/Win35Dict 290 dict def Win35Dict begin/bd{bind def}bind def/in{72
mul}bd/ed{exch def}bd/ld{load def}bd/tr/translate ld/gs/gsave ld/gr
/grestore ld/M/moveto ld/L/lineto ld/rmt/rmoveto ld/rlt/rlineto ld
/rct/rcurveto ld/st/stroke ld/n/newpath ld/sm/setmatrix ld/cm/currentmatrix
ld/cp/closepath ld/ARC/arcn ld/TR{65536 div}bd/lj/setlinejoin ld/lc
/setlinecap ld/ml/setmiterlimit ld/sl/setlinewidth ld/scignore false
def/sc{scignore{pop pop pop}{0 index 2 index eq 2 index 4 index eq
and{pop pop 255 div setgray}{3{255 div 3 1 roll}repeat setrgbcolor}ifelse}ifelse}bd
/FC{bR bG bB sc}bd/fC{/bB ed/bG ed/bR ed}bd/HC{hR hG hB sc}bd/hC{
/hB ed/hG ed/hR ed}bd/PC{pR pG pB sc}bd/pC{/pB ed/pG ed/pR ed}bd/sM
matrix def/PenW 1 def/iPen 5 def/mxF matrix def/mxE matrix def/mxUE
matrix def/mxUF matrix def/fBE false def/iDevRes 72 0 matrix defaultmatrix
dtransform dup mul exch dup mul add sqrt def/fPP false def/SS{fPP{
/SV save def}{gs}ifelse}bd/RS{fPP{SV restore}{gr}ifelse}bd/EJ{gsave
showpage grestore}bd/#C{userdict begin/#copies ed end}bd/FEbuf 2 string
def/FEglyph(G  )def/FE{1 exch{dup 16 FEbuf cvrs FEglyph exch 1 exch
putinterval 1 index exch FEglyph cvn put}for}bd/SM{/iRes ed/cyP ed
/cxPg ed/cyM ed/cxM ed 72 100 div dup scale dup 0 ne{90 eq{cyM exch
0 eq{cxM exch tr -90 rotate -1 1 scale}{cxM cxPg add exch tr +90 rotate}ifelse}{cyP
cyM sub exch 0 ne{cxM exch tr -90 rotate}{cxM cxPg add exch tr -90
rotate 1 -1 scale}ifelse}ifelse}{pop cyP cyM sub exch 0 ne{cxM cxPg
add exch tr 180 rotate}{cxM exch tr 1 -1 scale}ifelse}ifelse 100 iRes
div dup scale 0 0 transform .25 add round .25 sub exch .25 add round
.25 sub exch itransform translate}bd/SJ{1 index 0 eq{pop pop/fBE false
def}{1 index/Break ed div/dxBreak ed/fBE true def}ifelse}bd/ANSIVec[
16#0/grave 16#1/acute 16#2/circumflex 16#3/tilde 16#4/macron 16#5/breve
16#6/dotaccent 16#7/dieresis 16#8/ring 16#9/cedilla 16#A/hungarumlaut
16#B/ogonek 16#C/caron 16#D/dotlessi 16#27/quotesingle 16#60/grave
16#7C/bar 16#82/quotesinglbase 16#83/florin 16#84/quotedblbase 16#85
/ellipsis 16#86/dagger 16#87/daggerdbl 16#89/perthousand 16#8A/Scaron
16#8B/guilsinglleft 16#8C/OE 16#91/quoteleft 16#92/quoteright 16#93
/quotedblleft 16#94/quotedblright 16#95/bullet 16#96/endash 16#97
/emdash 16#99/trademark 16#9A/scaron 16#9B/guilsinglright 16#9C/oe
16#9F/Ydieresis 16#A0/space 16#A4/currency 16#A6/brokenbar 16#A7/section
16#A8/dieresis 16#A9/copyright 16#AA/ordfeminine 16#AB/guillemotleft
16#AC/logicalnot 16#AD/hyphen 16#AE/registered 16#AF/macron 16#B0/degree
16#B1/plusminus 16#B2/twosuperior 16#B3/threesuperior 16#B4/acute 16#B5
/mu 16#B6/paragraph 16#B7/periodcentered 16#B8/cedilla 16#B9/onesuperior
16#BA/ordmasculine 16#BB/guillemotright 16#BC/onequarter 16#BD/onehalf
16#BE/threequarters 16#BF/questiondown 16#C0/Agrave 16#C1/Aacute 16#C2
/Acircumflex 16#C3/Atilde 16#C4/Adieresis 16#C5/Aring 16#C6/AE 16#C7
/Ccedilla 16#C8/Egrave 16#C9/Eacute 16#CA/Ecircumflex 16#CB/Edieresis
16#CC/Igrave 16#CD/Iacute 16#CE/Icircumflex 16#CF/Idieresis 16#D0/Eth
16#D1/Ntilde 16#D2/Ograve 16#D3/Oacute 16#D4/Ocircumflex 16#D5/Otilde
16#D6/Odieresis 16#D7/multiply 16#D8/Oslash 16#D9/Ugrave 16#DA/Uacute
16#DB/Ucircumflex 16#DC/Udieresis 16#DD/Yacute 16#DE/Thorn 16#DF/germandbls
16#E0/agrave 16#E1/aacute 16#E2/acircumflex 16#E3/atilde 16#E4/adieresis
16#E5/aring 16#E6/ae 16#E7/ccedilla 16#E8/egrave 16#E9/eacute 16#EA
/ecircumflex 16#EB/edieresis 16#EC/igrave 16#ED/iacute 16#EE/icircumflex
16#EF/idieresis 16#F0/eth 16#F1/ntilde 16#F2/ograve 16#F3/oacute 16#F4
/ocircumflex 16#F5/otilde 16#F6/odieresis 16#F7/divide 16#F8/oslash
16#F9/ugrave 16#FA/uacute 16#FB/ucircumflex 16#FC/udieresis 16#FD/yacute
16#FE/thorn 16#FF/ydieresis ] def/reencdict 12 dict def/IsChar{basefontdict
/CharStrings get exch known}bd/MapCh{dup IsChar not{pop/bullet}if
newfont/Encoding get 3 1 roll put}bd/MapDegree{16#b0/degree IsChar{
/degree}{/ring}ifelse MapCh}bd/MapBB{16#a6/brokenbar IsChar{/brokenbar}{
/bar}ifelse MapCh}bd/ANSIFont{reencdict begin/newfontname ed/basefontname
ed FontDirectory newfontname known not{/basefontdict basefontname findfont
def/newfont basefontdict maxlength dict def basefontdict{exch dup/FID
ne{dup/Encoding eq{exch dup length array copy newfont 3 1 roll put}{exch
newfont 3 1 roll put}ifelse}{pop pop}ifelse}forall newfont/FontName
newfontname put 127 1 159{newfont/Encoding get exch/bullet put}for
ANSIVec aload pop ANSIVec length 2 idiv{MapCh}repeat MapDegree MapBB
newfontname newfont definefont pop}if newfontname end}bd/SB{FC/ULlen
ed/str ed str length fBE not{dup 1 gt{1 sub}if}if/cbStr ed/dxGdi ed
/y0 ed/x0 ed str stringwidth dup 0 ne{/y1 ed/x1 ed y1 y1 mul x1 x1
mul add sqrt dxGdi exch div 1 sub dup x1 mul cbStr div exch y1 mul
cbStr div}{exch abs neg dxGdi add cbStr div exch}ifelse/dyExtra ed
/dxExtra ed x0 y0 M fBE{dxBreak 0 BCh dxExtra dyExtra str awidthshow}{dxExtra
dyExtra str ashow}ifelse fUL{x0 y0 M dxUL dyUL rmt ULlen fBE{Break
add}if 0 mxUE transform gs rlt cyUL sl [] 0 setdash st gr}if fSO{x0
y0 M dxSO dySO rmt ULlen fBE{Break add}if 0 mxUE transform gs rlt cyUL
sl [] 0 setdash st gr}if n/fBE false def}bd/font{/name ed/Ascent ed
0 ne/fT3 ed 0 ne/fSO ed 0 ne/fUL ed/Sy ed/Sx ed 10.0 div/ori ed -10.0
div/esc ed/BCh ed name findfont/xAscent 0 def/yAscent Ascent def/ULesc
esc def ULesc mxUE rotate pop fT3{/esc 0 def xAscent yAscent mxUE transform
/yAscent ed/xAscent ed}if [Sx 0 0 Sy neg xAscent yAscent] esc mxE
rotate mxF concatmatrix makefont setfont [Sx 0 0 Sy neg 0 Ascent] mxUE
mxUF concatmatrix pop fUL{currentfont dup/FontInfo get/UnderlinePosition
known not{pop/Courier findfont}if/FontInfo get/UnderlinePosition get
1000 div 0 exch mxUF transform/dyUL ed/dxUL ed}if fSO{0 .3 mxUF transform
/dySO ed/dxSO ed}if fUL fSO or{currentfont dup/FontInfo get/UnderlineThickness
known not{pop/Courier findfont}if/FontInfo get/UnderlineThickness get
1000 div Sy mul/cyUL ed}if}bd/min{2 copy gt{exch}if pop}bd/max{2 copy
lt{exch}if pop}bd/CP{/ft ed{{ft 0 eq{clip}{eoclip}ifelse}stopped{currentflat
1 add setflat}{exit}ifelse}loop}bd/patfont 10 dict def patfont begin
/FontType 3 def/FontMatrix [1 0 0 -1 0 0] def/FontBBox [0 0 16 16]
def/Encoding StandardEncoding def/BuildChar{pop pop 16 0 0 0 16 16
setcachedevice 16 16 false [1 0 0 1 .25 .25]{pat}imagemask}bd end/p{
/pat 32 string def{}forall 0 1 7{dup 2 mul pat exch 3 index put dup
2 mul 1 add pat exch 3 index put dup 2 mul 16 add pat exch 3 index
put 2 mul 17 add pat exch 2 index put pop}for}bd/pfill{/PatFont patfont
definefont setfont/ch(AAAA)def X0 64 X1{Y1 -16 Y0{1 index exch M ch
show}for pop}for}bd/vert{X0 w X1{dup Y0 M Y1 L st}for}bd/horz{Y0 w
Y1{dup X0 exch M X1 exch L st}for}bd/fdiag{X0 w X1{Y0 M X1 X0 sub dup
rlt st}for Y0 w Y1{X0 exch M Y1 Y0 sub dup rlt st}for}bd/bdiag{X0 w
X1{Y1 M X1 X0 sub dup neg rlt st}for Y0 w Y1{X0 exch M Y1 Y0 sub dup
neg rlt st}for}bd/AU{1 add cvi 15 or}bd/AD{1 sub cvi -16 and}bd/SHR{pathbbox
AU/Y1 ed AU/X1 ed AD/Y0 ed AD/X0 ed}bd/hfill{/w iRes 37.5 div round
def 0.1 sl [] 0 setdash n dup 0 eq{horz}if dup 1 eq{vert}if dup 2 eq{fdiag}if
dup 3 eq{bdiag}if dup 4 eq{horz vert}if 5 eq{fdiag bdiag}if}bd/F{/ft
ed fm 256 and 0 ne{gs FC ft 0 eq{fill}{eofill}ifelse gr}if fm 1536
and 0 ne{SHR gs HC ft CP fm 1024 and 0 ne{/Tmp save def pfill Tmp restore}{fm
15 and hfill}ifelse gr}if}bd/S{PenW sl PC st}bd/m matrix def/GW{iRes
12 div PenW add cvi}bd/DoW{iRes 50 div PenW add cvi}bd/DW{iRes 8 div
PenW add cvi}bd/SP{/PenW ed/iPen ed iPen 0 eq iPen 6 eq or{[] 0 setdash}if
iPen 1 eq{[DW GW] 0 setdash}if iPen 2 eq{[DoW GW] 0 setdash}if iPen
3 eq{[DW GW DoW GW] 0 setdash}if iPen 4 eq{[DW GW DoW GW DoW GW] 0
setdash}if}bd/E{m cm pop tr scale 1 0 moveto 0 0 1 0 360 arc cp m sm}bd
/AG{/sy ed/sx ed sx div 4 1 roll sy div 4 1 roll sx div 4 1 roll sy
div 4 1 roll atan/a2 ed atan/a1 ed sx sy scale a1 a2 ARC}def/A{m cm
pop tr AG m sm}def/P{m cm pop tr 0 0 M AG cp m sm}def/RRect{n 4 copy
M 3 1 roll exch L 4 2 roll L L cp}bd/RRCC{/r ed/y1 ed/x1 ed/y0 ed/x0
ed x0 x1 add 2 div y0 M x1 y0 x1 y1 r arcto 4{pop}repeat x1 y1 x0 y1
r arcto 4{pop}repeat x0 y1 x0 y0 r arcto 4{pop}repeat x0 y0 x1 y0 r
arcto 4{pop}repeat cp}bd/RR{2 copy 0 eq exch 0 eq or{pop pop RRect}{2
copy eq{pop RRCC}{m cm pop/y2 ed/x2 ed/ys y2 x2 div 1 max def/xs x2
y2 div 1 max def/y1 exch ys div def/x1 exch xs div def/y0 exch ys div
def/x0 exch xs div def/r2 x2 y2 min def xs ys scale x0 x1 add 2 div
y0 M x1 y0 x1 y1 r2 arcto 4{pop}repeat x1 y1 x0 y1 r2 arcto 4{pop}repeat
x0 y1 x0 y0 r2 arcto 4{pop}repeat x0 y0 x1 y0 r2 arcto 4{pop}repeat
m sm cp}ifelse}ifelse}bd/PP{{rlt}repeat}bd/OB{gs 0 ne{7 3 roll/y ed
/x ed x y translate ULesc rotate x neg y neg translate x y 7 -3 roll}if
sc B fill gr}bd/B{M/dy ed/dx ed dx 0 rlt 0 dy rlt dx neg 0 rlt cp}bd
/CB{B clip n}bd/ErrHandler{errordict dup maxlength exch length gt
dup{errordict begin}if/errhelpdict 12 dict def errhelpdict begin/stackunderflow(operand stack underflow)def
/undefined(this name is not defined in a dictionary)def/VMerror(you have used up all the printer's memory)def
/typecheck(operator was expecting a different type of operand)def
/ioerror(input/output error occured)def end{end}if errordict begin
/handleerror{$error begin newerror{/newerror false def showpage 72
72 scale/x .25 def/y 9.6 def/Helvetica findfont .2 scalefont setfont
x y moveto(Offending Command = )show/command load{dup type/stringtype
ne{(max err string)cvs}if show}exec/y y .2 sub def x y moveto(Error = )show
errorname{dup type dup( max err string )cvs show( : )show/stringtype
ne{( max err string )cvs}if show}exec errordict begin errhelpdict errorname
known{x 1 add y .2 sub moveto errhelpdict errorname get show}if end
/y y .4 sub def x y moveto(Stack =)show ostack{/y y .2 sub def x 1
add y moveto dup type/stringtype ne{( max err string )cvs}if show}forall
showpage}if end}def end}bd end
/SVDoc save def
Win35Dict begin
ErrHandler
statusdict begin 0 setjobtimeout end
statusdict begin statusdict /jobname (Ventura - STELNEWS.CHP) put end
/setresolution where { pop 300 300 setresolution } if
gsave 
90 rotate 4 72 mul .55 -72 mul moveto /Times-Roman findfont
20 scalefont setfont 0.3 setgray (plasm-ph/9502002   26 Feb 1995) show grestore
SS
0 0 25 20 798 1100 300 SM
0 0 0 fC
/fm 256 def
2100 3 147 238 B
1 F
n
2100 3 147 3086 B
1 F
n
3 2850 147 238 B
1 F
n
3 2850 2245 238 B
1 F
n
2074 3 160 251 B
1 F
n
2074 3 160 3074 B
1 F
n
3 2825 160 251 B
1 F
n
3 2825 2232 251 B
1 F
n
2048 3 173 264 B
1 F
n
2048 3 173 3061 B
1 F
n
3 2799 173 264 B
1 F
n
3 2799 2219 264 B
1 F
n
2 2197 1196 666 B
1 F
n
255 255 255 fC
gs 1570 601 0 62 CB
1535 510 -11 107 B
1 F
n
gr
32 0 0 33 33 0 0 0 31 /Helvetica-Narrow /font15 ANSIFont font
0 0 0 fC
gs 2101 339 147 2788 CB
337 2908 38 (All ) 37 SB
374 2908 109 (opinions ) 108 SB
482 2908 82 (expres) 81 SB
563 2908 52 (sed ) 51 SB
614 2908 130 (herein are ) 129 SB
743 2908 52 (thos) 51 SB
794 2908 46 (e of) 45 SB
839 2908 16 ( t) 15 SB
854 2908 76 (he aut) 75 SB
929 2908 61 (hors ) 60 SB
989 2908 67 (and s) 66 SB
1055 2908 112 (hould not) 111 SB
1166 2908 46 ( be ) 45 SB
1211 2908 145 (reproduced,) 144 SB
1355 2908 61 ( quot) 60 SB
1415 2908 67 (ed in ) 66 SB
1481 2908 57 (publi) 57 SB
gr
gs 2101 339 147 2788 CB
1538 2908 37 (cat) 36 SB
1574 2908 58 (ions,) 57 SB
1631 2908 16 ( t) 15 SB
1646 2908 76 (ransm) 75 SB
1721 2908 22 (itt) 21 SB
1742 2908 46 (ed, ) 45 SB
1787 2908 61 (or us) 60 SB
1847 2908 67 (ed as) 66 SB
1913 2908 31 ( a ) 30 SB
1943 2908 100 (referenc) 99 SB
2042 2908 15 (e) 15 SB
gr
gs 2101 339 147 2788 CB
1025 2948 34 (wit) 33 SB
1058 2948 61 (hout ) 60 SB
1118 2948 46 (the ) 45 SB
1163 2948 97 (author\222s) 96 SB
1259 2948 22 ( c) 21 SB
1280 2948 82 (onsent) 81 SB
1361 2948 8 (.) 8 SB
gr
gs 2101 339 147 2788 CB
479 3003 50 (Oak) 49 SB
528 3003 87 ( Ridge ) 86 SB
614 3003 43 (Nat) 42 SB
656 3003 157 (ional Laborat) 156 SB
812 3003 46 (ory ) 45 SB
857 3003 28 (is ) 27 SB
884 3003 150 (managed by) 149 SB
1033 3003 31 ( M) 30 SB
1063 3003 61 (artin ) 60 SB
1123 3003 84 (Mariett) 83 SB
1206 3003 109 (a Energy) 108 SB
1314 3003 40 ( Sy) 39 SB
1353 3003 22 (st) 21 SB
1374 3003 52 (ems) 51 SB
1425 3003 16 (, ) 15 SB
1440 3003 37 (Inc) 36 SB
1476 3003 16 (.,) 15 SB
1491 3003 16 ( f) 15 SB
1506 3003 40 (or t) 39 SB
1545 3003 66 (he U.) 65 SB
1610 3003 34 (S. ) 33 SB
1643 3003 74 (Depar) 74 SB
gr
gs 2101 339 147 2788 CB
1717 3003 8 (t) 7 SB
1724 3003 69 (ment ) 68 SB
1792 3003 31 (of ) 30 SB
1822 3003 94 (Energy.) 93 SB
gr
255 255 255 fC
gs 1534 742 0 0 CB
1532 653 -74 13 B
1 F
n
gr
/PPT_ProcessAll true def

userdict /VPsave save put
userdict begin
/showpage{}def
-74 13 -74 667 1459 667 1459 13 newpath moveto lineto lineto lineto clip newpath
-44 576 translate 300 72 div dup neg scale
1533 300 div 615 72 div div 625 300 div 251 72 div div scale
311 -154 translate

/wCorelDict 200 dict def 
wCorelDict begin 
/wDocm matrix def 
/wSavm matrix def 
/wObjm matrix def 
/wBmp 512 string def 
/wScra 0.0 def 
statusdict /pageparams known 
{statusdict begin 
pageparams end 
1 ne {/wScra -90.0 def}if 
pop pop pop 
}if

/wDfreq 60 def 
/wDang 45 def 
/wDspot { } def 

/wSrgb { setrgbcolor } bind def 
/wScr { exch wScra add exch load setscreen } bind def 
/wSav { /wSobj save def } bind def 
/wRes { wSobj restore } bind def 
/wSg { setgray } bind def 
/wMto { moveto } bind def 
/wLto { lineto } bind def 
/wCto { curveto } bind def 
/wTl { translate } bind def 
/wCp { closepath } bind def 
/wGs { gsave } bind def 
/wGr { grestore } bind def 
/wCtm { matrix currentmatrix } bind def 
/wStm { setmatrix } bind def 
/wSld { [] 0 setdash } bind def 
/wDsh { [3 2] 0 setdash } bind def 
/wDtd { [1 2] 0 setdash } bind def 
/wFillp { } def 
/wCapp { } def 
/wLinep { } def 
/wPaintp { } def 

/InRange { 

3 -1 roll 
2 copy le {pop}{exch pop}ifelse 
2 copy ge {pop}{exch pop}ifelse 
} bind def 

/wDstChck { 
2 1 roll dup 3 -1 roll 
eq{ 1 add }if 
} bind def 

systemdict /setcmykcolor known 
{/wCMYK { 
setcmykcolor 
} bind def 
} 
{ 
/wCMYK { 

/wBlk exch def 
wBlk add 0 1 InRange 1 exch sub 3 1 roll 
wBlk add 0 1 InRange 1 exch sub 3 1 roll 
wBlk add 0 1 InRange 1 exch sub 3 1 roll 
setrgbcolor 
} bind def 
}ifelse 

/wSscr { 
currentscreen /wDspot exch def /wDang exch def /wDfreq exch def 
} bind def 

/wNeg { 
{1 exch sub} bind settransfer 
}bind def 

/wRscr { 
wDfreq wDang /wDspot wScr 
} bind def 

/wFnt { 
findfont exch scalefont setfont 
} bind def 

/wSvm { 
/wSavm wSavm currentmatrix def 
} bind def 

/wRsm { 
wSavm setmatrix 
} bind def 

/wSvp { 
currentpoint /wSavepy exch def /wSavepx exch def 
} bind def 

/wRsp { 
wSavepx wSavepy moveto 
} bind def 

/wSa { 

setlinejoin setlinecap 
} bind def 

/wPn { 

rotate scale 
} bind def 

/wStbb { 

/Bbury exch def 
/Bburx exch def 
/Bblly exch def 
/Bbllx exch def 
} bind def 

/wDot { 
dup mul exch dup mul add 1 exch sub 
} bind def 

/wLin { 
exch pop abs 1 exch sub 
} bind def 

/wRBMP { 
currentfile wBmp readhexstring pop 
} bind def 

/wR { 
0 0 moveto 
1 index 0 rlineto 
0 exch rlineto 
neg 0 rlineto 
wCp 
} bind def

/wE { 

wSvm 
scale 
0.5 0 moveto 
0.5 0.5 0.5 -90 270 arc 
wRsm 
} bind def 

/wCh { 
moveto 
true charpath 
Paintp 
} bind def 

/wACh { 
wSvm 
translate 
rotate 
0 0 moveto 
true charpath 
wRsm 
Paintp 
} bind def

/wChs { 
moveto 
show 
} bind def 

/wAChs { 
wSvm 
translate 
rotate 
0 0 moveto 
show 
wRsm 
} bind def

/ReEncodeSmall{ 
/bReEncode exch def 
/newfontname exch def 
/basefontname exch def /basefontdict basefontname findfont def 
/newfont basefontdict maxlength dict def basefontdict 
{1 index /FID ne 
{newfont 3 1 roll put} 
{pop pop}ifelse 
} forall 
bReEncode 
{ 
newfont /Encoding get 
256 array copy 
newfont exch /Encoding exch put 
newfont /FontName newfontname put gemenvec aload pop 
gemenvec length 2 idiv 
{newfont /Encoding get 3 1 roll put}repeat 
}if 
newfontname newfont definefont pop 
} bind def 
/gemenvec [ 
16#80 /grave 
16#81 /circumflex 
16#82 /tilde 
16#83 /dotlessi 
16#84 /florin 
16#85 /quotedblleft 
16#86 /quotedblright 
16#87 /guilsinglleft 
16#88 /guilsinglright 
16#89 /fi 
16#8a /fl 
16#8b /dagger 
16#8c /daggerdbl 
16#8d /endash 
16#8e /periodcentered 
16#8f /breve 
16#90 /quotedblbase 
16#91 /ellipsis 
16#92 /perthousand 
16#a1 /exclamdown 
16#a2 /cent 
16#a3 /sterling 
16#a4 /currency 
16#a5 /yen 
16#a6 /bar 
16#a7 /section 
16#a8 /dieresis 
16#a9 /copyright 
16#aa /ordfeminine 
16#ab /guillemotleft 
16#ac /logicalnot 
16#ad /emdash 
16#ae /registered 
16#af /space 
16#b0 /ring 
16#b1 /space 
16#b2 /space 
16#b3 /space 
16#b4 /acute 
16#b5 /space 
16#b6 /paragraph 
16#b7 /bullet 
16#b8 /cedilla 
16#b9 /space 
16#ba /ordmasculine 
16#bb /guillemotright 
16#bc /space 
16#bd /space 
16#be /space 
16#bf /questiondown 
16#c0 /Agrave 
16#c1 /Aacute 
16#c2 /Acircumflex 
16#c3 /Atilde 
16#c4 /Adieresis 
16#c5 /Aring 
16#c6 /AE 
16#c7 /Ccedilla 
16#c8 /Egrave 
16#c9 /Eacute 
16#ca /Ecircumflex 
16#cb /space 
16#cc /Igrave 
16#cd /Iacute 
16#ce /Icircumflex 
16#cf /Idieresis 
16#d0 /space 
16#d1 /Ntilde 
16#d2 /Ograve 
16#d3 /Oacute 
16#d4 /Ocircumflex 
16#d5 /Otilde 
16#d6 /Odieresis 
16#d7 /OE 
16#d8 /Oslash 
16#d9 /Ugrave 
16#da /Uacute 
16#db /Ucircumflex 
16#dc /Udieresis 
16#dd /space 
16#de /space 
16#df /germandbls 
16#e0 /agrave 
16#e1 /aacute 
16#e2 /acircumflex 
16#e3 /atilde 
16#e4 /adieresis 
16#e5 /aring 
16#e6 /ae 
16#e7 /ccedilla 
16#e8 /egrave 
16#e9 /eacute 
16#ea /ecircumflex 
16#eb /edieresis 
16#ec /igrave 
16#ed /iacute 
16#ee /icircumflex 
16#ef /idieresis 
16#f0 /space 
16#f1 /ntilde 
16#f2 /ograve 
16#f3 /oacute 
16#f4 /ocircumflex 
16#f5 /otilde 
16#f6 /odieresis 
16#f7 /oe 
16#f8 /oslash 
16#f9 /ugrave 
16#fa /uacute 
16#fb /ucircumflex 
16#fc /udieresis 
16#fd /space 
16#fe /space 
16#ff /ydieresis 
] def 

/wMax { 

2 copy ge{pop}{exch pop}ifelse 
} bind def 

/wMin { 

2 copy le{pop}{exch pop}ifelse 
} bind def 

/wFtn {

/Ang exch def 
/ToK exch 0 1 InRange def 
/ToY exch 0 1 InRange def 
/ToM exch 0 1 InRange def 
/ToC exch 0 1 InRange def 
/FrmK exch 0 1 InRange def 
/FrmY exch 0 1 InRange def 
/FrmM exch 0 1 InRange def 
/FrmC exch 0 1 InRange def 

ToC FrmC sub dup /dC exch def abs 
ToM FrmM sub dup /dM exch def abs 
ToY FrmY sub dup /dY exch def abs 
ToK FrmK sub dup /dK exch def abs

wMax wMax wMax 
200 mul cvi 200 wMin /fStp exch def 

/dC dC fStp div def 
/dM dM fStp div def 
/dY dY fStp div def 
/dK dK fStp div def 

eoclip 
newpath 
Bbllx Bblly moveto 
Bbllx Bbury lineto 
Bburx Bbury lineto 
Bburx Bblly lineto 
Ang rotate 
pathbbox 
/Stp 2 index 5 index sub fStp div def 
/Bllx1 4 index def 
/Bllx2 Bllx1 Stp add def 
/BburY 1 index def 
/BbllY 3 index def 
newpath 
4{ pop }repeat 

FrmC FrmM FrmY FrmK 
0 1 fStp 1 sub 
{ 
pop 
4 copy 
wCMYK 
dK add 4 1 roll 
dY add 4 1 roll 
dM add 4 1 roll 
dC add 4 1 roll 
Bllx1 BbllY moveto 
Bllx1 BburY lineto 
Bllx2 BburY lineto 
Bllx2 BbllY lineto 
fill 
/Bllx2 Bllx2 Stp add def 
/Bllx1 Bllx1 Stp add def 
} for 
4{ pop }repeat 
} bind def 

/wRfn {

/ToK exch 0 1 InRange def 
/ToY exch 0 1 InRange def 
/ToM exch 0 1 InRange def 
/ToC exch 0 1 InRange def 
/FrmK exch 0 1 InRange def 
/FrmY exch 0 1 InRange def 
/FrmM exch 0 1 InRange def 
/FrmC exch 0 1 InRange def 

ToC FrmC sub dup /dC exch def abs 
ToM FrmM sub dup /dM exch def abs 
ToY FrmY sub dup /dY exch def abs 
ToK FrmK sub dup /dK exch def abs

wMax wMax wMax 
200 mul cvi 200 wMin /fStp exch def 

/dC dC fStp div def 
/dM dM fStp div def 
/dY dY fStp div def 
/dK dK fStp div def 

4 setflat 
eoclip 
newpath 
Bbllx Bblly moveto 
Bbllx Bbury lineto 
Bburx Bbury lineto 
Bburx Bblly lineto 
pathbbox 
newpath 
3 -1 roll 
4 copy 
add 2 div 0 exch translate 
add 2 div 0 translate 
sub 2 div dup mul 
3 1 roll 
sub 2 div dup mul 
add sqrt 
dup /rad exch def 
fStp div /dr exch def 

FrmC FrmM FrmY FrmK 
0 1 fStp 1 sub 
{ 
pop 
4 copy 
wCMYK 
dK add 4 1 roll 
dY add 4 1 roll 
dM add 4 1 roll 
dC add 4 1 roll 
0 0 rad 0 360 arc fill 
/rad rad dr sub def 
} for 
4{ pop }repeat 
} bind def 

/wReg { 
0 wSg 4 setlinewidth 
2 copy 75 0 360 arc closepath 
2 copy moveto 125 0 rlineto 
2 copy moveto -125 0 rlineto 
2 copy moveto 0 125 rlineto 
moveto 0 -125 rlineto stroke 
} bind def 

/wArr { 

rotate 
-2 -1.5 rmoveto 
0.5 1.0 0.5 2.0 0 3 rcurveto 
3.5 -1.5 rlineto 
wCp 
eofill 
} bind def

0.072 dup scale
0.000 0.000 0.000 1.000 wCMYK 0 0 wSa wSld
1 setlinewidth
11.4738 setmiterlimit
wSscr

wSav
/wDocm wCtm def
wSav
-3331 4528 -2903 5014 wStbb
-3341 4515 -2892 5026 wStbb
-187 3485 wTl
[1.3665 0.0000 0.0000 1.3665 -3952.8768 1934.5945] concat /wObjm wCtm def
0 0 wSa
0.000 0.000 0.000 1.000 wCMYK wRscr
/Linep { 6 6 0.0 wPn stroke } bind def
/Fillp {  } bind def
/Paintp { wGs wDocm wStm wGs Fillp wGr Linep wGr newpath} bind def
 904 -298 wMto
804 -307 594 -411 718 -651 wCto
wGs wDocm wStm Linep wGr newpath
wRes
wSav
-3484 4204 -2881 5247 wStbb
-3499 4177 -2865 5273 wStbb
-187 3485 wTl
[1.3665 0.0000 0.0000 1.3665 -3952.8768 1934.5945] concat /wObjm wCtm def
0 0 wSa
0.000 0.000 0.000 1.000 wCMYK wRscr
/Linep { 6 6 0.0 wPn stroke } bind def
/Fillp {  } bind def
/Paintp { wGs wDocm wStm wGs Fillp wGr Linep wGr newpath} bind def
 920 -260 wMto
675 -128 588 -243 659 -519 wCto
730 -795 736 -739 736 -795 wCto
706 -854 682 -875 671 -888 wCto
612 -810 527 -560 522 -501 wCto
517 -442 482 -317 677 -232 wCto
wGs wDocm wStm Linep wGr newpath
wRes
wSav
-3730 4156 -3416 4730 wStbb
-3737 4141 -3408 4744 wStbb
-187 3485 wTl
[1.3665 0.0000 0.0000 1.3665 -3952.8768 1934.5945] concat /wObjm wCtm def
0 0 wSa
0.000 0.000 0.000 1.000 wCMYK wRscr
/Linep { 6 6 0.0 wPn stroke } bind def
/Fillp {  } bind def
/Paintp { wGs wDocm wStm wGs Fillp wGr Linep wGr newpath} bind def
 524 -506 wMto
441 -565 302 -778 529 -923 wCto
wGs wDocm wStm Linep wGr newpath
wRes
wSav
-3921 3850 -3282 4768 wStbb
-3936 3827 -3266 4790 wStbb
-187 3485 wTl
[1.3665 0.0000 0.0000 1.3665 -3952.8768 1934.5945] concat /wObjm wCtm def
0 0 wSa
0.000 0.000 0.000 1.000 wCMYK wRscr
/Linep { 6 6 0.0 wPn stroke } bind def
/Fillp {  } bind def
/Paintp { wGs wDocm wStm wGs Fillp wGr Linep wGr newpath} bind def
 515 -478 wMto
237 -486 217 -638 416 -842 wCto
616 -1045 595 -985 623 -1033 wCto
627 -1099 616 -1129 614 -1147 wCto
524 -1108 325 -934 291 -886 wCto
257 -837 162 -749 288 -578 wCto
wGs wDocm wStm Linep wGr newpath
wRes
wSav
-3823 3715 -3442 4198 wStbb
-3832 3702 -3432 4210 wStbb
-187 3485 wTl
[1.3665 0.0000 0.0000 1.3665 -3952.8768 1934.5945] concat /wObjm wCtm def
0 0 wSa
0.000 0.000 0.000 1.000 wCMYK wRscr
/Linep { 6 6 0.0 wPn stroke } bind def
/Fillp {  } bind def
/Paintp { wGs wDocm wStm wGs Fillp wGr Linep wGr newpath} bind def
 296 -895 wMto
256 -983 234 -1234 510 -1246 wCto
wGs wDocm wStm Linep wGr newpath
wRes
wSav
-4078 3489 -3192 4233 wStbb
-4100 3470 -3169 4251 wStbb
-187 3485 wTl
[1.3665 0.0000 0.0000 1.3665 -3952.8768 1934.5945] concat /wObjm wCtm def
0 0 wSa
0.000 0.000 0.000 1.000 wCMYK wRscr
/Linep { 6 6 0.0 wPn stroke } bind def
/Fillp {  } bind def
/Paintp { wGs wDocm wStm wGs Fillp wGr Linep wGr newpath} bind def
 273 -870 wMto
47 -1009 81 -1151 366 -1230 wCto
640 -1307 596 -1269 644 -1296 wCto
681 -1351 687 -1383 693 -1399 wCto
596 -1411 337 -1360 284 -1335 wCto
230 -1310 104 -1275 128 -1064 wCto
wGs wDocm wStm Linep wGr newpath
wRes
wSav
-3623 3319 -3134 3538 wStbb
-3635 3313 -3121 3543 wStbb
-187 3485 wTl
[1.3665 0.0000 0.0000 1.3665 -3952.8768 1934.5945] concat /wObjm wCtm def
0 0 wSa
0.000 0.000 0.000 1.000 wCMYK wRscr
/Linep { 6 6 0.0 wPn stroke } bind def
/Fillp {  } bind def
/Paintp { wGs wDocm wStm wGs Fillp wGr Linep wGr newpath} bind def
 616 -1401 wMto
734 -1423 694 -1409 708 -1451 wCto
734 -1531 735 -1521 684 -1531 wCto
657 -1536 630 -1527 616 -1512 wCto
574 -1465 585 -1468 539 -1444 wCto
505 -1425 462 -1384 435 -1378 wCto
411 -1422 397 -1433 380 -1506 wCto
wGs wDocm wStm Linep wGr newpath
wRes
wSav
-3441 3271 -3402 3448 wStbb
-3441 3266 -3401 3452 wStbb
-187 3485 wTl
[1.3665 0.0000 0.0000 1.3665 -3952.8768 1934.5945] concat /wObjm wCtm def
0 0 wSa
0.000 0.000 0.000 1.000 wCMYK wRscr
/Linep { 6 6 0.0 wPn stroke } bind def
/Fillp {  } bind def
/Paintp { wGs wDocm wStm wGs Fillp wGr Linep wGr newpath} bind def
 539 -1444 wMto
534 -1484 513 -1522 522 -1571 wCto
wGs wDocm wStm Linep wGr newpath
wRes
wSav
-3742 3146 -3232 3593 wStbb
-3754 3134 -3219 3604 wStbb
-187 3485 wTl
[1.3665 0.0000 0.0000 1.3665 -3952.8768 1934.5945] concat /wObjm wCtm def
0 0 wSa
0.000 0.000 0.000 1.000 wCMYK wRscr
/Linep { 6 6 0.0 wPn stroke } bind def
/Fillp {  } bind def
/Paintp { wGs wDocm wStm wGs Fillp wGr Linep wGr newpath} bind def
 293 -1338 wMto
301 -1429 415 -1662 663 -1532 wCto
wGs wDocm wStm Linep wGr newpath
wRes
wSav
-3968 3010 -2925 3612 wStbb
-3994 2994 -2898 3627 wStbb
-187 3485 wTl
[1.3665 0.0000 0.0000 1.3665 -3952.8768 1934.5945] concat /wObjm wCtm def
0 0 wSa
0.000 0.000 0.000 1.000 wCMYK wRscr
/Linep { 6 6 0.0 wPn stroke } bind def
/Fillp {  } bind def
/Paintp { wGs wDocm wStm wGs Fillp wGr Linep wGr newpath} bind def
 260 -1324 wMto
128 -1569 243 -1661 519 -1590 wCto
795 -1519 739 -1513 795 -1513 wCto
854 -1543 874 -1567 888 -1578 wCto
810 -1637 560 -1722 501 -1727 wCto
442 -1732 312 -1762 227 -1567 wCto
wGs wDocm wStm Linep wGr newpath
wRes
wSav
-3439 2735 -2881 3075 wStbb
-3452 2726 -2867 3083 wStbb
-187 3485 wTl
[1.3665 0.0000 0.0000 1.3665 -3952.8768 1934.5945] concat /wObjm wCtm def
0 0 wSa
0.000 0.000 0.000 1.000 wCMYK wRscr
/Linep { 6 6 0.0 wPn stroke } bind def
/Fillp {  } bind def
/Paintp { wGs wDocm wStm wGs Fillp wGr Linep wGr newpath} bind def
 515 -1725 wMto
553 -1779 763 -1963 920 -1717 wCto
wGs wDocm wStm Linep wGr newpath
wRes
wSav
-3491 2570 -2571 3205 wStbb
-3514 2554 -2548 3220 wStbb
-187 3485 wTl
[1.3665 0.0000 0.0000 1.3665 -3952.8768 1934.5945] concat /wObjm wCtm def
0 0 wSa
0.000 0.000 0.000 1.000 wCMYK wRscr
/Linep { 6 6 0.0 wPn stroke } bind def
/Fillp {  } bind def
/Paintp { wGs wDocm wStm wGs Fillp wGr Linep wGr newpath} bind def
 477 -1729 wMto
485 -2007 629 -2030 832 -1830 wCto
1039 -1627 989 -1652 1033 -1626 wCto
1099 -1622 1129 -1633 1147 -1635 wCto
1108 -1725 934 -1924 886 -1958 wCto
837 -1992 749 -2084 578 -1958 wCto
wGs wDocm wStm Linep wGr newpath
wRes
wSav
-2772 2735 -2398 3197 wStbb
-2781 2723 -2388 3208 wStbb
-187 3485 wTl
[1.3665 0.0000 0.0000 1.3665 -3952.8768 1934.5945] concat /wObjm wCtm def
0 0 wSa
0.000 0.000 0.000 1.000 wCMYK wRscr
/Linep { 6 6 0.0 wPn stroke } bind def
/Fillp {  } bind def
/Paintp { wGs wDocm wStm wGs Fillp wGr Linep wGr newpath} bind def
 1104 -1713 wMto
1185 -1628 1140 -1635 1178 -1644 wCto
1251 -1661 1274 -1637 1260 -1675 wCto
1251 -1698 1225 -1753 1205 -1759 wCto
1185 -1764 1135 -1770 1109 -1791 wCto
1078 -1815 1022 -1828 1003 -1848 wCto
1032 -1897 1038 -1916 1088 -1963 wCto
wGs wDocm wStm Linep wGr newpath
wRes
wSav
-2627 2849 -2493 2974 wStbb
-2630 2845 -2489 2977 wStbb
-187 3485 wTl
[1.3665 0.0000 0.0000 1.3665 -3952.8768 1934.5945] concat /wObjm wCtm def
0 0 wSa
0.000 0.000 0.000 1.000 wCMYK wRscr
/Linep { 6 6 0.0 wPn stroke } bind def
/Fillp {  } bind def
/Paintp { wGs wDocm wStm wGs Fillp wGr Linep wGr newpath} bind def
 1109 -1791 wMto
1142 -1817 1158 -1863 1204 -1880 wCto
wGs wDocm wStm Linep wGr newpath
wRes
wSav
-2914 2571 -2433 3088 wStbb
-2926 2558 -2420 3100 wStbb
-187 3485 wTl
[1.3665 0.0000 0.0000 1.3665 -3952.8768 1934.5945] concat /wObjm wCtm def
0 0 wSa
0.000 0.000 0.000 1.000 wCMYK wRscr
/Linep { 6 6 0.0 wPn stroke } bind def
/Fillp {  } bind def
/Paintp { wGs wDocm wStm wGs Fillp wGr Linep wGr newpath} bind def
 899 -1951 wMto
902 -1953 1232 -2083 1248 -1708 wCto
wGs wDocm wStm Linep wGr newpath
wRes
wSav
-2959 2400 -2210 3295 wStbb
-2977 2377 -2191 3317 wStbb
-187 3485 wTl
[1.3665 0.0000 0.0000 1.3665 -3952.8768 1934.5945] concat /wObjm wCtm def
0 0 wSa
0.000 0.000 0.000 1.000 wCMYK wRscr
/Linep { 6 6 0.0 wPn stroke } bind def
/Fillp {  } bind def
/Paintp { wGs wDocm wStm wGs Fillp wGr Linep wGr newpath} bind def
 866 -1972 wMto
1012 -2208 1148 -2157 1224 -1882 wCto
1301 -1607 1269 -1653 1296 -1605 wCto
1351 -1568 1383 -1562 1399 -1556 wCto
1411 -1653 1360 -1912 1335 -1965 wCto
1308 -2022 1280 -2144 1071 -2121 wCto
wGs wDocm wStm Linep wGr newpath
wRes
wSav
-2255 2864 -2044 3354 wStbb
-2260 2851 -2038 3366 wStbb
-187 3485 wTl
[1.3665 0.0000 0.0000 1.3665 -3952.8768 1934.5945] concat /wObjm wCtm def
0 0 wSa
0.000 0.000 0.000 1.000 wCMYK wRscr
/Linep { 6 6 0.0 wPn stroke } bind def
/Fillp {  } bind def
/Paintp { wGs wDocm wStm wGs Fillp wGr Linep wGr newpath} bind def
 1397 -1642 wMto
1423 -1513 1403 -1569 1444 -1555 wCto
1532 -1526 1519 -1524 1527 -1564 wCto
1533 -1588 1527 -1619 1512 -1633 wCto
1496 -1647 1456 -1678 1444 -1710 wCto
1430 -1746 1388 -1787 1381 -1814 wCto
1425 -1838 1432 -1852 1505 -1869 wCto
wGs wDocm wStm Linep wGr newpath
wRes
wSav
-2169 3044 -1992 3085 wStbb
-2173 3042 -1987 3086 wStbb
-187 3485 wTl
[1.3665 0.0000 0.0000 1.3665 -3952.8768 1934.5945] concat /wObjm wCtm def
0 0 wSa
0.000 0.000 0.000 1.000 wCMYK wRscr
/Linep { 6 6 0.0 wPn stroke } bind def
/Fillp {  } bind def
/Paintp { wGs wDocm wStm wGs Fillp wGr Linep wGr newpath} bind def
 1444 -1710 wMto
1477 -1714 1515 -1737 1571 -1727 wCto
wGs wDocm wStm Linep wGr newpath
wRes
wSav
-2311 2747 -1851 3256 wStbb
-2322 2734 -1839 3268 wStbb
-187 3485 wTl
[1.3665 0.0000 0.0000 1.3665 -3952.8768 1934.5945] concat /wObjm wCtm def
0 0 wSa
0.000 0.000 0.000 1.000 wCMYK wRscr
/Linep { 6 6 0.0 wPn stroke } bind def
/Fillp {  } bind def
/Paintp { wGs wDocm wStm wGs Fillp wGr Linep wGr newpath} bind def
 1340 -1954 wMto
1384 -1950 1674 -1860 1531 -1585 wCto
wGs wDocm wStm Linep wGr newpath
wRes
wSav
-2328 2508 -1721 3562 wStbb
-2343 2481 -1705 3588 wStbb
-187 3485 wTl
[1.3665 0.0000 0.0000 1.3665 -3952.8768 1934.5945] concat /wObjm wCtm def
0 0 wSa
0.000 0.000 0.000 1.000 wCMYK wRscr
/Linep { 6 6 0.0 wPn stroke } bind def
/Fillp {  } bind def
/Paintp { wGs wDocm wStm wGs Fillp wGr Linep wGr newpath} bind def
 1328 -1991 wMto
1583 -2129 1661 -1994 1592 -1723 wCto
1520 -1447 1513 -1510 1513 -1454 wCto
1543 -1395 1567 -1375 1578 -1361 wCto
1637 -1439 1722 -1689 1727 -1748 wCto
1730 -1780 1769 -1936 1572 -2022 wCto
wGs wDocm wStm Linep wGr newpath
wRes
wSav
-1975 3226 -1639 3690 wStbb
-1983 3214 -1630 3701 wStbb
-187 3485 wTl
[1.3665 0.0000 0.0000 1.3665 -3952.8768 1934.5945] concat /wObjm wCtm def
0 0 wSa
0.000 0.000 0.000 1.000 wCMYK wRscr
/Linep { 6 6 0.0 wPn stroke } bind def
/Fillp {  } bind def
/Paintp { wGs wDocm wStm wGs Fillp wGr Linep wGr newpath} bind def
 1623 -1435 wMto
1592 -1319 1586 -1365 1612 -1330 wCto
1659 -1269 1642 -1267 1702 -1312 wCto
1722 -1327 1723 -1360 1717 -1380 wCto
1710 -1400 1689 -1438 1694 -1471 wCto
1700 -1510 1684 -1568 1692 -1595 wCto
1742 -1594 1758 -1604 1829 -1583 wCto
wGs wDocm wStm Linep wGr newpath
wRes
wSav
-1828 3407 -1662 3477 wStbb
-1832 3405 -1657 3478 wStbb
-187 3485 wTl
[1.3665 0.0000 0.0000 1.3665 -3952.8768 1934.5945] concat /wObjm wCtm def
0 0 wSa
0.000 0.000 0.000 1.000 wCMYK wRscr
/Linep { 6 6 0.0 wPn stroke } bind def
/Fillp {  } bind def
/Paintp { wGs wDocm wStm wGs Fillp wGr Linep wGr newpath} bind def
 1694 -1471 wMto
1731 -1456 1774 -1455 1812 -1423 wCto
wGs wDocm wStm Linep wGr newpath
wRes
wSav
-1798 3048 -1452 3610 wStbb
-1806 3033 -1443 3624 wStbb
-187 3485 wTl
[1.3665 0.0000 0.0000 1.3665 -3952.8768 1934.5945] concat /wObjm wCtm def
0 0 wSa
0.000 0.000 0.000 1.000 wCMYK wRscr
/Linep { 6 6 0.0 wPn stroke } bind def
/Fillp {  } bind def
/Paintp { wGs wDocm wStm wGs Fillp wGr Linep wGr newpath} bind def
 1725 -1734 wMto
1776 -1698 1966 -1485 1716 -1326 wCto
wGs wDocm wStm Linep wGr newpath
wRes
wSav
-1926 2997 -1293 3916 wStbb
-1941 2974 -1277 3938 wStbb
-187 3485 wTl
[1.3665 0.0000 0.0000 1.3665 -3952.8768 1934.5945] concat /wObjm wCtm def
0 0 wSa
0.000 0.000 0.000 1.000 wCMYK wRscr
/Linep { 6 6 0.0 wPn stroke } bind def
/Fillp {  } bind def
/Paintp { wGs wDocm wStm wGs Fillp wGr Linep wGr newpath} bind def
 1734 -1771 wMto
2012 -1763 2024 -1609 1824 -1406 wCto
1624 -1202 1654 -1264 1626 -1216 wCto
1622 -1150 1633 -1120 1635 -1102 wCto
1725 -1141 1924 -1315 1958 -1363 wCto
1992 -1412 2082 -1506 1956 -1677 wCto
wGs wDocm wStm Linep wGr newpath
wRes
wSav
-1918 3712 -1449 4084 wStbb
-1929 3702 -1437 4093 wStbb
-187 3485 wTl
[1.3665 0.0000 0.0000 1.3665 -3952.8768 1934.5945] concat /wObjm wCtm def
0 0 wSa
0.000 0.000 0.000 1.000 wCMYK wRscr
/Linep { 6 6 0.0 wPn stroke } bind def
/Fillp {  } bind def
/Paintp { wGs wDocm wStm wGs Fillp wGr Linep wGr newpath} bind def
 1710 -1147 wMto
1628 -1071 1636 -1103 1650 -1059 wCto
1673 -980 1640 -979 1719 -998 wCto
1742 -1004 1754 -1023 1759 -1044 wCto
1763 -1064 1758 -1095 1791 -1140 wCto
1814 -1172 1833 -1229 1854 -1248 wCto
1897 -1222 1916 -1219 1968 -1164 wCto
wGs wDocm wStm Linep wGr newpath
wRes
wSav
-1695 3860 -1584 4002 wStbb
-1697 3856 -1581 4005 wStbb
-187 3485 wTl
[1.3665 0.0000 0.0000 1.3665 -3952.8768 1934.5945] concat /wObjm wCtm def
0 0 wSa
0.000 0.000 0.000 1.000 wCMYK wRscr
/Linep { 6 6 0.0 wPn stroke } bind def
/Fillp {  } bind def
/Paintp { wGs wDocm wStm wGs Fillp wGr Linep wGr newpath} bind def
 1791 -1140 wMto
1815 -1108 1853 -1086 1869 -1039 wCto
wGs wDocm wStm Linep wGr newpath
wRes
wSav
-1766 3570 -1396 4041 wStbb
-1775 3558 -1386 4052 wStbb
-187 3485 wTl
[1.3665 0.0000 0.0000 1.3665 -3952.8768 1934.5945] concat /wObjm wCtm def
0 0 wSa
0.000 0.000 0.000 1.000 wCMYK wRscr
/Linep { 6 6 0.0 wPn stroke } bind def
/Fillp {  } bind def
/Paintp { wGs wDocm wStm wGs Fillp wGr Linep wGr newpath} bind def
 1953 -1352 wMto
2001 -1246 2007 -1021 1739 -1010 wCto
wGs wDocm wStm Linep wGr newpath
wRes
wSav
-2016 3522 -1113 4276 wStbb
-2038 3503 -1090 4294 wStbb
-187 3485 wTl
[1.3665 0.0000 0.0000 1.3665 -3952.8768 1934.5945] concat /wObjm wCtm def
0 0 wSa
0.000 0.000 0.000 1.000 wCMYK wRscr
/Linep { 6 6 0.0 wPn stroke } bind def
/Fillp {  } bind def
/Paintp { wGs wDocm wStm wGs Fillp wGr Linep wGr newpath} bind def
 1977 -1387 wMto
2214 -1241 2157 -1101 1882 -1025 wCto
1607 -948 1653 -980 1605 -953 wCto
1568 -898 1562 -866 1556 -850 wCto
1653 -838 1912 -889 1965 -914 wCto
2019 -939 2148 -971 2125 -1182 wCto
wGs wDocm wStm Linep wGr newpath
wRes
wSav
-2093 4230 -1576 4450 wStbb
-2105 4224 -1563 4455 wStbb
-187 3485 wTl
[1.3665 0.0000 0.0000 1.3665 -3952.8768 1934.5945] concat /wObjm wCtm def
0 0 wSa
0.000 0.000 0.000 1.000 wCMYK wRscr
/Linep { 6 6 0.0 wPn stroke } bind def
/Fillp {  } bind def
/Paintp { wGs wDocm wStm wGs Fillp wGr Linep wGr newpath} bind def
 1642 -850 wMto
1538 -822 1567 -855 1546 -805 wCto
1512 -722 1500 -741 1576 -718 wCto
1600 -711 1617 -724 1633 -737 wCto
1651 -751 1678 -793 1710 -805 wCto
1746 -819 1792 -863 1819 -869 wCto
1843 -826 1857 -810 1875 -737 wCto
wGs wDocm wStm Linep wGr newpath
wRes
wSav
-1806 4317 -1762 4495 wStbb
-1807 4312 -1760 4499 wStbb
-187 3485 wTl
[1.3665 0.0000 0.0000 1.3665 -3952.8768 1934.5945] concat /wObjm wCtm def
0 0 wSa
0.000 0.000 0.000 1.000 wCMYK wRscr
/Linep { 6 6 0.0 wPn stroke } bind def
/Fillp {  } bind def
/Paintp { wGs wDocm wStm wGs Fillp wGr Linep wGr newpath} bind def
 1710 -805 wMto
1715 -765 1739 -727 1730 -678 wCto
wGs wDocm wStm Linep wGr newpath
wRes
wSav
-1963 4178 -1472 4659 wStbb
-1975 4165 -1459 4671 wStbb
-187 3485 wTl
[1.3665 0.0000 0.0000 1.3665 -3952.8768 1934.5945] concat /wObjm wCtm def
0 0 wSa
0.000 0.000 0.000 1.000 wCMYK wRscr
/Linep { 6 6 0.0 wPn stroke } bind def
/Fillp {  } bind def
/Paintp { wGs wDocm wStm wGs Fillp wGr Linep wGr newpath} bind def
 1951 -907 wMto
1951 -906 1899 -558 1595 -717 wCto
wGs wDocm wStm Linep wGr newpath
wRes
wSav
-2283 4154 -1188 4763 wStbb
-2310 4138 -1160 4778 wStbb
-187 3485 wTl
[1.3665 0.0000 0.0000 1.3665 -3952.8768 1934.5945] concat /wObjm wCtm def
0 0 wSa
0.000 0.000 0.000 1.000 wCMYK wRscr
/Linep { 6 6 0.0 wPn stroke } bind def
/Fillp {  } bind def
/Paintp { wGs wDocm wStm wGs Fillp wGr Linep wGr newpath} bind def
 1991 -925 wMto
2159 -612 1922 -604 1729 -654 wCto
1455 -725 1512 -736 1454 -736 wCto
1395 -706 1375 -682 1361 -671 wCto
1439 -612 1689 -527 1748 -522 wCto
1807 -517 1937 -482 2022 -677 wCto
wGs wDocm wStm Linep wGr newpath
wRes
wSav
-2423 4492 -1949 4856 wStbb
-2434 4482 -1937 4865 wStbb
-187 3485 wTl
[1.3665 0.0000 0.0000 1.3665 -3952.8768 1934.5945] concat /wObjm wCtm def
0 0 wSa
0.000 0.000 0.000 1.000 wCMYK wRscr
/Linep { 6 6 0.0 wPn stroke } bind def
/Fillp {  } bind def
/Paintp { wGs wDocm wStm wGs Fillp wGr Linep wGr newpath} bind def
 1440 -626 wMto
1329 -659 1374 -677 1330 -637 wCto
1258 -571 1270 -595 1312 -547 wCto
1329 -528 1352 -528 1371 -535 wCto
1391 -541 1438 -560 1471 -555 wCto
1510 -549 1564 -558 1591 -550 wCto
1590 -500 1602 -486 1580 -414 wCto
wGs wDocm wStm Linep wGr newpath
wRes
wSav
-2198 4659 -2128 4824 wStbb
-2199 4654 -2126 4828 wStbb
-187 3485 wTl
[1.3665 0.0000 0.0000 1.3665 -3952.8768 1934.5945] concat /wObjm wCtm def
0 0 wSa
0.000 0.000 0.000 1.000 wCMYK wRscr
/Linep { 6 6 0.0 wPn stroke } bind def
/Fillp {  } bind def
/Paintp { wGs wDocm wStm wGs Fillp wGr Linep wGr newpath} bind def
 1471 -555 wMto
1456 -518 1455 -475 1423 -437 wCto
wGs wDocm wStm Linep wGr newpath
wRes
wSav
-2321 4692 -1762 5007 wStbb
-2334 4684 -1748 5014 wStbb
-187 3485 wTl
[1.3665 0.0000 0.0000 1.3665 -3952.8768 1934.5945] concat /wObjm wCtm def
0 0 wSa
0.000 0.000 0.000 1.000 wCMYK wRscr
/Linep { 6 6 0.0 wPn stroke } bind def
/Fillp {  } bind def
/Paintp { wGs wDocm wStm wGs Fillp wGr Linep wGr newpath} bind def
 1739 -524 wMto
1681 -441 1478 -303 1333 -531 wCto
wGs wDocm wStm Linep wGr newpath
wRes
wSav
-2637 4561 -1718 5193 wStbb
-2659 4545 -1695 5208 wStbb
-187 3485 wTl
[1.3665 0.0000 0.0000 1.3665 -3952.8768 1934.5945] concat /wObjm wCtm def
0 0 wSa
0.000 0.000 0.000 1.000 wCMYK wRscr
/Linep { 6 6 0.0 wPn stroke } bind def
/Fillp {  } bind def
/Paintp { wGs wDocm wStm wGs Fillp wGr Linep wGr newpath} bind def
 1771 -515 wMto
1763 -237 1620 -219 1417 -419 wCto
1213 -618 1264 -595 1216 -623 wCto
1150 -627 1120 -616 1102 -614 wCto
1141 -524 1315 -325 1363 -291 wCto
1412 -257 1509 -167 1680 -293 wCto
wGs wDocm wStm Linep wGr newpath
wRes
wSav
-2817 4566 -2439 5028 wStbb
-2826 4554 -2429 5039 wStbb
-187 3485 wTl
[1.3665 0.0000 0.0000 1.3665 -3952.8768 1934.5945] concat /wObjm wCtm def
0 0 wSa
0.000 0.000 0.000 1.000 wCMYK wRscr
/Linep { 6 6 0.0 wPn stroke } bind def
/Fillp {  } bind def
/Paintp { wGs wDocm wStm wGs Fillp wGr Linep wGr newpath} bind def
 1145 -539 wMto
1064 -623 1106 -623 1057 -602 wCto
970 -565 984 -583 997 -533 wCto
1003 -509 1013 -505 1042 -493 wCto
1061 -485 1102 -484 1139 -461 wCto
1172 -440 1225 -421 1244 -401 wCto
1219 -357 1223 -339 1168 -288 wCto
wGs wDocm wStm Linep wGr newpath
wRes
wSav
-2724 4788 -2582 4900 wStbb
-2727 4785 -2578 4902 wStbb
-187 3485 wTl
[1.3665 0.0000 0.0000 1.3665 -3952.8768 1934.5945] concat /wObjm wCtm def
0 0 wSa
0.000 0.000 0.000 1.000 wCMYK wRscr
/Linep { 6 6 0.0 wPn stroke } bind def
/Fillp {  } bind def
/Paintp { wGs wDocm wStm wGs Fillp wGr Linep wGr newpath} bind def
 1139 -461 wMto
1107 -436 1084 -399 1038 -382 wCto
wGs wDocm wStm Linep wGr newpath
wRes
wSav
-2768 4716 -2294 5088 wStbb
-2779 4706 -2282 5097 wStbb
-187 3485 wTl
[1.3665 0.0000 0.0000 1.3665 -3952.8768 1934.5945] concat /wObjm wCtm def
0 0 wSa
0.000 0.000 0.000 1.000 wCMYK wRscr
/Linep { 6 6 0.0 wPn stroke } bind def
/Fillp {  } bind def
/Paintp { wGs wDocm wStm wGs Fillp wGr Linep wGr newpath} bind def
 1350 -302 wMto
1259 -259 1018 -244 1006 -513 wCto
wGs wDocm wStm Linep wGr newpath
wRes
wSav
-2997 4466 -2249 5371 wStbb
-3015 4443 -2230 5393 wStbb
-187 3485 wTl
[1.3665 0.0000 0.0000 1.3665 -3952.8768 1934.5945] concat /wObjm wCtm def
0 0 wSa
0.000 0.000 0.000 1.000 wCMYK wRscr
/Linep { 6 6 0.0 wPn stroke } bind def
/Fillp {  } bind def
/Paintp { wGs wDocm wStm wGs Fillp wGr Linep wGr newpath} bind def
 1383 -276 wMto
1236 -37 1100 -92 1023 -370 wCto
947 -645 979 -599 951 -647 wCto
896 -683 865 -689 849 -696 wCto
838 -611 891 -333 913 -286 wCto
938 -233 973 -102 1184 -126 wCto
wGs wDocm wStm Linep wGr newpath
wRes
wSav
-3172 4417 -2952 4901 wStbb
-3177 4404 -2946 4913 wStbb
-187 3485 wTl
[1.3665 0.0000 0.0000 1.3665 -3952.8768 1934.5945] concat /wObjm wCtm def
0 0 wSa
0.000 0.000 0.000 1.000 wCMYK wRscr
/Linep { 6 6 0.0 wPn stroke } bind def
/Fillp {  } bind def
/Paintp { wGs wDocm wStm wGs Fillp wGr Linep wGr newpath} bind def
 848 -619 wMto
819 -732 856 -702 803 -709 wCto
710 -720 730 -728 717 -679 wCto
710 -655 717 -646 736 -622 wCto
749 -605 783 -584 803 -545 wCto
822 -511 861 -471 868 -444 wCto
824 -419 809 -399 736 -381 wCto
wGs wDocm wStm Linep wGr newpath
wRes
wSav
-3208 4673 -3041 4711 wStbb
-3212 4672 -3036 4711 wStbb
-187 3485 wTl
[1.3665 0.0000 0.0000 1.3665 -3952.8768 1934.5945] concat /wObjm wCtm def
0 0 wSa
0.000 0.000 0.000 1.000 wCMYK wRscr
/Linep { 6 6 0.0 wPn stroke } bind def
/Fillp {  } bind def
/Paintp { wGs wDocm wStm wGs Fillp wGr Linep wGr newpath} bind def
 803 -545 wMto
764 -540 732 -520 684 -529 wCto
wGs wDocm wStm Linep wGr newpath
wRes
wSav
-3575 4072 -3211 4524 wStbb
-3584 4060 -3201 4535 wStbb
-187 3485 wTl
[1.3665 0.0000 0.0000 1.3665 -3952.8768 1934.5945] concat /wObjm wCtm def
0 0 wSa
0.000 0.000 0.000 1.000 wCMYK wRscr
/Linep { 6 6 0.0 wPn stroke } bind def
/Fillp {  } bind def
/Paintp { wGs wDocm wStm wGs Fillp wGr Linep wGr newpath} bind def
 638 -833 wMto
670 -945 679 -895 637 -928 wCto
562 -985 584 -981 547 -945 wCto
529 -928 531 -917 535 -886 wCto
538 -866 557 -830 555 -786 wCto
554 -752 570 -698 560 -666 wCto
510 -667 486 -657 415 -678 wCto
wGs wDocm wStm Linep wGr newpath
wRes
wSav
-3543 4275 -3380 4347 wStbb
-3547 4273 -3375 4348 wStbb
-187 3485 wTl
[1.3665 0.0000 0.0000 1.3665 -3952.8768 1934.5945] concat /wObjm wCtm def
0 0 wSa
0.000 0.000 0.000 1.000 wCMYK wRscr
/Linep { 6 6 0.0 wPn stroke } bind def
/Fillp {  } bind def
/Paintp { wGs wDocm wStm wGs Fillp wGr Linep wGr newpath} bind def
 555 -786 wMto
518 -802 476 -804 439 -836 wCto
wGs wDocm wStm Linep wGr newpath
wRes
wSav
-3753 3677 -3258 4035 wStbb
-3765 3668 -3245 4043 wStbb
-187 3485 wTl
[1.3665 0.0000 0.0000 1.3665 -3952.8768 1934.5945] concat /wObjm wCtm def
0 0 wSa
0.000 0.000 0.000 1.000 wCMYK wRscr
/Linep { 6 6 0.0 wPn stroke } bind def
/Fillp {  } bind def
/Paintp { wGs wDocm wStm wGs Fillp wGr Linep wGr newpath} bind def
 560 -1118 wMto
644 -1199 623 -1151 602 -1200 wCto
571 -1274 581 -1269 532 -1256 wCto
511 -1251 503 -1239 493 -1216 wCto
485 -1196 484 -1156 461 -1119 wCto
443 -1090 433 -1038 409 -1015 wCto
366 -1041 336 -1041 285 -1096 wCto
wGs wDocm wStm Linep wGr newpath
wRes
wSav
-3610 3754 -3509 3892 wStbb
-3612 3750 -3506 3895 wStbb
-187 3485 wTl
[1.3665 0.0000 0.0000 1.3665 -3952.8768 1934.5945] concat /wObjm wCtm def
0 0 wSa
0.000 0.000 0.000 1.000 wCMYK wRscr
/Linep { 6 6 0.0 wPn stroke } bind def
/Fillp {  } bind def
/Paintp { wGs wDocm wStm wGs Fillp wGr Linep wGr newpath} bind def
 461 -1119 wMto
436 -1151 407 -1170 390 -1217 wCto
wGs wDocm wStm Linep wGr newpath
wRes
wSav
-3258 2917 -2787 3269 wStbb
-3269 2908 -2775 3277 wStbb
-187 3485 wTl
[1.3665 0.0000 0.0000 1.3665 -3952.8768 1934.5945] concat /wObjm wCtm def
0 0 wSa
0.000 0.000 0.000 1.000 wCMYK wRscr
/Linep { 6 6 0.0 wPn stroke } bind def
/Fillp {  } bind def
/Paintp { wGs wDocm wStm wGs Fillp wGr Linep wGr newpath} bind def
 822 -1616 wMto
940 -1575 887 -1575 930 -1619 wCto
989 -1680 983 -1665 944 -1700 wCto
924 -1717 896 -1723 876 -1717 wCto
816 -1697 827 -1695 776 -1696 wCto
737 -1698 679 -1683 653 -1691 wCto
653 -1741 647 -1759 669 -1830 wCto
wGs wDocm wStm Linep wGr newpath
wRes
wSav
-3082 2937 -3012 3104 wStbb
-3083 2932 -3010 3108 wStbb
-187 3485 wTl
[1.3665 0.0000 0.0000 1.3665 -3952.8768 1934.5945] concat /wObjm wCtm def
0 0 wSa
0.000 0.000 0.000 1.000 wCMYK wRscr
/Linep { 6 6 0.0 wPn stroke } bind def
/Fillp {  } bind def
/Paintp { wGs wDocm wStm wGs Fillp wGr Linep wGr newpath} bind def
 776 -1696 wMto
791 -1734 792 -1777 824 -1815 wCto
wGs wDocm wStm Linep wGr newpath
wRes
wSav
-2044 3371 3965 4451 wStbb
-2194 3344 4115 4478 wStbb
-187 3485 wTl
[1.3665 0.0000 0.0000 1.6725 -3952.8768 2194.5491] concat /wObjm wCtm def
/Linep { } bind def
/Fillp { wRscr
0.000 0.000 0.000 0.000 wCMYK  eofill  } bind def
/Paintp { wGs wDocm wStm wGs Fillp wGr Linep wGr newpath} bind def
 5930 -1379 wMto
5930 -736 wLto
1536 -736 wLto
1536 -1379 wLto
5930 -1379 wLto
 wCp  Paintp
wRes
wSav
1627 3528 3798 4235 wStbb
1572 3510 3852 4252 wStbb
0 0 wTl
[1.0000 0.0000 0.0000 1.0000 0.0000 0.0000] concat /wObjm wCtm def
0 1 wSa
0.000 0.000 0.000 1.000 wCMYK wRscr
/Linep { 16 16 0.0 wPn stroke } bind def
/Fillp { wRscr
0.000 0.000 0.000 0.600 wCMYK  eofill  } bind def
/Paintp { wGs wDocm wStm wGs Fillp wGr Linep wGr newpath} bind def
 2307 4052 wMto
2334 4058 2378 4055 2384 4040 wCto
2380 4035 2260 3965 2271 3964 wCto
2255 3967 2252 3980 2260 3996 wCto
2286 4046 2307 4052 2307 4052 wCto
 wCp  1629 4142 wMto
1633 4157 1652 4185 1675 4192 wCto
1698 4200 1713 4200 1726 4193 wCto
1741 4186 1750 4167 1771 4177 wCto
1792 4188 1815 4192 1827 4193 wCto
1839 4193 1860 4200 1885 4193 wCto
1912 4186 2012 4139 2036 4050 wCto
2053 3981 2040 3896 2046 3877 wCto
2135 3878 wLto
2135 3878 2115 3969 2132 4048 wCto
2150 4127 2228 4182 2307 4193 wCto
2387 4204 2500 4151 2534 4055 wCto
2543 3967 2523 3946 2523 3946 wCto
2523 3946 2289 3837 2271 3825 wCto
2252 3812 2326 3743 2364 3741 wCto
2402 3740 2431 3866 2481 3867 wCto
2530 3867 2494 3866 2574 3867 wCto
2574 3867 2600 3849 2603 3858 wCto
2605 3867 2599 4126 2603 4137 wCto
2608 4149 2616 4192 2663 4192 wCto
2701 4193 2730 4170 2730 4151 wCto
2730 4109 2744 3851 2738 3814 wCto
2731 3776 2840 3711 2874 3807 wCto
2871 3912 2866 4095 2870 4146 wCto
2875 4194 2970 4222 3003 4142 wCto
3006 4013 3003 3859 3006 3818 wCto
3009 3788 3098 3689 3148 3815 wCto
3145 3976 3141 4090 3144 4146 wCto
3146 4185 3231 4233 3280 4137 wCto
3281 4035 3274 3883 3286 3875 wCto
3347 3877 3378 3878 3398 3875 wCto
3420 3874 3443 3843 3446 3825 wCto
3460 3725 3539 3739 3554 3741 wCto
3589 3747 3583 3770 3567 3786 wCto
3549 3802 3361 3958 3350 3980 wCto
3340 4000 3320 4076 3407 4146 wCto
3491 4214 3564 4200 3614 4184 wCto
3664 4169 3729 4091 3741 4054 wCto
3773 3937 3761 4008 3745 3941 wCto
3741 3922 3692 3871 3667 3926 wCto
3633 4001 3605 4127 3497 4047 wCto
3479 4033 3490 4028 3500 4019 wCto
3509 4009 3649 3882 3659 3873 wCto
3668 3863 3796 3733 3633 3628 wCto
3479 3530 3359 3717 3343 3740 wCto
3309 3739 3284 3751 3270 3739 wCto
3257 3727 3173 3530 2935 3661 wCto
2704 3534 2627 3703 2609 3740 wCto
2570 3739 2551 3741 2530 3740 wCto
2510 3740 2492 3612 2348 3605 wCto
2202 3598 2174 3735 2146 3739 wCto
2117 3743 2053 3744 2043 3735 wCto
2043 3703 2044 3661 2038 3649 wCto
2031 3637 1976 3583 1915 3665 wCto
1914 3819 1914 3972 1910 3996 wCto
1904 4031 1826 4106 1769 4005 wCto
1772 3857 1780 3688 1772 3668 wCto
1752 3618 1678 3574 1637 3673 wCto
1634 3831 1629 4142 1629 4142 wCto
 wCp  Paintp
wRes
wSav
-2991 3565 1205 4400 wStbb
-3095 3544 1309 4420 wStbb
0 0 wTl
[1.0000 0.0000 0.0000 1.0000 0.0000 0.0000] concat /wObjm wCtm def
0 1 wSa
0.000 0.000 0.000 1.000 wCMYK wRscr
/Linep { 16 16 0.0 wPn stroke } bind def
/Fillp { wRscr
0.000 0.000 0.000 0.600 wCMYK  eofill  } bind def
/Paintp { wGs wDocm wStm wGs Fillp wGr Linep wGr newpath} bind def
 currentflat 2 mul setflat
 568 4025 wMto
597 4025 631 4005 638 3979 wCto
643 3952 650 3780 641 3752 wCto
631 3725 596 3717 573 3717 wCto
565 3717 511 3751 510 3768 wCto
508 3795 495 3948 506 3980 wCto
519 4015 568 4025 568 4025 wCto
 wCp  -271 3996 wMto
-254 4016 -237 4038 -205 4035 wCto
-174 4033 -147 4009 -140 3979 wCto
-134 3946 -134 3786 -143 3756 wCto
-152 3727 -193 3717 -209 3720 wCto
-225 3724 -256 3736 -266 3754 wCto
-279 3784 -271 3996 -271 3996 wCto
 wCp  -2012 4020 wMto
-1985 4025 -1942 4023 -1935 4008 wCto
-1939 4003 -2059 3933 -2050 3930 wCto
-2065 3934 -2068 3948 -2059 3964 wCto
-2034 4013 -2012 4020 -2012 4020 wCto
 wCp  -1129 4001 wMto
-1118 4021 -1099 4046 -1064 4042 wCto
-1029 4036 -1025 4028 -1005 4003 wCto
-985 3977 -993 3796 -995 3775 wCto
-999 3752 -1047 3735 -1060 3736 wCto
-1074 3739 -1118 3744 -1126 3788 wCto
-1136 3842 -1129 4001 -1129 4001 wCto
 wCp  -643 4145 wMto
-693 4146 -708 4175 -731 4166 wCto
-753 4157 -777 4126 -778 4113 wCto
-780 4098 -771 3843 -777 3839 wCto
-812 3839 -870 3839 -870 3839 wCto
-867 4129 wLto
-867 4129 -873 4163 -908 4174 wCto
-943 4185 -959 4153 -967 4151 wCto
-1002 4149 -1021 4175 -1079 4174 wCto
-1079 4174 -1225 4177 -1265 4005 wCto
-1265 4005 -1268 3882 -1264 3839 wCto
-1304 3838 -1338 3838 -1338 3838 wCto
-1338 3838 -1339 4327 -1339 4323 wCto
-1345 4370 -1361 4397 -1404 4397 wCto
-1440 4398 -1480 4370 -1480 4323 wCto
-1480 4275 -1480 3839 -1480 3839 wCto
-1563 3839 wLto
-1563 3839 -1564 4331 -1563 4323 wCto
-1570 4360 -1604 4397 -1637 4397 wCto
-1671 4397 -1714 4364 -1714 4323 wCto
-1714 4281 -1714 3839 -1714 3839 wCto
-1714 3839 -1842 3842 -1864 3839 wCto
-1892 3835 -1907 3823 -1916 3775 wCto
-1926 3727 -1953 3719 -1979 3727 wCto
-1992 3729 -2043 3731 -2045 3779 wCto
-1942 3846 -1795 3912 -1780 3924 wCto
-1765 3936 -1776 3976 -1779 3988 wCto
-1782 4000 -1833 4158 -1988 4167 wCto
-2142 4175 -2197 4005 -2201 3987 wCto
-2206 3954 -2201 3839 -2201 3839 wCto
-2281 3838 wLto
-2281 3838 -2289 3984 -2278 4016 wCto
-2258 4070 -2209 4044 -2209 4103 wCto
-2209 4163 -2262 4155 -2277 4180 wCto
-2275 4204 -2281 4284 -2277 4256 wCto
-2275 4245 -2291 4304 -2336 4308 wCto
-2384 4313 -2407 4296 -2415 4267 wCto
-2421 4248 -2411 4180 -2415 4174 wCto
-2421 4167 -2484 4167 -2479 4107 wCto
-2473 4043 -2414 4054 -2415 4015 wCto
-2417 3988 -2415 3839 -2415 3839 wCto
-2556 3839 wLto
-2556 3839 -2685 3948 -2707 3969 wCto
-2727 3991 -2723 4020 -2672 4027 wCto
-2622 4035 -2575 3983 -2585 3937 wCto
-2556 3906 -2544 3889 -2544 3889 wCto
-2531 3889 -2467 3894 -2467 3956 wCto
-2467 4017 -2542 4166 -2688 4166 wCto
-2829 4166 -2861 4067 -2871 4031 wCto
-2887 3973 -2869 3992 -2862 3956 wCto
-2820 3922 -2669 3770 -2642 3752 wCto
-2645 3717 -2662 3713 -2673 3708 wCto
-2703 3696 -2742 3717 -2754 3776 wCto
-2765 3830 -2828 3837 -2828 3837 wCto
-2848 3837 -2943 3841 -2947 3839 wCto
-2952 3837 -2989 3820 -2985 3774 wCto
-2982 3727 -2951 3703 -2939 3700 wCto
-2928 3699 -2885 3708 -2866 3700 wCto
-2846 3691 -2804 3574 -2688 3571 wCto
-2572 3569 -2510 3707 -2472 3707 wCto
-2434 3707 -2422 3709 -2417 3708 wCto
-2417 3673 -2420 3646 -2414 3638 wCto
-2410 3629 -2391 3581 -2340 3583 wCto
-2290 3585 -2297 3597 -2278 3620 wCto
-2278 3656 -2274 3683 -2275 3708 wCto
-2275 3708 -2183 3715 -2174 3708 wCto
-2163 3703 -2107 3579 -1979 3585 wCto
-1852 3589 -1799 3683 -1798 3699 wCto
-1763 3699 -1726 3700 -1710 3699 wCto
-1710 3656 -1703 3650 -1701 3636 wCto
-1698 3621 -1683 3577 -1631 3579 wCto
-1583 3582 -1574 3602 -1562 3626 wCto
-1555 3658 -1563 3685 -1560 3700 wCto
-1535 3700 -1485 3703 -1470 3700 wCto
-1470 3666 -1465 3638 -1462 3629 wCto
-1458 3618 -1444 3567 -1389 3575 wCto
-1337 3582 -1355 3598 -1335 3621 wCto
-1335 3645 -1339 3688 -1333 3700 wCto
-1314 3700 -1254 3705 -1245 3700 wCto
-1235 3693 -1211 3618 -1098 3591 wCto
-1055 3581 -1028 3590 -1012 3597 wCto
-975 3612 -986 3614 -954 3590 wCto
-933 3575 -917 3586 -902 3590 wCto
-878 3595 -867 3605 -855 3636 wCto
-854 3677 -859 3691 -855 3708 wCto
-821 3708 -785 3716 -769 3708 wCto
-765 3684 -767 3626 -762 3614 wCto
-755 3603 -728 3573 -707 3571 wCto
-674 3570 -647 3598 -641 3621 wCto
-634 3642 -641 3695 -637 3708 wCto
-631 3723 -426 3723 -402 3713 wCto
-379 3704 -333 3597 -220 3578 wCto
-220 3578 -181 3574 -165 3586 wCto
-148 3599 -132 3609 -120 3605 wCto
-101 3605 -68 3583 -50 3579 wCto
-32 3577 -1 3603 8 3641 wCto
8 3681 4 3691 9 3708 wCto
53 3711 112 3712 142 3708 wCto
142 3673 139 3645 143 3633 wCto
148 3621 156 3579 206 3579 wCto
264 3579 275 3621 280 3629 wCto
284 3637 275 3695 282 3708 wCto
307 3708 367 3715 392 3708 wCto
422 3660 492 3575 577 3579 wCto
663 3585 748 3634 767 3700 wCto
810 3700 850 3704 870 3700 wCto
870 3661 870 3648 875 3637 wCto
881 3625 886 3577 944 3579 wCto
1003 3582 1006 3618 1006 3620 wCto
1015 3642 1021 3683 1018 3709 wCto
1014 3739 1006 3949 1011 3973 wCto
1015 4000 1028 4021 1046 4025 wCto
1064 4028 1138 4021 1162 4027 wCto
1188 4033 1196 4067 1199 4088 wCto
1203 4113 1181 4159 1142 4170 wCto
1102 4181 1073 4171 1046 4166 wCto
1018 4159 994 4142 984 4142 wCto
974 4142 967 4159 941 4163 wCto
908 4167 885 4153 871 4117 wCto
871 4074 871 3839 871 3839 wCto
778 3839 wLto
778 3839 781 4000 778 4025 wCto
774 4050 685 4165 572 4165 wCto
458 4165 387 4070 373 4024 wCto
364 3991 365 3839 365 3839 wCto
272 3839 wLto
272 3839 266 4003 271 4019 wCto
279 4046 351 4009 360 4088 wCto
369 4167 284 4141 283 4182 wCto
283 4222 279 4242 278 4264 wCto
276 4284 221 4317 198 4312 wCto
175 4307 162 4299 144 4265 wCto
140 4237 148 4192 137 4174 wCto
127 4157 65 4166 66 4088 wCto
67 4038 125 4038 136 4025 wCto
137 3991 151 3858 137 3838 wCto
96 3838 0 3839 0 3839 wCto
0 3839 -7 4130 -5 4126 wCto
-18 4157 -30 4173 -58 4174 wCto
-86 4175 -104 4149 -111 4142 wCto
-127 4147 -158 4175 -204 4174 wCto
-248 4173 -333 4165 -387 4078 wCto
-399 4056 -403 4033 -402 4025 wCto
-401 4017 -402 3839 -402 3839 wCto
-637 3839 wLto
-637 3839 -641 3964 -635 3979 wCto
-630 3992 -615 4023 -580 4025 wCto
-545 4029 -518 4024 -509 4017 wCto
-492 4013 -449 4048 -441 4088 wCto
-433 4130 -441 4173 -499 4170 wCto
-537 4169 -593 4145 -643 4145 wCto
 wCp  Paintp
wRes
wRes
 end

end
clear
userdict /VPsave get restore

/PPT_ProcessAll false def
32 0 0 46 46 0 0 0 43 /Helvetica-Bold /font13 ANSIFont font
0 0 0 fC
gs 1790 426 457 270 CB
532 345 87 (Pub) 86 SB
618 345 12 (l) 13 SB
631 345 37 (is) 38 SB
669 345 54 (he) 53 SB
722 345 41 (d ) 40 SB
762 345 53 (by) 54 SB
816 345 13 ( ) 12 SB
828 345 93 (Fusi) 94 SB
922 345 28 (o) 27 SB
949 345 72 (n E) 71 SB
1020 345 54 (ne) 53 SB
1073 345 154 (rgy Div) 155 SB
1228 345 37 (is) 38 SB
1266 345 81 (ion,) 80 SB
1346 345 49 ( O) 48 SB
1394 345 52 (ak) 51 SB
1445 345 140 ( Ridge) 139 SB
1584 345 45 ( N) 46 SB
1630 345 26 (a) 25 SB
1655 345 15 (t) 16 SB
1671 345 94 (iona) 93 SB
1764 345 12 (l) 13 SB
1777 345 13 ( ) 12 SB
1789 345 54 (La) 53 SB
1842 345 100 (bora) 99 SB
1941 345 86 (tory) 86 SB
gr
gs 1790 426 457 270 CB
532 400 73 (Bui) 74 SB
606 400 121 (lding ) 120 SB
726 400 52 (92) 51 SB
777 400 52 (01) 51 SB
828 400 15 (-) 16 SB
844 400 26 (2) 25 SB
869 400 26 (  ) 25 SB
894 400 26 (  ) 25 SB
919 400 26 (  ) 25 SB
944 400 44 ( P) 43 SB
987 400 49 (.O) 48 SB
1035 400 26 (. ) 25 SB
1060 400 87 (Box) 86 SB
1146 400 39 ( 2) 38 SB
1184 400 52 (00) 51 SB
1235 400 39 (9 ) 38 SB
1273 400 26 (  ) 25 SB
1298 400 26 (  ) 25 SB
1323 400 49 ( O) 48 SB
1371 400 52 (ak) 51 SB
1422 400 140 ( Ridge) 139 SB
1561 400 26 (, ) 25 SB
1586 400 29 (T) 28 SB
1614 400 32 (N) 33 SB
1647 400 39 ( 3) 38 SB
1685 400 52 (78) 51 SB
1736 400 52 (31) 51 SB
1787 400 67 (-80) 66 SB
1853 400 52 (71) 51 SB
1904 400 26 (, ) 25 SB
1929 400 32 (U) 33 SB
1962 400 62 (SA) 64 SB
gr
1638 1 532 455 B
1 F
n
32 0 0 46 46 0 0 0 42 /Helvetica /font12 ANSIFont font
gs 1790 426 457 270 CB
532 457 57 (Ed) 56 SB
588 457 49 (ito) 48 SB
636 457 15 (r) 16 SB
652 457 13 (:) 12 SB
664 457 36 ( J) 35 SB
699 457 90 (ame) 89 SB
788 457 67 (s A) 66 SB
854 457 26 (. ) 25 SB
879 457 123 (Rome) 122 SB
gr
gs 1790 426 457 270 CB
1265 457 12 (I) 13 SB
1278 457 23 (s) 22 SB
1300 457 75 (sue) 74 SB
1374 457 39 ( 3) 38 SB
1412 457 26 (7) 26 SB
gr
gs 1790 426 457 270 CB
1893 457 49 (Ja) 48 SB
1941 457 52 (nu) 51 SB
1992 457 77 (ary ) 76 SB
2068 457 52 (19) 51 SB
2119 457 52 (95) 51 SB
gr
32 0 0 33 33 0 0 0 31 /Helvetica /font12 ANSIFont font
gs 1790 426 457 270 CB
532 512 60 (E-M) 61 SB
593 512 17 (a) 19 SB
612 512 56 (il: ja) 58 SB
670 512 62 (r@o) 63 SB
733 512 29 (rn) 30 SB
763 512 51 (l.go) 53 SB
816 512 15 (v) 16 SB
832 512 319 (                               Ph) 320 SB
1152 512 17 (o) 18 SB
1170 512 18 (n) 19 SB
1189 512 17 (e) 18 SB
1207 512 47 (: \(6) 48 SB
1255 512 36 (15) 37 SB
1292 512 56 (\) 57) 57 SB
1349 512 47 (4-1) 48 SB
1397 512 36 (30) 37 SB
1434 512 18 (6) 18 SB
gr
945 6 222 663 B
1 F
n
32 0 0 58 58 0 0 0 53 /Helvetica-Bold /font13 ANSIFont font
gs 1096 1919 147 588 CB
237 701 196 (Electro) 197 SB
434 701 86 (n T) 87 SB
521 701 23 (r) 22 SB
543 701 32 (a) 33 SB
576 701 102 (nsp) 103 SB
679 701 186 (ort Stu) 187 SB
866 701 83 (die) 84 SB
950 701 134 (s in t) 135 SB
1085 701 67 (he) 67 SB
gr
gs 1096 1919 147 588 CB
315 771 77 (Co) 78 SB
393 771 151 (mpac) 152 SB
545 771 76 (t A) 77 SB
622 771 70 (ub) 71 SB
693 771 93 (urn) 94 SB
787 771 86 ( To) 87 SB
874 771 23 (r) 22 SB
896 771 32 (s) 33 SB
929 771 144 (atron) 145 SB
gr
32 0 0 42 42 0 0 0 38 /Times-Roman /font32 ANSIFont font
gs 1096 1919 147 588 CB
222 868 26 (T) 25 SB
247 868 40 (he) 39 SB
286 868 32 ( p) 31 SB
317 868 68 (robl) 67 SB
384 868 19 (e) 18 SB
402 868 44 (m ) 43 SB
445 868 35 (of) 34 SB
479 868 30 ( c) 29 SB
508 868 35 (ro) 34 SB
542 868 16 (s) 17 SB
559 868 44 (s-f) 43 SB
602 868 31 (ie) 30 SB
632 868 33 (ld) 32 SB
664 868 32 ( p) 31 SB
695 868 31 (la) 30 SB
725 868 68 (sma) 67 SB
792 868 11 ( ) 10 SB
802 868 26 (tr) 25 SB
827 868 40 (an) 39 SB
866 868 16 (s) 17 SB
883 868 21 (p) 20 SB
903 868 47 (ort) 46 SB
949 868 23 ( i) 22 SB
971 868 32 (n ) 31 SB
1002 868 33 (th) 32 SB
1034 868 30 (e ) 29 SB
1063 868 35 (pr) 34 SB
1097 868 35 (es) 35 SB
gr
gs 1096 1919 147 588 CB
1132 868 14 (-) 14 SB
gr
gs 1096 1919 147 588 CB
222 918 19 (e) 18 SB
240 918 40 (nc) 39 SB
279 918 30 (e ) 29 SB
308 918 35 (of) 34 SB
342 918 39 ( st) 38 SB
380 918 40 (oc) 39 SB
419 918 40 (ha) 39 SB
458 918 16 (s) 17 SB
475 918 12 (t) 11 SB
486 918 31 (ic) 30 SB
516 918 11 ( ) 10 SB
526 918 33 (m) 32 SB
558 918 40 (ag) 39 SB
597 918 40 (ne) 39 SB
636 918 24 (ti) 23 SB
659 918 30 (c ) 29 SB
688 918 26 (fi) 25 SB
713 918 19 (e) 18 SB
731 918 60 (lds ) 59 SB
790 918 40 (ha) 39 SB
829 918 41 (s r) 40 SB
869 918 38 (ec) 37 SB
906 918 19 (e) 18 SB
924 918 33 (iv) 32 SB
956 918 40 (ed) 39 SB
995 918 30 ( c) 29 SB
1024 918 42 (on) 41 SB
1065 918 16 (s) 17 SB
1082 918 12 (i) 11 SB
1093 918 40 (de) 39 SB
1132 918 14 (r) 14 SB
gr
gs 1096 1919 147 588 CB
1146 918 14 (-) 14 SB
gr
gs 1096 1919 147 588 CB
222 968 19 (a) 18 SB
240 968 52 (ble) 51 SB
291 968 11 ( ) 10 SB
301 968 19 (a) 18 SB
319 968 24 (tt) 23 SB
342 968 40 (en) 39 SB
381 968 24 (ti) 23 SB
404 968 53 (on ) 52 SB
456 968 33 (in) 32 SB
488 968 23 ( t) 22 SB
510 968 40 (he) 39 SB
549 968 11 ( ) 10 SB
559 968 24 (li) 23 SB
582 968 31 (te) 30 SB
612 968 33 (ra) 32 SB
644 968 33 (tu) 32 SB
676 968 33 (re) 32 SB
708 968 32 ( b) 31 SB
739 968 38 (ec) 37 SB
776 968 19 (a) 18 SB
794 968 56 (use) 55 SB
849 968 32 ( o) 31 SB
880 968 25 (f ) 24 SB
904 968 24 (it) 23 SB
927 968 16 (s) 17 SB
944 968 11 ( ) 10 SB
954 968 12 (i) 11 SB
965 968 54 (mp) 53 SB
1018 968 47 (ort) 46 SB
1064 968 40 (an) 39 SB
1103 968 38 (ce) 37 SB
gr
gs 1096 1919 147 588 CB
222 1018 33 (in) 32 SB
254 1018 32 ( p) 31 SB
285 1018 68 (rovi) 67 SB
352 1018 33 (di) 32 SB
384 1018 53 (ng ) 52 SB
436 1018 19 (a) 18 SB
454 1018 32 ( p) 31 SB
485 1018 52 (ote) 51 SB
536 1018 33 (nt) 32 SB
568 1018 31 (ia) 30 SB
598 1018 12 (l) 11 SB
609 1018 30 ( e) 29 SB
638 1018 42 (xp) 41 SB
679 1018 31 (la) 30 SB
709 1018 40 (na) 39 SB
748 1018 24 (ti) 23 SB
771 1018 53 (on ) 52 SB
823 1018 35 (of) 34 SB
857 1018 23 ( t) 22 SB
879 1018 40 (he) 39 SB
918 1018 11 ( ) 10 SB
928 1018 40 (an) 39 SB
967 1018 54 (om) 53 SB
1020 1018 31 (al) 30 SB
1050 1018 58 (ous) 58 SB
gr
gs 1096 1919 147 588 CB
222 1068 19 (e) 18 SB
240 1068 31 (le) 30 SB
270 1068 19 (c) 18 SB
288 1068 47 (tro) 46 SB
334 1068 32 (n ) 31 SB
365 1068 52 (the) 51 SB
416 1068 47 (rm) 46 SB
462 1068 19 (a) 18 SB
480 1068 23 (l ) 22 SB
502 1068 19 (c) 18 SB
520 1068 63 (ond) 62 SB
582 1068 40 (uc) 39 SB
621 1068 24 (ti) 23 SB
644 1068 45 (vit) 44 SB
688 1068 32 (y ) 31 SB
gr
32 0 0 42 42 0 0 0 42 /Symbol font
gs 1096 1919 147 588 CB
719 1064 23 (c) 23 SB
gr
32 0 0 33 33 0 0 0 30 /Times-Roman /font32 ANSIFont font
gs 1096 1919 147 588 CB
742 1085 15 (e) 15 SB
gr
32 0 0 42 42 0 0 0 38 /Times-Roman /font32 ANSIFont font
gs 1096 1919 147 588 CB
757 1068 11 ( ) 10 SB
767 1068 12 (i) 11 SB
778 1068 32 (n ) 31 SB
809 1068 52 (ma) 51 SB
860 1068 42 (gn) 41 SB
901 1068 31 (et) 30 SB
931 1068 31 (ic) 30 SB
961 1068 11 ( ) 10 SB
971 1068 19 (c) 18 SB
989 1068 68 (onfi) 67 SB
1056 1068 40 (ne) 39 SB
gr
gs 1096 1919 147 588 CB
1095 1068 14 (-) 14 SB
gr
gs 1096 1919 147 588 CB
222 1118 33 (m) 32 SB
254 1118 40 (en) 39 SB
293 1118 23 (t ) 22 SB
315 1118 40 (de) 39 SB
354 1118 33 (vi) 32 SB
386 1118 38 (ce) 37 SB
423 1118 27 (s ) 26 SB
449 1118 16 (s) 17 SB
466 1118 21 (u) 20 SB
486 1118 40 (ch) 39 SB
525 1118 30 ( a) 29 SB
554 1118 27 (s ) 26 SB
580 1118 54 (tok) 53 SB
633 1118 52 (am) 51 SB
684 1118 19 (a) 18 SB
702 1118 48 (ks ) 47 SB
749 1118 40 (an) 39 SB
788 1118 32 (d ) 31 SB
819 1118 16 (s) 17 SB
836 1118 12 (t) 11 SB
847 1118 19 (e) 18 SB
865 1118 24 (ll) 23 SB
888 1118 33 (ar) 32 SB
920 1118 31 (at) 30 SB
950 1118 62 (ors.) 61 SB
1011 1118 11 ( ) 10 SB
1021 1118 35 (In) 35 SB
gr
gs 1096 1919 147 588 CB
222 1168 31 (te) 30 SB
252 1168 47 (rm) 46 SB
298 1168 27 (s ) 26 SB
324 1168 46 (of ) 45 SB
369 1168 33 (th) 32 SB
401 1168 30 (e ) 29 SB
430 1168 33 (m) 32 SB
462 1168 19 (a) 18 SB
480 1168 61 (gne) 60 SB
540 1168 24 (ti) 23 SB
563 1168 19 (c) 18 SB
581 1168 25 ( f) 24 SB
605 1168 52 (luc) 51 SB
656 1168 12 (t) 11 SB
667 1168 40 (ua) 39 SB
706 1168 24 (ti) 23 SB
729 1168 53 (on ) 52 SB
781 1168 56 (spe) 55 SB
836 1168 31 (ct) 30 SB
866 1168 68 (rum) 67 SB
933 1168 11 (,) 10 SB
943 1168 11 ( ) 10 SB
953 1168 82 (whic) 81 SB
1034 1168 32 (h ) 31 SB
1065 1168 12 (i) 11 SB
1076 1168 16 (s) 17 SB
1093 1168 11 ( ) 10 SB
1103 1168 19 (a) 18 SB
1121 1168 16 (s) 16 SB
gr
gs 1096 1919 147 588 CB
1137 1168 14 (-) 14 SB
gr
gs 1096 1919 147 588 CB
222 1218 70 (sum) 69 SB
291 1218 40 (ed) 39 SB
330 1218 23 ( t) 22 SB
352 1218 32 (o ) 31 SB
383 1218 40 (be) 39 SB
422 1218 11 ( ) 10 SB
432 1218 114 (known) 113 SB
545 1218 22 (, ) 21 SB
566 1218 19 (a) 18 SB
584 1218 40 (na) 39 SB
623 1218 33 (ly) 32 SB
655 1218 24 (ti) 23 SB
678 1218 30 (c ) 29 SB
707 1218 35 (fo) 34 SB
741 1218 47 (rm) 46 SB
787 1218 52 (ula) 51 SB
838 1218 27 (s ) 26 SB
864 1218 40 (ha) 39 SB
903 1218 40 (ve) 39 SB
942 1218 32 ( b) 31 SB
973 1218 38 (ee) 37 SB
1010 1218 32 (n ) 31 SB
1041 1218 33 (gi) 32 SB
1073 1218 40 (ve) 39 SB
1112 1218 21 (n) 21 SB
gr
gs 1096 1919 147 588 CB
222 1268 49 (for) 48 SB
270 1268 11 ( ) 11 SB
gr
32 0 0 42 42 0 0 0 42 /Symbol font
gs 1096 1919 147 588 CB
281 1264 23 (c) 22 SB
gr
32 0 0 33 33 0 0 0 30 /Times-Roman /font32 ANSIFont font
gs 1096 1919 147 588 CB
303 1285 15 (e) 15 SB
gr
32 0 0 42 42 0 0 0 38 /Times-Roman /font32 ANSIFont font
gs 1096 1919 147 588 CB
318 1268 11 ( ) 10 SB
328 1268 44 (in ) 43 SB
371 1268 19 (e) 18 SB
389 1268 32 (ss) 33 SB
422 1268 19 (e) 18 SB
440 1268 33 (nt) 32 SB
472 1268 31 (ia) 30 SB
502 1268 12 (l) 11 SB
513 1268 44 (ly ) 43 SB
556 1268 19 (a) 18 SB
574 1268 24 (ll) 23 SB
597 1268 11 ( ) 10 SB
607 1268 40 (pa) 39 SB
646 1268 33 (ra) 32 SB
678 1268 52 (me) 51 SB
729 1268 12 (t) 11 SB
740 1268 33 (er) 32 SB
772 1268 25 ( r) 24 SB
796 1268 40 (eg) 39 SB
835 1268 45 (im) 44 SB
879 1268 19 (e) 18 SB
897 1268 16 (s) 17 SB
914 1268 11 ( ) 10 SB
924 1268 35 (of) 34 SB
958 1268 32 ( p) 31 SB
989 1268 53 (oss) 54 SB
1043 1268 12 (i) 11 SB
1054 1268 33 (bl) 32 SB
1086 1268 19 (e) 19 SB
gr
gs 1096 1919 147 588 CB
222 1318 63 (phy) 62 SB
284 1318 16 (s) 17 SB
301 1318 12 (i) 11 SB
312 1318 19 (c) 18 SB
330 1318 31 (al) 30 SB
360 1318 11 ( ) 10 SB
370 1318 33 (re) 32 SB
402 1318 31 (le) 30 SB
432 1318 40 (va) 39 SB
471 1318 40 (nc) 39 SB
510 1318 19 (e) 18 SB
528 1318 25 ( [) 24 SB
552 1318 46 (1].) 45 SB
597 1318 11 ( ) 10 SB
607 1318 100 (Howe) 99 SB
706 1318 40 (ve) 39 SB
745 1318 14 (r) 12 SB
757 1318 22 (, ) 21 SB
778 1318 19 (c) 18 SB
796 1318 56 (onf) 55 SB
851 1318 59 (irm) 58 SB
909 1318 19 (a) 18 SB
927 1318 24 (ti) 23 SB
950 1318 53 (on ) 52 SB
1002 1318 35 (of) 34 SB
1036 1318 23 ( t) 22 SB
1058 1318 40 (he) 39 SB
gr
gs 1096 1919 147 588 CB
222 1368 33 (th) 32 SB
254 1368 40 (eo) 39 SB
293 1368 33 (re) 32 SB
325 1368 24 (ti) 23 SB
348 1368 38 (ca) 37 SB
385 1368 12 (l) 11 SB
396 1368 44 ( m) 43 SB
439 1368 42 (od) 41 SB
480 1368 31 (el) 30 SB
510 1368 48 (s h) 47 SB
557 1368 46 (as ) 45 SB
602 1368 54 (not) 53 SB
655 1368 11 ( ) 10 SB
665 1368 40 (be) 39 SB
704 1368 40 (en) 39 SB
743 1368 32 ( d) 31 SB
774 1368 61 (one) 60 SB
834 1368 11 ( ) 10 SB
844 1368 40 (ex) 39 SB
883 1368 40 (pe) 39 SB
922 1368 59 (rim) 58 SB
980 1368 19 (e) 18 SB
998 1368 33 (nt) 32 SB
1030 1368 31 (al) 30 SB
1060 1368 33 (ly) 29 SB
1089 1368 22 (. ) 21 SB
gr
gs 1096 1919 147 588 CB
222 1441 62 (On ) 61 SB
283 1441 12 (t) 11 SB
294 1441 40 (he) 39 SB
333 1441 11 ( ) 10 SB
343 1441 82 (Com) 81 SB
424 1441 40 (pa) 39 SB
463 1441 19 (c) 18 SB
481 1441 23 (t ) 21 SB
502 1441 107 (Aubur) 106 SB
608 1441 32 (n ) 31 SB
639 1441 26 (T) 22 SB
661 1441 35 (or) 34 SB
695 1441 16 (s) 17 SB
712 1441 19 (a) 18 SB
730 1441 12 (t) 11 SB
741 1441 67 (ron ) 65 SB
806 1441 72 (\(CA) 67 SB
873 1441 26 (T) 25 SB
898 1441 25 (\),) 24 SB
922 1441 11 ( ) 10 SB
932 1441 19 (a) 18 SB
950 1441 11 ( ) 10 SB
960 1441 19 (\223) 18 SB
978 1441 47 (ste) 46 SB
1024 1441 24 (ll) 23 SB
1047 1441 19 (a) 18 SB
1065 1441 33 (ra) 32 SB
1097 1441 47 (tor) 46 SB
gr
gs 1096 1919 147 588 CB
222 1491 33 (di) 32 SB
254 1491 61 (ode) 60 SB
314 1491 19 (\224) 18 SB
332 1491 76 ( syst) 75 SB
407 1491 52 (em) 51 SB
458 1491 11 ( ) 10 SB
468 1491 49 ([2]) 48 SB
516 1491 60 ( wa) 59 SB
575 1491 27 (s ) 26 SB
601 1491 40 (ad) 39 SB
640 1491 54 (opt) 53 SB
693 1491 40 (ed) 39 SB
732 1491 30 ( a) 29 SB
761 1491 27 (s ) 26 SB
787 1491 30 (a ) 29 SB
816 1491 12 (t) 11 SB
827 1491 54 (ool) 53 SB
880 1491 23 ( t) 22 SB
902 1491 32 (o ) 31 SB
933 1491 56 (pro) 55 SB
988 1491 54 (vid) 53 SB
1041 1491 30 (e ) 29 SB
1070 1491 19 (a) 18 SB
gr
gs 1096 1919 147 588 CB
222 1541 47 (ste) 46 SB
268 1541 19 (a) 18 SB
286 1541 84 (dy-st) 83 SB
369 1541 31 (at) 30 SB
399 1541 19 (e) 18 SB
417 1541 11 ( ) 10 SB
427 1541 31 (el) 30 SB
457 1541 19 (e) 18 SB
475 1541 31 (ct) 30 SB
505 1541 56 (ron) 55 SB
560 1541 11 ( ) 10 SB
570 1541 16 (s) 17 SB
587 1541 21 (o) 20 SB
607 1541 54 (urc) 53 SB
660 1541 19 (e) 18 SB
678 1541 22 (. ) 21 SB
699 1541 30 (A) 27 SB
726 1541 11 ( ) 10 SB
736 1541 47 (set) 46 SB
782 1541 11 ( ) 10 SB
792 1541 46 (of ) 45 SB
837 1541 21 (h) 20 SB
857 1541 31 (el) 30 SB
887 1541 31 (ic) 30 SB
917 1541 19 (a) 18 SB
935 1541 23 (l ) 21 SB
956 1541 40 (co) 39 SB
995 1541 24 (il) 23 SB
1018 1541 16 (s) 17 SB
1035 1541 11 ( ) 9 SB
1044 1541 65 (was) 65 SB
gr
gs 1096 1919 147 588 CB
222 1591 33 (in) 32 SB
254 1591 16 (s) 17 SB
271 1591 12 (t) 11 SB
282 1591 19 (a) 18 SB
300 1591 24 (ll) 23 SB
323 1591 40 (ed) 39 SB
362 1591 23 ( i) 22 SB
384 1591 70 (nsid) 69 SB
453 1591 30 (e ) 29 SB
482 1591 12 (t) 11 SB
493 1591 40 (he) 39 SB
532 1591 39 ( C) 38 SB
570 1591 30 (A) 25 SB
595 1591 26 (T) 25 SB
620 1591 11 ( ) 10 SB
630 1591 40 (va) 39 SB
669 1591 19 (c) 18 SB
687 1591 75 (uum) 74 SB
761 1591 11 ( ) 10 SB
771 1591 40 (ve) 39 SB
810 1591 16 (s) 17 SB
827 1591 35 (se) 34 SB
861 1591 23 (l ) 22 SB
883 1591 12 (t) 11 SB
894 1591 32 (o ) 31 SB
925 1591 33 (cr) 32 SB
957 1591 38 (ea) 37 SB
994 1591 12 (t) 11 SB
1005 1591 30 (e ) 29 SB
1034 1591 33 (re) 32 SB
1066 1591 37 (so) 37 SB
gr
gs 1096 1919 147 588 CB
1103 1591 14 (-) 14 SB
gr
gs 1096 1919 147 588 CB
222 1641 40 (na) 39 SB
261 1641 33 (nt) 32 SB
293 1641 44 ( m) 43 SB
336 1641 19 (a) 18 SB
354 1641 61 (gne) 60 SB
414 1641 12 (t) 11 SB
425 1641 31 (ic) 30 SB
455 1641 11 ( ) 10 SB
465 1641 40 (pe) 39 SB
504 1641 47 (rtu) 46 SB
550 1641 54 (rba) 53 SB
603 1641 24 (ti) 23 SB
626 1641 53 (on ) 52 SB
678 1641 26 (fi) 25 SB
703 1641 19 (e) 18 SB
721 1641 60 (lds.) 59 SB
780 1641 11 ( ) 10 SB
790 1641 26 (T) 25 SB
815 1641 40 (he) 39 SB
854 1641 11 ( ) 10 SB
864 1641 52 (ma) 51 SB
915 1641 12 (i) 11 SB
926 1641 32 (n ) 31 SB
957 1641 19 (c) 18 SB
975 1641 54 (om) 53 SB
1028 1641 42 (po) 42 SB
gr
gs 1096 1919 147 588 CB
1070 1641 14 (-) 14 SB
gr
gs 1096 1919 147 588 CB
222 1691 40 (ne) 39 SB
261 1691 33 (nt) 32 SB
293 1691 16 (s) 17 SB
310 1691 11 ( ) 10 SB
320 1691 35 (of) 34 SB
354 1691 23 ( t) 22 SB
376 1691 40 (he) 39 SB
415 1691 11 ( ) 10 SB
425 1691 40 (pe) 39 SB
464 1691 47 (rtu) 46 SB
510 1691 54 (rba) 53 SB
563 1691 24 (ti) 23 SB
586 1691 53 (on ) 52 SB
638 1691 33 (m) 32 SB
670 1691 19 (a) 18 SB
688 1691 61 (gne) 60 SB
748 1691 24 (ti) 23 SB
771 1691 19 (c) 18 SB
789 1691 25 ( f) 24 SB
813 1691 31 (ie) 30 SB
843 1691 33 (ld) 32 SB
875 1691 67 ( spe) 66 SB
941 1691 19 (c) 18 SB
959 1691 26 (tr) 25 SB
984 1691 54 (um) 53 SB
1037 1691 30 ( a) 29 SB
1066 1691 33 (re) 32 SB
gr
32 0 0 42 42 0 0 0 39 /Times-Italic /font31 ANSIFont font
gs 1096 1919 147 588 CB
222 1740 21 (n) 21 SB
gr
32 0 0 42 42 0 0 0 38 /Times-Roman /font32 ANSIFont font
gs 1096 1919 147 588 CB
243 1741 12 (/) 11 SB
gr
32 0 0 42 42 0 0 0 39 /Times-Italic /font31 ANSIFont font
gs 1096 1919 147 588 CB
254 1740 30 (m) 30 SB
gr
32 0 0 42 42 0 0 0 38 /Times-Roman /font32 ANSIFont font
gs 1096 1919 147 588 CB
284 1741 35 ( =) 34 SB
318 1741 11 ( ) 10 SB
328 1741 33 (1/) 32 SB
360 1741 32 (3 ) 31 SB
391 1741 40 (an) 39 SB
430 1741 32 (d ) 31 SB
gr
32 0 0 42 42 0 0 0 39 /Times-Italic /font31 ANSIFont font
gs 1096 1919 147 588 CB
461 1740 21 (n) 21 SB
gr
32 0 0 42 42 0 0 0 38 /Times-Roman /font32 ANSIFont font
gs 1096 1919 147 588 CB
482 1741 12 (/) 12 SB
gr
32 0 0 42 42 0 0 0 39 /Times-Italic /font31 ANSIFont font
gs 1096 1919 147 588 CB
494 1740 30 (m) 30 SB
gr
32 0 0 42 42 0 0 0 38 /Times-Roman /font32 ANSIFont font
gs 1096 1919 147 588 CB
524 1741 11 ( ) 10 SB
534 1741 24 (=) 23 SB
557 1741 11 ( ) 10 SB
567 1741 54 (1/4) 53 SB
620 1741 44 ( m) 43 SB
663 1741 42 (od) 41 SB
704 1741 46 (es,) 45 SB
749 1741 11 ( ) 10 SB
759 1741 82 (whic) 81 SB
840 1741 32 (h ) 31 SB
871 1741 42 (wi) 41 SB
912 1741 24 (ll) 23 SB
935 1741 30 ( c) 29 SB
964 1741 33 (re) 32 SB
996 1741 19 (a) 18 SB
1014 1741 31 (te) 30 SB
1044 1741 11 ( ) 10 SB
1054 1741 33 (m) 32 SB
1086 1741 40 (ag) 39 SB
gr
gs 1096 1919 147 588 CB
1125 1741 14 (-) 14 SB
gr
gs 1096 1919 147 588 CB
222 1791 40 (ne) 39 SB
261 1791 24 (ti) 23 SB
284 1791 19 (c) 18 SB
302 1791 23 ( i) 22 SB
324 1791 47 (sla) 46 SB
370 1791 69 (nds ) 68 SB
438 1791 53 (on ) 52 SB
490 1791 12 (t) 11 SB
501 1791 40 (he) 39 SB
540 1791 25 ( r) 24 SB
564 1791 31 (at) 30 SB
594 1791 33 (io) 32 SB
626 1791 40 (na) 39 SB
665 1791 23 (l ) 22 SB
687 1791 84 (surfa) 83 SB
770 1791 19 (c) 18 SB
788 1791 19 (e) 18 SB
806 1791 16 (s) 17 SB
823 1791 11 ( ) 10 SB
833 1791 35 (of) 34 SB
867 1791 25 ( r) 24 SB
891 1791 52 (ota) 51 SB
942 1791 12 (t) 11 SB
953 1791 54 (ion) 53 SB
1006 1791 31 (al) 30 SB
1036 1791 11 ( ) 10 SB
1046 1791 45 (tra) 44 SB
1090 1791 37 (ns) 37 SB
gr
gs 1096 1919 147 588 CB
1127 1791 14 (-) 14 SB
gr
gs 1096 1919 147 588 CB
222 1841 49 (for) 48 SB
270 1841 44 (m ) 43 SB
313 1841 19 (a) 18 SB
331 1841 23 (t ) 22 SB
gr
14 2 353 1867 B
1 F
n
32 0 0 42 42 0 0 0 42 /Symbol font
gs 1096 1919 147 588 CB
353 1837 14 (i) 14 SB
gr
32 0 0 42 42 0 0 0 38 /Times-Roman /font32 ANSIFont font
gs 1096 1919 147 588 CB
367 1841 11 ( ) 10 SB
377 1841 24 (=) 23 SB
400 1841 33 (1/) 32 SB
432 1841 32 (3 ) 31 SB
463 1841 40 (an) 39 SB
502 1841 32 (d ) 31 SB
533 1841 54 (1/4) 53 SB
586 1841 22 (. ) 21 SB
607 1841 26 (T) 25 SB
632 1841 40 (he) 39 SB
671 1841 11 ( ) 10 SB
681 1841 40 (va) 39 SB
720 1841 19 (c) 18 SB
738 1841 75 (uum) 74 SB
812 1841 11 ( ) 10 SB
822 1841 33 (m) 32 SB
854 1841 40 (ag) 39 SB
893 1841 40 (ne) 39 SB
932 1841 24 (ti) 23 SB
955 1841 30 (c ) 29 SB
984 1841 14 (f) 13 SB
997 1841 31 (ie) 30 SB
1027 1841 33 (ld) 32 SB
gr
gs 1096 1919 147 588 CB
222 1891 19 (c) 18 SB
240 1891 68 (onfi) 67 SB
307 1891 56 (gur) 55 SB
362 1891 31 (at) 30 SB
392 1891 33 (io) 32 SB
424 1891 32 (n ) 31 SB
455 1891 76 (was ) 75 SB
530 1891 19 (c) 18 SB
548 1891 77 (hose) 76 SB
624 1891 32 (n ) 31 SB
655 1891 16 (s) 17 SB
672 1891 21 (o) 20 SB
692 1891 23 ( t) 22 SB
714 1891 40 (ha) 39 SB
753 1891 23 (t ) 22 SB
775 1891 12 (t) 11 SB
786 1891 40 (he) 39 SB
825 1891 30 ( a) 29 SB
854 1891 40 (ve) 39 SB
893 1891 33 (ra) 32 SB
925 1891 40 (ge) 39 SB
964 1891 11 ( ) 10 SB
974 1891 33 (m) 32 SB
1006 1891 54 (ino) 53 SB
1059 1891 25 (r ) 24 SB
1083 1891 33 (ra) 32 SB
gr
gs 1096 1919 147 588 CB
1115 1891 14 (-) 14 SB
gr
gs 1096 1919 147 588 CB
222 1941 33 (di) 32 SB
254 1941 62 (us f) 61 SB
315 1941 46 (or ) 45 SB
360 1941 33 (th) 32 SB
392 1941 30 (e ) 29 SB
gr
32 0 0 42 42 0 0 0 39 /Times-Italic /font31 ANSIFont font
gs 1096 1919 147 588 CB
421 1940 21 (n) 21 SB
gr
32 0 0 42 42 0 0 0 38 /Times-Roman /font32 ANSIFont font
gs 1096 1919 147 588 CB
442 1941 12 (/) 11 SB
gr
32 0 0 42 42 0 0 0 39 /Times-Italic /font31 ANSIFont font
gs 1096 1919 147 588 CB
453 1940 30 (m) 30 SB
gr
32 0 0 42 42 0 0 0 38 /Times-Roman /font32 ANSIFont font
gs 1096 1919 147 588 CB
483 1941 11 ( ) 10 SB
493 1941 35 (= ) 34 SB
527 1941 33 (1/) 32 SB
559 1941 32 (4 ) 31 SB
590 1941 84 (surfa) 83 SB
673 1941 19 (c) 18 SB
691 1941 19 (e) 18 SB
709 1941 23 ( i) 22 SB
731 1941 46 (s a) 45 SB
776 1941 12 (t) 11 SB
787 1941 30 ( a) 29 SB
816 1941 42 (bo) 41 SB
857 1941 44 (ut ) 43 SB
900 1941 32 (5 ) 31 SB
931 1941 19 (c) 18 SB
949 1941 33 (m) 32 SB
981 1941 22 (, ) 21 SB
1002 1941 19 (a) 18 SB
1020 1941 53 (nd ) 52 SB
1072 1941 12 (t) 11 SB
1083 1941 40 (he) 39 SB
gr
gs 1096 1919 147 588 CB
222 1991 19 (a) 18 SB
240 1991 40 (ve) 39 SB
279 1991 33 (ra) 32 SB
311 1991 40 (ge) 39 SB
350 1991 44 ( m) 43 SB
393 1991 12 (i) 11 SB
404 1991 67 (nor ) 66 SB
470 1991 33 (ra) 32 SB
502 1991 33 (di) 32 SB
534 1991 48 (us ) 47 SB
581 1991 46 (of ) 45 SB
626 1991 33 (th) 32 SB
658 1991 30 (e ) 29 SB
gr
32 0 0 42 42 0 0 0 39 /Times-Italic /font31 ANSIFont font
gs 1096 1919 147 588 CB
687 1990 21 (n) 21 SB
gr
32 0 0 42 42 0 0 0 38 /Times-Roman /font32 ANSIFont font
gs 1096 1919 147 588 CB
708 1991 12 (/) 11 SB
gr
32 0 0 42 42 0 0 0 39 /Times-Italic /font31 ANSIFont font
gs 1096 1919 147 588 CB
719 1990 30 (m) 30 SB
gr
32 0 0 42 42 0 0 0 38 /Times-Roman /font32 ANSIFont font
gs 1096 1919 147 588 CB
749 1991 11 ( ) 10 SB
759 1991 35 (= ) 34 SB
793 1991 33 (1/) 32 SB
825 1991 32 (3 ) 31 SB
856 1991 84 (surfa) 83 SB
939 1991 19 (c) 18 SB
957 1991 19 (e) 18 SB
975 1991 23 ( i) 22 SB
997 1991 46 (s a) 45 SB
1042 1991 12 (t) 11 SB
1053 1991 30 ( a) 29 SB
1082 1991 42 (bo) 41 SB
1123 1991 33 (ut) 33 SB
gr
gs 1096 1919 147 588 CB
222 2041 32 (8 ) 31 SB
253 2041 19 (c) 18 SB
271 2041 44 (m.) 43 SB
314 2041 11 ( ) 10 SB
324 2041 40 (W) 38 SB
362 2041 12 (i) 11 SB
373 2041 33 (th) 32 SB
405 2041 62 ( suf) 61 SB
466 2041 14 (f) 13 SB
479 2041 31 (ic) 30 SB
509 2041 31 (ie) 30 SB
539 2041 33 (nt) 32 SB
571 2041 11 ( ) 10 SB
581 2041 40 (ex) 39 SB
620 2041 31 (te) 30 SB
650 2041 54 (rna) 53 SB
703 2041 12 (l) 11 SB
714 2041 44 (ly ) 43 SB
757 2041 19 (a) 18 SB
775 2041 42 (pp) 41 SB
816 2041 24 (li) 23 SB
839 2041 40 (ed) 39 SB
878 2041 11 ( ) 10 SB
888 2041 33 (re) 32 SB
920 2041 16 (s) 17 SB
937 2041 21 (o) 20 SB
957 2041 40 (na) 39 SB
996 2041 44 (nt ) 43 SB
1039 2041 14 (f) 13 SB
1052 2041 31 (ie) 30 SB
1082 2041 33 (ld) 32 SB
1114 2041 16 (s) 17 SB
1131 2041 11 (,) 10 SB
gr
gs 1096 1919 147 588 CB
222 2091 33 (th) 32 SB
254 2091 30 (e ) 29 SB
283 2091 12 (i) 11 SB
294 2091 16 (s) 17 SB
311 2091 12 (l) 11 SB
322 2091 19 (a) 18 SB
340 2091 69 (nds ) 68 SB
408 2091 54 (wil) 53 SB
461 2091 23 (l ) 22 SB
483 2091 61 (ove) 60 SB
543 2091 26 (rl) 25 SB
568 2091 19 (a) 18 SB
586 2091 32 (p ) 31 SB
617 2091 40 (an) 39 SB
656 2091 32 (d ) 31 SB
687 2091 33 (cr) 32 SB
719 2091 38 (ea) 37 SB
756 2091 12 (t) 11 SB
767 2091 30 (e ) 29 SB
796 2091 49 (sto) 48 SB
844 2091 40 (ch) 39 SB
883 2091 47 (ast) 46 SB
929 2091 31 (ic) 30 SB
959 2091 11 ( ) 10 SB
969 2091 33 (re) 32 SB
1001 2091 54 (gio) 53 SB
1054 2091 60 (ns i) 59 SB
1113 2091 21 (n) 21 SB
gr
gs 1096 1919 147 588 CB
222 2141 33 (th) 32 SB
254 2141 30 (e ) 29 SB
283 2141 33 (m) 32 SB
315 2141 19 (a) 18 SB
333 2141 61 (gne) 60 SB
393 2141 24 (ti) 23 SB
416 2141 19 (c) 18 SB
434 2141 25 ( f) 24 SB
458 2141 31 (ie) 30 SB
488 2141 33 (ld) 32 SB
520 2141 16 (s) 17 SB
537 2141 11 ( ) 10 SB
547 2141 63 (whi) 62 SB
609 2141 19 (c) 18 SB
627 2141 32 (h ) 31 SB
658 2141 40 (en) 39 SB
697 2141 40 (ab) 39 SB
736 2141 31 (le) 30 SB
766 2141 11 ( ) 10 SB
776 2141 16 (s) 17 SB
793 2141 12 (t) 11 SB
804 2141 63 (udy) 62 SB
866 2141 32 ( o) 31 SB
897 2141 25 (f ) 24 SB
921 2141 52 (the) 51 SB
972 2141 11 ( ) 10 SB
982 2141 68 (stoc) 67 SB
1049 2141 40 (ha) 39 SB
1088 2141 40 (sti) 39 SB
1127 2141 19 (c) 18 SB
gr
gs 1096 1919 147 588 CB
222 2191 26 (tr) 25 SB
247 2191 40 (an) 39 SB
286 2191 16 (s) 17 SB
303 2191 21 (p) 20 SB
323 2191 47 (ort) 46 SB
369 2191 32 ( o) 31 SB
400 2191 25 (f ) 24 SB
424 2191 52 (the) 51 SB
475 2191 11 ( ) 10 SB
485 2191 19 (e) 18 SB
503 2191 31 (le) 30 SB
533 2191 19 (c) 18 SB
551 2191 47 (tro) 46 SB
597 2191 48 (ns.) 48 SB
gr
gs 1096 1919 147 588 CB
222 2263 26 (T) 25 SB
247 2263 40 (he) 39 SB
286 2263 69 ( sub) 68 SB
354 2263 16 (s) 17 SB
371 2263 19 (e) 18 SB
389 2263 42 (qu) 41 SB
430 2263 40 (en) 39 SB
469 2263 23 (t ) 22 SB
491 2263 33 (m) 32 SB
523 2263 45 (oti) 44 SB
567 2263 53 (on ) 52 SB
619 2263 35 (of) 34 SB
653 2263 23 ( t) 22 SB
675 2263 40 (he) 39 SB
714 2263 11 ( ) 10 SB
724 2263 31 (el) 30 SB
754 2263 19 (e) 18 SB
772 2263 31 (ct) 30 SB
802 2263 56 (ron) 55 SB
857 2263 16 (s) 17 SB
874 2263 11 ( ) 10 SB
884 2263 12 (l) 11 SB
895 2263 38 (ea) 37 SB
932 2263 33 (vi) 32 SB
964 2263 53 (ng ) 52 SB
1016 2263 12 (t) 11 SB
1027 2263 40 (he) 39 SB
gr
gs 1096 1919 147 588 CB
222 2313 56 (spa) 55 SB
277 2313 38 (ce) 37 SB
314 2313 11 ( ) 10 SB
324 2313 19 (c) 18 SB
342 2313 40 (ha) 39 SB
381 2313 14 (r) 13 SB
394 2313 40 (ge) 39 SB
433 2313 67 ( she) 66 SB
499 2313 19 (a) 18 SB
517 2313 33 (th) 32 SB
549 2313 32 ( n) 31 SB
580 2313 38 (ea) 37 SB
617 2313 25 (r ) 24 SB
641 2313 12 (t) 11 SB
652 2313 40 (he) 39 SB
691 2313 32 ( h) 31 SB
722 2313 44 (ot ) 43 SB
765 2313 26 (fi) 25 SB
790 2313 31 (la) 30 SB
820 2313 33 (m) 32 SB
852 2313 19 (e) 18 SB
870 2313 44 (nt ) 43 SB
913 2313 12 (i) 11 SB
924 2313 16 (s) 17 SB
941 2313 11 ( ) 10 SB
951 2313 33 (di) 32 SB
983 2313 14 (f) 13 SB
996 2313 63 (fusi) 62 SB
1058 2313 40 (ve) 39 SB
gr
gs 1096 1919 147 588 CB
222 2363 19 (a) 18 SB
240 2363 54 (lon) 53 SB
293 2363 32 (g ) 31 SB
324 2363 40 (an) 39 SB
363 2363 32 (d ) 31 SB
394 2363 38 (ac) 37 SB
431 2363 35 (ro) 34 SB
465 2363 16 (s) 17 SB
482 2363 27 (s ) 26 SB
508 2363 33 (th) 32 SB
540 2363 30 (e ) 29 SB
569 2363 33 (m) 32 SB
601 2363 19 (a) 18 SB
619 2363 61 (gne) 60 SB
679 2363 24 (ti) 23 SB
702 2363 19 (c) 18 SB
720 2363 25 ( f) 24 SB
744 2363 31 (ie) 30 SB
774 2363 33 (ld) 32 SB
806 2363 22 (, ) 21 SB
827 2363 33 (m) 32 SB
859 2363 19 (a) 18 SB
877 2363 45 (inl) 44 SB
921 2363 32 (y ) 31 SB
952 2363 33 (in) 32 SB
984 2363 54 (vol) 53 SB
1037 2363 54 (vin) 53 SB
1090 2363 21 (g) 21 SB
gr
gs 1096 1042 1152 597 CB
1227 672 19 (c) 18 SB
1245 672 45 (oll) 44 SB
1289 672 40 (isi) 39 SB
1328 672 69 (ons ) 68 SB
1396 672 46 (of ) 45 SB
1441 672 19 (e) 18 SB
1459 672 31 (le) 30 SB
1489 672 19 (c) 18 SB
1507 672 47 (tro) 46 SB
1553 672 90 (ns wi) 89 SB
1642 672 33 (th) 32 SB
1674 672 32 ( n) 31 SB
1705 672 40 (eu) 39 SB
1744 672 45 (tra) 44 SB
1788 672 12 (l) 11 SB
1799 672 32 ( p) 31 SB
1830 672 33 (ar) 32 SB
1862 672 24 (ti) 23 SB
1885 672 31 (cl) 30 SB
1915 672 19 (e) 18 SB
1933 672 16 (s) 17 SB
1950 672 11 ( ) 10 SB
1960 672 35 (of) 34 SB
1994 672 25 ( r) 24 SB
2018 672 47 (esi) 46 SB
2064 672 61 (dua) 60 SB
2124 672 12 (l) 12 SB
gr
gs 1096 1042 1152 597 CB
1227 722 40 (ga) 39 SB
1266 722 38 (s. ) 37 SB
1303 722 35 (In) 34 SB
1337 722 30 ( c) 29 SB
1366 722 33 (yl) 32 SB
1398 722 33 (in) 32 SB
1430 722 47 (dri) 46 SB
1476 722 38 (ca) 37 SB
1513 722 12 (l) 11 SB
1524 722 30 ( a) 29 SB
1553 722 42 (pp) 41 SB
1594 722 68 (roxi) 67 SB
1661 722 33 (m) 32 SB
1693 722 31 (at) 30 SB
1723 722 33 (io) 32 SB
1755 722 32 (n,) 31 SB
1786 722 23 ( t) 22 SB
1808 722 40 (he) 39 SB
1847 722 11 ( ) 10 SB
1857 722 31 (el) 30 SB
1887 722 19 (e) 18 SB
1905 722 31 (ct) 30 SB
1935 722 56 (ron) 55 SB
1990 722 25 ( f) 24 SB
2014 722 54 (lux) 53 SB
gr
gs 1096 1042 1152 597 CB
1227 772 19 (a) 18 SB
1245 772 33 (cr) 32 SB
1277 772 53 (oss) 54 SB
1331 772 11 ( ) 10 SB
1341 772 12 (t) 11 SB
1352 772 40 (he) 39 SB
1391 772 44 ( m) 43 SB
1434 772 19 (a) 18 SB
1452 772 61 (gne) 60 SB
1512 772 12 (t) 11 SB
1523 772 31 (ic) 30 SB
1553 772 11 ( ) 10 SB
1563 772 16 (s) 17 SB
1580 772 21 (u) 20 SB
1600 772 47 (rfa) 46 SB
1646 772 19 (c) 18 SB
1664 772 46 (es ) 45 SB
1709 772 39 (is ) 38 SB
1747 772 33 (gi) 32 SB
1779 772 40 (ve) 39 SB
1818 772 32 (n ) 31 SB
1849 772 42 (by) 42 SB
gr
32 0 0 42 42 0 0 0 42 /Symbol font
gs 1096 1042 1152 597 CB
1310 869 25 (G) 25 SB
gr
32 0 0 42 42 0 0 0 38 /Times-Roman /font32 ANSIFont font
gs 1096 1042 1152 597 CB
1335 873 11 ( ) 11 SB
gr
32 0 0 42 42 0 0 0 42 /Symbol font
gs 1096 1042 1152 597 CB
1346 869 23 (=) 22 SB
gr
32 0 0 42 42 0 0 0 38 /Times-Roman /font32 ANSIFont font
gs 1096 1042 1152 597 CB
1368 873 11 ( ) 11 SB
gr
32 0 0 42 42 0 0 0 39 /Times-Italic /font31 ANSIFont font
gs 1096 1042 1152 597 CB
1414 837 14 (I) 14 SB
gr
32 0 0 33 33 0 0 0 30 /Times-Roman /font32 ANSIFont font
gs 1096 1042 1152 597 CB
1428 858 15 (e) 14 SB
1442 858 26 (m) 26 SB
gr
32 0 0 42 42 0 0 0 38 /Times-Roman /font32 ANSIFont font
gs 1096 1042 1152 597 CB
1381 909 21 (4) 21 SB
gr
32 0 0 42 42 0 0 0 42 /Symbol font
gs 1096 1042 1152 597 CB
1402 905 23 (p) 23 SB
gr
32 0 0 33 33 0 0 0 30 /Times-Roman /font32 ANSIFont font
gs 1096 1042 1152 597 CB
1425 899 17 (2) 16 SB
gr
32 0 0 42 42 0 0 0 39 /Times-Italic /font31 ANSIFont font
gs 1096 1042 1152 597 CB
1441 908 26 (R) 25 SB
gr
gs 1096 1042 1152 597 CB
1466 908 16 (r) 17 SB
gr
gs 1096 1042 1152 597 CB
1483 908 19 (e) 18 SB
gr
124 1 1379 896 B
1 F
n
32 0 0 42 42 0 0 0 38 /Times-Roman /font32 ANSIFont font
gs 1096 1042 1152 597 CB
1503 873 22 ( ,) 21 SB
gr
gs 1096 1042 1152 597 CB
1227 995 70 (whe) 69 SB
1296 995 33 (re) 32 SB
1328 995 11 ( ) 10 SB
gr
32 0 0 42 42 0 0 0 39 /Times-Italic /font31 ANSIFont font
gs 1096 1042 1152 597 CB
1338 994 16 (r) 17 SB
gr
32 0 0 42 42 0 0 0 38 /Times-Roman /font32 ANSIFont font
gs 1096 1042 1152 597 CB
1355 995 11 ( ) 10 SB
1365 995 12 (i) 11 SB
1376 995 16 (s) 17 SB
1393 995 11 ( ) 10 SB
1403 995 12 (t) 11 SB
1414 995 40 (he) 39 SB
1453 995 44 ( m) 43 SB
1496 995 12 (i) 11 SB
1507 995 67 (nor ) 66 SB
1573 995 33 (ra) 32 SB
1605 995 33 (di) 32 SB
1637 995 48 (us,) 47 SB
1684 995 11 ( ) 11 SB
gr
32 0 0 42 42 0 0 0 39 /Times-Italic /font31 ANSIFont font
gs 1096 1042 1152 597 CB
1695 994 26 (R) 25 SB
gr
32 0 0 42 42 0 0 0 38 /Times-Roman /font32 ANSIFont font
gs 1096 1042 1152 597 CB
1720 995 11 ( ) 10 SB
1730 995 39 (is ) 38 SB
1768 995 33 (th) 32 SB
1800 995 30 (e ) 29 SB
1829 995 33 (m) 32 SB
1861 995 19 (a) 18 SB
1879 995 47 (jor) 46 SB
1925 995 25 ( r) 24 SB
1949 995 40 (ad) 39 SB
1988 995 60 (ius,) 59 SB
2047 995 11 ( ) 10 SB
2057 995 40 (an) 39 SB
2096 995 21 (d) 21 SB
gr
32 0 0 42 42 0 0 0 39 /Times-Italic /font31 ANSIFont font
gs 1096 1042 1152 597 CB
1227 1044 14 (I) 14 SB
gr
32 0 0 33 33 0 0 0 30 /Times-Roman /font32 ANSIFont font
gs 1096 1042 1152 597 CB
1241 1062 41 (em) 40 SB
gr
32 0 0 42 42 0 0 0 38 /Times-Roman /font32 ANSIFont font
gs 1096 1042 1152 597 CB
1281 1045 23 ( i) 22 SB
1303 1045 39 (s t) 38 SB
1341 1045 40 (he) 39 SB
1380 1045 11 ( ) 10 SB
1390 1045 52 (em) 51 SB
1441 1045 12 (i) 11 SB
1452 1045 16 (s) 17 SB
1469 1045 28 (si) 27 SB
1496 1045 53 (on ) 52 SB
1548 1045 19 (c) 18 SB
1566 1045 68 (urre) 67 SB
1633 1045 33 (nt) 32 SB
1665 1045 11 ( ) 10 SB
1675 1045 46 (of ) 45 SB
1720 1045 12 (t) 11 SB
1731 1045 40 (he) 39 SB
1770 1045 11 ( ) 10 SB
1780 1045 54 (hot) 53 SB
1833 1045 11 ( ) 10 SB
1843 1045 26 (fi) 25 SB
1868 1045 31 (la) 30 SB
1898 1045 33 (m) 32 SB
1930 1045 40 (en) 39 SB
1969 1045 23 (t.) 22 SB
1991 1045 11 ( ) 10 SB
2001 1045 26 (T) 25 SB
2026 1045 40 (he) 39 SB
2065 1045 11 ( ) 10 SB
2075 1045 47 (dif) 46 SB
gr
gs 1096 1042 1152 597 CB
2121 1045 14 (-) 14 SB
gr
gs 1096 1042 1152 597 CB
1227 1095 63 (fusi) 62 SB
1289 1095 45 (vit) 44 SB
1333 1095 32 (y ) 31 SB
1364 1095 46 (of ) 45 SB
1409 1095 12 (t) 11 SB
1420 1095 40 (he) 39 SB
1459 1095 30 ( e) 29 SB
1488 1095 12 (l) 11 SB
1499 1095 38 (ec) 37 SB
1536 1095 12 (t) 11 SB
1547 1095 83 (rons ) 82 SB
1629 1095 39 (is ) 38 SB
1667 1095 40 (de) 39 SB
1706 1095 31 (te) 30 SB
1736 1095 47 (rm) 46 SB
1782 1095 12 (i) 11 SB
1793 1095 40 (ne) 39 SB
1832 1095 32 (d ) 31 SB
1863 1095 53 (by ) 52 SB
1915 1095 49 ([3]) 48 SB
gr
32 0 0 42 42 0 0 0 39 /Times-Italic /font31 ANSIFont font
gs 1096 1042 1152 597 CB
1310 1196 30 (D) 30 SB
gr
32 0 0 33 33 0 0 0 33 /Symbol font
gs 1096 1042 1152 597 CB
1340 1214 22 (^) 22 SB
gr
32 0 0 42 42 0 0 0 38 /Times-Roman /font32 ANSIFont font
gs 1096 1042 1152 597 CB
1362 1197 11 ( ) 10 SB
gr
32 0 0 42 42 0 0 0 42 /Symbol font
gs 1096 1042 1152 597 CB
1372 1193 23 (=) 23 SB
gr
32 0 0 42 42 0 0 0 38 /Times-Roman /font32 ANSIFont font
gs 1096 1042 1152 597 CB
1395 1197 11 ( ) 10 SB
gr
32 0 0 42 42 0 0 0 42 /Symbol font
gs 1096 1042 1152 597 CB
1547 1165 25 (G) 25 SB
gr
32 0 0 42 42 0 0 0 38 /Times-Roman /font32 ANSIFont font
gs 1096 1042 1152 597 CB
1408 1227 8 (|) 8 SB
gr
32 0 0 42 42 0 0 0 42 /Symbol font
gs 1096 1042 1152 597 CB
1416 1223 14 (\() 14 SB
gr
gs 1096 1042 1152 597 CB
1430 1223 30 (\321) 29 SB
gr
32 0 0 42 42 0 0 0 39 /Times-Italic /font31 ANSIFont font
gs 1096 1042 1152 597 CB
1459 1226 21 (n) 21 SB
gr
32 0 0 33 33 0 0 0 30 /Times-Roman /font32 ANSIFont font
gs 1096 1042 1152 597 CB
1480 1247 15 (e) 15 SB
gr
32 0 0 42 42 0 0 0 38 /Times-Roman /font32 ANSIFont font
gs 1096 1042 1152 597 CB
1495 1227 11 ( ) 10 SB
gr
32 0 0 42 42 0 0 0 42 /Symbol font
gs 1096 1042 1152 597 CB
1505 1223 23 (+) 23 SB
gr
32 0 0 42 42 0 0 0 38 /Times-Roman /font32 ANSIFont font
gs 1096 1042 1152 597 CB
1528 1227 11 ( ) 10 SB
gr
32 0 0 42 42 0 0 0 39 /Times-Italic /font31 ANSIFont font
gs 1096 1042 1152 597 CB
1538 1226 19 (e) 18 SB
1556 1226 47 (nE) 46 SB
gr
32 0 0 33 33 0 0 0 33 /Symbol font
gs 1096 1042 1152 597 CB
1602 1244 22 (^) 22 SB
gr
32 0 0 33 33 0 0 0 30 /Times-Roman /font32 ANSIFont font
gs 1096 1042 1152 597 CB
1624 1223 8 ( ) 8 SB
gr
32 0 0 42 42 0 0 0 42 /Symbol font
gs 1096 1042 1152 597 CB
1632 1223 7 (\244) 7 SB
gr
32 0 0 33 33 0 0 0 30 /Times-Roman /font32 ANSIFont font
gs 1096 1042 1152 597 CB
1639 1240 8 ( ) 8 SB
gr
32 0 0 42 42 0 0 0 39 /Times-Italic /font31 ANSIFont font
gs 1096 1042 1152 597 CB
1647 1226 23 (T) 19 SB
1666 1226 19 (e) 18 SB
gr
32 0 0 42 42 0 0 0 42 /Symbol font
gs 1096 1042 1152 597 CB
1689 1223 14 (\)) 13 SB
gr
32 0 0 42 42 0 0 0 38 /Times-Roman /font32 ANSIFont font
gs 1096 1042 1152 597 CB
1702 1227 8 (|) 9 SB
gr
307 1 1405 1220 B
1 F
n
32 0 0 42 42 0 0 0 38 /Times-Roman /font32 ANSIFont font
gs 1096 1042 1152 597 CB
1713 1197 11 ( ) 10 SB
1723 1197 35 (= ) 34 SB
gr
32 0 0 42 42 0 0 0 39 /Times-Italic /font31 ANSIFont font
gs 1096 1042 1152 597 CB
1833 1161 14 (I) 14 SB
gr
32 0 0 33 33 0 0 0 30 /Times-Roman /font32 ANSIFont font
gs 1096 1042 1152 597 CB
1847 1182 15 (e) 14 SB
1861 1182 26 (m) 26 SB
gr
32 0 0 42 42 0 0 0 38 /Times-Roman /font32 ANSIFont font
gs 1096 1042 1152 597 CB
1759 1233 21 (4) 21 SB
gr
32 0 0 42 42 0 0 0 42 /Symbol font
gs 1096 1042 1152 597 CB
1780 1229 23 (p) 23 SB
gr
32 0 0 33 33 0 0 0 30 /Times-Roman /font32 ANSIFont font
gs 1096 1042 1152 597 CB
1803 1223 17 (2) 16 SB
gr
32 0 0 42 42 0 0 0 39 /Times-Italic /font31 ANSIFont font
gs 1096 1042 1152 597 CB
1819 1232 42 (Rr) 40 SB
1859 1232 19 (e) 18 SB
gr
32 0 0 44 44 0 0 0 39 /Times-Roman /font32 ANSIFont font
gs 1096 1042 1152 597 CB
1879 1232 9 (|) 10 SB
gr
32 0 0 42 42 0 0 0 42 /Symbol font
gs 1096 1042 1152 597 CB
1887 1229 30 (\321) 30 SB
gr
32 0 0 44 44 0 0 0 41 /Times-Italic /font31 ANSIFont font
gs 1096 1042 1152 597 CB
1917 1230 22 (n) 24 SB
gr
32 0 0 35 35 0 0 0 31 /Times-Roman /font32 ANSIFont font
gs 1096 1042 1152 597 CB
1938 1252 16 (e) 15 SB
gr
32 0 0 44 44 0 0 0 39 /Times-Roman /font32 ANSIFont font
gs 1096 1042 1152 597 CB
1952 1232 9 (|) 10 SB
gr
205 1 1757 1220 B
1 F
n
32 0 0 42 42 0 0 0 38 /Times-Roman /font32 ANSIFont font
gs 1096 1042 1152 597 CB
1963 1197 11 ( ) 10 SB
gr
gs 1096 1042 1152 597 CB
1310 1314 60 (for ) 59 SB
gr
32 0 0 42 42 0 0 0 38 /Times-Roman /font32 ANSIFont font
gs 1096 1042 1152 597 CB
1369 1314 8 (|) 8 SB
gr
32 0 0 44 44 0 0 0 44 /Symbol font
gs 1096 1042 1152 597 CB
1377 1308 31 (\321) 30 SB
gr
32 0 0 42 42 0 0 0 39 /Times-Italic /font31 ANSIFont font
gs 1096 1042 1152 597 CB
1407 1313 21 (n) 20 SB
gr
32 0 0 33 33 0 0 0 30 /Times-Roman /font32 ANSIFont font
gs 1096 1042 1152 597 CB
1427 1334 15 (e) 15 SB
gr
32 0 0 42 42 0 0 0 38 /Times-Roman /font32 ANSIFont font
gs 1096 1042 1152 597 CB
1442 1314 8 (|) 8 SB
gr
32 0 0 42 42 0 0 0 38 /Times-Roman /font32 ANSIFont font
gs 1096 1042 1152 597 CB
1450 1314 35 ( >) 34 SB
1484 1314 24 (>) 23 SB
1507 1314 11 ( ) 10 SB
gr
32 0 0 42 42 0 0 0 39 /Times-Italic /font31 ANSIFont font
gs 1096 1042 1152 597 CB
1517 1313 40 (en) 39 SB
1556 1313 26 (E) 26 SB
gr
32 0 0 33 33 0 0 0 33 /Symbol font
gs 1096 1042 1152 597 CB
1582 1331 22 (^) 21 SB
gr
32 0 0 33 33 0 0 0 30 /Times-Roman /font32 ANSIFont font
gs 1096 1042 1152 597 CB
1603 1310 8 ( ) 9 SB
gr
32 0 0 42 42 0 0 0 42 /Symbol font
gs 1096 1042 1152 597 CB
1612 1310 7 (\244) 6 SB
gr
32 0 0 35 35 0 0 0 31 /Times-Roman /font32 ANSIFont font
gs 1096 1042 1152 597 CB
1618 1326 9 ( ) 9 SB
gr
32 0 0 44 44 0 0 0 41 /Times-Italic /font31 ANSIFont font
gs 1096 1042 1152 597 CB
1627 1311 24 (T) 26 SB
gr
32 0 0 33 33 0 0 0 30 /Times-Roman /font32 ANSIFont font
gs 1096 1042 1152 597 CB
1650 1334 15 (e) 14 SB
gr
32 0 0 42 42 0 0 0 38 /Times-Roman /font32 ANSIFont font
gs 1096 1042 1152 597 CB
1664 1314 22 ( .) 21 SB
gr
gs 1096 1042 1152 597 CB
1227 1410 49 (He) 48 SB
1275 1410 33 (re) 32 SB
1307 1410 11 ( ) 11 SB
gr
32 0 0 42 42 0 0 0 39 /Times-Italic /font31 ANSIFont font
gs 1096 1042 1152 597 CB
1318 1409 21 (n) 20 SB
gr
32 0 0 33 33 0 0 0 30 /Times-Roman /font32 ANSIFont font
gs 1096 1042 1152 597 CB
1338 1427 15 (e) 15 SB
gr
32 0 0 42 42 0 0 0 38 /Times-Roman /font32 ANSIFont font
gs 1096 1042 1152 597 CB
1353 1410 11 ( ) 10 SB
1363 1410 39 (is ) 38 SB
1401 1410 52 (the) 51 SB
1452 1410 11 ( ) 10 SB
1462 1410 19 (e) 18 SB
1480 1410 31 (le) 30 SB
1510 1410 19 (c) 18 SB
1528 1410 47 (tro) 46 SB
1574 1410 32 (n ) 31 SB
1605 1410 40 (de) 39 SB
1644 1410 61 (nsit) 60 SB
1704 1410 21 (y) 18 SB
1722 1410 11 (,) 10 SB
1732 1410 11 ( ) 11 SB
gr
32 0 0 42 42 0 0 0 39 /Times-Italic /font31 ANSIFont font
gs 1096 1042 1152 597 CB
1743 1409 23 (T) 23 SB
gr
32 0 0 33 33 0 0 0 30 /Times-Roman /font32 ANSIFont font
gs 1096 1042 1152 597 CB
1766 1427 15 (e) 14 SB
gr
32 0 0 42 42 0 0 0 38 /Times-Roman /font32 ANSIFont font
gs 1096 1042 1152 597 CB
1780 1410 23 ( i) 22 SB
1802 1410 39 (s t) 38 SB
1840 1410 40 (he) 39 SB
1879 1410 11 ( ) 10 SB
1889 1410 31 (el) 30 SB
1919 1410 19 (e) 18 SB
1937 1410 31 (ct) 30 SB
1967 1410 56 (ron) 55 SB
2022 1410 23 ( t) 22 SB
2044 1410 19 (e) 18 SB
2062 1410 33 (m) 33 SB
gr
gs 1096 1042 1152 597 CB
2095 1410 14 (-) 13 SB
gr
gs 1096 1042 1152 597 CB
1227 1460 40 (pe) 39 SB
1266 1460 33 (ra) 32 SB
1298 1460 33 (tu) 32 SB
1330 1460 33 (re) 32 SB
1362 1460 22 (, ) 21 SB
1383 1460 19 (a) 18 SB
1401 1460 53 (nd ) 52 SB
gr
32 0 0 42 42 0 0 0 39 /Times-Italic /font31 ANSIFont font
gs 1096 1042 1152 597 CB
1453 1459 26 (E) 25 SB
gr
32 0 0 33 33 0 0 0 33 /Symbol font
gs 1096 1042 1152 597 CB
1478 1474 22 (^) 22 SB
gr
32 0 0 42 42 0 0 0 38 /Times-Roman /font32 ANSIFont font
gs 1096 1042 1152 597 CB
1500 1460 11 ( ) 10 SB
1510 1460 39 (is ) 38 SB
1548 1460 33 (th) 32 SB
1580 1460 30 (e ) 29 SB
1609 1460 40 (pe) 39 SB
1648 1460 35 (rp) 34 SB
1682 1460 40 (en) 39 SB
1721 1460 52 (dic) 51 SB
1772 1460 33 (ul) 32 SB
1804 1460 19 (a) 18 SB
1822 1460 25 (r ) 24 SB
1846 1460 31 (el) 30 SB
1876 1460 19 (e) 18 SB
1894 1460 31 (ct) 30 SB
1924 1460 26 (ri) 25 SB
1949 1460 30 (c ) 29 SB
1978 1460 26 (fi) 25 SB
2003 1460 19 (e) 18 SB
2021 1460 44 (ld.) 43 SB
2064 1460 11 ( ) 10 SB
gr
255 255 128 fC
945 410 222 2454 B
1 F
n
32 0 0 67 67 0 0 0 61 /Helvetica-BoldOblique /font14 ANSIFont font
0 0 0 fC
gs 1096 560 147 2379 CB
448 2476 18 (I) 19 SB
467 2476 41 (n) 40 SB
507 2476 19 ( ) 18 SB
525 2476 82 (thi) 81 SB
606 2476 56 (s ) 55 SB
661 2476 134 (issu) 133 SB
794 2476 56 (e ) 55 SB
849 2476 38 (. ) 37 SB
886 2476 19 (.) 18 SB
904 2476 19 ( ) 18 SB
922 2476 19 (.) 19 SB
gr
32 0 0 38 38 0 0 0 36 /Helvetica-Bold /font13 ANSIFont font
gs 1096 560 147 2379 CB
237 2599 36 (El) 35 SB
272 2599 55 (ect) 54 SB
326 2599 61 (ron) 60 SB
386 2599 24 ( t) 23 SB
409 2599 36 (ra) 35 SB
444 2599 90 (nspo) 89 SB
533 2599 28 (rt) 27 SB
560 2599 32 ( s) 31 SB
591 2599 59 (tud) 58 SB
649 2599 32 (ie) 31 SB
680 2599 32 (s ) 31 SB
711 2599 34 (in) 33 SB
744 2599 24 ( t) 23 SB
767 2599 55 (he ) 54 SB
821 2599 85 (Com) 83 SB
904 2599 44 (pa) 43 SB
947 2599 45 (ct ) 44 SB
991 2599 73 (Aub) 72 SB
1063 2599 61 (urn) 60 SB
gr
gs 1096 560 147 2379 CB
237 2644 61 (Tor) 60 SB
297 2644 55 (sat) 54 SB
351 2644 38 (ro) 37 SB
388 2644 23 (n) 23 SB
gr
32 0 0 38 38 0 0 0 34 /Helvetica /font12 ANSIFont font
gs 1096 560 147 2379 CB
237 2691 94 (The s) 93 SB
330 2691 49 (tell) 48 SB
378 2691 34 (ar) 33 SB
411 2691 32 (at) 31 SB
442 2691 34 (or) 33 SB
475 2691 32 ( d) 31 SB
506 2691 30 (io) 29 SB
535 2691 41 (de) 42 SB
577 2691 11 ( ) 10 SB
587 2691 33 (m) 31 SB
618 2691 20 (e) 21 SB
639 2691 11 (t) 10 SB
649 2691 74 (hod ) 73 SB
722 2691 9 (i) 8 SB
730 2691 30 (s ) 29 SB
759 2691 92 (used ) 91 SB
850 2691 32 (to) 31 SB
881 2691 44 ( m) 42 SB
923 2691 20 (e) 21 SB
944 2691 21 (a) 20 SB
964 2691 53 (sur) 52 SB
1016 2691 20 (e) 21 SB
1037 2691 11 ( ) 10 SB
1047 2691 24 (tr) 23 SB
1070 2691 61 (ans) 60 SB
gr
gs 1096 560 147 2379 CB
1130 2691 13 (-) 13 SB
gr
gs 1096 560 147 2379 CB
237 2736 55 (por) 54 SB
291 2736 11 (t) 10 SB
301 2736 20 ( i) 19 SB
320 2736 32 (n ) 31 SB
351 2736 30 (st) 29 SB
380 2736 40 (oc) 39 SB
419 2736 72 (hast) 71 SB
490 2736 9 (i) 8 SB
498 2736 30 (c ) 29 SB
527 2736 33 (m) 31 SB
558 2736 103 (agneti) 102 SB
660 2736 30 (c ) 29 SB
689 2736 11 (f) 10 SB
699 2736 9 (i) 8 SB
707 2736 20 (e) 21 SB
728 2736 9 (l) 8 SB
736 2736 51 (ds ) 50 SB
786 2736 9 (i) 8 SB
794 2736 32 (n ) 31 SB
825 2736 86 (CAT.) 85 SB
910 2736 11 ( ) 11 SB
gr
gs 1096 560 147 2379 CB
927 2736 11 (.) 10 SB
937 2736 11 (.) 10 SB
947 2736 22 (..) 21 SB
968 2736 11 (.) 10 SB
978 2736 11 (.) 10 SB
988 2736 22 (..) 21 SB
1009 2736 11 (.) 10 SB
1019 2736 11 (.) 10 SB
1029 2736 22 (..) 21 SB
1050 2736 11 (.) 10 SB
1060 2736 11 (.) 10 SB
1070 2736 22 (..) 21 SB
1091 2736 11 (.) 10 SB
1101 2736 11 (.) 10 SB
gr
gs 1096 560 147 2379 CB
1121 2736 32 ( 1) 31 SB
gr
/PPT_ProcessAll true def

userdict /VPsave save put
userdict begin
/showpage{}def
1002 1488 1002 2768 2398 2768 2398 1488 newpath moveto lineto lineto lineto clip newpath
1227 2467 translate 300 72 div dup neg scale
946 300 div 339 72 div div 680 300 div 244 72 div div scale
-14 -22 translate
/AutoFlatness true def
/wCorel5Dict 300 dict def wCorel5Dict begin/bd{bind def}bind def/ld{load def}
bd/xd{exch def}bd/_ null def/rp{{pop}repeat}bd/@cp/closepath ld/@gs/gsave ld
/@gr/grestore ld/@np/newpath ld/Tl/translate ld/$sv 0 def/@sv{/$sv save def}bd
/@rs{$sv restore}bd/spg/showpage ld/showpage{}bd currentscreen/@dsp xd/$dsp
/@dsp def/$dsa xd/$dsf xd/$sdf false def/$SDF false def/$Scra 0 def/SetScr
/setscreen ld/setscreen{3 rp}bd/@ss{2 index 0 eq{$dsf 3 1 roll 4 -1 roll pop}
if exch $Scra add exch load SetScr}bd/SepMode_5 where{pop}{/SepMode_5 0 def}
ifelse/CurrentInkName_5 where{pop}{/CurrentInkName_5(Composite)def}ifelse
/$ink_5 where{pop}{/$ink_5 -1 def}ifelse/$c 0 def/$m 0 def/$y 0 def/$k 0 def
/$t 1 def/$n _ def/$o 0 def/$fil 0 def/$C 0 def/$M 0 def/$Y 0 def/$K 0 def/$T 1
def/$N _ def/$O 0 def/$PF false def/s1c 0 def/s1m 0 def/s1y 0 def/s1k 0 def
/s1t 0 def/s1n _ def/$bkg false def/SK 0 def/SM 0 def/SY 0 def/SC 0 def/$op
false def matrix currentmatrix/$ctm xd/$ptm matrix def/$ttm matrix def/$stm
matrix def/$fst 128 def/$pad 0 def/$rox 0 def/$roy 0 def/$ffpnt true def
/CorelDrawReencodeVect[16#0/grave 16#5/breve 16#6/dotaccent 16#8/ring
16#A/hungarumlaut 16#B/ogonek 16#C/caron 16#D/dotlessi 16#27/quotesingle
16#60/grave 16#7C/bar
16#82/quotesinglbase/florin/quotedblbase/ellipsis/dagger/daggerdbl
16#88/circumflex/perthousand/Scaron/guilsinglleft/OE
16#91/quoteleft/quoteright/quotedblleft/quotedblright/bullet/endash/emdash
16#98/tilde/trademark/scaron/guilsinglright/oe 16#9F/Ydieresis
16#A1/exclamdown/cent/sterling/currency/yen/brokenbar/section
16#a8/dieresis/copyright/ordfeminine/guillemotleft/logicalnot/minus/registered/macron
16#b0/degree/plusminus/twosuperior/threesuperior/acute/mu/paragraph/periodcentered
16#b8/cedilla/onesuperior/ordmasculine/guillemotright/onequarter/onehalf/threequarters/questiondown
16#c0/Agrave/Aacute/Acircumflex/Atilde/Adieresis/Aring/AE/Ccedilla
16#c8/Egrave/Eacute/Ecircumflex/Edieresis/Igrave/Iacute/Icircumflex/Idieresis
16#d0/Eth/Ntilde/Ograve/Oacute/Ocircumflex/Otilde/Odieresis/multiply
16#d8/Oslash/Ugrave/Uacute/Ucircumflex/Udieresis/Yacute/Thorn/germandbls
16#e0/agrave/aacute/acircumflex/atilde/adieresis/aring/ae/ccedilla
16#e8/egrave/eacute/ecircumflex/edieresis/igrave/iacute/icircumflex/idieresis
16#f0/eth/ntilde/ograve/oacute/ocircumflex/otilde/odieresis/divide
16#f8/oslash/ugrave/uacute/ucircumflex/udieresis/yacute/thorn/ydieresis]def
/@BeginSysCorelDict{systemdict/Corel20Dict known{systemdict/Corel20Dict get
exec}if}bd/@EndSysCorelDict{systemdict/Corel20Dict known{end}if}bd AutoFlatness
{/@ifl{dup currentflat exch sub 10 gt{
([Error: PathTooComplex; OffendingCommand: AnyPaintingOperator]\n)print flush
@np exit}{currentflat 2 add setflat}ifelse}bd/@fill/fill ld/fill{currentflat{
{@fill}stopped{@ifl}{exit}ifelse}bind loop setflat}bd/@eofill/eofill ld/eofill
{currentflat{{@eofill}stopped{@ifl}{exit}ifelse}bind loop setflat}bd/@clip
/clip ld/clip{currentflat{{@clip}stopped{@ifl}{exit}ifelse}bind loop setflat}
bd/@eoclip/eoclip ld/eoclip{currentflat{{@eoclip}stopped{@ifl}{exit}ifelse}
bind loop setflat}bd/@stroke/stroke ld/stroke{currentflat{{@stroke}stopped
{@ifl}{exit}ifelse}bind loop setflat}bd}if/d/setdash ld/j/setlinejoin ld/J
/setlinecap ld/M/setmiterlimit ld/w/setlinewidth ld/O{/$o xd}bd/R{/$O xd}bd/W
/eoclip ld/c/curveto ld/C/c ld/l/lineto ld/L/l ld/rl/rlineto ld/m/moveto ld/n
/newpath ld/N/newpath ld/P{11 rp}bd/u{}bd/U{}bd/A{pop}bd/q/@gs ld/Q/@gr ld/`
{}bd/~{}bd/@{}bd/&{}bd/@j{@sv @np}bd/@J{@rs}bd/g{1 exch sub/$k xd/$c 0 def/$m 0
def/$y 0 def/$t 1 def/$n _ def/$fil 0 def}bd/G{1 sub neg/$K xd _ 1 0 0 0/$C xd
/$M xd/$Y xd/$T xd/$N xd}bd/k{1 index type/stringtype eq{/$t xd/$n xd}{/$t 0
def/$n _ def}ifelse/$k xd/$y xd/$m xd/$c xd/$fil 0 def}bd/K{1 index type
/stringtype eq{/$T xd/$N xd}{/$T 0 def/$N _ def}ifelse/$K xd/$Y xd/$M xd/$C xd
}bd/x/k ld/X/K ld/sf{1 index type/stringtype eq{/s1t xd/s1n xd}{/s1t 0 def/s1n
_ def}ifelse/s1k xd/s1y xd/s1m xd/s1c xd}bd/i{dup 0 ne{setflat}{pop}ifelse}bd
/v{4 -2 roll 2 copy 6 -2 roll c}bd/V/v ld/y{2 copy c}bd/Y/y ld/@w{matrix rotate
/$ptm xd matrix scale $ptm dup concatmatrix/$ptm xd 1 eq{$ptm exch dup
concatmatrix/$ptm xd}if 1 w}bd/@g{1 eq dup/$sdf xd{/$scp xd/$sca xd/$scf xd}if
}bd/@G{1 eq dup/$SDF xd{/$SCP xd/$SCA xd/$SCF xd}if}bd/@D{2 index 0 eq{$dsf 3 1
roll 4 -1 roll pop}if 3 copy exch $Scra add exch load SetScr/$dsp xd/$dsa xd
/$dsf xd}bd/$ngx{$SDF{$SCF SepMode_5 0 eq{$SCA}{$dsa}ifelse $SCP @ss}if}bd/p{
/$pm xd 7 rp/$pyf xd/$pxf xd/$pn xd/$fil 1 def}bd/@MN{2 copy le{pop}{exch pop}
ifelse}bd/@MX{2 copy ge{pop}{exch pop}ifelse}bd/InRange{3 -1 roll @MN @MX}bd
/wDstChck{2 1 roll dup 3 -1 roll eq{1 add}if}bd/@dot{dup mul exch dup mul add 1
exch sub}bd/@lin{exch pop abs 1 exch sub}bd/cmyk2rgb{3{dup 5 -1 roll add 1 exch
sub dup 0 lt{pop 0}if exch}repeat pop}bd/rgb2cmyk{3{1 exch sub 3 1 roll}repeat
3 copy @MN @MN 3{dup 5 -1 roll sub neg exch}repeat}bd/rgb2g{2 index .299 mul 2
index .587 mul add 1 index .114 mul add 4 1 roll 3 rp}bd/WaldoColor_5 where{
pop}{/SetRgb/setrgbcolor ld/GetRgb/currentrgbcolor ld/SetGry/setgray ld/GetGry
/currentgray ld/SetRgb2 systemdict/setrgbcolor get def/GetRgb2 systemdict
/currentrgbcolor get def/SetHsb systemdict/sethsbcolor get def/GetHsb
systemdict/currenthsbcolor get def/rgb2hsb{SetRgb2 GetHsb}bd/hsb2rgb{3 -1 roll
dup floor sub 3 1 roll SetHsb GetRgb2}bd/setcmykcolor where{pop/SetCmyk_5
/setcmykcolor ld}{/SetCmyk_5{cmyk2rgb SetRgb}bd}ifelse/currentcmykcolor where{
pop/GetCmyk/currentcmykcolor ld}{/GetCmyk{GetRgb rgb2cmyk}bd}ifelse
/setoverprint where{pop}{/setoverprint{/$op xd}bd}ifelse/currentoverprint where
{pop}{/currentoverprint{$op}bd}ifelse/@tc_5{5 -1 roll dup 1 ge{pop}{4{dup 6 -1
roll mul exch}repeat pop}ifelse}bd/@trp{exch pop 5 1 roll @tc_5}bd
/setprocesscolor_5{SepMode_5 0 eq{SetCmyk_5}{0 4 $ink_5 sub index exch pop 5 1
roll 4 rp SepsColor true eq{$ink_5 3 gt{1 sub neg SetGry}{0 0 0 4 $ink_5 roll
SetCmyk_5}ifelse}{1 sub neg SetGry}ifelse}ifelse}bd/findcmykcustomcolor where
{pop}{/findcmykcustomcolor{5 array astore}bd}ifelse/setcustomcolor where{pop}{
/setcustomcolor{exch aload pop SepMode_5 0 eq{pop @tc_5 setprocesscolor_5}{
CurrentInkName_5 eq{4 index}{0}ifelse 6 1 roll 5 rp 1 sub neg SetGry}ifelse}bd
}ifelse/@scc_5{dup type/booleantype eq{setoverprint}{1 eq setoverprint}ifelse
dup _ eq{pop setprocesscolor_5 pop}{findcmykcustomcolor exch setcustomcolor}
ifelse SepMode_5 0 eq{true}{GetGry 1 eq currentoverprint and not}ifelse}bd
/colorimage where{pop/ColorImage/colorimage ld}{/ColorImage{/ncolors xd pop
/dataaq xd{dataaq ncolors dup 3 eq{/$dat xd 0 1 $dat length 3 div 1 sub{dup 3
mul $dat 1 index get 255 div $dat 2 index 1 add get 255 div $dat 3 index 2 add
get 255 div rgb2g 255 mul cvi exch pop $dat 3 1 roll put}for $dat 0 $dat length
3 idiv getinterval pop}{4 eq{/$dat xd 0 1 $dat length 4 div 1 sub{dup 4 mul
$dat 1 index get 255 div $dat 2 index 1 add get 255 div $dat 3 index 2 add get
255 div $dat 4 index 3 add get 255 div cmyk2rgb rgb2g 255 mul cvi exch pop $dat
3 1 roll put}for $dat 0 $dat length ncolors idiv getinterval}if}ifelse}image}
bd}ifelse/setcmykcolor{1 5 1 roll _ currentoverprint @scc_5/$ffpnt xd}bd
/currentcmykcolor{0 0 0 0}bd/setrgbcolor{rgb2cmyk setcmykcolor}bd
/currentrgbcolor{currentcmykcolor cmyk2rgb}bd/sethsbcolor{hsb2rgb setrgbcolor}
bd/currenthsbcolor{currentrgbcolor rgb2hsb}bd/setgray{dup dup setrgbcolor}bd
/currentgray{currentrgbcolor rgb2g}bd}ifelse/WaldoColor_5 true def/@sft{$tllx
$pxf add dup $tllx gt{$pwid sub}if/$tx xd $tury $pyf sub dup $tury lt{$phei
add}if/$ty xd}bd/@stb{pathbbox/$ury xd/$urx xd/$lly xd/$llx xd}bd/@ep{{cvx exec
}forall}bd/@tp{@sv/$in true def 2 copy dup $lly le{/$in false def}if $phei sub
$ury ge{/$in false def}if dup $urx ge{/$in false def}if $pwid add $llx le{/$in
false def}if $in{@np 2 copy m $pwid 0 rl 0 $phei neg rl $pwid neg 0 rl 0 $phei
rl clip @np $pn cvlit load aload pop 7 -1 roll 5 index sub 7 -1 roll 3 index
sub Tl matrix currentmatrix/$ctm xd @ep 4 rp}{2 rp}ifelse @rs}bd/@th{@sft 0 1
$tly 1 sub{dup $psx mul $tx add{dup $llx gt{$pwid sub}{exit}ifelse}loop exch
$phei mul $ty exch sub 0 1 $tlx 1 sub{$pwid mul 3 copy 3 -1 roll add exch @tp
pop}for 2 rp}for}bd/@tv{@sft 0 1 $tlx 1 sub{dup $pwid mul $tx add exch $psy mul
$ty exch sub{dup $ury lt{$phei add}{exit}ifelse}loop 0 1 $tly 1 sub{$phei mul 3
copy sub @tp pop}for 2 rp}for}bd/@pf{@gs $ctm setmatrix $pm concat @stb eoclip
Bburx Bbury $pm itransform/$tury xd/$turx xd Bbllx Bblly $pm itransform/$tlly
xd/$tllx xd/$wid $turx $tllx sub def/$hei $tury $tlly sub def @gs $vectpat{1 0
0 0 0 _ $o @scc_5{eofill}if}{$t $c $m $y $k $n $o @scc_5{SepMode_5 0 eq $pfrg
or{$tllx $tlly Tl $wid $hei scale <00> 8 1 false[8 0 0 1 0 0]{}imagemask}{
/$bkg true def}ifelse}if}ifelse @gr $wid 0 gt $hei 0 gt and{$pn cvlit load
aload pop/$pd xd 3 -1 roll sub/$phei xd exch sub/$pwid xd $wid $pwid div
ceiling 1 add/$tlx xd $hei $phei div ceiling 1 add/$tly xd $psx 0 eq{@tv}{@th}
ifelse}if @gr @np/$bkg false def}bd/@dlt{$fse $fss sub/nff xd $frb dup 1 eq
exch 2 eq or{$frt dup $frc $frm $fry $frk @tc_5 4 copy cmyk2rgb rgb2hsb 3 copy
/myb xd/mys xd/myh xd $tot $toc $tom $toy $tok @tc_5 cmyk2rgb rgb2hsb 3 1 roll
4 1 roll 5 1 roll sub neg nff div/kdb xd sub neg nff div/kds xd sub neg dup 0
eq{pop $frb 2 eq{.99}{-.99}ifelse}if dup $frb 2 eq exch 0 lt and{1 add}if dup
$frb 1 eq exch 0 gt and{1 sub}if nff div/kdh xd}{$frt dup $frc $frm $fry $frk
@tc_5 5 copy $tot dup $toc $tom $toy $tok @tc_5 5 1 roll 6 1 roll 7 1 roll 8 1
roll 9 1 roll sub neg nff dup 1 gt{1 sub}if div/$dk xd sub neg nff dup 1 gt{1
sub}if div/$dy xd sub neg nff dup 1 gt{1 sub}if div/$dm xd sub neg nff dup 1 gt
{1 sub}if div/$dc xd sub neg nff dup 1 gt{1 sub}if div/$dt xd}ifelse}bd/ffcol{
5 copy $fsit 0 eq{setcmykcolor pop}{SepMode_5 0 ne{$frn findcmykcustomcolor
exch setcustomcolor}{4 rp $frc $frm $fry $frk $frn findcmykcustomcolor exch
setcustomcolor}ifelse}ifelse}bd/@ftl{1 index 4 index sub dup $pad mul dup/$pdw
xd 2 mul sub $fst div/$wid xd 2 index sub/$hei xd pop Tl @dlt $fss 0 eq{ffcol n
0 0 m 0 $hei l $pdw $hei l $pdw 0 l @cp $ffpnt{fill}{@np}ifelse}if $fss $wid
mul $pdw add 0 Tl nff{ffcol n 0 0 m 0 $hei l $wid $hei l $wid 0 l @cp $ffpnt
{fill}{@np}ifelse $wid 0 Tl $frb dup 1 eq exch 2 eq or{4 rp myh mys myb kdb add
3 1 roll kds add 3 1 roll kdh add 3 1 roll 3 copy/myb xd/mys xd/myh xd hsb2rgb
rgb2cmyk}{$dk add 5 1 roll $dy add 5 1 roll $dm add 5 1 roll $dc add 5 1 roll
$dt add 5 1 roll}ifelse}repeat 5 rp $tot dup $toc $tom $toy $tok @tc_5 ffcol n
0 0 m 0 $hei l $pdw $hei l $pdw 0 l @cp $ffpnt{fill}{@np}ifelse 5 rp}bd/@ftrs{
1 index 4 index sub dup $rox mul/$row xd 2 div 1 index 4 index sub dup $roy mul
/$roh xd 2 div 2 copy dup mul exch dup mul add sqrt $row dup mul $roh dup mul
add sqrt add dup/$hei xd $fst div/$wid xd 4 index add $roh add exch 5 index add
$row add exch Tl $fan rotate 4 rp @dlt $fss 0 eq{ffcol $fty 3 eq{$hei dup neg
dup m 2 mul @sqr}{0 0 m 0 0 $hei 0 360 arc}ifelse $ffpnt{fill}{@np}ifelse}if
1.0 $pad 2 mul sub dup scale $hei $fss $wid mul sub/$hei xd nff{ffcol $fty 3 eq
{n $hei dup neg dup m 2 mul @sqr}{n 0 0 m 0 0 $hei 0 360 arc}ifelse $ffpnt
{fill}{@np}ifelse/$hei $hei $wid sub def $frb dup 1 eq exch 2 eq or{4 rp myh
mys myb kdb add 3 1 roll kds add 3 1 roll kdh add 3 1 roll 3 copy/myb xd/mys xd
/myh xd hsb2rgb rgb2cmyk}{$dk add 5 1 roll $dy add 5 1 roll $dm add 5 1 roll
$dc add 5 1 roll $dt add 5 1 roll}ifelse}repeat 5 rp}bd/@ftc{1 index 4 index
sub dup $rox mul/$row xd 2 div 1 index 4 index sub dup $roy mul/$roh xd 2 div 2
copy dup mul exch dup mul add sqrt $row dup mul $roh dup mul add sqrt add dup
/$hei xd $fst div/$wid xd 4 index add $roh add exch 5 index add $row add exch
Tl 4 rp @dlt $fss 0 eq{ffcol $ffpnt{fill}{@np}ifelse}{n}ifelse/$dang 180 $fst 1
sub div def/$sang $dang -2 div 180 add def/$eang $dang 2 div 180 add def/$sang
$sang $dang $fss mul add def/$eang $eang $dang $fss mul add def/$sang $eang
$dang sub def nff{ffcol n 0 0 m 0 0 $hei $sang $fan add $eang $fan add arc
$ffpnt{fill}{@np}ifelse 0 0 m 0 0 $hei $eang neg $fan add $sang neg $fan add
arc $ffpnt{fill}{@np}ifelse/$sang $eang def/$eang $eang $dang add def $frb dup
1 eq exch 2 eq or{4 rp myh mys myb kdb add 3 1 roll kds add 3 1 roll kdh add 3
1 roll 3 copy/myb xd/mys xd/myh xd hsb2rgb rgb2cmyk}{$dk add 5 1 roll $dy add 5
1 roll $dm add 5 1 roll $dc add 5 1 roll $dt add 5 1 roll}ifelse}repeat 5 rp}
bd/@ff{/$fss 0 def $o 1 eq setoverprint 1 1 $fsc 1 sub{dup 1 sub $fsit 0 eq{
$fsa exch 5 mul 5 getinterval aload 2 rp/$frk xd/$fry xd/$frm xd/$frc xd/$frn _
def/$frt 1 def $fsa exch 5 mul 5 getinterval aload pop $fss add/$fse xd/$tok xd
/$toy xd/$tom xd/$toc xd/$ton _ def/$tot 1 def}{$fsa exch 7 mul 7 getinterval
aload 2 rp/$frt xd/$frn xd/$frk xd/$fry xd/$frm xd/$frc xd $fsa exch 7 mul 7
getinterval aload pop $fss add/$fse xd/$tot xd/$ton xd/$tok xd/$toy xd/$tom xd
/$toc xd}ifelse $fsit 0 eq SepMode_5 0 eq or dup not CurrentInkName_5 $frn eq
and or{@sv $ctm setmatrix eoclip Bbllx Bblly Bburx Bbury $fty 2 eq{@ftc}{1
index 3 index m 2 copy l 3 index 1 index l 3 index 3 index l @cp $fty dup 1 eq
exch 3 eq or{@ftrs}{4 rp $fan rotate pathbbox @ftl}ifelse}ifelse @rs/$fss $fse
def}{1 0 0 0 0 _ $o @scc_5{fill}if}ifelse}for @np}bd/@Pf{@sv SepMode_5 0 eq
$ink_5 3 eq or{0 J 0 j[]0 d $t $c $m $y $k $n $o @scc_5 pop $ctm setmatrix 72
1000 div dup matrix scale dup concat dup Bburx exch Bbury exch itransform
ceiling cvi/Bbury xd ceiling cvi/Bburx xd Bbllx exch Bblly exch itransform
floor cvi/Bblly xd floor cvi/Bbllx xd $Prm aload pop $Psn load exec}{1 SetGry
eofill}ifelse @rs @np}bd/F{matrix currentmatrix $sdf{$scf $sca $scp @ss}if $fil
1 eq{@pf}{$fil 2 eq{@ff}{$fil 3 eq{@Pf}{$t $c $m $y $k $n $o @scc_5{eofill}
{@np}ifelse}ifelse}ifelse}ifelse $sdf{$dsf $dsa $dsp @ss}if setmatrix}bd/f{@cp
F}bd/S{matrix currentmatrix $ctm setmatrix $SDF{$SCF $SCA $SCP @ss}if $T $C $M
$Y $K $N $O @scc_5{matrix currentmatrix $ptm concat stroke setmatrix}
{@np}ifelse $SDF{$dsf $dsa $dsp @ss}if setmatrix}bd/s{@cp S}bd/B{@gs F @gr S}
bd/b{@cp B}bd/E{5 array astore exch cvlit xd}bd/@cc{currentfile $dat
readhexstring pop}bd/@sm{/$ctm $ctm currentmatrix def}bd/@E{/Bbury xd/Bburx xd
/Bblly xd/Bbllx xd}bd/@c{@cp}bd/@p{/$fil 1 def 1 eq dup/$vectpat xd{/$pfrg true
def}{@gs $t $c $m $y $k $n $o @scc_5/$pfrg xd @gr}ifelse/$pm xd/$psy xd/$psx xd
/$pyf xd/$pxf xd/$pn xd}bd/@P{/$fil 3 def/$Psn xd array astore/$Prm xd}bd/@k{
/$fil 2 def/$roy xd/$rox xd/$pad xd/$fty xd/$fan xd $fty 1 eq{/$fan 0 def}if
/$frb xd/$fst xd/$fsc xd/$fsa xd/$fsit 0 def}bd/@x{/$fil 2 def/$roy xd/$rox xd
/$pad xd/$fty xd/$fan xd $fty 1 eq{/$fan 0 def}if/$frb xd/$fst xd/$fsc xd/$fsa
xd/$fsit 1 def}bd/@ii{concat 3 index 3 index m 3 index 1 index l 2 copy l 1
index 3 index l 3 index 3 index l clip 4 rp}bd/tcc{@cc}def/@i{@sm @gs @ii 6
index 1 ne{/$frg true def 2 rp}{1 eq{s1t s1c s1m s1y s1k s1n $O @scc_5/$frg xd
}{/$frg false def}ifelse 1 eq{@gs $ctm setmatrix F @gr}if}ifelse @np/$ury xd
/$urx xd/$lly xd/$llx xd/$bts xd/$hei xd/$wid xd/$dat $wid $bts mul 8 div
ceiling cvi string def $bkg $frg or{$SDF{$SCF $SCA $SCP @ss}if $llx $lly Tl
$urx $llx sub $ury $lly sub scale $bkg{$t $c $m $y $k $n $o @scc_5 pop}if $wid
$hei abs $bts 1 eq{$bkg}{$bts}ifelse[$wid 0 0 $hei neg 0 $hei 0
gt{$hei}{0}ifelse]/tcc load $bts 1 eq{imagemask}{image}ifelse $SDF{$dsf $dsa
$dsp @ss}if}{$hei abs{tcc pop}repeat}ifelse @gr $ctm setmatrix}bd/@M{@sv}bd/@N
{/@cc{}def 1 eq{12 -1 roll neg 12 1 roll @I}{13 -1 roll neg 13 1 roll @i}
ifelse @rs}bd/@I{@sm @gs @ii @np/$ury xd/$urx xd/$lly xd/$llx xd/$ncl xd/$bts
xd/$hei xd/$wid xd/$dat $wid $bts mul $ncl mul 8 div ceiling cvi string def
$ngx $llx $lly Tl $urx $llx sub $ury $lly sub scale $wid $hei abs $bts[$wid 0 0
$hei neg 0 $hei 0 gt{$hei}{0}ifelse]/@cc load false $ncl ColorImage $SDF{$dsf
$dsa $dsp @ss}if @gr $ctm setmatrix}bd/z{exch findfont exch scalefont setfont}
bd/ZB{9 dict dup begin 4 1 roll/FontType 3 def/FontMatrix xd/FontBBox xd
/Encoding 256 array def 0 1 255{Encoding exch/.notdef put}for/CharStrings 256
dict def CharStrings/.notdef{}put/Metrics 256 dict def Metrics/.notdef 3 -1
roll put/BuildChar{exch dup/$char exch/Encoding get 3 index get def dup
/Metrics get $char get aload pop setcachedevice begin Encoding exch get
CharStrings exch get end exec}def end definefont pop}bd/ZBAddChar{findfont
begin dup 4 1 roll dup 6 1 roll Encoding 3 1 roll put CharStrings 3 1 roll put
Metrics 3 1 roll put end}bd/Z{findfont dup maxlength 2 add dict exch dup{1
index/FID ne{3 index 3 1 roll put}{2 rp}ifelse}forall pop dup dup/Encoding get
256 array copy dup/$fe xd/Encoding exch put dup/Fontname 3 index put 3 -1 roll
dup length 0 ne{0 exch{dup type 0 type eq{exch pop}{$fe exch 2 index exch put 1
add}ifelse}forall pop}if dup 256 dict dup/$met xd/Metrics exch put dup
/FontMatrix get 0 get 1000 mul 1 exch div 3 index length 256 eq{0 1 255{dup $fe
exch get dup/.notdef eq{2 rp}{5 index 3 -1 roll get 2 index mul $met 3 1 roll
put}ifelse}for}if pop definefont pop pop}bd/@ftx{{currentpoint 3 -1 roll(0)dup
3 -1 roll 0 exch put dup @gs true charpath $ctm setmatrix @@txt @gr @np
stringwidth pop 3 -1 roll add exch moveto}forall}bd/@ft{matrix currentmatrix
exch $sdf{$scf $sca $scp @ss}if $fil 1 eq{/@@txt/@pf ld @ftx}{$fil 2 eq{/@@txt
/@ff ld @ftx}{$fil 3 eq{/@@txt/@Pf ld @ftx}{$t $c $m $y $k $n $o @scc_5{show}
{pop}ifelse}ifelse}ifelse}ifelse $sdf{$dsf $dsa $dsp @ss}if setmatrix}bd/@st{
matrix currentmatrix exch $SDF{$SCF $SCA $SCP @ss}if $T $C $M $Y $K $N $O
@scc_5{{currentpoint 3 -1 roll(0)dup 3 -1 roll 0 exch put dup @gs true charpath
$ctm setmatrix $ptm concat stroke @gr @np stringwidth pop 3 -1 roll add exch
moveto}forall}{pop}ifelse $SDF{$dsf $dsa $dsp @ss}if setmatrix}bd/@te{@ft}bd
/@tr{@st}bd/@ta{dup @gs @ft @gr @st}bd/@t@a{dup @gs @st @gr @ft}bd/@tm{@sm
concat}bd/e{/t{@te}def}bd/r{/t{@tr}def}bd/o{/t{pop}def}bd/a{/t{@ta}def}bd/@a{
/t{@t@a}def}bd/t{@te}def/T{@np $ctm setmatrix/$ttm matrix def}bd/ddt{t}def/@t{
/$stm $stm currentmatrix def 3 1 roll moveto $ttm concat ddt $stm setmatrix}bd
/@n{/$ttm exch matrix rotate def}bd/@s{}bd/@l{}bd/@B{@gs S @gr F}bd/@b{@cp @B}
bd/@sep{CurrentInkName_5(Composite)eq{/$ink_5 -1 def}{CurrentInkName_5(Cyan)eq
{/$ink_5 0 def}{CurrentInkName_5(Magenta)eq{/$ink_5 1 def}{CurrentInkName_5
(Yellow)eq{/$ink_5 2 def}{CurrentInkName_5(Black)eq{/$ink_5 3 def}{/$ink_5 4
def}ifelse}ifelse}ifelse}ifelse}ifelse}bd/@whi{@gs -72000 dup moveto -72000
72000 lineto 72000 dup lineto 72000 -72000 lineto @cp 1 SetGry fill @gr}bd
/@neg{[{1 exch sub}/exec cvx currenttransfer/exec cvx]cvx settransfer @whi}bd
/currentscale{1 0 dtransform matrix defaultmatrix idtransform dup mul exch dup
mul add sqrt 0 1 dtransform matrix defaultmatrix idtransform dup mul exch dup
mul add sqrt}bd/@unscale{currentscale 1 exch div exch 1 exch div exch scale}bd
/@sqr{dup 0 rlineto dup 0 exch rlineto neg 0 rlineto @cp}bd/corelsym{@gs @np Tl
-90 rotate 7{45 rotate -.75 2 moveto 1.5 @sqr fill}repeat @gr}bd/@reg_cor{@gs
@np Tl -6 -6 moveto 12 @sqr @gs 1 GetGry sub SetGry fill @gr 4{90 rotate 0 4 m
0 4 rl}repeat stroke 0 0 corelsym @gr}bd/@reg_std{@gs @np Tl .3 w 0 0 5 0 360
arc @cp @gs 1 GetGry sub SetGry fill @gr 4{90 rotate 0 0 m 0 8 rl}repeat stroke
@gr}bd/@reg_inv{@gs @np Tl .3 w 0 0 5 0 360 arc @cp @gs 1 GetGry sub SetGry
fill @gr 4{90 rotate 0 0 m 0 8 rl}repeat stroke 0 0 m 0 0 5 90 180 arc @cp 0 0
5 270 360 arc @cp GetGry fill @gr}bd/$corelmeter[1 .95 .75 .50 .25 .05 0]def
/@colormeter{@gs @np 0 SetGry 0.3 w/Courier findfont 5 scalefont setfont/yy xd
/xx xd 0 1 6{dup xx 20 sub yy m 20 @sqr @gs $corelmeter exch get dup SetGry
fill @gr stroke xx 18 sub yy 2 add m exch dup 3 ge{1 SetGry}{0 SetGry}ifelse 3
eq{pop}{100 mul 100 exch sub cvi 20 string cvs show}ifelse/yy yy 20 add def}
for @gr}bd/@calbar{@gs Tl @gs @np 0 0 m @gs 20 @sqr 1 1 0 0 0 _ 0 @scc_5 pop
fill @gr 20 0 Tl 0 0 m @gs 20 @sqr 1 1 0 1 0 _ 0 @scc_5 pop fill @gr 20 0 Tl 0
0 m @gs 20 @sqr 1 0 0 1 0 _ 0 @scc_5 pop fill @gr 20 0 Tl 0 0 m @gs 20 @sqr 1 0
1 1 0 _ 0 @scc_5 pop fill @gr 20 0 Tl 0 0 m @gs 20 @sqr 1 0 1 0 0 _ 0 @scc_5
pop fill @gr 20 0 Tl 0 0 m @gs 20 @sqr 1 1 1 0 0 _ 0 @scc_5 pop fill @gr 20 0
Tl 0 0 m @gs 20 @sqr 1 1 1 1 0 _ 0 @scc_5 pop fill @gr @gr @np -84 0 Tl @gs 0 0
m 20 @sqr clip 1 1 0 0 0 _ 0 @scc_5 pop @gain @gr 20 0 Tl @gs 0 0 m 20 @sqr
clip 1 0 1 0 0 _ 0 @scc_5 pop @gain @gr 20 0 Tl @gs 0 0 m 20 @sqr clip 1 0 0 1
0 _ 0 @scc_5 pop @gain @gr 20 0 Tl @gs 0 0 m 20 @sqr clip 1 0 0 0 1 _ 0 @scc_5
pop @gain @gr @gr}bd/@gain{10 10 Tl @np 0 0 m 0 10 360{0 0 15 4 -1 roll dup 5
add arc @cp}for fill}bd/@crop{@gs 10 div/$croplen xd .3 w 0 SetGry Tl rotate 0
0 m 0 $croplen neg rl stroke @gr}bd/@colorbox{@gs @np Tl 100 exch sub 100 div
SetGry -8 -8 moveto 16 @sqr fill 0 SetGry 10 -2 moveto show @gr}bd/deflevel 0
def/@sax{/deflevel deflevel 1 add def}bd/@eax{/deflevel deflevel dup 0 gt{1
sub}if def deflevel 0 gt{/eax load}{eax}ifelse}bd/eax{{exec}forall}bd/@rax{
deflevel 0 eq{@rs @sv}if}bd/@daq{dup type/arraytype eq{{}forall}if}bd/@BMP{
/@cc xd 11 index 1 eq{12 -1 roll pop @i}{7 -2 roll 2 rp @I}ifelse}bd systemdict
/pdfmark known not{/pdfmark/cleartomark ld}if end
wCorel5Dict begin
@BeginSysCorelDict
2.6131 setmiterlimit
1.00 setflat
/$fst 14 def
[ 0 0 0 0 0 0 0 0 0 0 0 0 0 0 0 0 0 
0 0 0 0 0 0 0 0 0 0 0 0 0 0 0 278 
278 355 556 556 889 667 191 333 333 389 584 278 333 278 278 556 
556 556 556 556 556 556 556 556 556 278 278 584 584 584 556 1015 
667 667 722 722 667 611 778 722 278 500 667 556 833 722 778 667 
778 722 667 611 722 667 944 667 667 611 278 278 278 469 556 333 
556 556 500 556 556 278 556 556 222 222 500 222 833 556 556 556 
556 333 500 278 556 500 722 500 500 500 334 260 334 584 350 350 
350 222 556 333 1000 556 556 333 1000 667 333 1000 350 350 350 350 
222 222 333 333 350 556 1000 333 1000 500 333 944 350 350 667 278 
333 556 556 556 556 260 556 333 737 370 556 584 584 737 333 400 
584 333 333 333 556 537 278 333 333 365 556 834 834 834 611 667 
667 667 667 667 667 1000 722 667 667 667 667 278 278 278 278 722 
722 778 778 778 778 778 584 778 722 722 722 722 667 667 611 556 
556 556 556 556 556 889 500 556 556 556 556 278 278 278 278 556 
556 556 556 556 556 556 584 611 556 556 556 556 500 556 500 ]
CorelDrawReencodeVect /_R44-Helvetica /Helvetica Z

@sv
@sm
@sv
@sm @sv @sv
@rax 
104.54 163.37 105.55 163.66 @E
0 J 2 j [] 0 d 0 R 0 @G
0.00 0.00 0.00 1.00 K
0 1.008 1.008 0.000 @w
104.54 163.51 m
105.55 163.51 L
S

@rax 
108.50 165.38 109.51 165.67 @E
0 J 2 j [] 0 d 0 R 0 @G
0.00 0.00 0.00 1.00 K
0 1.008 1.008 0.000 @w
108.50 165.53 m
109.51 165.53 L
S

@rax 
112.54 167.40 113.54 167.69 @E
0 J 2 j [] 0 d 0 R 0 @G
0.00 0.00 0.00 1.00 K
0 1.008 1.008 0.000 @w
112.54 167.54 m
113.54 167.54 L
S

@rax 
116.57 169.42 117.50 169.70 @E
0 J 2 j [] 0 d 0 R 0 @G
0.00 0.00 0.00 1.00 K
0 1.008 1.008 0.000 @w
116.57 169.56 m
117.50 169.56 L
S

@rax 
120.53 171.36 121.54 171.65 @E
0 J 2 j [] 0 d 0 R 0 @G
0.00 0.00 0.00 1.00 K
0 1.008 1.008 0.000 @w
120.53 171.50 m
121.54 171.50 L
S

@rax 
124.56 172.51 125.57 173.52 @E
0 J 2 j [] 0 d 0 R 0 @G
0.00 0.00 0.00 1.00 K
0 1.008 1.008 0.000 @w
124.56 172.51 m
125.57 173.52 L
S

@rax 
128.52 174.53 129.53 175.54 @E
0 J 2 j [] 0 d 0 R 0 @G
0.00 0.00 0.00 1.00 K
0 1.008 1.008 0.000 @w
128.52 174.53 m
129.53 175.54 L
S

@rax 
132.55 176.54 133.56 177.55 @E
0 J 2 j [] 0 d 0 R 0 @G
0.00 0.00 0.00 1.00 K
0 1.008 1.008 0.000 @w
132.55 176.54 m
133.56 177.55 L
S

@rax 
136.51 178.42 137.52 178.70 @E
0 J 0 j [] 0 d 0 R 0 @G
0.00 0.00 0.00 1.00 K
0 1.008 1.008 0.000 @w
136.51 178.56 m
136.51 178.56 L
137.52 178.56 L
S

@rax 
140.54 180.36 141.55 180.65 @E
0 J 2 j [] 0 d 0 R 0 @G
0.00 0.00 0.00 1.00 K
0 1.008 1.008 0.000 @w
140.54 180.50 m
141.55 180.50 L
S

@rax 
144.50 182.38 145.51 182.66 @E
0 J 2 j [] 0 d 0 R 0 @G
0.00 0.00 0.00 1.00 K
0 1.008 1.008 0.000 @w
144.50 182.52 m
145.51 182.52 L
S

@rax 
148.54 184.39 149.54 184.68 @E
0 J 2 j [] 0 d 0 R 0 @G
0.00 0.00 0.00 1.00 K
0 1.008 1.008 0.000 @w
148.54 184.54 m
149.54 184.54 L
S

@rax 
152.57 186.41 153.50 186.70 @E
0 J 2 j [] 0 d 0 R 0 @G
0.00 0.00 0.00 1.00 K
0 1.008 1.008 0.000 @w
152.57 186.55 m
153.50 186.55 L
S

@rax 
156.53 188.42 157.54 188.71 @E
0 J 2 j [] 0 d 0 R 0 @G
0.00 0.00 0.00 1.00 K
0 1.008 1.008 0.000 @w
156.53 188.57 m
157.54 188.57 L
S

@rax 
160.56 189.50 161.57 190.51 @E
0 J 2 j [] 0 d 0 R 0 @G
0.00 0.00 0.00 1.00 K
0 1.008 1.008 0.000 @w
160.56 189.50 m
161.57 190.51 L
S

@rax 
164.52 191.52 165.53 192.53 @E
0 J 2 j [] 0 d 0 R 0 @G
0.00 0.00 0.00 1.00 K
0 1.008 1.008 0.000 @w
164.52 191.52 m
165.53 192.53 L
S

@rax 
167.54 193.39 168.55 193.68 @E
0 J 2 j [] 0 d 0 R 0 @G
0.00 0.00 0.00 1.00 K
0 1.008 1.008 0.000 @w
167.54 193.54 m
168.55 193.54 L
S

@rax 
172.51 195.41 173.52 195.70 @E
0 J 2 j [] 0 d 0 R 0 @G
0.00 0.00 0.00 1.00 K
0 1.008 1.008 0.000 @w
172.51 195.55 m
173.52 195.55 L
S

@rax 
176.54 197.42 177.55 197.71 @E
0 J 2 j [] 0 d 0 R 0 @G
0.00 0.00 0.00 1.00 K
0 1.008 1.008 0.000 @w
176.54 197.57 m
177.55 197.57 L
S

@rax 
180.50 198.36 181.51 198.65 @E
0 J 2 j [] 0 d 0 R 0 @G
0.00 0.00 0.00 1.00 K
0 1.008 1.008 0.000 @w
180.50 198.50 m
181.51 198.50 L
S

@rax 
184.54 199.37 185.54 199.66 @E
0 J 2 j [] 0 d 0 R 0 @G
0.00 0.00 0.00 1.00 K
0 1.008 1.008 0.000 @w
184.54 199.51 m
185.54 199.51 L
S

@rax 
188.57 199.37 189.50 199.66 @E
0 J 2 j [] 0 d 0 R 0 @G
0.00 0.00 0.00 1.00 K
0 1.008 1.008 0.000 @w
188.57 199.51 m
189.50 199.51 L
S

@rax 
192.53 199.37 193.54 199.66 @E
0 J 2 j [] 0 d 0 R 0 @G
0.00 0.00 0.00 1.00 K
0 1.008 1.008 0.000 @w
192.53 199.51 m
193.54 199.51 L
S

@rax 
195.55 199.37 196.56 199.66 @E
0 J 2 j [] 0 d 0 R 0 @G
0.00 0.00 0.00 1.00 K
0 1.008 1.008 0.000 @w
195.55 199.51 m
196.56 199.51 L
S

@rax 
199.51 199.37 201.53 199.66 @E
0 J 2 j [] 0 d 0 R 0 @G
0.00 0.00 0.00 1.00 K
0 1.008 1.008 0.000 @w
199.51 199.51 m
201.53 199.51 L
S

@rax 
204.55 199.37 205.56 199.66 @E
0 J 2 j [] 0 d 0 R 0 @G
0.00 0.00 0.00 1.00 K
0 1.008 1.008 0.000 @w
204.55 199.51 m
205.56 199.51 L
S

@rax 
208.51 199.37 209.52 199.66 @E
0 J 2 j [] 0 d 0 R 0 @G
0.00 0.00 0.00 1.00 K
0 1.008 1.008 0.000 @w
208.51 199.51 m
209.52 199.51 L
S

@rax 
212.54 199.37 213.55 199.66 @E
0 J 2 j [] 0 d 0 R 0 @G
0.00 0.00 0.00 1.00 K
0 1.008 1.008 0.000 @w
212.54 199.51 m
213.55 199.51 L
S

@rax 
216.50 199.37 217.51 199.66 @E
0 J 2 j [] 0 d 0 R 0 @G
0.00 0.00 0.00 1.00 K
0 1.008 1.008 0.000 @w
216.50 199.51 m
217.51 199.51 L
S

@rax 
220.54 199.37 221.54 199.66 @E
0 J 2 j [] 0 d 0 R 0 @G
0.00 0.00 0.00 1.00 K
0 1.008 1.008 0.000 @w
220.54 199.51 m
221.54 199.51 L
S

@rax 
223.56 199.37 224.57 199.66 @E
0 J 2 j [] 0 d 0 R 0 @G
0.00 0.00 0.00 1.00 K
0 1.008 1.008 0.000 @w
223.56 199.51 m
224.57 199.51 L
S

@rax 
227.52 199.37 228.53 199.66 @E
0 J 2 j [] 0 d 0 R 0 @G
0.00 0.00 0.00 1.00 K
0 1.008 1.008 0.000 @w
227.52 199.51 m
228.53 199.51 L
S

@rax 
231.55 199.37 232.56 199.66 @E
0 J 2 j [] 0 d 0 R 0 @G
0.00 0.00 0.00 1.00 K
0 1.008 1.008 0.000 @w
231.55 199.51 m
232.56 199.51 L
S

@rax 
235.51 199.37 236.52 199.66 @E
0 J 2 j [] 0 d 0 R 0 @G
0.00 0.00 0.00 1.00 K
0 1.008 1.008 0.000 @w
235.51 199.51 m
236.52 199.51 L
S

@rax 
239.54 199.37 240.55 199.66 @E
0 J 2 j [] 0 d 0 R 0 @G
0.00 0.00 0.00 1.00 K
0 1.008 1.008 0.000 @w
239.54 199.51 m
240.55 199.51 L
S

@rax 
244.37 199.37 244.66 199.66 @E
0 J 2 j [] 0 d 0 R 0 @G
0.00 0.00 0.00 1.00 K
0 1.008 1.008 0.000 @w
244.51 199.51 m
244.51 199.51 L
S

@rax 
247.54 199.37 248.54 199.66 @E
0 J 2 j [] 0 d 0 R 0 @G
0.00 0.00 0.00 1.00 K
0 1.008 1.008 0.000 @w
247.54 199.51 m
248.54 199.51 L
S

@rax 
251.57 199.37 252.50 199.66 @E
0 J 2 j [] 0 d 0 R 0 @G
0.00 0.00 0.00 1.00 K
0 1.008 1.008 0.000 @w
251.57 199.51 m
252.50 199.51 L
S

@rax 
255.53 199.37 256.54 199.66 @E
0 J 2 j [] 0 d 0 R 0 @G
0.00 0.00 0.00 1.00 K
0 1.008 1.008 0.000 @w
255.53 199.51 m
256.54 199.51 L
S

@rax 
260.57 199.37 261.50 199.66 @E
0 J 2 j [] 0 d 0 R 0 @G
0.00 0.00 0.00 1.00 K
0 1.008 1.008 0.000 @w
260.57 199.51 m
261.50 199.51 L
S

@rax 
263.52 200.38 264.53 200.66 @E
0 J 2 j [] 0 d 0 R 0 @G
0.00 0.00 0.00 1.00 K
0 1.008 1.008 0.000 @w
263.52 200.52 m
264.53 200.52 L
S

@rax 
267.55 201.53 268.56 202.54 @E
0 J 2 j [] 0 d 0 R 0 @G
0.00 0.00 0.00 1.00 K
0 1.008 1.008 0.000 @w
267.55 201.53 m
268.56 202.54 L
S

@rax 
272.38 203.54 272.66 204.55 @E
0 J 2 j [] 0 d 0 R 0 @G
0.00 0.00 0.00 1.00 K
0 1.008 1.008 0.000 @w
272.52 203.54 m
272.52 204.55 L
S

@rax 
276.55 206.42 277.56 206.71 @E
0 J 2 j [] 0 d 0 R 0 @G
0.00 0.00 0.00 1.00 K
0 1.008 1.008 0.000 @w
276.55 206.57 m
277.56 206.57 L
S

@rax 
279.50 207.50 280.51 208.51 @E
0 J 2 j [] 0 d 0 R 0 @G
0.00 0.00 0.00 1.00 K
0 1.008 1.008 0.000 @w
279.50 207.50 m
280.51 208.51 L
S

@rax 
283.54 209.38 284.54 209.66 @E
0 J 2 j [] 0 d 0 R 0 @G
0.00 0.00 0.00 1.00 K
0 1.008 1.008 0.000 @w
283.54 209.52 m
284.54 209.52 L
S

@rax 
287.57 211.39 288.50 211.68 @E
0 J 2 j [] 0 d 0 R 0 @G
0.00 0.00 0.00 1.00 K
0 1.008 1.008 0.000 @w
287.57 211.54 m
288.50 211.54 L
S

@rax 
292.39 213.41 292.68 213.70 @E
0 J 2 j [] 0 d 0 R 0 @G
0.00 0.00 0.00 1.00 K
0 1.008 1.008 0.000 @w
292.54 213.55 m
292.54 213.55 L
S

@rax 
295.56 214.56 296.57 215.57 @E
0 J 2 j [] 0 d 0 R 0 @G
0.00 0.00 0.00 1.00 K
0 1.008 1.008 0.000 @w
295.56 214.56 m
296.57 215.57 L
S

@rax 
299.52 217.37 300.53 217.66 @E
0 J 2 j [] 0 d 0 R 0 @G
0.00 0.00 0.00 1.00 K
0 1.008 1.008 0.000 @w
299.52 217.51 m
300.53 217.51 L
S

@rax 
303.55 219.38 304.56 219.67 @E
0 J 2 j [] 0 d 0 R 0 @G
0.00 0.00 0.00 1.00 K
0 1.008 1.008 0.000 @w
303.55 219.53 m
304.56 219.53 L
S

@rax 
307.44 220.54 308.45 221.54 @E
0 J 2 j [] 0 d 0 R 0 @G
0.00 0.00 0.00 1.00 K
0 1.008 1.008 0.000 @w
307.44 220.54 m
308.45 221.54 L
S

@rax 
311.47 222.55 312.48 223.56 @E
0 J 2 j [] 0 d 0 R 0 @G
0.00 0.00 0.00 1.00 K
0 1.008 1.008 0.000 @w
311.47 222.55 m
312.48 223.56 L
S

@rax 
315.43 224.57 316.44 225.50 @E
0 J 2 j [] 0 d 0 R 0 @G
0.00 0.00 0.00 1.00 K
0 1.008 1.008 0.000 @w
315.43 224.57 m
316.44 225.50 L
S

@rax 
319.46 226.51 320.47 227.52 @E
0 J 2 j [] 0 d 0 R 0 @G
0.00 0.00 0.00 1.00 K
0 1.008 1.008 0.000 @w
319.46 226.51 m
320.47 227.52 L
S

@rax 
323.50 228.38 324.43 228.67 @E
0 J 2 j [] 0 d 0 R 0 @G
0.00 0.00 0.00 1.00 K
0 1.008 1.008 0.000 @w
323.50 228.53 m
324.43 228.53 L
S

@rax 
327.46 230.40 328.46 230.69 @E
0 J 2 j [] 0 d 0 R 0 @G
0.00 0.00 0.00 1.00 K
0 1.008 1.008 0.000 @w
327.46 230.54 m
328.46 230.54 L
S

@rax 
331.49 232.42 332.50 232.70 @E
0 J 2 j [] 0 d 0 R 0 @G
0.00 0.00 0.00 1.00 K
0 1.008 1.008 0.000 @w
331.49 232.56 m
332.50 232.56 L
S

@rax 
335.45 234.36 336.46 234.65 @E
0 J 2 j [] 0 d 0 R 0 @G
0.00 0.00 0.00 1.00 K
0 1.008 1.008 0.000 @w
335.45 234.50 m
336.46 234.50 L
S

@rax 
339.48 236.38 340.49 236.66 @E
0 J 2 j [] 0 d 0 R 0 @G
0.00 0.00 0.00 1.00 K
0 1.008 1.008 0.000 @w
339.48 236.52 m
340.49 236.52 L
S

@rax 
104.54 86.54 340.49 198.50 @E
0 J 0 j [] 0 d 0 R 0 @G
0.00 0.00 0.00 1.00 K
0 1.800 1.800 0.000 @w
104.54 86.54 m
105.55 87.55 L
116.57 92.52 L
127.51 97.56 L
136.51 102.53 L
142.56 104.54 L
151.56 109.51 L
156.53 111.53 L
162.50 114.55 L
165.53 115.56 L
173.52 119.52 L
177.55 121.54 L
183.53 124.56 L
189.50 127.51 L
193.54 129.53 L
201.53 132.55 L
204.55 134.57 L
208.51 136.51 L
214.56 138.53 L
216.50 139.54 L
221.54 142.56 L
225.50 144.50 L
230.54 146.52 L
236.52 149.54 L
244.51 153.50 L
252.50 157.54 L
256.54 159.55 L
261.50 161.57 L
266.54 163.51 L
272.52 166.54 L
277.56 168.55 L
282.53 171.50 L
286.56 173.52 L
292.54 175.54 L
300.53 179.57 L
312.48 185.54 L
321.48 189.50 L
329.47 193.54 L
340.49 198.50 L
S

@rax 
77.54 67.39 83.52 67.68 @E
0 J 2 j [] 0 d 0 R 0 @G
0.00 0.00 0.00 1.00 K
0 1.008 1.008 0.000 @w
77.54 67.54 m
83.52 67.54 L
S

@rax 
77.54 98.42 83.52 98.71 @E
0 J 2 j [] 0 d 0 R 0 @G
0.00 0.00 0.00 1.00 K
0 1.008 1.008 0.000 @w
77.54 98.57 m
83.52 98.57 L
S

@rax 
77.54 130.39 83.52 130.68 @E
0 J 2 j [] 0 d 0 R 0 @G
0.00 0.00 0.00 1.00 K
0 1.008 1.008 0.000 @w
77.54 130.54 m
83.52 130.54 L
S

@rax 
77.54 161.42 83.52 161.71 @E
0 J 2 j [] 0 d 0 R 0 @G
0.00 0.00 0.00 1.00 K
0 1.008 1.008 0.000 @w
77.54 161.57 m
83.52 161.57 L
S

@rax 
77.54 192.38 83.52 192.67 @E
0 J 2 j [] 0 d 0 R 0 @G
0.00 0.00 0.00 1.00 K
0 1.008 1.008 0.000 @w
77.54 192.53 m
83.52 192.53 L
S

@rax 
77.54 224.42 83.52 224.71 @E
0 J 2 j [] 0 d 0 R 0 @G
0.00 0.00 0.00 1.00 K
0 1.008 1.008 0.000 @w
77.54 224.57 m
83.52 224.57 L
S

@rax 
77.54 255.38 83.52 255.67 @E
0 J 2 j [] 0 d 0 R 0 @G
0.00 0.00 0.00 1.00 K
0 1.008 1.008 0.000 @w
77.54 255.53 m
83.52 255.53 L
S

@rax 
77.54 76.39 80.57 76.68 @E
0 J 2 j [] 0 d 0 R 0 @G
0.00 0.00 0.00 1.00 K
0 1.008 1.008 0.000 @w
77.54 76.54 m
80.57 76.54 L
S

@rax 
77.54 82.37 80.57 82.66 @E
0 J 2 j [] 0 d 0 R 0 @G
0.00 0.00 0.00 1.00 K
0 1.008 1.008 0.000 @w
77.54 82.51 m
80.57 82.51 L
S

@rax 
77.54 86.40 80.57 86.69 @E
0 J 2 j [] 0 d 0 R 0 @G
0.00 0.00 0.00 1.00 K
0 1.008 1.008 0.000 @w
77.54 86.54 m
80.57 86.54 L
S

@rax 
77.54 89.42 80.57 89.71 @E
0 J 2 j [] 0 d 0 R 0 @G
0.00 0.00 0.00 1.00 K
0 1.008 1.008 0.000 @w
77.54 89.57 m
80.57 89.57 L
S

@rax 
77.54 91.37 80.57 91.66 @E
0 J 2 j [] 0 d 0 R 0 @G
0.00 0.00 0.00 1.00 K
0 1.008 1.008 0.000 @w
77.54 91.51 m
80.57 91.51 L
S

@rax 
77.54 94.39 80.57 94.68 @E
0 J 2 j [] 0 d 0 R 0 @G
0.00 0.00 0.00 1.00 K
0 1.008 1.008 0.000 @w
77.54 94.54 m
80.57 94.54 L
S

@rax 
77.54 95.40 80.57 95.69 @E
0 J 2 j [] 0 d 0 R 0 @G
0.00 0.00 0.00 1.00 K
0 1.008 1.008 0.000 @w
77.54 95.54 m
80.57 95.54 L
S

@rax 
77.54 97.42 80.57 97.70 @E
0 J 2 j [] 0 d 0 R 0 @G
0.00 0.00 0.00 1.00 K
0 1.008 1.008 0.000 @w
77.54 97.56 m
80.57 97.56 L
S

@rax 
77.54 108.36 80.57 108.65 @E
0 J 2 j [] 0 d 0 R 0 @G
0.00 0.00 0.00 1.00 K
0 1.008 1.008 0.000 @w
77.54 108.50 m
80.57 108.50 L
S

@rax 
77.54 113.40 80.57 113.69 @E
0 J 2 j [] 0 d 0 R 0 @G
0.00 0.00 0.00 1.00 K
0 1.008 1.008 0.000 @w
77.54 113.54 m
80.57 113.54 L
S

@rax 
77.54 117.36 80.57 117.65 @E
0 J 2 j [] 0 d 0 R 0 @G
0.00 0.00 0.00 1.00 K
0 1.008 1.008 0.000 @w
77.54 117.50 m
80.57 117.50 L
S

@rax 
77.54 120.38 80.57 120.67 @E
0 J 2 j [] 0 d 0 R 0 @G
0.00 0.00 0.00 1.00 K
0 1.008 1.008 0.000 @w
77.54 120.53 m
80.57 120.53 L
S

@rax 
77.54 123.41 80.57 123.70 @E
0 J 2 j [] 0 d 0 R 0 @G
0.00 0.00 0.00 1.00 K
0 1.008 1.008 0.000 @w
77.54 123.55 m
80.57 123.55 L
S

@rax 
77.54 125.42 80.57 125.71 @E
0 J 2 j [] 0 d 0 R 0 @G
0.00 0.00 0.00 1.00 K
0 1.008 1.008 0.000 @w
77.54 125.57 m
80.57 125.57 L
S

@rax 
77.54 127.37 80.57 127.66 @E
0 J 2 j [] 0 d 0 R 0 @G
0.00 0.00 0.00 1.00 K
0 1.008 1.008 0.000 @w
77.54 127.51 m
80.57 127.51 L
S

@rax 
77.54 128.38 80.57 128.66 @E
0 J 2 j [] 0 d 0 R 0 @G
0.00 0.00 0.00 1.00 K
0 1.008 1.008 0.000 @w
77.54 128.52 m
80.57 128.52 L
S

@rax 
77.54 139.39 80.57 139.68 @E
0 J 2 j [] 0 d 0 R 0 @G
0.00 0.00 0.00 1.00 K
0 1.008 1.008 0.000 @w
77.54 139.54 m
80.57 139.54 L
S

@rax 
77.54 145.37 80.57 145.66 @E
0 J 2 j [] 0 d 0 R 0 @G
0.00 0.00 0.00 1.00 K
0 1.008 1.008 0.000 @w
77.54 145.51 m
80.57 145.51 L
S

@rax 
77.54 148.39 80.57 148.68 @E
0 J 2 j [] 0 d 0 R 0 @G
0.00 0.00 0.00 1.00 K
0 1.008 1.008 0.000 @w
77.54 148.54 m
80.57 148.54 L
S

@rax 
77.54 152.42 80.57 152.71 @E
0 J 2 j [] 0 d 0 R 0 @G
0.00 0.00 0.00 1.00 K
0 1.008 1.008 0.000 @w
77.54 152.57 m
80.57 152.57 L
S

@rax 
77.54 154.37 80.57 154.66 @E
0 J 2 j [] 0 d 0 R 0 @G
0.00 0.00 0.00 1.00 K
0 1.008 1.008 0.000 @w
77.54 154.51 m
80.57 154.51 L
S

@rax 
77.54 156.38 80.57 156.67 @E
0 J 2 j [] 0 d 0 R 0 @G
0.00 0.00 0.00 1.00 K
0 1.008 1.008 0.000 @w
77.54 156.53 m
80.57 156.53 L
S

@rax 
77.54 158.40 80.57 158.69 @E
0 J 2 j [] 0 d 0 R 0 @G
0.00 0.00 0.00 1.00 K
0 1.008 1.008 0.000 @w
77.54 158.54 m
80.57 158.54 L
S

@rax 
77.54 159.41 80.57 159.70 @E
0 J 2 j [] 0 d 0 R 0 @G
0.00 0.00 0.00 1.00 K
0 1.008 1.008 0.000 @w
77.54 159.55 m
80.57 159.55 L
S

@rax 
77.54 170.42 80.57 170.71 @E
0 J 2 j [] 0 d 0 R 0 @G
0.00 0.00 0.00 1.00 K
0 1.008 1.008 0.000 @w
77.54 170.57 m
80.57 170.57 L
S

@rax 
77.54 176.40 80.57 176.69 @E
0 J 2 j [] 0 d 0 R 0 @G
0.00 0.00 0.00 1.00 K
0 1.008 1.008 0.000 @w
77.54 176.54 m
80.57 176.54 L
S

@rax 
77.54 180.36 80.57 180.65 @E
0 J 2 j [] 0 d 0 R 0 @G
0.00 0.00 0.00 1.00 K
0 1.008 1.008 0.000 @w
77.54 180.50 m
80.57 180.50 L
S

@rax 
77.54 183.38 80.57 183.67 @E
0 J 2 j [] 0 d 0 R 0 @G
0.00 0.00 0.00 1.00 K
0 1.008 1.008 0.000 @w
77.54 183.53 m
80.57 183.53 L
S

@rax 
77.54 185.40 80.57 185.69 @E
0 J 2 j [] 0 d 0 R 0 @G
0.00 0.00 0.00 1.00 K
0 1.008 1.008 0.000 @w
77.54 185.54 m
80.57 185.54 L
S

@rax 
77.54 188.42 80.57 188.71 @E
0 J 2 j [] 0 d 0 R 0 @G
0.00 0.00 0.00 1.00 K
0 1.008 1.008 0.000 @w
77.54 188.57 m
80.57 188.57 L
S

@rax 
77.54 189.36 80.57 189.65 @E
0 J 2 j [] 0 d 0 R 0 @G
0.00 0.00 0.00 1.00 K
0 1.008 1.008 0.000 @w
77.54 189.50 m
80.57 189.50 L
S

@rax 
77.54 191.38 80.57 191.66 @E
0 J 2 j [] 0 d 0 R 0 @G
0.00 0.00 0.00 1.00 K
0 1.008 1.008 0.000 @w
77.54 191.52 m
80.57 191.52 L
S

@rax 
77.54 202.39 80.57 202.68 @E
0 J 2 j [] 0 d 0 R 0 @G
0.00 0.00 0.00 1.00 K
0 1.008 1.008 0.000 @w
77.54 202.54 m
80.57 202.54 L
S

@rax 
77.54 207.36 80.57 207.65 @E
0 J 2 j [] 0 d 0 R 0 @G
0.00 0.00 0.00 1.00 K
0 1.008 1.008 0.000 @w
77.54 207.50 m
80.57 207.50 L
S

@rax 
77.54 211.39 80.57 211.68 @E
0 J 2 j [] 0 d 0 R 0 @G
0.00 0.00 0.00 1.00 K
0 1.008 1.008 0.000 @w
77.54 211.54 m
80.57 211.54 L
S

@rax 
77.54 214.42 80.57 214.70 @E
0 J 2 j [] 0 d 0 R 0 @G
0.00 0.00 0.00 1.00 K
0 1.008 1.008 0.000 @w
77.54 214.56 m
80.57 214.56 L
S

@rax 
77.54 217.37 80.57 217.66 @E
0 J 2 j [] 0 d 0 R 0 @G
0.00 0.00 0.00 1.00 K
0 1.008 1.008 0.000 @w
77.54 217.51 m
80.57 217.51 L
S

@rax 
77.54 219.38 80.57 219.67 @E
0 J 2 j [] 0 d 0 R 0 @G
0.00 0.00 0.00 1.00 K
0 1.008 1.008 0.000 @w
77.54 219.53 m
80.57 219.53 L
S

@rax 
77.54 221.40 80.57 221.69 @E
0 J 2 j [] 0 d 0 R 0 @G
0.00 0.00 0.00 1.00 K
0 1.008 1.008 0.000 @w
77.54 221.54 m
80.57 221.54 L
S

@rax 
77.54 222.41 80.57 222.70 @E
0 J 2 j [] 0 d 0 R 0 @G
0.00 0.00 0.00 1.00 K
0 1.008 1.008 0.000 @w
77.54 222.55 m
80.57 222.55 L
S

@rax 
77.54 233.42 80.57 233.71 @E
0 J 2 j [] 0 d 0 R 0 @G
0.00 0.00 0.00 1.00 K
0 1.008 1.008 0.000 @w
77.54 233.57 m
80.57 233.57 L
S

@rax 
77.54 239.40 80.57 239.69 @E
0 J 2 j [] 0 d 0 R 0 @G
0.00 0.00 0.00 1.00 K
0 1.008 1.008 0.000 @w
77.54 239.54 m
80.57 239.54 L
S

@rax 
77.54 243.36 80.57 243.65 @E
0 J 2 j [] 0 d 0 R 0 @G
0.00 0.00 0.00 1.00 K
0 1.008 1.008 0.000 @w
77.54 243.50 m
80.57 243.50 L
S

@rax 
77.54 246.38 80.57 246.67 @E
0 J 2 j [] 0 d 0 R 0 @G
0.00 0.00 0.00 1.00 K
0 1.008 1.008 0.000 @w
77.54 246.53 m
80.57 246.53 L
S

@rax 
77.54 248.40 80.57 248.69 @E
0 J 2 j [] 0 d 0 R 0 @G
0.00 0.00 0.00 1.00 K
0 1.008 1.008 0.000 @w
77.54 248.54 m
80.57 248.54 L
S

@rax 
77.54 250.42 80.57 250.70 @E
0 J 2 j [] 0 d 0 R 0 @G
0.00 0.00 0.00 1.00 K
0 1.008 1.008 0.000 @w
77.54 250.56 m
80.57 250.56 L
S

@rax 
77.54 252.36 80.57 252.65 @E
0 J 2 j [] 0 d 0 R 0 @G
0.00 0.00 0.00 1.00 K
0 1.008 1.008 0.000 @w
77.54 252.50 m
80.57 252.50 L
S

@rax 
77.54 254.38 80.57 254.66 @E
0 J 2 j [] 0 d 0 R 0 @G
0.00 0.00 0.00 1.00 K
0 1.008 1.008 0.000 @w
77.54 254.52 m
80.57 254.52 L
S

@rax 
335.45 67.39 341.50 67.68 @E
0 J 2 j [] 0 d 0 R 0 @G
0.00 0.00 0.00 1.00 K
0 1.008 1.008 0.000 @w
341.50 67.54 m
335.45 67.54 L
S

@rax 
335.45 98.42 341.50 98.71 @E
0 J 2 j [] 0 d 0 R 0 @G
0.00 0.00 0.00 1.00 K
0 1.008 1.008 0.000 @w
341.50 98.57 m
335.45 98.57 L
S

@rax 
335.45 130.39 341.50 130.68 @E
0 J 2 j [] 0 d 0 R 0 @G
0.00 0.00 0.00 1.00 K
0 1.008 1.008 0.000 @w
341.50 130.54 m
335.45 130.54 L
S

@rax 
335.45 161.42 341.50 161.71 @E
0 J 2 j [] 0 d 0 R 0 @G
0.00 0.00 0.00 1.00 K
0 1.008 1.008 0.000 @w
341.50 161.57 m
335.45 161.57 L
S

@rax 
335.45 192.38 341.50 192.67 @E
0 J 2 j [] 0 d 0 R 0 @G
0.00 0.00 0.00 1.00 K
0 1.008 1.008 0.000 @w
341.50 192.53 m
335.45 192.53 L
S

@rax 
335.45 224.42 341.50 224.71 @E
0 J 2 j [] 0 d 0 R 0 @G
0.00 0.00 0.00 1.00 K
0 1.008 1.008 0.000 @w
341.50 224.57 m
335.45 224.57 L
S

@rax 
335.45 255.38 341.50 255.67 @E
0 J 2 j [] 0 d 0 R 0 @G
0.00 0.00 0.00 1.00 K
0 1.008 1.008 0.000 @w
341.50 255.53 m
335.45 255.53 L
S

@rax 
338.47 76.39 341.50 76.68 @E
0 J 2 j [] 0 d 0 R 0 @G
0.00 0.00 0.00 1.00 K
0 1.008 1.008 0.000 @w
341.50 76.54 m
338.47 76.54 L
S

@rax 
338.47 82.37 341.50 82.66 @E
0 J 2 j [] 0 d 0 R 0 @G
0.00 0.00 0.00 1.00 K
0 1.008 1.008 0.000 @w
341.50 82.51 m
338.47 82.51 L
S

@rax 
338.47 86.40 341.50 86.69 @E
0 J 2 j [] 0 d 0 R 0 @G
0.00 0.00 0.00 1.00 K
0 1.008 1.008 0.000 @w
341.50 86.54 m
338.47 86.54 L
S

@rax 
338.47 89.42 341.50 89.71 @E
0 J 2 j [] 0 d 0 R 0 @G
0.00 0.00 0.00 1.00 K
0 1.008 1.008 0.000 @w
341.50 89.57 m
338.47 89.57 L
S

@rax 
338.47 91.37 341.50 91.66 @E
0 J 2 j [] 0 d 0 R 0 @G
0.00 0.00 0.00 1.00 K
0 1.008 1.008 0.000 @w
341.50 91.51 m
338.47 91.51 L
S

@rax 
338.47 94.39 341.50 94.68 @E
0 J 2 j [] 0 d 0 R 0 @G
0.00 0.00 0.00 1.00 K
0 1.008 1.008 0.000 @w
341.50 94.54 m
338.47 94.54 L
S

@rax 
338.47 95.40 341.50 95.69 @E
0 J 2 j [] 0 d 0 R 0 @G
0.00 0.00 0.00 1.00 K
0 1.008 1.008 0.000 @w
341.50 95.54 m
338.47 95.54 L
S

@rax 
338.47 97.42 341.50 97.70 @E
0 J 2 j [] 0 d 0 R 0 @G
0.00 0.00 0.00 1.00 K
0 1.008 1.008 0.000 @w
341.50 97.56 m
338.47 97.56 L
S

@rax 
338.47 108.36 341.50 108.65 @E
0 J 2 j [] 0 d 0 R 0 @G
0.00 0.00 0.00 1.00 K
0 1.008 1.008 0.000 @w
341.50 108.50 m
338.47 108.50 L
S

@rax 
338.47 113.40 341.50 113.69 @E
0 J 2 j [] 0 d 0 R 0 @G
0.00 0.00 0.00 1.00 K
0 1.008 1.008 0.000 @w
341.50 113.54 m
338.47 113.54 L
S

@rax 
338.47 117.36 341.50 117.65 @E
0 J 2 j [] 0 d 0 R 0 @G
0.00 0.00 0.00 1.00 K
0 1.008 1.008 0.000 @w
341.50 117.50 m
338.47 117.50 L
S

@rax 
338.47 120.38 341.50 120.67 @E
0 J 2 j [] 0 d 0 R 0 @G
0.00 0.00 0.00 1.00 K
0 1.008 1.008 0.000 @w
341.50 120.53 m
338.47 120.53 L
S

@rax 
338.47 123.41 341.50 123.70 @E
0 J 2 j [] 0 d 0 R 0 @G
0.00 0.00 0.00 1.00 K
0 1.008 1.008 0.000 @w
341.50 123.55 m
338.47 123.55 L
S

@rax 
338.47 125.42 341.50 125.71 @E
0 J 2 j [] 0 d 0 R 0 @G
0.00 0.00 0.00 1.00 K
0 1.008 1.008 0.000 @w
341.50 125.57 m
338.47 125.57 L
S

@rax 
338.47 127.37 341.50 127.66 @E
0 J 2 j [] 0 d 0 R 0 @G
0.00 0.00 0.00 1.00 K
0 1.008 1.008 0.000 @w
341.50 127.51 m
338.47 127.51 L
S

@rax 
338.47 128.38 341.50 128.66 @E
0 J 2 j [] 0 d 0 R 0 @G
0.00 0.00 0.00 1.00 K
0 1.008 1.008 0.000 @w
341.50 128.52 m
338.47 128.52 L
S

@rax 
338.47 139.39 341.50 139.68 @E
0 J 2 j [] 0 d 0 R 0 @G
0.00 0.00 0.00 1.00 K
0 1.008 1.008 0.000 @w
341.50 139.54 m
338.47 139.54 L
S

@rax 
338.47 145.37 341.50 145.66 @E
0 J 2 j [] 0 d 0 R 0 @G
0.00 0.00 0.00 1.00 K
0 1.008 1.008 0.000 @w
341.50 145.51 m
338.47 145.51 L
S

@rax 
338.47 148.39 341.50 148.68 @E
0 J 2 j [] 0 d 0 R 0 @G
0.00 0.00 0.00 1.00 K
0 1.008 1.008 0.000 @w
341.50 148.54 m
338.47 148.54 L
S

@rax 
338.47 152.42 341.50 152.71 @E
0 J 2 j [] 0 d 0 R 0 @G
0.00 0.00 0.00 1.00 K
0 1.008 1.008 0.000 @w
341.50 152.57 m
338.47 152.57 L
S

@rax 
338.47 154.37 341.50 154.66 @E
0 J 2 j [] 0 d 0 R 0 @G
0.00 0.00 0.00 1.00 K
0 1.008 1.008 0.000 @w
341.50 154.51 m
338.47 154.51 L
S

@rax 
338.47 156.38 341.50 156.67 @E
0 J 2 j [] 0 d 0 R 0 @G
0.00 0.00 0.00 1.00 K
0 1.008 1.008 0.000 @w
341.50 156.53 m
338.47 156.53 L
S

@rax 
338.47 158.40 341.50 158.69 @E
0 J 2 j [] 0 d 0 R 0 @G
0.00 0.00 0.00 1.00 K
0 1.008 1.008 0.000 @w
341.50 158.54 m
338.47 158.54 L
S

@rax 
338.47 159.41 341.50 159.70 @E
0 J 2 j [] 0 d 0 R 0 @G
0.00 0.00 0.00 1.00 K
0 1.008 1.008 0.000 @w
341.50 159.55 m
338.47 159.55 L
S

@rax 
338.47 170.42 341.50 170.71 @E
0 J 2 j [] 0 d 0 R 0 @G
0.00 0.00 0.00 1.00 K
0 1.008 1.008 0.000 @w
341.50 170.57 m
338.47 170.57 L
S

@rax 
338.47 176.40 341.50 176.69 @E
0 J 2 j [] 0 d 0 R 0 @G
0.00 0.00 0.00 1.00 K
0 1.008 1.008 0.000 @w
341.50 176.54 m
338.47 176.54 L
S

@rax 
338.47 180.36 341.50 180.65 @E
0 J 2 j [] 0 d 0 R 0 @G
0.00 0.00 0.00 1.00 K
0 1.008 1.008 0.000 @w
341.50 180.50 m
338.47 180.50 L
S

@rax 
338.47 183.38 341.50 183.67 @E
0 J 2 j [] 0 d 0 R 0 @G
0.00 0.00 0.00 1.00 K
0 1.008 1.008 0.000 @w
341.50 183.53 m
338.47 183.53 L
S

@rax 
338.47 185.40 341.50 185.69 @E
0 J 2 j [] 0 d 0 R 0 @G
0.00 0.00 0.00 1.00 K
0 1.008 1.008 0.000 @w
341.50 185.54 m
338.47 185.54 L
S

@rax 
338.47 188.42 341.50 188.71 @E
0 J 2 j [] 0 d 0 R 0 @G
0.00 0.00 0.00 1.00 K
0 1.008 1.008 0.000 @w
341.50 188.57 m
338.47 188.57 L
S

@rax 
338.47 189.36 341.50 189.65 @E
0 J 2 j [] 0 d 0 R 0 @G
0.00 0.00 0.00 1.00 K
0 1.008 1.008 0.000 @w
341.50 189.50 m
338.47 189.50 L
S

@rax 
338.47 191.38 341.50 191.66 @E
0 J 2 j [] 0 d 0 R 0 @G
0.00 0.00 0.00 1.00 K
0 1.008 1.008 0.000 @w
341.50 191.52 m
338.47 191.52 L
S

@rax 
338.47 202.39 341.50 202.68 @E
0 J 2 j [] 0 d 0 R 0 @G
0.00 0.00 0.00 1.00 K
0 1.008 1.008 0.000 @w
341.50 202.54 m
338.47 202.54 L
S

@rax 
338.47 207.36 341.50 207.65 @E
0 J 2 j [] 0 d 0 R 0 @G
0.00 0.00 0.00 1.00 K
0 1.008 1.008 0.000 @w
341.50 207.50 m
338.47 207.50 L
S

@rax 
338.47 211.39 341.50 211.68 @E
0 J 2 j [] 0 d 0 R 0 @G
0.00 0.00 0.00 1.00 K
0 1.008 1.008 0.000 @w
341.50 211.54 m
338.47 211.54 L
S

@rax 
338.47 214.42 341.50 214.70 @E
0 J 2 j [] 0 d 0 R 0 @G
0.00 0.00 0.00 1.00 K
0 1.008 1.008 0.000 @w
341.50 214.56 m
338.47 214.56 L
S

@rax 
338.47 217.37 341.50 217.66 @E
0 J 2 j [] 0 d 0 R 0 @G
0.00 0.00 0.00 1.00 K
0 1.008 1.008 0.000 @w
341.50 217.51 m
338.47 217.51 L
S

@rax 
338.47 219.38 341.50 219.67 @E
0 J 2 j [] 0 d 0 R 0 @G
0.00 0.00 0.00 1.00 K
0 1.008 1.008 0.000 @w
341.50 219.53 m
338.47 219.53 L
S

@rax 
338.47 221.40 341.50 221.69 @E
0 J 2 j [] 0 d 0 R 0 @G
0.00 0.00 0.00 1.00 K
0 1.008 1.008 0.000 @w
341.50 221.54 m
338.47 221.54 L
S

@rax 
338.47 222.41 341.50 222.70 @E
0 J 2 j [] 0 d 0 R 0 @G
0.00 0.00 0.00 1.00 K
0 1.008 1.008 0.000 @w
341.50 222.55 m
338.47 222.55 L
S

@rax 
338.47 233.42 341.50 233.71 @E
0 J 2 j [] 0 d 0 R 0 @G
0.00 0.00 0.00 1.00 K
0 1.008 1.008 0.000 @w
341.50 233.57 m
338.47 233.57 L
S

@rax 
338.47 239.40 341.50 239.69 @E
0 J 2 j [] 0 d 0 R 0 @G
0.00 0.00 0.00 1.00 K
0 1.008 1.008 0.000 @w
341.50 239.54 m
338.47 239.54 L
S

@rax 
338.47 243.36 341.50 243.65 @E
0 J 2 j [] 0 d 0 R 0 @G
0.00 0.00 0.00 1.00 K
0 1.008 1.008 0.000 @w
341.50 243.50 m
338.47 243.50 L
S

@rax 
338.47 246.38 341.50 246.67 @E
0 J 2 j [] 0 d 0 R 0 @G
0.00 0.00 0.00 1.00 K
0 1.008 1.008 0.000 @w
341.50 246.53 m
338.47 246.53 L
S

@rax 
338.47 248.40 341.50 248.69 @E
0 J 2 j [] 0 d 0 R 0 @G
0.00 0.00 0.00 1.00 K
0 1.008 1.008 0.000 @w
341.50 248.54 m
338.47 248.54 L
S

@rax 
338.47 250.42 341.50 250.70 @E
0 J 2 j [] 0 d 0 R 0 @G
0.00 0.00 0.00 1.00 K
0 1.008 1.008 0.000 @w
341.50 250.56 m
338.47 250.56 L
S

@rax 
338.47 252.36 341.50 252.65 @E
0 J 2 j [] 0 d 0 R 0 @G
0.00 0.00 0.00 1.00 K
0 1.008 1.008 0.000 @w
341.50 252.50 m
338.47 252.50 L
S

@rax 
338.47 254.38 341.50 254.66 @E
0 J 2 j [] 0 d 0 R 0 @G
0.00 0.00 0.00 1.00 K
0 1.008 1.008 0.000 @w
341.50 254.52 m
338.47 254.52 L
S

@rax 
341.35 67.54 341.64 73.51 @E
0 J 2 j [] 0 d 0 R 0 @G
0.00 0.00 0.00 1.00 K
0 1.008 1.008 0.000 @w
341.50 67.54 m
341.50 73.51 L
S

@rax 
275.40 67.54 275.69 73.51 @E
0 J 2 j [] 0 d 0 R 0 @G
0.00 0.00 0.00 1.00 K
0 1.008 1.008 0.000 @w
275.54 67.54 m
275.54 73.51 L
S

@rax 
209.38 67.54 209.66 73.51 @E
0 J 2 j [] 0 d 0 R 0 @G
0.00 0.00 0.00 1.00 K
0 1.008 1.008 0.000 @w
209.52 67.54 m
209.52 73.51 L
S

@rax 
143.42 67.54 143.71 73.51 @E
0 J 2 j [] 0 d 0 R 0 @G
0.00 0.00 0.00 1.00 K
0 1.008 1.008 0.000 @w
143.57 67.54 m
143.57 73.51 L
S

@rax 
77.40 67.54 77.69 73.51 @E
0 J 2 j [] 0 d 0 R 0 @G
0.00 0.00 0.00 1.00 K
0 1.008 1.008 0.000 @w
77.54 67.54 m
77.54 73.51 L
S

@rax 
321.34 67.54 321.62 70.56 @E
0 J 2 j [] 0 d 0 R 0 @G
0.00 0.00 0.00 1.00 K
0 1.008 1.008 0.000 @w
321.48 67.54 m
321.48 70.56 L
S

@rax 
310.32 67.54 310.61 70.56 @E
0 J 2 j [] 0 d 0 R 0 @G
0.00 0.00 0.00 1.00 K
0 1.008 1.008 0.000 @w
310.46 67.54 m
310.46 70.56 L
S

@rax 
301.39 67.54 301.68 70.56 @E
0 J 2 j [] 0 d 0 R 0 @G
0.00 0.00 0.00 1.00 K
0 1.008 1.008 0.000 @w
301.54 67.54 m
301.54 70.56 L
S

@rax 
295.42 67.54 295.70 70.56 @E
0 J 2 j [] 0 d 0 R 0 @G
0.00 0.00 0.00 1.00 K
0 1.008 1.008 0.000 @w
295.56 67.54 m
295.56 70.56 L
S

@rax 
290.38 67.54 290.66 70.56 @E
0 J 2 j [] 0 d 0 R 0 @G
0.00 0.00 0.00 1.00 K
0 1.008 1.008 0.000 @w
290.52 67.54 m
290.52 70.56 L
S

@rax 
285.41 67.54 285.70 70.56 @E
0 J 2 j [] 0 d 0 R 0 @G
0.00 0.00 0.00 1.00 K
0 1.008 1.008 0.000 @w
285.55 67.54 m
285.55 70.56 L
S

@rax 
282.38 67.54 282.67 70.56 @E
0 J 2 j [] 0 d 0 R 0 @G
0.00 0.00 0.00 1.00 K
0 1.008 1.008 0.000 @w
282.53 67.54 m
282.53 70.56 L
S

@rax 
278.42 67.54 278.71 70.56 @E
0 J 2 j [] 0 d 0 R 0 @G
0.00 0.00 0.00 1.00 K
0 1.008 1.008 0.000 @w
278.57 67.54 m
278.57 70.56 L
S

@rax 
255.38 67.54 255.67 70.56 @E
0 J 2 j [] 0 d 0 R 0 @G
0.00 0.00 0.00 1.00 K
0 1.008 1.008 0.000 @w
255.53 67.54 m
255.53 70.56 L
S

@rax 
244.37 67.54 244.66 70.56 @E
0 J 2 j [] 0 d 0 R 0 @G
0.00 0.00 0.00 1.00 K
0 1.008 1.008 0.000 @w
244.51 67.54 m
244.51 70.56 L
S

@rax 
235.37 67.54 235.66 70.56 @E
0 J 2 j [] 0 d 0 R 0 @G
0.00 0.00 0.00 1.00 K
0 1.008 1.008 0.000 @w
235.51 67.54 m
235.51 70.56 L
S

@rax 
229.39 67.54 229.68 70.56 @E
0 J 2 j [] 0 d 0 R 0 @G
0.00 0.00 0.00 1.00 K
0 1.008 1.008 0.000 @w
229.54 67.54 m
229.54 70.56 L
S

@rax 
224.42 67.54 224.71 70.56 @E
0 J 2 j [] 0 d 0 R 0 @G
0.00 0.00 0.00 1.00 K
0 1.008 1.008 0.000 @w
224.57 67.54 m
224.57 70.56 L
S

@rax 
219.38 67.54 219.67 70.56 @E
0 J 2 j [] 0 d 0 R 0 @G
0.00 0.00 0.00 1.00 K
0 1.008 1.008 0.000 @w
219.53 67.54 m
219.53 70.56 L
S

@rax 
216.36 67.54 216.65 70.56 @E
0 J 2 j [] 0 d 0 R 0 @G
0.00 0.00 0.00 1.00 K
0 1.008 1.008 0.000 @w
216.50 67.54 m
216.50 70.56 L
S

@rax 
212.40 67.54 212.69 70.56 @E
0 J 2 j [] 0 d 0 R 0 @G
0.00 0.00 0.00 1.00 K
0 1.008 1.008 0.000 @w
212.54 67.54 m
212.54 70.56 L
S

@rax 
189.36 67.54 189.65 70.56 @E
0 J 2 j [] 0 d 0 R 0 @G
0.00 0.00 0.00 1.00 K
0 1.008 1.008 0.000 @w
189.50 67.54 m
189.50 70.56 L
S

@rax 
177.41 67.54 177.70 70.56 @E
0 J 2 j [] 0 d 0 R 0 @G
0.00 0.00 0.00 1.00 K
0 1.008 1.008 0.000 @w
177.55 67.54 m
177.55 70.56 L
S

@rax 
169.42 67.54 169.70 70.56 @E
0 J 2 j [] 0 d 0 R 0 @G
0.00 0.00 0.00 1.00 K
0 1.008 1.008 0.000 @w
169.56 67.54 m
169.56 70.56 L
S

@rax 
163.37 67.54 163.66 70.56 @E
0 J 2 j [] 0 d 0 R 0 @G
0.00 0.00 0.00 1.00 K
0 1.008 1.008 0.000 @w
163.51 67.54 m
163.51 70.56 L
S

@rax 
158.40 67.54 158.69 70.56 @E
0 J 2 j [] 0 d 0 R 0 @G
0.00 0.00 0.00 1.00 K
0 1.008 1.008 0.000 @w
158.54 67.54 m
158.54 70.56 L
S

@rax 
153.36 67.54 153.65 70.56 @E
0 J 2 j [] 0 d 0 R 0 @G
0.00 0.00 0.00 1.00 K
0 1.008 1.008 0.000 @w
153.50 67.54 m
153.50 70.56 L
S

@rax 
149.40 67.54 149.69 70.56 @E
0 J 2 j [] 0 d 0 R 0 @G
0.00 0.00 0.00 1.00 K
0 1.008 1.008 0.000 @w
149.54 67.54 m
149.54 70.56 L
S

@rax 
146.38 67.54 146.66 70.56 @E
0 J 2 j [] 0 d 0 R 0 @G
0.00 0.00 0.00 1.00 K
0 1.008 1.008 0.000 @w
146.52 67.54 m
146.52 70.56 L
S

@rax 
123.41 67.54 123.70 70.56 @E
0 J 2 j [] 0 d 0 R 0 @G
0.00 0.00 0.00 1.00 K
0 1.008 1.008 0.000 @w
123.55 67.54 m
123.55 70.56 L
S

@rax 
111.38 67.54 111.67 70.56 @E
0 J 2 j [] 0 d 0 R 0 @G
0.00 0.00 0.00 1.00 K
0 1.008 1.008 0.000 @w
111.53 67.54 m
111.53 70.56 L
S

@rax 
103.39 67.54 103.68 70.56 @E
0 J 2 j [] 0 d 0 R 0 @G
0.00 0.00 0.00 1.00 K
0 1.008 1.008 0.000 @w
103.54 67.54 m
103.54 70.56 L
S

@rax 
97.42 67.54 97.70 70.56 @E
0 J 2 j [] 0 d 0 R 0 @G
0.00 0.00 0.00 1.00 K
0 1.008 1.008 0.000 @w
97.56 67.54 m
97.56 70.56 L
S

@rax 
91.37 67.54 91.66 70.56 @E
0 J 2 j [] 0 d 0 R 0 @G
0.00 0.00 0.00 1.00 K
0 1.008 1.008 0.000 @w
91.51 67.54 m
91.51 70.56 L
S

@rax 
87.41 67.54 87.70 70.56 @E
0 J 2 j [] 0 d 0 R 0 @G
0.00 0.00 0.00 1.00 K
0 1.008 1.008 0.000 @w
87.55 67.54 m
87.55 70.56 L
S

@rax 
83.38 67.54 83.66 70.56 @E
0 J 2 j [] 0 d 0 R 0 @G
0.00 0.00 0.00 1.00 K
0 1.008 1.008 0.000 @w
83.52 67.54 m
83.52 70.56 L
S

@rax 
80.42 67.54 80.71 70.56 @E
0 J 2 j [] 0 d 0 R 0 @G
0.00 0.00 0.00 1.00 K
0 1.008 1.008 0.000 @w
80.57 67.54 m
80.57 70.56 L
S

@rax 
341.35 249.55 341.64 255.53 @E
0 J 2 j [] 0 d 0 R 0 @G
0.00 0.00 0.00 1.00 K
0 1.008 1.008 0.000 @w
341.50 255.53 m
341.50 249.55 L
S

@rax 
275.40 249.55 275.69 255.53 @E
0 J 2 j [] 0 d 0 R 0 @G
0.00 0.00 0.00 1.00 K
0 1.008 1.008 0.000 @w
275.54 255.53 m
275.54 249.55 L
S

@rax 
209.38 249.55 209.66 255.53 @E
0 J 2 j [] 0 d 0 R 0 @G
0.00 0.00 0.00 1.00 K
0 1.008 1.008 0.000 @w
209.52 255.53 m
209.52 249.55 L
S

@rax 
143.42 249.55 143.71 255.53 @E
0 J 2 j [] 0 d 0 R 0 @G
0.00 0.00 0.00 1.00 K
0 1.008 1.008 0.000 @w
143.57 255.53 m
143.57 249.55 L
S

@rax 
77.40 249.55 77.69 255.53 @E
0 J 2 j [] 0 d 0 R 0 @G
0.00 0.00 0.00 1.00 K
0 1.008 1.008 0.000 @w
77.54 255.53 m
77.54 249.55 L
S

@rax 
321.34 252.50 321.62 255.53 @E
0 J 2 j [] 0 d 0 R 0 @G
0.00 0.00 0.00 1.00 K
0 1.008 1.008 0.000 @w
321.48 255.53 m
321.48 252.50 L
S

@rax 
310.32 252.50 310.61 255.53 @E
0 J 2 j [] 0 d 0 R 0 @G
0.00 0.00 0.00 1.00 K
0 1.008 1.008 0.000 @w
310.46 255.53 m
310.46 252.50 L
S

@rax 
301.39 252.50 301.68 255.53 @E
0 J 2 j [] 0 d 0 R 0 @G
0.00 0.00 0.00 1.00 K
0 1.008 1.008 0.000 @w
301.54 255.53 m
301.54 252.50 L
S

@rax 
295.42 252.50 295.70 255.53 @E
0 J 2 j [] 0 d 0 R 0 @G
0.00 0.00 0.00 1.00 K
0 1.008 1.008 0.000 @w
295.56 255.53 m
295.56 252.50 L
S

@rax 
290.38 252.50 290.66 255.53 @E
0 J 2 j [] 0 d 0 R 0 @G
0.00 0.00 0.00 1.00 K
0 1.008 1.008 0.000 @w
290.52 255.53 m
290.52 252.50 L
S

@rax 
285.41 252.50 285.70 255.53 @E
0 J 2 j [] 0 d 0 R 0 @G
0.00 0.00 0.00 1.00 K
0 1.008 1.008 0.000 @w
285.55 255.53 m
285.55 252.50 L
S

@rax 
282.38 252.50 282.67 255.53 @E
0 J 2 j [] 0 d 0 R 0 @G
0.00 0.00 0.00 1.00 K
0 1.008 1.008 0.000 @w
282.53 255.53 m
282.53 252.50 L
S

@rax 
278.42 252.50 278.71 255.53 @E
0 J 2 j [] 0 d 0 R 0 @G
0.00 0.00 0.00 1.00 K
0 1.008 1.008 0.000 @w
278.57 255.53 m
278.57 252.50 L
S

@rax 
255.38 252.50 255.67 255.53 @E
0 J 2 j [] 0 d 0 R 0 @G
0.00 0.00 0.00 1.00 K
0 1.008 1.008 0.000 @w
255.53 255.53 m
255.53 252.50 L
S

@rax 
244.37 252.50 244.66 255.53 @E
0 J 2 j [] 0 d 0 R 0 @G
0.00 0.00 0.00 1.00 K
0 1.008 1.008 0.000 @w
244.51 255.53 m
244.51 252.50 L
S

@rax 
235.37 252.50 235.66 255.53 @E
0 J 2 j [] 0 d 0 R 0 @G
0.00 0.00 0.00 1.00 K
0 1.008 1.008 0.000 @w
235.51 255.53 m
235.51 252.50 L
S

@rax 
229.39 252.50 229.68 255.53 @E
0 J 2 j [] 0 d 0 R 0 @G
0.00 0.00 0.00 1.00 K
0 1.008 1.008 0.000 @w
229.54 255.53 m
229.54 252.50 L
S

@rax 
224.42 252.50 224.71 255.53 @E
0 J 2 j [] 0 d 0 R 0 @G
0.00 0.00 0.00 1.00 K
0 1.008 1.008 0.000 @w
224.57 255.53 m
224.57 252.50 L
S

@rax 
219.38 252.50 219.67 255.53 @E
0 J 2 j [] 0 d 0 R 0 @G
0.00 0.00 0.00 1.00 K
0 1.008 1.008 0.000 @w
219.53 255.53 m
219.53 252.50 L
S

@rax 
216.36 252.50 216.65 255.53 @E
0 J 2 j [] 0 d 0 R 0 @G
0.00 0.00 0.00 1.00 K
0 1.008 1.008 0.000 @w
216.50 255.53 m
216.50 252.50 L
S

@rax 
212.40 252.50 212.69 255.53 @E
0 J 2 j [] 0 d 0 R 0 @G
0.00 0.00 0.00 1.00 K
0 1.008 1.008 0.000 @w
212.54 255.53 m
212.54 252.50 L
S

@rax 
189.36 252.50 189.65 255.53 @E
0 J 2 j [] 0 d 0 R 0 @G
0.00 0.00 0.00 1.00 K
0 1.008 1.008 0.000 @w
189.50 255.53 m
189.50 252.50 L
S

@rax 
177.41 252.50 177.70 255.53 @E
0 J 2 j [] 0 d 0 R 0 @G
0.00 0.00 0.00 1.00 K
0 1.008 1.008 0.000 @w
177.55 255.53 m
177.55 252.50 L
S

@rax 
169.42 252.50 169.70 255.53 @E
0 J 2 j [] 0 d 0 R 0 @G
0.00 0.00 0.00 1.00 K
0 1.008 1.008 0.000 @w
169.56 255.53 m
169.56 252.50 L
S

@rax 
163.37 252.50 163.66 255.53 @E
0 J 2 j [] 0 d 0 R 0 @G
0.00 0.00 0.00 1.00 K
0 1.008 1.008 0.000 @w
163.51 255.53 m
163.51 252.50 L
S

@rax 
158.40 252.50 158.69 255.53 @E
0 J 2 j [] 0 d 0 R 0 @G
0.00 0.00 0.00 1.00 K
0 1.008 1.008 0.000 @w
158.54 255.53 m
158.54 252.50 L
S

@rax 
153.36 252.50 153.65 255.53 @E
0 J 2 j [] 0 d 0 R 0 @G
0.00 0.00 0.00 1.00 K
0 1.008 1.008 0.000 @w
153.50 255.53 m
153.50 252.50 L
S

@rax 
149.40 252.50 149.69 255.53 @E
0 J 2 j [] 0 d 0 R 0 @G
0.00 0.00 0.00 1.00 K
0 1.008 1.008 0.000 @w
149.54 255.53 m
149.54 252.50 L
S

@rax 
146.38 252.50 146.66 255.53 @E
0 J 2 j [] 0 d 0 R 0 @G
0.00 0.00 0.00 1.00 K
0 1.008 1.008 0.000 @w
146.52 255.53 m
146.52 252.50 L
S

@rax 
123.41 252.50 123.70 255.53 @E
0 J 2 j [] 0 d 0 R 0 @G
0.00 0.00 0.00 1.00 K
0 1.008 1.008 0.000 @w
123.55 255.53 m
123.55 252.50 L
S

@rax 
111.38 252.50 111.67 255.53 @E
0 J 2 j [] 0 d 0 R 0 @G
0.00 0.00 0.00 1.00 K
0 1.008 1.008 0.000 @w
111.53 255.53 m
111.53 252.50 L
S

@rax 
103.39 252.50 103.68 255.53 @E
0 J 2 j [] 0 d 0 R 0 @G
0.00 0.00 0.00 1.00 K
0 1.008 1.008 0.000 @w
103.54 255.53 m
103.54 252.50 L
S

@rax 
97.42 252.50 97.70 255.53 @E
0 J 2 j [] 0 d 0 R 0 @G
0.00 0.00 0.00 1.00 K
0 1.008 1.008 0.000 @w
97.56 255.53 m
97.56 252.50 L
S

@rax 
91.37 252.50 91.66 255.53 @E
0 J 2 j [] 0 d 0 R 0 @G
0.00 0.00 0.00 1.00 K
0 1.008 1.008 0.000 @w
91.51 255.53 m
91.51 252.50 L
S

@rax 
87.41 252.50 87.70 255.53 @E
0 J 2 j [] 0 d 0 R 0 @G
0.00 0.00 0.00 1.00 K
0 1.008 1.008 0.000 @w
87.55 255.53 m
87.55 252.50 L
S

@rax 
83.38 252.50 83.66 255.53 @E
0 J 2 j [] 0 d 0 R 0 @G
0.00 0.00 0.00 1.00 K
0 1.008 1.008 0.000 @w
83.52 255.53 m
83.52 252.50 L
S

@rax 
80.42 252.50 80.71 255.53 @E
0 J 2 j [] 0 d 0 R 0 @G
0.00 0.00 0.00 1.00 K
0 1.008 1.008 0.000 @w
80.57 255.53 m
80.57 252.50 L
S

@rax 
77.40 67.54 77.69 255.53 @E
0 J 2 j [] 0 d 0 R 0 @G
0.00 0.00 0.00 1.00 K
0 1.440 1.440 0.000 @w
77.54 67.54 m
77.54 255.53 L
S

@rax 
341.35 67.54 341.64 255.53 @E
0 J 2 j [] 0 d 0 R 0 @G
0.00 0.00 0.00 1.00 K
0 1.440 1.440 0.000 @w
341.50 67.54 m
341.50 255.53 L
S

@rax 
77.54 67.39 341.50 67.68 @E
0 J 2 j [] 0 d 0 R 0 @G
0.00 0.00 0.00 1.00 K
0 1.440 1.440 0.000 @w
77.54 67.54 m
341.50 67.54 L
S

@rax 
77.54 255.38 341.50 255.67 @E
0 J 2 j [] 0 d 0 R 0 @G
0.00 0.00 0.00 1.00 K
0 1.440 1.440 0.000 @w
77.54 255.53 m
341.50 255.53 L
S

@rax 
102.53 169.56 105.55 172.51 @E
0 J 0 j [] 0 d 0 R 0 @G
0.00 0.00 0.00 1.00 K
0 1.008 1.008 0.000 @w
105.55 171.00 m
105.55 170.21 104.83 169.56 104.04 169.56 c
103.25 169.56 102.53 170.21 102.53 171.00 c
102.53 171.86 103.25 172.51 104.04 172.51 c
104.83 172.51 105.55 171.86 105.55 171.00 c
@c
S

@rax 
103.54 170.57 106.56 173.52 @E
0 J 0 j [] 0 d 0 R 0 @G
0.00 0.00 0.00 1.00 K
0 1.008 1.008 0.000 @w
106.56 172.01 m
106.56 171.22 105.84 170.57 105.05 170.57 c
104.18 170.57 103.54 171.22 103.54 172.01 c
103.54 172.87 104.18 173.52 105.05 173.52 c
105.84 173.52 106.56 172.87 106.56 172.01 c
@c
S

@rax 
114.55 173.52 118.51 176.54 @E
0 J 0 j [] 0 d 0 R 0 @G
0.00 0.00 0.00 1.00 K
0 1.008 1.008 0.000 @w
118.51 175.03 m
118.51 174.24 117.65 173.52 116.57 173.52 c
115.42 173.52 114.55 174.24 114.55 175.03 c
114.55 175.82 115.42 176.54 116.57 176.54 c
117.65 176.54 118.51 175.82 118.51 175.03 c
@c
S

@rax 
125.57 173.52 128.52 177.55 @E
0 J 0 j [] 0 d 0 R 0 @G
0.00 0.00 0.00 1.00 K
0 1.008 1.008 0.000 @w
128.52 175.54 m
128.52 174.46 127.87 173.52 127.01 173.52 c
126.22 173.52 125.57 174.46 125.57 175.54 c
125.57 176.62 126.22 177.55 127.01 177.55 c
127.87 177.55 128.52 176.62 128.52 175.54 c
@c
S

@rax 
134.57 175.54 137.52 178.56 @E
0 J 0 j [] 0 d 0 R 0 @G
0.00 0.00 0.00 1.00 K
0 1.008 1.008 0.000 @w
137.52 177.05 m
137.52 176.18 136.87 175.54 136.01 175.54 c
135.22 175.54 134.57 176.18 134.57 177.05 c
134.57 177.84 135.22 178.56 136.01 178.56 c
136.87 178.56 137.52 177.84 137.52 177.05 c
@c
S

@rax 
140.54 174.53 143.57 177.55 @E
0 J 0 j [] 0 d 0 R 0 @G
0.00 0.00 0.00 1.00 K
0 1.008 1.008 0.000 @w
143.57 176.04 m
143.57 175.25 142.85 174.53 142.06 174.53 c
141.19 174.53 140.54 175.25 140.54 176.04 c
140.54 176.83 141.19 177.55 142.06 177.55 c
142.85 177.55 143.57 176.83 143.57 176.04 c
@c
S

@rax 
149.54 176.54 152.57 179.57 @E
0 J 0 j [] 0 d 0 R 0 @G
0.00 0.00 0.00 1.00 K
0 1.008 1.008 0.000 @w
152.57 178.06 m
152.57 177.19 151.85 176.54 151.06 176.54 c
150.19 176.54 149.54 177.19 149.54 178.06 c
149.54 178.85 150.19 179.57 151.06 179.57 c
151.85 179.57 152.57 178.85 152.57 178.06 c
@c
S

@rax 
154.51 178.56 158.54 182.52 @E
0 J 0 j [] 0 d 0 R 0 @G
0.00 0.00 0.00 1.00 K
0 1.008 1.008 0.000 @w
158.54 180.50 m
158.54 179.42 157.61 178.56 156.53 178.56 c
155.45 178.56 154.51 179.42 154.51 180.50 c
154.51 181.66 155.45 182.52 156.53 182.52 c
157.61 182.52 158.54 181.66 158.54 180.50 c
@c
S

@rax 
160.56 179.57 163.51 182.52 @E
0 J 0 j [] 0 d 0 R 0 @G
0.00 0.00 0.00 1.00 K
0 1.008 1.008 0.000 @w
163.51 181.01 m
163.51 180.22 162.86 179.57 162.00 179.57 c
161.21 179.57 160.56 180.22 160.56 181.01 c
160.56 181.87 161.21 182.52 162.00 182.52 c
162.86 182.52 163.51 181.87 163.51 181.01 c
@c
S

@rax 
163.51 183.53 166.54 186.55 @E
0 J 0 j [] 0 d 0 R 0 @G
0.00 0.00 0.00 1.00 K
0 1.008 1.008 0.000 @w
166.54 185.04 m
166.54 184.25 165.89 183.53 165.02 183.53 c
164.23 183.53 163.51 184.25 163.51 185.04 c
163.51 185.83 164.23 186.55 165.02 186.55 c
165.89 186.55 166.54 185.83 166.54 185.04 c
@c
S

@rax 
171.50 182.52 174.53 185.54 @E
0 J 0 j [] 0 d 0 R 0 @G
0.00 0.00 0.00 1.00 K
0 1.008 1.008 0.000 @w
174.53 184.03 m
174.53 183.24 173.88 182.52 173.02 182.52 c
172.22 182.52 171.50 183.24 171.50 184.03 c
171.50 184.82 172.22 185.54 173.02 185.54 c
173.88 185.54 174.53 184.82 174.53 184.03 c
@c
S

@rax 
175.54 183.53 178.56 186.55 @E
0 J 0 j [] 0 d 0 R 0 @G
0.00 0.00 0.00 1.00 K
0 1.008 1.008 0.000 @w
178.56 185.04 m
178.56 184.25 177.84 183.53 177.05 183.53 c
176.18 183.53 175.54 184.25 175.54 185.04 c
175.54 185.83 176.18 186.55 177.05 186.55 c
177.84 186.55 178.56 185.83 178.56 185.04 c
@c
S

@rax 
181.51 184.54 184.54 188.57 @E
0 J 0 j [] 0 d 0 R 0 @G
0.00 0.00 0.00 1.00 K
0 1.008 1.008 0.000 @w
184.54 186.55 m
184.54 185.47 183.89 184.54 183.02 184.54 c
182.23 184.54 181.51 185.47 181.51 186.55 c
181.51 187.63 182.23 188.57 183.02 188.57 c
183.89 188.57 184.54 187.63 184.54 186.55 c
@c
S

@rax 
187.56 188.57 190.51 191.52 @E
0 J 0 j [] 0 d 0 R 0 @G
0.00 0.00 0.00 1.00 K
0 1.008 1.008 0.000 @w
190.51 190.01 m
190.51 189.22 189.86 188.57 189.00 188.57 c
188.21 188.57 187.56 189.22 187.56 190.01 c
187.56 190.87 188.21 191.52 189.00 191.52 c
189.86 191.52 190.51 190.87 190.51 190.01 c
@c
S

@rax 
192.53 190.51 195.55 193.54 @E
0 J 0 j [] 0 d 0 R 0 @G
0.00 0.00 0.00 1.00 K
0 1.008 1.008 0.000 @w
195.55 192.02 m
195.55 191.23 194.83 190.51 194.04 190.51 c
193.25 190.51 192.53 191.23 192.53 192.02 c
192.53 192.89 193.25 193.54 194.04 193.54 c
194.83 193.54 195.55 192.89 195.55 192.02 c
@c
S

@rax 
199.51 190.51 202.54 193.54 @E
0 J 0 j [] 0 d 0 R 0 @G
0.00 0.00 0.00 1.00 K
0 1.008 1.008 0.000 @w
202.54 192.02 m
202.54 191.23 201.89 190.51 201.02 190.51 c
200.23 190.51 199.51 191.23 199.51 192.02 c
199.51 192.89 200.23 193.54 201.02 193.54 c
201.89 193.54 202.54 192.89 202.54 192.02 c
@c
S

@rax 
202.54 191.52 205.56 194.54 @E
0 J 0 j [] 0 d 0 R 0 @G
0.00 0.00 0.00 1.00 K
0 1.008 1.008 0.000 @w
205.56 193.03 m
205.56 192.24 204.84 191.52 204.05 191.52 c
203.18 191.52 202.54 192.24 202.54 193.03 c
202.54 193.82 203.18 194.54 204.05 194.54 c
204.84 194.54 205.56 193.82 205.56 193.03 c
@c
S

@rax 
206.57 191.52 209.52 194.54 @E
0 J 0 j [] 0 d 0 R 0 @G
0.00 0.00 0.00 1.00 K
0 1.008 1.008 0.000 @w
209.52 193.03 m
209.52 192.24 208.87 191.52 208.01 191.52 c
207.22 191.52 206.57 192.24 206.57 193.03 c
206.57 193.82 207.22 194.54 208.01 194.54 c
208.87 194.54 209.52 193.82 209.52 193.03 c
@c
S

@rax 
212.54 193.54 215.57 196.56 @E
0 J 0 j [] 0 d 0 R 0 @G
0.00 0.00 0.00 1.00 K
0 1.008 1.008 0.000 @w
215.57 195.05 m
215.57 194.18 214.85 193.54 214.06 193.54 c
213.19 193.54 212.54 194.18 212.54 195.05 c
212.54 195.84 213.19 196.56 214.06 196.56 c
214.85 196.56 215.57 195.84 215.57 195.05 c
@c
S

@rax 
214.56 194.54 217.51 197.57 @E
0 J 0 j [] 0 d 0 R 0 @G
0.00 0.00 0.00 1.00 K
0 1.008 1.008 0.000 @w
217.51 196.06 m
217.51 195.19 216.86 194.54 216.00 194.54 c
215.21 194.54 214.56 195.19 214.56 196.06 c
214.56 196.85 215.21 197.57 216.00 197.57 c
216.86 197.57 217.51 196.85 217.51 196.06 c
@c
S

@rax 
219.53 195.55 222.55 198.50 @E
0 J 0 j [] 0 d 0 R 0 @G
0.00 0.00 0.00 1.00 K
0 1.008 1.008 0.000 @w
222.55 197.06 m
222.55 196.20 221.83 195.55 221.04 195.55 c
220.25 195.55 219.53 196.20 219.53 197.06 c
219.53 197.86 220.25 198.50 221.04 198.50 c
221.83 198.50 222.55 197.86 222.55 197.06 c
@c
S

@rax 
223.56 196.56 226.51 199.51 @E
0 J 0 j [] 0 d 0 R 0 @G
0.00 0.00 0.00 1.00 K
0 1.008 1.008 0.000 @w
226.51 198.00 m
226.51 197.21 225.86 196.56 225.00 196.56 c
224.21 196.56 223.56 197.21 223.56 198.00 c
223.56 198.86 224.21 199.51 225.00 199.51 c
225.86 199.51 226.51 198.86 226.51 198.00 c
@c
S

@rax 
228.53 198.50 231.55 201.53 @E
0 J 0 j [] 0 d 0 R 0 @G
0.00 0.00 0.00 1.00 K
0 1.008 1.008 0.000 @w
231.55 200.02 m
231.55 199.22 230.83 198.50 230.04 198.50 c
229.25 198.50 228.53 199.22 228.53 200.02 c
228.53 200.88 229.25 201.53 230.04 201.53 c
230.83 201.53 231.55 200.88 231.55 200.02 c
@c
S

@rax 
234.50 199.51 237.53 202.54 @E
0 J 0 j [] 0 d 0 R 0 @G
0.00 0.00 0.00 1.00 K
0 1.008 1.008 0.000 @w
237.53 201.02 m
237.53 200.23 236.88 199.51 236.02 199.51 c
235.22 199.51 234.50 200.23 234.50 201.02 c
234.50 201.89 235.22 202.54 236.02 202.54 c
236.88 202.54 237.53 201.89 237.53 201.02 c
@c
S

@rax 
243.50 202.54 246.53 205.56 @E
0 J 0 j [] 0 d 0 R 0 @G
0.00 0.00 0.00 1.00 K
0 1.008 1.008 0.000 @w
246.53 204.05 m
246.53 203.18 245.88 202.54 245.02 202.54 c
244.22 202.54 243.50 203.18 243.50 204.05 c
243.50 204.84 244.22 205.56 245.02 205.56 c
245.88 205.56 246.53 204.84 246.53 204.05 c
@c
S

@rax 
250.56 204.55 253.51 207.50 @E
0 J 0 j [] 0 d 0 R 0 @G
0.00 0.00 0.00 1.00 K
0 1.008 1.008 0.000 @w
253.51 206.06 m
253.51 205.20 252.86 204.55 252.00 204.55 c
251.21 204.55 250.56 205.20 250.56 206.06 c
250.56 206.86 251.21 207.50 252.00 207.50 c
252.86 207.50 253.51 206.86 253.51 206.06 c
@c
S

@rax 
255.53 205.56 258.55 208.51 @E
0 J 0 j [] 0 d 0 R 0 @G
0.00 0.00 0.00 1.00 K
0 1.008 1.008 0.000 @w
258.55 207.00 m
258.55 206.21 257.83 205.56 257.04 205.56 c
256.25 205.56 255.53 206.21 255.53 207.00 c
255.53 207.86 256.25 208.51 257.04 208.51 c
257.83 208.51 258.55 207.86 258.55 207.00 c
@c
S

@rax 
259.56 207.50 262.51 210.53 @E
0 J 0 j [] 0 d 0 R 0 @G
0.00 0.00 0.00 1.00 K
0 1.008 1.008 0.000 @w
262.51 209.02 m
262.51 208.22 261.86 207.50 261.00 207.50 c
260.21 207.50 259.56 208.22 259.56 209.02 c
259.56 209.88 260.21 210.53 261.00 210.53 c
261.86 210.53 262.51 209.88 262.51 209.02 c
@c
S

@rax 
264.53 208.51 267.55 211.54 @E
0 J 0 j [] 0 d 0 R 0 @G
0.00 0.00 0.00 1.00 K
0 1.008 1.008 0.000 @w
267.55 210.02 m
267.55 209.23 266.83 208.51 266.04 208.51 c
265.25 208.51 264.53 209.23 264.53 210.02 c
264.53 210.89 265.25 211.54 266.04 211.54 c
266.83 211.54 267.55 210.89 267.55 210.02 c
@c
S

@rax 
270.50 210.53 273.53 213.55 @E
0 J 0 j [] 0 d 0 R 0 @G
0.00 0.00 0.00 1.00 K
0 1.008 1.008 0.000 @w
273.53 212.04 m
273.53 211.25 272.88 210.53 272.02 210.53 c
271.22 210.53 270.50 211.25 270.50 212.04 c
270.50 212.83 271.22 213.55 272.02 213.55 c
272.88 213.55 273.53 212.83 273.53 212.04 c
@c
S

@rax 
275.54 211.54 278.57 214.56 @E
0 J 0 j [] 0 d 0 R 0 @G
0.00 0.00 0.00 1.00 K
0 1.008 1.008 0.000 @w
278.57 213.05 m
278.57 212.18 277.85 211.54 277.06 211.54 c
276.19 211.54 275.54 212.18 275.54 213.05 c
275.54 213.84 276.19 214.56 277.06 214.56 c
277.85 214.56 278.57 213.84 278.57 213.05 c
@c
S

@rax 
280.51 212.54 283.54 215.57 @E
0 J 0 j [] 0 d 0 R 0 @G
0.00 0.00 0.00 1.00 K
0 1.008 1.008 0.000 @w
283.54 214.06 m
283.54 213.19 282.89 212.54 282.02 212.54 c
281.23 212.54 280.51 213.19 280.51 214.06 c
280.51 214.85 281.23 215.57 282.02 215.57 c
282.89 215.57 283.54 214.85 283.54 214.06 c
@c
S

@rax 
284.54 213.55 287.57 216.50 @E
0 J 0 j [] 0 d 0 R 0 @G
0.00 0.00 0.00 1.00 K
0 1.008 1.008 0.000 @w
287.57 215.06 m
287.57 214.20 286.85 213.55 286.06 213.55 c
285.19 213.55 284.54 214.20 284.54 215.06 c
284.54 215.86 285.19 216.50 286.06 216.50 c
286.85 216.50 287.57 215.86 287.57 215.06 c
@c
S

@rax 
290.52 216.50 293.54 219.53 @E
0 J 0 j [] 0 d 0 R 0 @G
0.00 0.00 0.00 1.00 K
0 1.008 1.008 0.000 @w
293.54 218.02 m
293.54 217.22 292.82 216.50 292.03 216.50 c
291.24 216.50 290.52 217.22 290.52 218.02 c
290.52 218.88 291.24 219.53 292.03 219.53 c
292.82 219.53 293.54 218.88 293.54 218.02 c
@c
S

@rax 
298.51 217.51 301.54 220.54 @E
0 J 0 j [] 0 d 0 R 0 @G
0.00 0.00 0.00 1.00 K
0 1.008 1.008 0.000 @w
301.54 219.02 m
301.54 218.23 300.89 217.51 300.02 217.51 c
299.23 217.51 298.51 218.23 298.51 219.02 c
298.51 219.89 299.23 220.54 300.02 220.54 c
300.89 220.54 301.54 219.89 301.54 219.02 c
@c
S

@rax 
310.46 220.54 313.49 223.56 @E
0 J 0 j [] 0 d 0 R 0 @G
0.00 0.00 0.00 1.00 K
0 1.008 1.008 0.000 @w
313.49 222.05 m
313.49 221.18 312.77 220.54 311.98 220.54 c
311.11 220.54 310.46 221.18 310.46 222.05 c
310.46 222.84 311.11 223.56 311.98 223.56 c
312.77 223.56 313.49 222.84 313.49 222.05 c
@c
S

@rax 
319.46 222.55 322.49 225.50 @E
0 J 0 j [] 0 d 0 R 0 @G
0.00 0.00 0.00 1.00 K
0 1.008 1.008 0.000 @w
322.49 224.06 m
322.49 223.20 321.77 222.55 320.98 222.55 c
320.11 222.55 319.46 223.20 319.46 224.06 c
319.46 224.86 320.11 225.50 320.98 225.50 c
321.77 225.50 322.49 224.86 322.49 224.06 c
@c
S

@rax 
327.46 227.52 330.48 231.55 @E
0 J 0 j [] 0 d 0 R 0 @G
0.00 0.00 0.00 1.00 K
0 1.008 1.008 0.000 @w
330.48 229.54 m
330.48 228.46 329.76 227.52 328.97 227.52 c
328.18 227.52 327.46 228.46 327.46 229.54 c
327.46 230.62 328.18 231.55 328.97 231.55 c
329.76 231.55 330.48 230.62 330.48 229.54 c
@c
S

@rax 
338.47 231.55 341.50 234.50 @E
0 J 0 j [] 0 d 0 R 0 @G
0.00 0.00 0.00 1.00 K
0 1.008 1.008 0.000 @w
341.50 233.06 m
341.50 232.20 340.78 231.55 339.98 231.55 c
339.12 231.55 338.47 232.20 338.47 233.06 c
338.47 233.86 339.12 234.50 339.98 234.50 c
340.78 234.50 341.50 233.86 341.50 233.06 c
@c
S

@rax 49.54 64.73 62.35 73.44 @E
[0.07200 0.00000 0.00000 0.07200 49.53600 65.01600] @tm
 0 O 0 @g
0.00 0.00 0.00 1.00 k
e
/_R44-Helvetica 166.00 z
0 0 (10) @t
T
@rax 49.54 64.73 62.35 73.44 @E
[0.07200 0.00000 0.00000 0.07200 49.53600 65.01600] @tm
0 J 0 j [] 0 d 0 R 0 @G
0.00 0.00 0.00 1.00 K
0 0.216 0.216 0.000 @w
r
/_R44-Helvetica 166.00 z
0 0 (10) @t
T
@rax 64.51 70.85 68.62 76.61 @E
[0.07200 0.00000 0.00000 0.07200 64.51200 71.06400] @tm
 0 O 0 @g
0.00 0.00 0.00 1.00 k
e
/_R44-Helvetica 110.00 z
0 0 (0) @t
T
@rax 64.51 70.85 68.62 76.61 @E
[0.07200 0.00000 0.00000 0.07200 64.51200 71.06400] @tm
0 J 0 j [] 0 d 0 R 0 @G
0.00 0.00 0.00 1.00 K
0 0.216 0.216 0.000 @w
r
/_R44-Helvetica 110.00 z
0 0 (0) @t
T
@rax 49.54 95.76 62.35 104.47 @E
[0.07200 0.00000 0.00000 0.07200 49.53600 96.04800] @tm
 0 O 0 @g
0.00 0.00 0.00 1.00 k
e
/_R44-Helvetica 166.00 z
0 0 (10) @t
T
@rax 49.54 95.76 62.35 104.47 @E
[0.07200 0.00000 0.00000 0.07200 49.53600 96.04800] @tm
0 J 0 j [] 0 d 0 R 0 @G
0.00 0.00 0.00 1.00 K
0 0.216 0.216 0.000 @w
r
/_R44-Helvetica 166.00 z
0 0 (10) @t
T
@rax 64.51 103.03 67.32 108.58 @E
[0.07200 0.00000 0.00000 0.07200 64.51200 103.03200] @tm
 0 O 0 @g
0.00 0.00 0.00 1.00 k
e
/_R44-Helvetica 110.00 z
0 0 (1) @t
T
@rax 64.51 103.03 67.32 108.58 @E
[0.07200 0.00000 0.00000 0.07200 64.51200 103.03200] @tm
0 J 0 j [] 0 d 0 R 0 @G
0.00 0.00 0.00 1.00 K
0 0.216 0.216 0.000 @w
r
/_R44-Helvetica 110.00 z
0 0 (1) @t
T
@rax 49.54 127.73 62.35 136.44 @E
[0.07200 0.00000 0.00000 0.07200 49.53600 128.01599] @tm
 0 O 0 @g
0.00 0.00 0.00 1.00 k
e
/_R44-Helvetica 166.00 z
0 0 (10) @t
T
@rax 49.54 127.73 62.35 136.44 @E
[0.07200 0.00000 0.00000 0.07200 49.53600 128.01599] @tm
0 J 0 j [] 0 d 0 R 0 @G
0.00 0.00 0.00 1.00 K
0 0.216 0.216 0.000 @w
r
/_R44-Helvetica 166.00 z
0 0 (10) @t
T
@rax 64.51 134.06 68.54 139.61 @E
[0.07200 0.00000 0.00000 0.07200 64.51200 134.06399] @tm
 0 O 0 @g
0.00 0.00 0.00 1.00 k
e
/_R44-Helvetica 110.00 z
0 0 (2) @t
T
@rax 64.51 134.06 68.54 139.61 @E
[0.07200 0.00000 0.00000 0.07200 64.51200 134.06399] @tm
0 J 0 j [] 0 d 0 R 0 @G
0.00 0.00 0.00 1.00 K
0 0.216 0.216 0.000 @w
r
/_R44-Helvetica 110.00 z
0 0 (2) @t
T
@rax 49.54 158.76 62.35 167.47 @E
[0.07200 0.00000 0.00000 0.07200 49.53600 159.04799] @tm
 0 O 0 @g
0.00 0.00 0.00 1.00 k
e
/_R44-Helvetica 166.00 z
0 0 (10) @t
T
@rax 49.54 158.76 62.35 167.47 @E
[0.07200 0.00000 0.00000 0.07200 49.53600 159.04799] @tm
0 J 0 j [] 0 d 0 R 0 @G
0.00 0.00 0.00 1.00 K
0 0.216 0.216 0.000 @w
r
/_R44-Helvetica 166.00 z
0 0 (10) @t
T
@rax 64.51 164.81 68.62 170.57 @E
[0.07200 0.00000 0.00000 0.07200 64.51200 165.02399] @tm
 0 O 0 @g
0.00 0.00 0.00 1.00 k
e
/_R44-Helvetica 110.00 z
0 0 (3) @t
T
@rax 64.51 164.81 68.62 170.57 @E
[0.07200 0.00000 0.00000 0.07200 64.51200 165.02399] @tm
0 J 0 j [] 0 d 0 R 0 @G
0.00 0.00 0.00 1.00 K
0 0.216 0.216 0.000 @w
r
/_R44-Helvetica 110.00 z
0 0 (3) @t
T
@rax 49.54 189.72 62.35 198.43 @E
[0.07200 0.00000 0.00000 0.07200 49.53600 190.00799] @tm
 0 O 0 @g
0.00 0.00 0.00 1.00 k
e
/_R44-Helvetica 166.00 z
0 0 (10) @t
T
@rax 49.54 189.72 62.35 198.43 @E
[0.07200 0.00000 0.00000 0.07200 49.53600 190.00799] @tm
0 J 0 j [] 0 d 0 R 0 @G
0.00 0.00 0.00 1.00 K
0 0.216 0.216 0.000 @w
r
/_R44-Helvetica 166.00 z
0 0 (10) @t
T
@rax 64.51 196.06 68.69 201.60 @E
[0.07200 0.00000 0.00000 0.07200 64.51200 196.05599] @tm
 0 O 0 @g
0.00 0.00 0.00 1.00 k
e
/_R44-Helvetica 110.00 z
0 0 (4) @t
T
@rax 64.51 196.06 68.69 201.60 @E
[0.07200 0.00000 0.00000 0.07200 64.51200 196.05599] @tm
0 J 0 j [] 0 d 0 R 0 @G
0.00 0.00 0.00 1.00 K
0 0.216 0.216 0.000 @w
r
/_R44-Helvetica 110.00 z
0 0 (4) @t
T
@rax 49.54 221.76 62.35 230.47 @E
[0.07200 0.00000 0.00000 0.07200 49.53600 222.04799] @tm
 0 O 0 @g
0.00 0.00 0.00 1.00 k
e
/_R44-Helvetica 166.00 z
0 0 (10) @t
T
@rax 49.54 221.76 62.35 230.47 @E
[0.07200 0.00000 0.00000 0.07200 49.53600 222.04799] @tm
0 J 0 j [] 0 d 0 R 0 @G
0.00 0.00 0.00 1.00 K
0 0.216 0.216 0.000 @w
r
/_R44-Helvetica 166.00 z
0 0 (10) @t
T
@rax 64.51 227.81 68.62 233.50 @E
[0.07200 0.00000 0.00000 0.07200 64.51200 228.02399] @tm
 0 O 0 @g
0.00 0.00 0.00 1.00 k
e
/_R44-Helvetica 110.00 z
0 0 (5) @t
T
@rax 64.51 227.81 68.62 233.50 @E
[0.07200 0.00000 0.00000 0.07200 64.51200 228.02399] @tm
0 J 0 j [] 0 d 0 R 0 @G
0.00 0.00 0.00 1.00 K
0 0.216 0.216 0.000 @w
r
/_R44-Helvetica 110.00 z
0 0 (5) @t
T
@rax 49.54 252.72 62.35 261.43 @E
[0.07200 0.00000 0.00000 0.07200 49.53600 253.00799] @tm
 0 O 0 @g
0.00 0.00 0.00 1.00 k
e
/_R44-Helvetica 166.00 z
0 0 (10) @t
T
@rax 49.54 252.72 62.35 261.43 @E
[0.07200 0.00000 0.00000 0.07200 49.53600 253.00799] @tm
0 J 0 j [] 0 d 0 R 0 @G
0.00 0.00 0.00 1.00 K
0 0.216 0.216 0.000 @w
r
/_R44-Helvetica 166.00 z
0 0 (10) @t
T
@rax 64.51 258.84 68.62 264.60 @E
[0.07200 0.00000 0.00000 0.07200 64.51200 259.05599] @tm
 0 O 0 @g
0.00 0.00 0.00 1.00 k
e
/_R44-Helvetica 110.00 z
0 0 (6) @t
T
@rax 64.51 258.84 68.62 264.60 @E
[0.07200 0.00000 0.00000 0.07200 64.51200 259.05599] @tm
0 J 0 j [] 0 d 0 R 0 @G
0.00 0.00 0.00 1.00 K
0 0.216 0.216 0.000 @w
r
/_R44-Helvetica 110.00 z
0 0 (6) @t
T
@rax 330.48 48.74 343.30 57.46 @E
[0.07200 0.00000 0.00000 0.07200 330.47999 49.03200] @tm
 0 O 0 @g
0.00 0.00 0.00 1.00 k
e
/_R44-Helvetica 166.00 z
0 0 (10) @t
T
@rax 330.48 48.74 343.30 57.46 @E
[0.07200 0.00000 0.00000 0.07200 330.47999 49.03200] @tm
0 J 0 j [] 0 d 0 R 0 @G
0.00 0.00 0.00 1.00 K
0 0.216 0.216 0.000 @w
r
/_R44-Helvetica 166.00 z
0 0 (10) @t
T
@rax 345.46 56.02 350.93 61.56 @E
[0.07200 0.00000 0.00000 0.07200 345.45599 56.01600] @tm
 0 O 0 @g
0.00 0.00 0.00 1.00 k
e
/_R44-Helvetica 110.00 z
0 0 (-1) @t
T
@rax 345.46 56.02 350.93 61.56 @E
[0.07200 0.00000 0.00000 0.07200 345.45599 56.01600] @tm
0 J 0 j [] 0 d 0 R 0 @G
0.00 0.00 0.00 1.00 K
0 0.216 0.216 0.000 @w
r
/_R44-Helvetica 110.00 z
0 0 (-1) @t
T
@rax 266.54 48.74 279.36 57.46 @E
[0.07200 0.00000 0.00000 0.07200 266.54399 49.03200] @tm
 0 O 0 @g
0.00 0.00 0.00 1.00 k
e
/_R44-Helvetica 166.00 z
0 0 (10) @t
T
@rax 266.54 48.74 279.36 57.46 @E
[0.07200 0.00000 0.00000 0.07200 266.54399 49.03200] @tm
0 J 0 j [] 0 d 0 R 0 @G
0.00 0.00 0.00 1.00 K
0 0.216 0.216 0.000 @w
r
/_R44-Helvetica 166.00 z
0 0 (10) @t
T
@rax 281.52 55.80 285.62 61.56 @E
[0.07200 0.00000 0.00000 0.07200 281.51999 56.01600] @tm
 0 O 0 @g
0.00 0.00 0.00 1.00 k
e
/_R44-Helvetica 110.00 z
0 0 (0) @t
T
@rax 281.52 55.80 285.62 61.56 @E
[0.07200 0.00000 0.00000 0.07200 281.51999 56.01600] @tm
0 J 0 j [] 0 d 0 R 0 @G
0.00 0.00 0.00 1.00 K
0 0.216 0.216 0.000 @w
r
/_R44-Helvetica 110.00 z
0 0 (0) @t
T
@rax 200.52 48.74 213.34 57.46 @E
[0.07200 0.00000 0.00000 0.07200 200.51999 49.03200] @tm
 0 O 0 @g
0.00 0.00 0.00 1.00 k
e
/_R44-Helvetica 166.00 z
0 0 (10) @t
T
@rax 200.52 48.74 213.34 57.46 @E
[0.07200 0.00000 0.00000 0.07200 200.51999 49.03200] @tm
0 J 0 j [] 0 d 0 R 0 @G
0.00 0.00 0.00 1.00 K
0 0.216 0.216 0.000 @w
r
/_R44-Helvetica 166.00 z
0 0 (10) @t
T
@rax 215.57 56.02 218.38 61.56 @E
[0.07200 0.00000 0.00000 0.07200 215.56799 56.01600] @tm
 0 O 0 @g
0.00 0.00 0.00 1.00 k
e
/_R44-Helvetica 110.00 z
0 0 (1) @t
T
@rax 215.57 56.02 218.38 61.56 @E
[0.07200 0.00000 0.00000 0.07200 215.56799 56.01600] @tm
0 J 0 j [] 0 d 0 R 0 @G
0.00 0.00 0.00 1.00 K
0 0.216 0.216 0.000 @w
r
/_R44-Helvetica 110.00 z
0 0 (1) @t
T
@rax 134.57 48.74 147.38 57.46 @E
[0.07200 0.00000 0.00000 0.07200 134.56799 49.03200] @tm
 0 O 0 @g
0.00 0.00 0.00 1.00 k
e
/_R44-Helvetica 166.00 z
0 0 (10) @t
T
@rax 134.57 48.74 147.38 57.46 @E
[0.07200 0.00000 0.00000 0.07200 134.56799 49.03200] @tm
0 J 0 j [] 0 d 0 R 0 @G
0.00 0.00 0.00 1.00 K
0 0.216 0.216 0.000 @w
r
/_R44-Helvetica 166.00 z
0 0 (10) @t
T
@rax 149.54 56.02 153.58 61.56 @E
[0.07200 0.00000 0.00000 0.07200 149.54399 56.01600] @tm
 0 O 0 @g
0.00 0.00 0.00 1.00 k
e
/_R44-Helvetica 110.00 z
0 0 (2) @t
T
@rax 149.54 56.02 153.58 61.56 @E
[0.07200 0.00000 0.00000 0.07200 149.54399 56.01600] @tm
0 J 0 j [] 0 d 0 R 0 @G
0.00 0.00 0.00 1.00 K
0 0.216 0.216 0.000 @w
r
/_R44-Helvetica 110.00 z
0 0 (2) @t
T
@rax 68.54 48.74 81.36 57.46 @E
[0.07200 0.00000 0.00000 0.07200 68.54400 49.03200] @tm
 0 O 0 @g
0.00 0.00 0.00 1.00 k
e
/_R44-Helvetica 166.00 z
0 0 (10) @t
T
@rax 68.54 48.74 81.36 57.46 @E
[0.07200 0.00000 0.00000 0.07200 68.54400 49.03200] @tm
0 J 0 j [] 0 d 0 R 0 @G
0.00 0.00 0.00 1.00 K
0 0.216 0.216 0.000 @w
r
/_R44-Helvetica 166.00 z
0 0 (10) @t
T
@rax 82.51 55.80 86.62 61.56 @E
[0.07200 0.00000 0.00000 0.07200 82.51200 56.01600] @tm
 0 O 0 @g
0.00 0.00 0.00 1.00 k
e
/_R44-Helvetica 110.00 z
0 0 (3) @t
T
@rax 82.51 55.80 86.62 61.56 @E
[0.07200 0.00000 0.00000 0.07200 82.51200 56.01600] @tm
0 J 0 j [] 0 d 0 R 0 @G
0.00 0.00 0.00 1.00 K
0 0.216 0.216 0.000 @w
r
/_R44-Helvetica 110.00 z
0 0 (3) @t
T
@rax 189.50 29.66 197.14 40.90 @E
[0.07200 0.00000 0.00000 0.07200 189.50399 30.02400] @tm
 0 O 0 @g
0.00 0.00 0.00 1.00 k
e
/Symbol 193.00 z
0 0 (l) @t
T
@rax 189.50 29.66 197.14 40.90 @E
[0.07200 0.00000 0.00000 0.07200 189.50399 30.02400] @tm
0 J 0 j [] 0 d 0 R 0 @G
0.00 0.00 0.00 1.00 K
0 0.216 0.216 0.000 @w
r
/Symbol 193.00 z
0 0 (l) @t
T
@rax 197.42 23.83 201.89 28.66 @E
[0.07200 0.00000 0.00000 0.07200 197.56799 24.04800] @tm
 0 O 0 @g
0.00 0.00 0.00 1.00 k
e
/Symbol 126.00 z
0 0 (n) @t
T
@rax 197.42 23.83 201.89 28.66 @E
[0.07200 0.00000 0.00000 0.07200 197.56799 24.04800] @tm
0 J 0 j [] 0 d 0 R 0 @G
0.00 0.00 0.00 1.00 K
0 0.216 0.216 0.000 @w
r
/Symbol 126.00 z
0 0 (n) @t
T
@rax 207.14 30.67 211.54 41.26 @E
[0.07200 0.00000 0.00000 0.07200 203.54399 31.03200] @tm
 0 O 0 @g
0.00 0.00 0.00 1.00 k
e
/_R44-Helvetica 193.00 z
54 0 (/) @t
T
@rax 207.14 30.67 211.54 41.26 @E
[0.07200 0.00000 0.00000 0.07200 203.54399 31.03200] @tm
0 J 0 j [] 0 d 0 R 0 @G
0.00 0.00 0.00 1.00 K
0 0.216 0.216 0.000 @w
r
/_R44-Helvetica 193.00 z
54 0 (/) @t
T
@rax 206.64 25.85 222.05 43.99 @E
[0.07200 0.00000 0.00000 0.07200 210.52799 30.02400] @tm
 0 O 0 @g
0.00 0.00 0.00 1.00 k
e
/Symbol 193.00 z
0 0 (p) @t
T
@rax 206.64 25.85 222.05 43.99 @E
[0.07200 0.00000 0.00000 0.07200 210.52799 30.02400] @tm
0 J 0 j [] 0 d 0 R 0 @G
0.00 0.00 0.00 1.00 K
0 0.216 0.216 0.000 @w
r
/Symbol 193.00 z
0 0 (p) @t
T
@rax 218.52 31.03 228.02 41.04 @E
[0.07200 0.00000 0.00000 0.07200 218.51999 31.03200] @tm
 0 O 0 @g
0.00 0.00 0.00 1.00 k
e
/_R44-Helvetica 193.00 z
0 0 (R) @t
T
@rax 218.52 31.03 228.02 41.04 @E
[0.07200 0.00000 0.00000 0.07200 218.51999 31.03200] @tm
0 J 0 j [] 0 d 0 R 0 @G
0.00 0.00 0.00 1.00 K
0 0.216 0.216 0.000 @w
r
/_R44-Helvetica 193.00 z
0 0 (R) @t
T
@rax 270.50 136.01 278.57 144.58 @E
[0.07200 0.00000 0.00000 0.07200 270.50399 136.00799] @tm
 0 O 0 @g
0.00 0.00 0.00 1.00 k
e
/_R44-Helvetica 166.00 z
0 0 (D) @t
T
@rax 270.50 136.01 278.57 144.58 @E
[0.07200 0.00000 0.00000 0.07200 270.50399 136.00799] @tm
0 J 0 j [] 0 d 0 R 0 @G
0.00 0.00 0.00 1.00 K
0 0.216 0.216 0.000 @w
r
/_R44-Helvetica 166.00 z
0 0 (D) @t
T
@rax 279.50 131.83 283.25 136.30 @E
[0.07200 0.00000 0.00000 0.07200 279.50399 132.04799] @tm
 0 O 0 @g
0.00 0.00 0.00 1.00 k
e
/_R44-Helvetica 110.00 z
0 0 (c) @t
T
@rax 279.50 131.83 283.25 136.30 @E
[0.07200 0.00000 0.00000 0.07200 279.50399 132.04799] @tm
0 J 0 j [] 0 d 0 R 0 @G
0.00 0.00 0.00 1.00 K
0 0.216 0.216 0.000 @w
r
/_R44-Helvetica 110.00 z
0 0 (c) @t
T
@rax 280.01 132.41 294.05 148.03 @E
[0.07200 0.00000 0.00000 0.07200 283.53599 136.00799] @tm
 0 O 0 @g
0.00 0.00 0.00 1.00 k
e
/_R44-Helvetica 166.00 z
0 0 (=) @t
T
@rax 280.01 132.41 294.05 148.03 @E
[0.07200 0.00000 0.00000 0.07200 283.53599 136.00799] @tm
0 J 0 j [] 0 d 0 R 0 @G
0.00 0.00 0.00 1.00 K
0 0.216 0.216 0.000 @w
r
/_R44-Helvetica 166.00 z
0 0 (=) @t
T
@rax 290.52 134.21 296.42 143.06 @E
[0.07200 0.00000 0.00000 0.07200 290.51999 137.01599] @tm
 0 O 0 @g
0.00 0.00 0.00 1.00 k
e
/Symbol 166.00 z
0 0 (r) @t
T
@rax 290.52 134.21 296.42 143.06 @E
[0.07200 0.00000 0.00000 0.07200 290.51999 137.01599] @tm
0 J 0 j [] 0 d 0 R 0 @G
0.00 0.00 0.00 1.00 K
0 0.216 0.216 0.000 @w
r
/Symbol 166.00 z
0 0 (r) @t
T
@rax 297.50 131.83 301.61 136.30 @E
[0.07200 0.00000 0.00000 0.07200 297.50399 132.04799] @tm
 0 O 0 @g
0.00 0.00 0.00 1.00 k
e
/_R44-Helvetica 110.00 z
0 0 (e) @t
T
@rax 297.50 131.83 301.61 136.30 @E
[0.07200 0.00000 0.00000 0.07200 297.50399 132.04799] @tm
0 J 0 j [] 0 d 0 R 0 @G
0.00 0.00 0.00 1.00 K
0 0.216 0.216 0.000 @w
r
/_R44-Helvetica 110.00 z
0 0 (e) @t
T
@rax 301.54 143.06 305.57 148.61 @E
[0.07200 0.00000 0.00000 0.07200 301.53599 143.06399] @tm
 0 O 0 @g
0.00 0.00 0.00 1.00 k
e
/_R44-Helvetica 110.00 z
0 0 (2) @t
T
@rax 301.54 143.06 305.57 148.61 @E
[0.07200 0.00000 0.00000 0.07200 301.53599 143.06399] @tm
0 J 0 j [] 0 d 0 R 0 @G
0.00 0.00 0.00 1.00 K
0 0.216 0.216 0.000 @w
r
/_R44-Helvetica 110.00 z
0 0 (2) @t
T
@rax 306.29 136.73 312.12 143.14 @E
[0.07200 0.00000 0.00000 0.07200 306.43199 137.01599] @tm
 0 O 0 @g
0.00 0.00 0.00 1.00 k
e
/Symbol 166.00 z
0 0 (n) @t
T
@rax 306.29 136.73 312.12 143.14 @E
[0.07200 0.00000 0.00000 0.07200 306.43199 137.01599] @tm
0 J 0 j [] 0 d 0 R 0 @G
0.00 0.00 0.00 1.00 K
0 0.216 0.216 0.000 @w
r
/Symbol 166.00 z
0 0 (n) @t
T
@rax 312.48 131.83 320.76 136.30 @E
[0.07200 0.00000 0.00000 0.07200 312.47999 132.04799] @tm
 0 O 0 @g
0.00 0.00 0.00 1.00 k
e
/_R44-Helvetica 110.00 z
0 0 (en) @t
T
@rax 312.48 131.83 320.76 136.30 @E
[0.07200 0.00000 0.00000 0.07200 312.47999 132.04799] @tm
0 J 0 j [] 0 d 0 R 0 @G
0.00 0.00 0.00 1.00 K
0 0.216 0.216 0.000 @w
r
/_R44-Helvetica 110.00 z
0 0 (en) @t
T
@rax 220.54 172.01 228.60 180.58 @E
[0.07200 0.00000 0.00000 0.07200 220.53599 172.00799] @tm
 0 O 0 @g
0.00 0.00 0.00 1.00 k
e
/_R44-Helvetica 166.00 z
0 0 (D) @t
T
@rax 220.54 172.01 228.60 180.58 @E
[0.07200 0.00000 0.00000 0.07200 220.53599 172.00799] @tm
0 J 0 j [] 0 d 0 R 0 @G
0.00 0.00 0.00 1.00 K
0 0.216 0.216 0.000 @w
r
/_R44-Helvetica 166.00 z
0 0 (D) @t
T
@rax 229.54 177.34 241.78 183.31 @E
[0.07200 0.00000 0.00000 0.07200 229.53599 179.06399] @tm
 0 O 0 @g
0.00 0.00 0.00 1.00 k
e
/_R44-Helvetica 110.00 z
0 0 (e) @t
58 0 (xp) @t
T
@rax 229.54 177.34 241.78 183.31 @E
[0.07200 0.00000 0.00000 0.07200 229.53599 179.06399] @tm
0 J 0 j [] 0 d 0 R 0 @G
0.00 0.00 0.00 1.00 K
0 0.216 0.216 0.000 @w
r
/_R44-Helvetica 110.00 z
0 0 (e) @t
58 0 (xp) @t
T
@rax 94.54 148.03 102.60 156.60 @E
[0.07200 0.00000 0.00000 0.07200 94.53600 148.03199] @tm
 0 O 0 @g
0.00 0.00 0.00 1.00 k
e
/_R44-Helvetica 166.00 z
0 0 (D) @t
T
@rax 94.54 148.03 102.60 156.60 @E
[0.07200 0.00000 0.00000 0.07200 94.53600 148.03199] @tm
0 J 0 j [] 0 d 0 R 0 @G
0.00 0.00 0.00 1.00 K
0 0.216 0.216 0.000 @w
r
/_R44-Helvetica 166.00 z
0 0 (D) @t
T
@rax 104.54 143.78 119.59 149.83 @E
[0.07200 0.00000 0.00000 0.07200 104.54400 143.99999] @tm
 0 O 0 @g
0.00 0.00 0.00 1.00 k
e
/_R44-Helvetica 110.00 z
0 0 (G.S) @t
188 0 (.) @t
T
@rax 104.54 143.78 119.59 149.83 @E
[0.07200 0.00000 0.00000 0.07200 104.54400 143.99999] @tm
0 J 0 j [] 0 d 0 R 0 @G
0.00 0.00 0.00 1.00 K
0 0.216 0.216 0.000 @w
r
/_R44-Helvetica 110.00 z
0 0 (G.S) @t
188 0 (.) @t
T
@rax 297.50 239.04 305.57 247.61 @E
[0.07200 0.00000 0.00000 0.07200 297.50399 239.03999] @tm
 0 O 0 @g
0.00 0.00 0.00 1.00 k
e
/_R44-Helvetica 166.00 z
0 0 (D) @t
T
@rax 297.50 239.04 305.57 247.61 @E
[0.07200 0.00000 0.00000 0.07200 297.50399 239.03999] @tm
0 J 0 j [] 0 d 0 R 0 @G
0.00 0.00 0.00 1.00 K
0 0.216 0.216 0.000 @w
r
/_R44-Helvetica 166.00 z
0 0 (D) @t
T
@rax 306.43 234.79 319.10 240.84 @E
[0.07200 0.00000 0.00000 0.07200 306.43199 235.00799] @tm
 0 O 0 @g
0.00 0.00 0.00 1.00 k
e
/_R44-Helvetica 110.00 z
0 0 (P) @t
53 0 (.S) @t
155 0 (.) @t
T
@rax 306.43 234.79 319.10 240.84 @E
[0.07200 0.00000 0.00000 0.07200 306.43199 235.00799] @tm
0 J 0 j [] 0 d 0 R 0 @G
0.00 0.00 0.00 1.00 K
0 0.216 0.216 0.000 @w
r
/_R44-Helvetica 110.00 z
0 0 (P) @t
53 0 (.S) @t
155 0 (.) @t
T
@rax 212.54 210.02 220.61 218.59 @E
[0.07200 0.00000 0.00000 0.07200 212.54399 210.02399] @tm
 0 O 0 @g
0.00 0.00 0.00 1.00 k
e
/_R44-Helvetica 166.00 z
0 0 (D) @t
T
@rax 212.54 210.02 220.61 218.59 @E
[0.07200 0.00000 0.00000 0.07200 212.54399 210.02399] @tm
0 J 0 j [] 0 d 0 R 0 @G
0.00 0.00 0.00 1.00 K
0 0.216 0.216 0.000 @w
r
/_R44-Helvetica 166.00 z
0 0 (D) @t
T
@rax 221.54 204.34 225.65 210.31 @E
[0.07200 0.00000 0.00000 0.07200 221.54399 206.06399] @tm
 0 O 0 @g
0.00 0.00 0.00 1.00 k
e
/_R44-Helvetica 110.00 z
0 0 (p) @t
T
@rax 221.54 204.34 225.65 210.31 @E
[0.07200 0.00000 0.00000 0.07200 221.54399 206.06399] @tm
0 J 0 j [] 0 d 0 R 0 @G
0.00 0.00 0.00 1.00 K
0 0.216 0.216 0.000 @w
r
/_R44-Helvetica 110.00 z
0 0 (p) @t
T
@rax 
196.06 180.00 207.00 188.06 @E
 0 O 0 @g
0.00 0.00 0.00 1.00 k
0 J 0 j [] 0 d 0 R 0 @G
0.00 0.00 0.00 1.00 K
0 0.216 0.216 0.000 @w
196.06 188.06 m
207.00 187.06 L
202.03 185.04 L
204.05 180.00 L
196.06 188.06 L
@c
B

@rax 
202.54 177.55 217.51 184.54 @E
0 J 2 j [] 0 d 0 R 0 @G
0.00 0.00 0.00 1.00 K
0 1.008 1.008 0.000 @w
202.54 184.54 m
217.51 177.55 L
S

@rax 
124.56 180.36 129.53 180.65 @E
0 J 2 j [] 0 d 0 R 0 @G
0.00 0.00 0.00 1.00 K
0 1.008 1.008 0.000 @w
124.56 180.50 m
129.53 180.50 L
S

@rax 
124.56 165.38 129.53 165.67 @E
0 J 2 j [] 0 d 0 R 0 @G
0.00 0.00 0.00 1.00 K
0 1.008 1.008 0.000 @w
124.56 165.53 m
129.53 165.53 L
S

@rax 
127.37 165.53 127.66 179.57 @E
0 J 2 j [] 0 d 0 R 0 @G
0.00 0.00 0.00 1.00 K
0 1.008 1.008 0.000 @w
127.51 179.57 m
127.51 165.53 L
S

@rax 15.19 58.54 44.06 260.21 @E
[0.00001 0.07200 -0.07200 0.00001 39.24000 64.36800] @tm
 0 O 0 @g
0.00 0.00 0.00 1.00 k
e
/_R44-Helvetica 222.00 z
0 0 (Diffusion Coefficient \133cm /s\135) @t
T
@rax 15.19 58.54 44.06 260.21 @E
[0.00001 0.07200 -0.07200 0.00001 39.24000 64.36800] @tm
 0 O 0 @g
0.00 0.00 0.00 1.00 k
e
/_R44-Helvetica 111.00 z
2391 111 (2) @t
T
@rs @rs @sm
@rs
@rs
@EndSysCorelDict
end

end
clear
userdict /VPsave get restore

/PPT_ProcessAll false def
32 0 0 38 38 0 0 0 36 /Helvetica-Bold /font13 ANSIFont font
gs 1096 480 1152 2467 CB
1227 2541 34 (Fi) 33 SB
1260 2541 34 (g.) 33 SB
1293 2541 11 ( ) 10 SB
1303 2541 32 (1.) 31 SB
gr
32 0 0 38 38 0 0 0 34 /Helvetica /font12 ANSIFont font
gs 1096 480 1152 2467 CB
1334 2543 45 ( El) 44 SB
1378 2543 20 (e) 21 SB
1399 2543 19 (c) 18 SB
1417 2543 24 (tr) 23 SB
1440 2543 53 (on ) 52 SB
1492 2543 30 (di) 29 SB
1521 2543 11 (f) 10 SB
1531 2543 11 (f) 10 SB
1541 2543 49 (usi) 48 SB
1589 2543 53 (on ) 52 SB
1641 2543 40 (co) 39 SB
1680 2543 20 (e) 21 SB
1701 2543 11 (f) 10 SB
1711 2543 20 (fi) 19 SB
1730 2543 28 (ci) 27 SB
1757 2543 20 (e) 21 SB
1778 2543 21 (n) 20 SB
1798 2543 22 (t ) 21 SB
1819 2543 40 (ac) 39 SB
1858 2543 53 (ros) 52 SB
1910 2543 30 (s ) 29 SB
1939 2543 84 (good) 83 SB
2022 2543 44 ( m) 42 SB
2064 2543 42 (ag) 41 SB
gr
gs 1096 480 1152 2467 CB
2105 2543 13 (-) 13 SB
gr
gs 1096 480 1152 2467 CB
1227 2588 41 (ne) 42 SB
1269 2588 11 (t) 10 SB
1279 2588 9 (i) 8 SB
1287 2588 30 (c ) 29 SB
1316 2588 53 (sur) 52 SB
1368 2588 11 (f) 10 SB
1378 2588 60 (ace) 61 SB
1439 2588 19 (s) 18 SB
1457 2588 32 ( a) 31 SB
1488 2588 30 (s ) 29 SB
1517 2588 32 (a ) 31 SB
1548 2588 32 (fu) 31 SB
1579 2588 51 (nct) 50 SB
1629 2588 9 (i) 8 SB
1637 2588 53 (on ) 52 SB
1689 2588 32 (of) 31 SB
1720 2588 11 ( ) 10 SB
1730 2588 32 (th) 31 SB
1761 2588 20 (e) 21 SB
1782 2588 11 ( ) 10 SB
1792 2588 20 (e) 21 SB
1813 2588 9 (l) 8 SB
1821 2588 20 (e) 21 SB
1842 2588 30 (ct) 29 SB
1871 2588 34 (ro) 33 SB
1904 2588 55 (n-n) 54 SB
1958 2588 20 (e) 21 SB
1979 2588 32 (ut) 31 SB
2010 2588 34 (ra) 33 SB
2043 2588 20 (l ) 19 SB
2062 2588 40 (co) 39 SB
2101 2588 18 (ll) 17 SB
2118 2588 9 (i) 8 SB
gr
gs 1096 480 1152 2467 CB
2126 2588 13 (-) 13 SB
gr
gs 1096 480 1152 2467 CB
1227 2633 28 (si) 27 SB
1254 2633 63 (ona) 62 SB
1316 2633 18 (li) 17 SB
1333 2633 11 (t) 10 SB
1343 2633 30 (y.) 29 SB
1372 2633 11 ( ) 10 SB
1382 2633 53 (Als) 52 SB
1434 2633 32 (o ) 31 SB
1465 2633 30 (pl) 29 SB
1494 2633 32 (ot) 31 SB
1525 2633 63 (ted ) 62 SB
1587 2633 76 (are t) 75 SB
1662 2633 71 (he c) 70 SB
1732 2633 30 (la) 29 SB
1761 2633 47 (ssi) 46 SB
1807 2633 49 (cal) 48 SB
1855 2633 11 ( ) 10 SB
1865 2633 32 (\(s) 31 SB
1896 2633 30 (ol) 29 SB
1925 2633 30 (id) 29 SB
1954 2633 20 ( l) 19 SB
1973 2633 9 (i) 8 SB
1981 2633 41 (ne) 42 SB
2023 2633 13 (\)) 12 SB
2035 2633 11 ( ) 10 SB
2045 2633 63 (and) 63 SB
gr
gs 1096 480 1152 2467 CB
1227 2678 11 (t) 10 SB
1237 2678 41 (he) 42 SB
1279 2678 11 ( ) 10 SB
1289 2678 61 (con) 60 SB
1349 2678 39 (ve) 40 SB
1389 2678 32 (nt) 31 SB
1420 2678 9 (i) 8 SB
1428 2678 63 (ona) 62 SB
1490 2678 20 (l ) 19 SB
1509 2678 40 (ca) 39 SB
1548 2678 28 (lc) 27 SB
1575 2678 30 (ul) 29 SB
1604 2678 32 (at) 31 SB
1635 2678 30 (io) 29 SB
1664 2678 51 (ns ) 50 SB
1714 2678 32 (of) 31 SB
1745 2678 11 ( ) 10 SB
1755 2678 32 (th) 31 SB
1786 2678 20 (e) 21 SB
1807 2678 11 ( ) 10 SB
1817 2678 41 (ne) 42 SB
1859 2678 40 (oc) 39 SB
1898 2678 30 (la) 29 SB
1927 2678 47 (ssi) 46 SB
1973 2678 49 (cal) 48 SB
gr
gs 1096 480 1152 2467 CB
1227 2723 11 (t) 10 SB
1237 2723 94 (okam) 92 SB
1329 2723 40 (ak) 39 SB
1368 2723 22 ( t) 21 SB
1389 2723 13 (r) 12 SB
1401 2723 82 (ansp) 81 SB
1482 2723 45 (ort) 44 SB
1526 2723 11 ( ) 10 SB
1536 2723 30 (di) 29 SB
1565 2723 11 (f) 10 SB
1575 2723 32 (fu) 31 SB
1606 2723 28 (si) 27 SB
1633 2723 53 (on ) 52 SB
1685 2723 82 (coeff) 81 SB
1766 2723 9 (i) 8 SB
1774 2723 28 (ci) 27 SB
1801 2723 20 (e) 21 SB
1822 2723 32 (nt) 31 SB
1853 2723 11 ( ) 10 SB
1863 2723 11 (f) 10 SB
1873 2723 45 (or ) 44 SB
1917 2723 11 (t) 10 SB
1927 2723 41 (he) 42 SB
1969 2723 11 ( ) 10 SB
1979 2723 20 (e) 21 SB
2000 2723 9 (l) 8 SB
2008 2723 20 (e) 21 SB
2029 2723 19 (c) 18 SB
2047 2723 24 (tr) 23 SB
2070 2723 61 (ons) 60 SB
gr
gs 1096 480 1152 2467 CB
1227 2768 34 (\(d) 33 SB
1260 2768 32 (ot) 31 SB
1291 2768 63 (ted ) 62 SB
1353 2768 18 (li) 17 SB
1370 2768 65 (ne\).) 64 SB
gr
1 #C
statusdict begin /manualfeed false store end
EJ RS
SVDoc restore
end


/Win35Dict 290 dict def Win35Dict begin/bd{bind def}bind def/in{72
mul}bd/ed{exch def}bd/ld{load def}bd/tr/translate ld/gs/gsave ld/gr
/grestore ld/M/moveto ld/L/lineto ld/rmt/rmoveto ld/rlt/rlineto ld
/rct/rcurveto ld/st/stroke ld/n/newpath ld/sm/setmatrix ld/cm/currentmatrix
ld/cp/closepath ld/ARC/arcn ld/TR{65536 div}bd/lj/setlinejoin ld/lc
/setlinecap ld/ml/setmiterlimit ld/sl/setlinewidth ld/scignore false
def/sc{scignore{pop pop pop}{0 index 2 index eq 2 index 4 index eq
and{pop pop 255 div setgray}{3{255 div 3 1 roll}repeat setrgbcolor}ifelse}ifelse}bd
/FC{bR bG bB sc}bd/fC{/bB ed/bG ed/bR ed}bd/HC{hR hG hB sc}bd/hC{
/hB ed/hG ed/hR ed}bd/PC{pR pG pB sc}bd/pC{/pB ed/pG ed/pR ed}bd/sM
matrix def/PenW 1 def/iPen 5 def/mxF matrix def/mxE matrix def/mxUE
matrix def/mxUF matrix def/fBE false def/iDevRes 72 0 matrix defaultmatrix
dtransform dup mul exch dup mul add sqrt def/fPP false def/SS{fPP{
/SV save def}{gs}ifelse}bd/RS{fPP{SV restore}{gr}ifelse}bd/EJ{gsave
showpage grestore}bd/#C{userdict begin/#copies ed end}bd/FEbuf 2 string
def/FEglyph(G  )def/FE{1 exch{dup 16 FEbuf cvrs FEglyph exch 1 exch
putinterval 1 index exch FEglyph cvn put}for}bd/SM{/iRes ed/cyP ed
/cxPg ed/cyM ed/cxM ed 72 100 div dup scale dup 0 ne{90 eq{cyM exch
0 eq{cxM exch tr -90 rotate -1 1 scale}{cxM cxPg add exch tr +90 rotate}ifelse}{cyP
cyM sub exch 0 ne{cxM exch tr -90 rotate}{cxM cxPg add exch tr -90
rotate 1 -1 scale}ifelse}ifelse}{pop cyP cyM sub exch 0 ne{cxM cxPg
add exch tr 180 rotate}{cxM exch tr 1 -1 scale}ifelse}ifelse 100 iRes
div dup scale 0 0 transform .25 add round .25 sub exch .25 add round
.25 sub exch itransform translate}bd/SJ{1 index 0 eq{pop pop/fBE false
def}{1 index/Break ed div/dxBreak ed/fBE true def}ifelse}bd/ANSIVec[
16#0/grave 16#1/acute 16#2/circumflex 16#3/tilde 16#4/macron 16#5/breve
16#6/dotaccent 16#7/dieresis 16#8/ring 16#9/cedilla 16#A/hungarumlaut
16#B/ogonek 16#C/caron 16#D/dotlessi 16#27/quotesingle 16#60/grave
16#7C/bar 16#82/quotesinglbase 16#83/florin 16#84/quotedblbase 16#85
/ellipsis 16#86/dagger 16#87/daggerdbl 16#89/perthousand 16#8A/Scaron
16#8B/guilsinglleft 16#8C/OE 16#91/quoteleft 16#92/quoteright 16#93
/quotedblleft 16#94/quotedblright 16#95/bullet 16#96/endash 16#97
/emdash 16#99/trademark 16#9A/scaron 16#9B/guilsinglright 16#9C/oe
16#9F/Ydieresis 16#A0/space 16#A4/currency 16#A6/brokenbar 16#A7/section
16#A8/dieresis 16#A9/copyright 16#AA/ordfeminine 16#AB/guillemotleft
16#AC/logicalnot 16#AD/hyphen 16#AE/registered 16#AF/macron 16#B0/degree
16#B1/plusminus 16#B2/twosuperior 16#B3/threesuperior 16#B4/acute 16#B5
/mu 16#B6/paragraph 16#B7/periodcentered 16#B8/cedilla 16#B9/onesuperior
16#BA/ordmasculine 16#BB/guillemotright 16#BC/onequarter 16#BD/onehalf
16#BE/threequarters 16#BF/questiondown 16#C0/Agrave 16#C1/Aacute 16#C2
/Acircumflex 16#C3/Atilde 16#C4/Adieresis 16#C5/Aring 16#C6/AE 16#C7
/Ccedilla 16#C8/Egrave 16#C9/Eacute 16#CA/Ecircumflex 16#CB/Edieresis
16#CC/Igrave 16#CD/Iacute 16#CE/Icircumflex 16#CF/Idieresis 16#D0/Eth
16#D1/Ntilde 16#D2/Ograve 16#D3/Oacute 16#D4/Ocircumflex 16#D5/Otilde
16#D6/Odieresis 16#D7/multiply 16#D8/Oslash 16#D9/Ugrave 16#DA/Uacute
16#DB/Ucircumflex 16#DC/Udieresis 16#DD/Yacute 16#DE/Thorn 16#DF/germandbls
16#E0/agrave 16#E1/aacute 16#E2/acircumflex 16#E3/atilde 16#E4/adieresis
16#E5/aring 16#E6/ae 16#E7/ccedilla 16#E8/egrave 16#E9/eacute 16#EA
/ecircumflex 16#EB/edieresis 16#EC/igrave 16#ED/iacute 16#EE/icircumflex
16#EF/idieresis 16#F0/eth 16#F1/ntilde 16#F2/ograve 16#F3/oacute 16#F4
/ocircumflex 16#F5/otilde 16#F6/odieresis 16#F7/divide 16#F8/oslash
16#F9/ugrave 16#FA/uacute 16#FB/ucircumflex 16#FC/udieresis 16#FD/yacute
16#FE/thorn 16#FF/ydieresis ] def/reencdict 12 dict def/IsChar{basefontdict
/CharStrings get exch known}bd/MapCh{dup IsChar not{pop/bullet}if
newfont/Encoding get 3 1 roll put}bd/MapDegree{16#b0/degree IsChar{
/degree}{/ring}ifelse MapCh}bd/MapBB{16#a6/brokenbar IsChar{/brokenbar}{
/bar}ifelse MapCh}bd/ANSIFont{reencdict begin/newfontname ed/basefontname
ed FontDirectory newfontname known not{/basefontdict basefontname findfont
def/newfont basefontdict maxlength dict def basefontdict{exch dup/FID
ne{dup/Encoding eq{exch dup length array copy newfont 3 1 roll put}{exch
newfont 3 1 roll put}ifelse}{pop pop}ifelse}forall newfont/FontName
newfontname put 127 1 159{newfont/Encoding get exch/bullet put}for
ANSIVec aload pop ANSIVec length 2 idiv{MapCh}repeat MapDegree MapBB
newfontname newfont definefont pop}if newfontname end}bd/SB{FC/ULlen
ed/str ed str length fBE not{dup 1 gt{1 sub}if}if/cbStr ed/dxGdi ed
/y0 ed/x0 ed str stringwidth dup 0 ne{/y1 ed/x1 ed y1 y1 mul x1 x1
mul add sqrt dxGdi exch div 1 sub dup x1 mul cbStr div exch y1 mul
cbStr div}{exch abs neg dxGdi add cbStr div exch}ifelse/dyExtra ed
/dxExtra ed x0 y0 M fBE{dxBreak 0 BCh dxExtra dyExtra str awidthshow}{dxExtra
dyExtra str ashow}ifelse fUL{x0 y0 M dxUL dyUL rmt ULlen fBE{Break
add}if 0 mxUE transform gs rlt cyUL sl [] 0 setdash st gr}if fSO{x0
y0 M dxSO dySO rmt ULlen fBE{Break add}if 0 mxUE transform gs rlt cyUL
sl [] 0 setdash st gr}if n/fBE false def}bd/font{/name ed/Ascent ed
0 ne/fT3 ed 0 ne/fSO ed 0 ne/fUL ed/Sy ed/Sx ed 10.0 div/ori ed -10.0
div/esc ed/BCh ed name findfont/xAscent 0 def/yAscent Ascent def/ULesc
esc def ULesc mxUE rotate pop fT3{/esc 0 def xAscent yAscent mxUE transform
/yAscent ed/xAscent ed}if [Sx 0 0 Sy neg xAscent yAscent] esc mxE
rotate mxF concatmatrix makefont setfont [Sx 0 0 Sy neg 0 Ascent] mxUE
mxUF concatmatrix pop fUL{currentfont dup/FontInfo get/UnderlinePosition
known not{pop/Courier findfont}if/FontInfo get/UnderlinePosition get
1000 div 0 exch mxUF transform/dyUL ed/dxUL ed}if fSO{0 .3 mxUF transform
/dySO ed/dxSO ed}if fUL fSO or{currentfont dup/FontInfo get/UnderlineThickness
known not{pop/Courier findfont}if/FontInfo get/UnderlineThickness get
1000 div Sy mul/cyUL ed}if}bd/min{2 copy gt{exch}if pop}bd/max{2 copy
lt{exch}if pop}bd/CP{/ft ed{{ft 0 eq{clip}{eoclip}ifelse}stopped{currentflat
1 add setflat}{exit}ifelse}loop}bd/patfont 10 dict def patfont begin
/FontType 3 def/FontMatrix [1 0 0 -1 0 0] def/FontBBox [0 0 16 16]
def/Encoding StandardEncoding def/BuildChar{pop pop 16 0 0 0 16 16
setcachedevice 16 16 false [1 0 0 1 .25 .25]{pat}imagemask}bd end/p{
/pat 32 string def{}forall 0 1 7{dup 2 mul pat exch 3 index put dup
2 mul 1 add pat exch 3 index put dup 2 mul 16 add pat exch 3 index
put 2 mul 17 add pat exch 2 index put pop}for}bd/pfill{/PatFont patfont
definefont setfont/ch(AAAA)def X0 64 X1{Y1 -16 Y0{1 index exch M ch
show}for pop}for}bd/vert{X0 w X1{dup Y0 M Y1 L st}for}bd/horz{Y0 w
Y1{dup X0 exch M X1 exch L st}for}bd/fdiag{X0 w X1{Y0 M X1 X0 sub dup
rlt st}for Y0 w Y1{X0 exch M Y1 Y0 sub dup rlt st}for}bd/bdiag{X0 w
X1{Y1 M X1 X0 sub dup neg rlt st}for Y0 w Y1{X0 exch M Y1 Y0 sub dup
neg rlt st}for}bd/AU{1 add cvi 15 or}bd/AD{1 sub cvi -16 and}bd/SHR{pathbbox
AU/Y1 ed AU/X1 ed AD/Y0 ed AD/X0 ed}bd/hfill{/w iRes 37.5 div round
def 0.1 sl [] 0 setdash n dup 0 eq{horz}if dup 1 eq{vert}if dup 2 eq{fdiag}if
dup 3 eq{bdiag}if dup 4 eq{horz vert}if 5 eq{fdiag bdiag}if}bd/F{/ft
ed fm 256 and 0 ne{gs FC ft 0 eq{fill}{eofill}ifelse gr}if fm 1536
and 0 ne{SHR gs HC ft CP fm 1024 and 0 ne{/Tmp save def pfill Tmp restore}{fm
15 and hfill}ifelse gr}if}bd/S{PenW sl PC st}bd/m matrix def/GW{iRes
12 div PenW add cvi}bd/DoW{iRes 50 div PenW add cvi}bd/DW{iRes 8 div
PenW add cvi}bd/SP{/PenW ed/iPen ed iPen 0 eq iPen 6 eq or{[] 0 setdash}if
iPen 1 eq{[DW GW] 0 setdash}if iPen 2 eq{[DoW GW] 0 setdash}if iPen
3 eq{[DW GW DoW GW] 0 setdash}if iPen 4 eq{[DW GW DoW GW DoW GW] 0
setdash}if}bd/E{m cm pop tr scale 1 0 moveto 0 0 1 0 360 arc cp m sm}bd
/AG{/sy ed/sx ed sx div 4 1 roll sy div 4 1 roll sx div 4 1 roll sy
div 4 1 roll atan/a2 ed atan/a1 ed sx sy scale a1 a2 ARC}def/A{m cm
pop tr AG m sm}def/P{m cm pop tr 0 0 M AG cp m sm}def/RRect{n 4 copy
M 3 1 roll exch L 4 2 roll L L cp}bd/RRCC{/r ed/y1 ed/x1 ed/y0 ed/x0
ed x0 x1 add 2 div y0 M x1 y0 x1 y1 r arcto 4{pop}repeat x1 y1 x0 y1
r arcto 4{pop}repeat x0 y1 x0 y0 r arcto 4{pop}repeat x0 y0 x1 y0 r
arcto 4{pop}repeat cp}bd/RR{2 copy 0 eq exch 0 eq or{pop pop RRect}{2
copy eq{pop RRCC}{m cm pop/y2 ed/x2 ed/ys y2 x2 div 1 max def/xs x2
y2 div 1 max def/y1 exch ys div def/x1 exch xs div def/y0 exch ys div
def/x0 exch xs div def/r2 x2 y2 min def xs ys scale x0 x1 add 2 div
y0 M x1 y0 x1 y1 r2 arcto 4{pop}repeat x1 y1 x0 y1 r2 arcto 4{pop}repeat
x0 y1 x0 y0 r2 arcto 4{pop}repeat x0 y0 x1 y0 r2 arcto 4{pop}repeat
m sm cp}ifelse}ifelse}bd/PP{{rlt}repeat}bd/OB{gs 0 ne{7 3 roll/y ed
/x ed x y translate ULesc rotate x neg y neg translate x y 7 -3 roll}if
sc B fill gr}bd/B{M/dy ed/dx ed dx 0 rlt 0 dy rlt dx neg 0 rlt cp}bd
/CB{B clip n}bd/ErrHandler{errordict dup maxlength exch length gt
dup{errordict begin}if/errhelpdict 12 dict def errhelpdict begin/stackunderflow(operand stack underflow)def
/undefined(this name is not defined in a dictionary)def/VMerror(you have used up all the printer's memory)def
/typecheck(operator was expecting a different type of operand)def
/ioerror(input/output error occured)def end{end}if errordict begin
/handleerror{$error begin newerror{/newerror false def showpage 72
72 scale/x .25 def/y 9.6 def/Helvetica findfont .2 scalefont setfont
x y moveto(Offending Command = )show/command load{dup type/stringtype
ne{(max err string)cvs}if show}exec/y y .2 sub def x y moveto(Error = )show
errorname{dup type dup( max err string )cvs show( : )show/stringtype
ne{( max err string )cvs}if show}exec errordict begin errhelpdict errorname
known{x 1 add y .2 sub moveto errhelpdict errorname get show}if end
/y y .4 sub def x y moveto(Stack =)show ostack{/y y .2 sub def x 1
add y moveto dup type/stringtype ne{( max err string )cvs}if show}forall
showpage}if end}def end}bd end
/SVDoc save def
Win35Dict begin
ErrHandler
statusdict begin 0 setjobtimeout end
statusdict begin statusdict /jobname (Ventura - STELNEWS.CHP) put end
/setresolution where { pop 300 300 setresolution } if
SS
0 0 25 20 798 1100 300 SM
0 0 0 fC
/fm 256 def
2100 3 147 238 B
1 F
n
2100 3 147 3086 B
1 F
n
3 2850 147 238 B
1 F
n
3 2850 2245 238 B
1 F
n
2074 3 160 251 B
1 F
n
2074 3 160 3074 B
1 F
n
3 2825 160 251 B
1 F
n
3 2825 2232 251 B
1 F
n
2048 3 173 264 B
1 F
n
2048 3 173 3061 B
1 F
n
3 2799 173 264 B
1 F
n
3 2799 2219 264 B
1 F
n
2 2700 1196 313 B
1 F
n
32 0 0 42 42 0 0 0 38 /Times-Roman /font32 ANSIFont font
gs 1096 2851 147 238 CB
222 312 26 (T) 25 SB
247 312 40 (he) 39 SB
286 312 67 ( spa) 66 SB
352 312 12 (t) 11 SB
363 312 31 (ia) 30 SB
393 312 23 (l ) 22 SB
415 312 33 (di) 32 SB
447 312 54 (stri) 53 SB
500 312 54 (but) 53 SB
553 312 33 (io) 32 SB
585 312 32 (n ) 31 SB
616 312 46 (of ) 45 SB
661 312 33 (th) 32 SB
693 312 30 (e ) 29 SB
722 312 19 (e) 18 SB
740 312 31 (le) 30 SB
770 312 19 (c) 18 SB
788 312 26 (tr) 25 SB
813 312 69 (ons ) 68 SB
881 312 44 (in ) 43 SB
924 312 12 (t) 11 SB
935 312 40 (he) 39 SB
974 312 23 ( t) 22 SB
996 312 70 (orsa) 69 SB
1065 312 26 (tr) 25 SB
1090 312 42 (on) 42 SB
gr
gs 1096 2851 147 238 CB
222 362 49 (wa) 48 SB
270 362 16 (s) 17 SB
287 362 11 ( ) 10 SB
297 362 33 (m) 32 SB
329 362 54 (oni) 53 SB
382 362 33 (to) 32 SB
414 362 33 (re) 32 SB
446 362 32 (d ) 31 SB
477 362 53 (by ) 52 SB
529 362 19 (a) 18 SB
547 362 44 ( m) 43 SB
590 362 12 (i) 11 SB
601 362 52 (nia) 51 SB
652 362 12 (t) 11 SB
663 362 54 (ure) 53 SB
716 362 11 ( ) 10 SB
726 362 54 (mo) 53 SB
779 362 40 (va) 39 SB
818 362 52 (ble) 51 SB
869 362 11 ( ) 10 SB
879 362 56 (pro) 55 SB
934 362 40 (be) 39 SB
973 362 53 ( wi) 52 SB
1025 362 33 (th) 32 SB
1057 362 30 ( a) 29 SB
1086 362 11 ( ) 10 SB
1096 362 33 (di) 32 SB
gr
gs 1096 2851 147 238 CB
1128 362 14 (-) 14 SB
gr
gs 1096 2851 147 238 CB
222 412 19 (a) 18 SB
240 412 52 (me) 51 SB
291 412 12 (t) 11 SB
302 412 33 (er) 32 SB
334 412 32 ( o) 31 SB
365 412 25 (f ) 24 SB
389 412 32 (0.) 31 SB
420 412 53 (25 ) 52 SB
472 412 33 (m) 32 SB
504 412 44 (m ) 43 SB
547 412 19 (a) 18 SB
565 412 53 (nd ) 52 SB
617 412 19 (a) 18 SB
635 412 11 ( ) 10 SB
645 412 31 (le) 30 SB
675 412 54 (ngt) 53 SB
728 412 32 (h ) 31 SB
759 412 46 (of ) 45 SB
804 412 32 (1 ) 31 SB
835 412 33 (m) 32 SB
867 412 33 (m) 32 SB
899 412 11 (.) 10 SB
909 412 11 ( ) 10 SB
919 412 47 (Th) 46 SB
965 412 30 (e ) 29 SB
994 412 19 (e) 18 SB
1012 412 42 (qu) 41 SB
1053 412 31 (al) 30 SB
gr
gs 1096 2851 147 238 CB
222 462 19 (e) 18 SB
240 462 31 (le) 30 SB
270 462 19 (c) 18 SB
288 462 47 (tro) 46 SB
334 462 32 (n ) 31 SB
365 462 40 (de) 39 SB
404 462 61 (nsit) 60 SB
464 462 32 (y ) 31 SB
495 462 40 (co) 39 SB
534 462 54 (nto) 53 SB
587 462 46 (ur ) 45 SB
632 462 84 (surfa) 83 SB
715 462 19 (c) 18 SB
733 462 19 (e) 18 SB
751 462 16 (s) 17 SB
768 462 11 ( ) 10 SB
778 462 19 (a) 18 SB
796 462 33 (re) 32 SB
828 462 11 ( ) 10 SB
838 462 77 (foun) 76 SB
914 462 32 (d ) 31 SB
945 462 44 (in ) 43 SB
988 462 42 (go) 41 SB
1029 462 42 (od) 42 SB
gr
gs 1096 2851 147 238 CB
222 512 19 (a) 18 SB
240 512 54 (gre) 53 SB
293 512 19 (e) 18 SB
311 512 52 (me) 51 SB
362 512 33 (nt) 32 SB
394 512 11 ( ) 10 SB
404 512 54 (wit) 53 SB
457 512 32 (h ) 31 SB
488 512 33 (m) 32 SB
520 512 19 (a) 18 SB
538 512 61 (gne) 60 SB
598 512 24 (ti) 23 SB
621 512 19 (c) 18 SB
639 512 11 ( ) 10 SB
649 512 47 (flu) 46 SB
695 512 32 (x ) 31 SB
726 512 84 (surfa) 83 SB
809 512 19 (c) 18 SB
827 512 19 (e) 18 SB
845 512 16 (s) 17 SB
862 512 11 (.) 10 SB
872 512 11 ( ) 10 SB
882 512 30 (A) 27 SB
909 512 11 ( ) 10 SB
919 512 49 (sm) 48 SB
967 512 31 (al) 30 SB
997 512 23 (l ) 21 SB
1018 512 47 (Re) 46 SB
1064 512 31 (ta) 30 SB
1094 512 35 (rd) 35 SB
gr
gs 1096 2851 147 238 CB
1129 512 14 (-) 13 SB
gr
gs 1096 2851 147 238 CB
222 562 33 (in) 32 SB
254 562 32 (g ) 31 SB
285 562 54 (Fie) 53 SB
338 562 12 (l) 11 SB
349 562 32 (d ) 31 SB
380 562 26 (E) 25 SB
405 562 40 (ne) 39 SB
444 562 14 (r) 13 SB
457 562 53 (gy ) 52 SB
509 562 70 (Ana) 69 SB
578 562 12 (l) 11 SB
589 562 40 (yz) 39 SB
628 562 33 (er) 32 SB
660 562 11 ( ) 10 SB
670 562 91 (\(RFE) 90 SB
760 562 55 (A\) ) 54 SB
814 562 35 ([4) 34 SB
848 562 25 (] ) 24 SB
872 562 49 (wa) 48 SB
920 562 16 (s) 17 SB
937 562 11 ( ) 9 SB
946 562 77 (used) 76 SB
1022 562 11 ( ) 10 SB
1032 562 33 (to) 32 SB
gr
gs 1096 2851 147 238 CB
222 612 33 (m) 32 SB
254 612 38 (ea) 37 SB
291 612 70 (sure) 69 SB
360 612 11 ( ) 10 SB
370 612 33 (th) 32 SB
402 612 30 (e ) 29 SB
431 612 19 (e) 18 SB
449 612 31 (le) 30 SB
479 612 19 (c) 18 SB
497 612 26 (tr) 25 SB
522 612 53 (on ) 52 SB
574 612 40 (pa) 39 SB
613 612 33 (ra) 32 SB
645 612 24 (ll) 23 SB
668 612 19 (e) 18 SB
686 612 23 (l ) 22 SB
708 612 19 (e) 18 SB
726 612 40 (ne) 39 SB
765 612 14 (r) 13 SB
778 612 53 (gy ) 52 SB
830 612 33 (di) 32 SB
862 612 16 (s) 17 SB
879 612 12 (t) 11 SB
890 612 26 (ri) 25 SB
915 612 54 (but) 53 SB
968 612 54 (ion) 53 SB
1021 612 22 (. ) 21 SB
gr
gs 1096 2851 147 238 CB
222 685 26 (T) 25 SB
247 685 40 (he) 39 SB
286 685 30 ( e) 29 SB
315 685 12 (l) 11 SB
326 685 38 (ec) 37 SB
363 685 12 (t) 11 SB
374 685 67 (ron ) 66 SB
440 685 33 (di) 32 SB
472 685 14 (f) 13 SB
485 685 63 (fusi) 62 SB
547 685 53 (on ) 52 SB
599 685 19 (c) 18 SB
617 685 40 (oe) 39 SB
656 685 14 (f) 13 SB
669 685 45 (fic) 44 SB
713 685 12 (i) 11 SB
724 685 40 (en) 39 SB
763 685 23 (t ) 22 SB
785 685 19 (a) 18 SB
803 685 33 (cr) 32 SB
835 685 53 (oss) 54 SB
889 685 11 ( ) 10 SB
899 685 12 (t) 11 SB
910 685 40 (he) 39 SB
949 685 32 ( g) 31 SB
980 685 74 (ood ) 73 SB
1053 685 33 (m) 32 SB
1085 685 19 (a) 18 SB
1103 685 21 (g) 21 SB
gr
gs 1096 2851 147 238 CB
1124 685 14 (-) 14 SB
gr
gs 1096 2851 147 238 CB
222 735 40 (ne) 39 SB
261 735 24 (ti) 23 SB
284 735 19 (c) 18 SB
302 735 62 ( sur) 61 SB
363 735 33 (fa) 32 SB
395 735 38 (ce) 37 SB
432 735 27 (s ) 26 SB
458 735 39 (is ) 38 SB
496 735 54 (plo) 53 SB
549 735 24 (tt) 23 SB
572 735 40 (ed) 39 SB
611 735 23 ( i) 22 SB
633 735 32 (n ) 31 SB
664 735 56 (Fig) 55 SB
719 735 22 (. ) 21 SB
740 735 32 (1 ) 31 SB
771 735 19 (a) 18 SB
789 735 46 (s a) 45 SB
834 735 11 ( ) 10 SB
844 735 56 (fun) 55 SB
899 735 31 (ct) 30 SB
929 735 33 (io) 32 SB
961 735 32 (n ) 31 SB
992 735 46 (of ) 45 SB
1037 735 33 (th) 32 SB
1069 735 19 (e) 19 SB
gr
gs 1096 2851 147 238 CB
222 785 19 (e) 18 SB
240 785 31 (le) 30 SB
270 785 19 (c) 18 SB
288 785 47 (tro) 46 SB
334 785 75 (n-ne) 74 SB
408 785 33 (ut) 32 SB
440 785 33 (ra) 32 SB
472 785 23 (l ) 22 SB
494 785 40 (pa) 39 SB
533 785 26 (rt) 25 SB
558 785 31 (ic) 30 SB
588 785 12 (l) 11 SB
599 785 30 (e ) 29 SB
628 785 19 (c) 18 SB
646 785 33 (ol) 32 SB
678 785 24 (li) 23 SB
701 785 16 (s) 17 SB
718 785 12 (i) 11 SB
729 785 61 (ona) 60 SB
789 785 12 (l) 11 SB
800 785 24 (it) 23 SB
823 785 21 (y) 18 SB
841 785 11 (,) 10 SB
851 785 23 ( t) 22 SB
873 785 40 (ha) 39 SB
912 785 23 (t ) 22 SB
934 785 12 (i) 11 SB
945 785 16 (s) 17 SB
962 785 11 ( ) 10 SB
972 785 33 (re) 32 SB
1004 785 35 (pr) 34 SB
1038 785 19 (e) 19 SB
gr
gs 1096 2851 147 238 CB
1057 785 14 (-) 13 SB
gr
gs 1096 2851 147 238 CB
222 835 56 (sen) 55 SB
277 835 31 (te) 30 SB
307 835 32 (d ) 31 SB
338 835 53 (by ) 52 SB
390 835 12 (t) 11 SB
401 835 40 (he) 39 SB
440 835 25 ( r) 24 SB
464 835 31 (at) 30 SB
494 835 33 (io) 32 SB
526 835 32 ( o) 31 SB
557 835 25 (f ) 24 SB
581 835 52 (the) 51 SB
632 835 11 ( ) 10 SB
642 835 19 (e) 18 SB
660 835 31 (le) 30 SB
690 835 19 (c) 18 SB
708 835 47 (tro) 46 SB
754 835 32 (n ) 31 SB
785 835 52 (me) 51 SB
836 835 19 (a) 18 SB
854 835 32 (n ) 31 SB
885 835 47 (fre) 46 SB
931 835 19 (e) 18 SB
949 835 11 ( ) 10 SB
959 835 40 (pa) 39 SB
998 835 44 (th ) 43 SB
gr
32 0 0 42 42 0 0 0 42 /Symbol font
gs 1096 2851 147 238 CB
1042 831 23 (l) 23 SB
gr
gs 1096 2851 147 238 CB
1041 831 23 (l) 23 SB
gr
32 0 0 33 33 0 0 0 33 /Symbol font
gs 1096 2851 147 238 CB
1065 849 17 (n) 17 SB
gr
gs 1096 2851 147 238 CB
1064 849 17 (n) 17 SB
gr
32 0 0 42 42 0 0 0 38 /Times-Roman /font32 ANSIFont font
gs 1096 2851 147 238 CB
222 885 61 (ove) 60 SB
282 885 25 (r ) 24 SB
gr
32 0 0 42 42 0 0 0 42 /Symbol font
gs 1096 2851 147 238 CB
306 881 23 (p) 23 SB
gr
32 0 0 42 42 0 0 0 38 /Times-Roman /font32 ANSIFont font
gs 1096 2851 147 238 CB
329 885 28 (R) 27 SB
356 885 22 (. ) 21 SB
377 885 26 (T) 25 SB
402 885 40 (he) 39 SB
441 885 11 ( ) 10 SB
451 885 40 (ch) 39 SB
490 885 40 (an) 39 SB
529 885 40 (ge) 39 SB
568 885 32 ( o) 31 SB
599 885 25 (f ) 24 SB
623 885 33 (th) 32 SB
655 885 30 (e ) 29 SB
684 885 19 (e) 18 SB
702 885 12 (l) 11 SB
713 885 38 (ec) 37 SB
750 885 12 (t) 11 SB
761 885 70 (ron-) 69 SB
830 885 40 (ne) 39 SB
869 885 47 (utr) 46 SB
915 885 31 (al) 30 SB
945 885 11 ( ) 10 SB
955 885 19 (c) 18 SB
973 885 45 (oll) 44 SB
1017 885 40 (isi) 39 SB
1056 885 61 (ona) 60 SB
1116 885 12 (l) 12 SB
gr
gs 1096 2851 147 238 CB
1128 885 14 (-) 13 SB
gr
gs 1096 2851 147 238 CB
222 935 24 (it) 23 SB
245 935 32 (y ) 31 SB
276 935 39 (is ) 38 SB
314 935 19 (a) 18 SB
332 935 40 (ch) 39 SB
371 935 31 (ie) 30 SB
401 935 40 (ve) 39 SB
440 935 32 (d ) 31 SB
471 935 53 (by ) 52 SB
523 935 19 (c) 18 SB
541 935 40 (ha) 39 SB
580 935 54 (ngi) 53 SB
633 935 53 (ng ) 52 SB
685 935 35 (pr) 34 SB
719 935 105 (essure) 104 SB
823 935 11 ( ) 10 SB
833 935 46 (of ) 45 SB
878 935 12 (t) 11 SB
889 935 40 (he) 39 SB
928 935 32 ( h) 31 SB
959 935 31 (el) 30 SB
989 935 33 (iu) 32 SB
1021 935 44 (m ) 43 SB
1064 935 40 (ga) 39 SB
1103 935 16 (s) 16 SB
gr
gs 1096 2851 147 238 CB
222 985 33 (in) 32 SB
254 985 23 ( t) 22 SB
276 985 40 (he) 39 SB
315 985 11 ( ) 10 SB
325 985 40 (va) 39 SB
364 985 40 (cu) 39 SB
403 985 54 (um) 53 SB
456 985 32 ( v) 31 SB
487 985 70 (esse) 69 SB
556 985 23 (l ) 22 SB
578 985 19 (c) 18 SB
596 985 40 (ha) 39 SB
635 985 54 (mb) 53 SB
688 985 33 (er) 30 SB
718 985 11 (.) 10 SB
728 985 11 ( ) 10 SB
738 985 68 (Plot) 67 SB
805 985 31 (te) 30 SB
835 985 32 (d ) 31 SB
866 985 19 (a) 18 SB
884 985 60 (lso ) 59 SB
943 985 12 (i) 11 SB
954 985 32 (n ) 31 SB
985 985 67 (Fig.) 66 SB
1051 985 11 ( ) 10 SB
1061 985 32 (1 ) 31 SB
1092 985 28 (is) 28 SB
gr
gs 1096 2851 147 238 CB
222 1035 33 (th) 32 SB
254 1035 30 (e ) 29 SB
283 1035 19 (c) 18 SB
301 1035 31 (la) 30 SB
331 1035 63 (ssic) 62 SB
393 1035 19 (a) 18 SB
411 1035 23 (l ) 22 SB
433 1035 14 ([) 14 SB
gr
32 0 0 42 42 0 0 0 39 /Times-Italic /font31 ANSIFont font
gs 1096 2851 147 238 CB
447 1034 30 (D) 30 SB
gr
32 0 0 42 42 0 0 0 38 /Times-Roman /font32 ANSIFont font
gs 1096 2851 147 238 CB
477 1035 11 ( ) 10 SB
gr
32 0 0 44 44 0 0 0 44 /Symbol font
gs 1096 2851 147 238 CB
487 1029 24 (=) 23 SB
gr
32 0 0 42 42 0 0 0 38 /Times-Roman /font32 ANSIFont font
gs 1096 2851 147 238 CB
510 1035 11 ( ) 10 SB
gr
32 0 0 44 44 0 0 0 44 /Symbol font
gs 1096 2851 147 238 CB
520 1029 15 (\() 15 SB
gr
32 0 0 42 42 0 0 0 42 /Symbol font
gs 1096 2851 147 238 CB
534 1031 23 (r) 23 SB
gr
32 0 0 33 33 0 0 0 30 /Times-Roman /font32 ANSIFont font
gs 1096 2851 147 238 CB
557 1052 15 (e) 14 SB
gr
32 0 0 42 42 0 0 0 42 /Symbol font
gs 1096 2851 147 238 CB
571 1031 14 (\)) 14 SB
gr
32 0 0 35 35 0 0 0 31 /Times-Roman /font32 ANSIFont font
gs 1096 2851 147 238 CB
585 1022 18 (2) 17 SB
gr
32 0 0 44 44 0 0 0 44 /Symbol font
gs 1096 2851 147 238 CB
602 1029 23 (n) 21 SB
gr
32 0 0 35 35 0 0 0 31 /Times-Roman /font32 ANSIFont font
gs 1096 2851 147 238 CB
623 1051 16 (e) 15 SB
638 1051 18 (n) 16 SB
gr
32 0 0 42 42 0 0 0 38 /Times-Roman /font32 ANSIFont font
gs 1096 2851 147 238 CB
654 1035 25 (] ) 24 SB
678 1035 40 (an) 39 SB
717 1035 32 (d ) 31 SB
748 1035 40 (co) 39 SB
787 1035 61 (nve) 60 SB
847 1035 33 (nt) 32 SB
879 1035 54 (ion) 53 SB
932 1035 31 (al) 30 SB
962 1035 11 ( ) 10 SB
972 1035 38 (ca) 37 SB
1009 1035 12 (l) 11 SB
1020 1035 40 (cu) 39 SB
1059 1035 31 (la) 30 SB
1089 1035 24 (ti) 23 SB
1112 1035 42 (on) 42 SB
gr
gs 1096 2851 147 238 CB
222 1085 46 (of ) 45 SB
267 1085 12 (t) 11 SB
278 1085 40 (he) 39 SB
317 1085 32 ( n) 31 SB
348 1085 40 (eo) 39 SB
387 1085 31 (cl) 30 SB
417 1085 19 (a) 18 SB
435 1085 16 (s) 17 SB
452 1085 28 (si) 27 SB
479 1085 38 (ca) 37 SB
516 1085 12 (l) 11 SB
527 1085 23 ( t) 22 SB
549 1085 61 (oka) 60 SB
609 1085 33 (m) 32 SB
641 1085 19 (a) 18 SB
659 1085 32 (k ) 31 SB
690 1085 45 (tra) 44 SB
734 1085 79 (nspo) 78 SB
812 1085 37 (rt ) 36 SB
848 1085 33 (di) 32 SB
880 1085 14 (f) 13 SB
893 1085 63 (fusi) 62 SB
955 1085 53 (on ) 52 SB
1007 1085 19 (c) 18 SB
1025 1085 40 (oe) 39 SB
1064 1085 14 (f) 13 SB
1077 1085 26 (fi) 26 SB
gr
gs 1096 2851 147 238 CB
1103 1085 14 (-) 13 SB
gr
gs 1096 2851 147 238 CB
222 1135 19 (c) 18 SB
240 1135 31 (ie) 30 SB
270 1135 33 (nt) 32 SB
302 1135 32 ( o) 31 SB
333 1135 25 (f ) 24 SB
357 1135 52 (the) 51 SB
408 1135 11 ( ) 10 SB
418 1135 19 (e) 18 SB
436 1135 31 (le) 30 SB
466 1135 19 (c) 18 SB
484 1135 47 (tro) 46 SB
530 1135 59 (ns, ) 58 SB
588 1135 70 (whe) 69 SB
657 1135 33 (re) 32 SB
689 1135 11 ( ) 10 SB
699 1135 33 (th) 32 SB
731 1135 30 (e ) 29 SB
760 1135 19 (c) 18 SB
778 1135 54 (oul) 53 SB
831 1135 54 (om) 53 SB
884 1135 32 (b ) 31 SB
915 1135 19 (c) 18 SB
933 1135 45 (oll) 44 SB
977 1135 40 (isi) 39 SB
1016 1135 53 (on ) 52 SB
1068 1135 47 (fre) 46 SB
gr
gs 1096 2851 147 238 CB
1114 1135 14 (-) 14 SB
gr
gs 1096 2851 147 238 CB
222 1185 61 (que) 60 SB
282 1185 40 (nc) 39 SB
321 1185 32 (y ) 31 SB
gr
32 0 0 42 42 0 0 0 42 /Symbol font
gs 1096 2851 147 238 CB
352 1181 22 (n) 21 SB
gr
32 0 0 42 42 0 0 0 38 /Times-Roman /font32 ANSIFont font
gs 1096 2851 147 238 CB
373 1185 32 ( h) 31 SB
404 1185 46 (as ) 45 SB
449 1185 40 (be) 39 SB
488 1185 19 (e) 18 SB
506 1185 32 (n ) 31 SB
537 1185 33 (re) 32 SB
569 1185 52 (pla) 51 SB
620 1185 19 (c) 18 SB
638 1185 19 (e) 18 SB
656 1185 32 (d ) 31 SB
687 1185 53 (by ) 52 SB
739 1185 33 (th) 32 SB
771 1185 30 (e ) 29 SB
800 1185 19 (e) 18 SB
818 1185 31 (le) 30 SB
848 1185 19 (c) 18 SB
866 1185 26 (tr) 25 SB
891 1185 77 (on-n) 76 SB
967 1185 40 (eu) 39 SB
1006 1185 45 (tra) 44 SB
1050 1185 12 (l) 11 SB
gr
gs 1096 2851 147 238 CB
222 1235 19 (c) 18 SB
240 1235 45 (oll) 44 SB
284 1235 40 (isi) 39 SB
323 1235 53 (on ) 52 SB
375 1235 47 (fre) 46 SB
421 1235 42 (qu) 41 SB
462 1235 40 (en) 39 SB
501 1235 40 (cy) 39 SB
540 1235 11 ( ) 11 SB
gr
32 0 0 42 42 0 0 0 42 /Symbol font
gs 1096 2851 147 238 CB
551 1231 22 (n) 21 SB
gr
32 0 0 33 33 0 0 0 30 /Times-Roman /font32 ANSIFont font
gs 1096 2851 147 238 CB
572 1252 32 (en) 31 SB
gr
32 0 0 42 42 0 0 0 38 /Times-Roman /font32 ANSIFont font
gs 1096 2851 147 238 CB
603 1235 22 (. ) 21 SB
624 1235 49 (He) 48 SB
672 1235 33 (re) 32 SB
gr
32 0 0 42 42 0 0 0 39 /Times-Italic /font31 ANSIFont font
gs 1096 2851 147 238 CB
222 1284 30 (D) 30 SB
gr
32 0 0 33 33 0 0 0 30 /Times-Roman /font32 ANSIFont font
gs 1096 2851 147 238 CB
252 1302 50 (G.S) 51 SB
303 1302 8 (.) 8 SB
gr
32 0 0 42 42 0 0 0 38 /Times-Roman /font32 ANSIFont font
gs 1096 2851 147 238 CB
311 1285 11 ( ) 10 SB
gr
32 0 0 42 42 0 0 0 42 /Symbol font
gs 1096 2851 147 238 CB
321 1281 23 (=) 23 SB
gr
32 0 0 42 42 0 0 0 38 /Times-Roman /font32 ANSIFont font
gs 1096 2851 147 238 CB
344 1285 11 ( ) 10 SB
gr
32 0 0 42 42 0 0 0 42 /Symbol font
gs 1096 2851 147 238 CB
354 1281 14 (\() 14 SB
gr
32 0 0 42 42 0 0 0 39 /Times-Italic /font31 ANSIFont font
gs 1096 2851 147 238 CB
368 1284 26 (R) 25 SB
gr
32 0 0 35 35 0 0 0 31 /Times-Roman /font32 ANSIFont font
gs 1096 2851 147 238 CB
393 1280 9 ( ) 9 SB
gr
32 0 0 42 42 0 0 0 42 /Symbol font
gs 1096 2851 147 238 CB
402 1281 7 (\244) 6 SB
gr
32 0 0 33 33 0 0 0 30 /Times-Roman /font32 ANSIFont font
gs 1096 2851 147 238 CB
408 1298 8 ( ) 9 SB
gr
32 0 0 42 42 0 0 0 39 /Times-Italic /font31 ANSIFont font
gs 1096 2851 147 238 CB
417 1284 16 (r) 16 SB
gr
32 0 0 42 42 0 0 0 42 /Symbol font
gs 1096 2851 147 238 CB
433 1281 14 (\)) 14 SB
gr
32 0 0 25 25 0 0 0 22 /Times-Roman /font32 ANSIFont font
gs 1096 2851 147 238 CB
447 1270 13 (3) 12 SB
gr
32 0 0 33 33 0 0 0 33 /Symbol font
gs 1096 2851 147 238 CB
459 1270 6 (\244) 6 SB
gr
32 0 0 25 25 0 0 0 22 /Times-Roman /font32 ANSIFont font
gs 1096 2851 147 238 CB
465 1286 13 (2) 12 SB
gr
32 0 0 42 42 0 0 0 42 /Symbol font
gs 1096 2851 147 238 CB
477 1281 14 (\() 14 SB
gr
32 0 0 42 42 0 0 0 38 /Times-Roman /font32 ANSIFont font
gs 1096 2851 147 238 CB
491 1285 21 (2) 21 SB
gr
32 0 0 42 42 0 0 0 42 /Symbol font
gs 1096 2851 147 238 CB
512 1281 23 (p) 22 SB
gr
32 0 0 33 33 0 0 0 30 /Times-Roman /font32 ANSIFont font
gs 1096 2851 147 238 CB
534 1281 8 ( ) 9 SB
gr
32 0 0 42 42 0 0 0 42 /Symbol font
gs 1096 2851 147 238 CB
543 1281 7 (\244) 6 SB
gr
32 0 0 33 33 0 0 0 30 /Times-Roman /font32 ANSIFont font
gs 1096 2851 147 238 CB
549 1298 8 ( ) 9 SB
gr
32 0 0 42 42 0 0 0 39 /Times-Italic /font31 ANSIFont font
gs 1096 2851 147 238 CB
558 1284 12 (t) 11 SB
gr
32 0 0 42 42 0 0 0 42 /Symbol font
gs 1096 2851 147 238 CB
569 1281 14 (\)) 14 SB
gr
32 0 0 33 33 0 0 0 30 /Times-Roman /font32 ANSIFont font
gs 1096 2851 147 238 CB
583 1273 17 (2) 16 SB
gr
32 0 0 42 42 0 0 0 42 /Symbol font
gs 1096 2851 147 238 CB
599 1281 14 (\() 14 SB
gr
gs 1096 2851 147 238 CB
613 1281 23 (r) 23 SB
gr
32 0 0 33 33 0 0 0 30 /Times-Roman /font32 ANSIFont font
gs 1096 2851 147 238 CB
636 1302 15 (e) 15 SB
gr
32 0 0 42 42 0 0 0 42 /Symbol font
gs 1096 2851 147 238 CB
651 1281 14 (\)) 13 SB
gr
32 0 0 33 33 0 0 0 30 /Times-Roman /font32 ANSIFont font
gs 1096 2851 147 238 CB
664 1273 17 (2) 17 SB
gr
32 0 0 42 42 0 0 0 42 /Symbol font
gs 1096 2851 147 238 CB
681 1281 22 (n) 21 SB
gr
32 0 0 33 33 0 0 0 30 /Times-Roman /font32 ANSIFont font
gs 1096 2851 147 238 CB
702 1302 32 (en) 32 SB
gr
32 0 0 42 42 0 0 0 38 /Times-Roman /font32 ANSIFont font
gs 1096 2851 147 238 CB
734 1285 11 ( ) 10 SB
744 1285 11 ( ) 10 SB
754 1285 39 (is ) 38 SB
792 1285 33 (th) 32 SB
824 1285 30 (e ) 29 SB
853 1285 49 (Ga) 48 SB
901 1285 31 (le) 30 SB
931 1285 19 (e) 18 SB
949 1285 77 (v-Sa) 76 SB
1025 1285 42 (gd) 41 SB
1066 1285 38 (ee) 37 SB
1103 1285 21 (v) 21 SB
gr
gs 1096 2851 147 238 CB
222 1335 33 (di) 32 SB
254 1335 14 (f) 13 SB
267 1335 84 (fusio) 83 SB
350 1335 32 (n ) 31 SB
381 1335 40 (co) 39 SB
420 1335 33 (ef) 32 SB
452 1335 14 (f) 13 SB
465 1335 31 (ic) 30 SB
495 1335 31 (ie) 30 SB
525 1335 33 (nt) 32 SB
557 1335 11 (,) 10 SB
567 1335 30 ( a) 29 SB
596 1335 42 (nd) 41 SB
637 1335 11 ( ) 11 SB
gr
32 0 0 42 42 0 0 0 39 /Times-Italic /font31 ANSIFont font
gs 1096 2851 147 238 CB
648 1334 30 (D) 30 SB
gr
32 0 0 33 33 0 0 0 30 /Times-Roman /font32 ANSIFont font
gs 1096 2851 147 238 CB
678 1352 18 (P) 14 SB
692 1352 26 (.S) 27 SB
719 1352 8 (.) 8 SB
gr
32 0 0 42 42 0 0 0 38 /Times-Roman /font32 ANSIFont font
gs 1096 2851 147 238 CB
727 1335 11 ( ) 10 SB
737 1335 35 (= ) 34 SB
gr
32 0 0 42 42 0 0 0 42 /Symbol font
gs 1096 2851 147 238 CB
771 1331 14 (\() 14 SB
gr
32 0 0 42 42 0 0 0 38 /Times-Roman /font32 ANSIFont font
gs 1096 2851 147 238 CB
785 1335 21 (2) 20 SB
gr
32 0 0 42 42 0 0 0 42 /Symbol font
gs 1096 2851 147 238 CB
805 1331 23 (p) 23 SB
gr
32 0 0 33 33 0 0 0 30 /Times-Roman /font32 ANSIFont font
gs 1096 2851 147 238 CB
828 1331 8 ( ) 8 SB
gr
32 0 0 42 42 0 0 0 42 /Symbol font
gs 1096 2851 147 238 CB
836 1331 7 (\244) 7 SB
gr
32 0 0 33 33 0 0 0 30 /Times-Roman /font32 ANSIFont font
gs 1096 2851 147 238 CB
843 1348 8 ( ) 8 SB
gr
32 0 0 42 42 0 0 0 39 /Times-Italic /font31 ANSIFont font
gs 1096 2851 147 238 CB
851 1334 12 (t) 12 SB
gr
32 0 0 42 42 0 0 0 42 /Symbol font
gs 1096 2851 147 238 CB
863 1331 14 (\)) 14 SB
gr
32 0 0 33 33 0 0 0 30 /Times-Roman /font32 ANSIFont font
gs 1096 2851 147 238 CB
877 1323 17 (2) 16 SB
gr
32 0 0 42 42 0 0 0 42 /Symbol font
gs 1096 2851 147 238 CB
893 1331 14 (\() 14 SB
gr
gs 1096 2851 147 238 CB
907 1331 23 (r) 23 SB
gr
32 0 0 33 33 0 0 0 30 /Times-Roman /font32 ANSIFont font
gs 1096 2851 147 238 CB
930 1352 15 (e) 14 SB
gr
32 0 0 44 44 0 0 0 44 /Symbol font
gs 1096 2851 147 238 CB
944 1329 15 (\)) 14 SB
gr
32 0 0 35 35 0 0 0 31 /Times-Roman /font32 ANSIFont font
gs 1096 2851 147 238 CB
958 1322 18 (2) 17 SB
gr
32 0 0 44 44 0 0 0 44 /Symbol font
gs 1096 2851 147 238 CB
975 1329 23 (n) 21 SB
gr
32 0 0 33 33 0 0 0 30 /Times-Roman /font32 ANSIFont font
gs 1096 2851 147 238 CB
996 1352 32 (en) 31 SB
gr
32 0 0 42 42 0 0 0 38 /Times-Roman /font32 ANSIFont font
gs 1096 2851 147 238 CB
1027 1335 23 ( i) 22 SB
1049 1335 39 (s t) 38 SB
1087 1335 40 (he) 39 SB
gr
gs 1096 2851 147 238 CB
222 1385 49 (Pfi) 48 SB
270 1385 70 (rsch) 69 SB
339 1385 56 (-Sc) 55 SB
394 1385 54 (hlu) 53 SB
447 1385 31 (te) 30 SB
477 1385 25 (r ) 24 SB
501 1385 33 (di) 32 SB
533 1385 14 (f) 13 SB
546 1385 84 (fusio) 83 SB
629 1385 32 (n ) 31 SB
660 1385 40 (co) 39 SB
699 1385 33 (ef) 32 SB
731 1385 14 (f) 13 SB
744 1385 31 (ic) 30 SB
774 1385 31 (ie) 30 SB
804 1385 33 (nt) 32 SB
836 1385 11 (.) 10 SB
846 1385 25 ( I) 24 SB
870 1385 32 (n ) 31 SB
901 1385 52 (the) 51 SB
952 1385 35 (se) 34 SB
986 1385 30 ( e) 29 SB
1015 1385 42 (xp) 41 SB
1056 1385 33 (re) 32 SB
1088 1385 16 (s) 17 SB
gr
gs 1096 2851 147 238 CB
1105 1385 14 (-) 13 SB
gr
gs 1096 2851 147 238 CB
222 1435 70 (sion) 69 SB
291 1435 16 (s) 17 SB
308 1435 11 (,) 10 SB
318 1435 11 ( ) 10 SB
gr
32 0 0 42 42 0 0 0 42 /Symbol font
gs 1096 2851 147 238 CB
328 1431 23 (r) 23 SB
gr
32 0 0 33 33 0 0 0 30 /Times-Roman /font32 ANSIFont font
gs 1096 2851 147 238 CB
351 1452 15 (e) 15 SB
gr
32 0 0 42 42 0 0 0 38 /Times-Roman /font32 ANSIFont font
gs 1096 2851 147 238 CB
366 1435 11 ( ) 10 SB
gr
32 0 0 42 42 0 0 0 42 /Symbol font
gs 1096 2851 147 238 CB
376 1431 23 (=) 23 SB
gr
32 0 0 42 42 0 0 0 38 /Times-Roman /font32 ANSIFont font
gs 1096 2851 147 238 CB
399 1435 11 ( ) 10 SB
gr
32 0 0 42 42 0 0 0 41 /NewCenturySchlbk-Italic /font22 ANSIFont font
gs 1096 2851 147 238 CB
409 1432 22 (v) 23 SB
gr
32 0 0 33 33 0 0 0 33 /Symbol font
gs 1096 2851 147 238 CB
430 1449 22 (^) 22 SB
gr
32 0 0 33 33 0 0 0 30 /Times-Roman /font32 ANSIFont font
gs 1096 2851 147 238 CB
452 1431 8 ( ) 8 SB
gr
32 0 0 42 42 0 0 0 42 /Symbol font
gs 1096 2851 147 238 CB
460 1431 7 (\244) 7 SB
gr
32 0 0 33 33 0 0 0 30 /Times-Roman /font32 ANSIFont font
gs 1096 2851 147 238 CB
467 1448 8 ( ) 8 SB
gr
32 0 0 42 42 0 0 0 42 /Symbol font
gs 1096 2851 147 238 CB
475 1431 29 (w) 29 SB
gr
32 0 0 33 33 0 0 0 30 /Times-Roman /font32 ANSIFont font
gs 1096 2851 147 238 CB
504 1452 30 (ce) 29 SB
gr
32 0 0 42 42 0 0 0 38 /Times-Roman /font32 ANSIFont font
gs 1096 2851 147 238 CB
533 1435 23 ( i) 22 SB
555 1435 39 (s t) 38 SB
593 1435 40 (he) 39 SB
632 1435 11 ( ) 10 SB
642 1435 31 (el) 30 SB
672 1435 19 (e) 18 SB
690 1435 31 (ct) 30 SB
720 1435 56 (ron) 55 SB
775 1435 37 ( L) 36 SB
811 1435 19 (a) 18 SB
829 1435 47 (rm) 46 SB
875 1435 46 (or ) 45 SB
920 1435 33 (ra) 32 SB
952 1435 33 (di) 32 SB
984 1435 48 (us,) 47 SB
1031 1435 30 ( a) 29 SB
1060 1435 42 (nd) 41 SB
gr
32 0 0 42 42 0 0 0 42 /Symbol font
gs 1096 2851 147 238 CB
222 1481 29 (w) 29 SB
gr
32 0 0 33 33 0 0 0 30 /Times-Roman /font32 ANSIFont font
gs 1096 2851 147 238 CB
251 1502 15 (c) 14 SB
265 1502 15 (e) 15 SB
gr
32 0 0 42 42 0 0 0 38 /Times-Roman /font32 ANSIFont font
gs 1096 2851 147 238 CB
280 1485 11 ( ) 10 SB
290 1485 39 (is ) 38 SB
328 1485 52 (the) 51 SB
379 1485 11 ( ) 10 SB
389 1485 19 (e) 18 SB
407 1485 31 (le) 30 SB
437 1485 19 (c) 18 SB
455 1485 47 (tro) 46 SB
501 1485 32 (n ) 31 SB
532 1485 77 (gyro) 76 SB
608 1485 47 (fre) 46 SB
654 1485 61 (que) 60 SB
714 1485 40 (nc) 39 SB
753 1485 21 (y) 18 SB
771 1485 11 (.) 10 SB
781 1485 11 ( ) 10 SB
791 1485 40 (W) 35 SB
826 1485 30 (e ) 29 SB
855 1485 19 (a) 18 SB
873 1485 86 (ssum) 85 SB
958 1485 40 (ed) 39 SB
997 1485 11 ( ) 10 SB
gr
32 0 0 42 42 0 0 0 41 /NewCenturySchlbk-Italic /font22 ANSIFont font
gs 1096 2851 147 238 CB
1007 1482 22 (v) 22 SB
gr
32 0 0 33 33 0 0 0 33 /Symbol font
gs 1096 2851 147 238 CB
1029 1499 22 (^) 22 SB
gr
32 0 0 42 42 0 0 0 38 /Times-Roman /font32 ANSIFont font
gs 1096 2851 147 238 CB
1051 1485 11 ( ) 10 SB
gr
32 0 0 42 42 0 0 0 42 /Symbol font
gs 1096 2851 147 238 CB
1061 1481 23 (=) 23 SB
gr
32 0 0 42 42 0 0 0 38 /Times-Roman /font32 ANSIFont font
gs 1096 2851 147 238 CB
1084 1485 11 ( ) 10 SB
gr
32 0 0 42 42 0 0 0 41 /NewCenturySchlbk-Italic /font22 ANSIFont font
gs 1096 2851 147 238 CB
1094 1482 22 (v) 22 SB
gr
32 0 0 44 44 0 0 0 44 /Symbol font
gs 1096 2851 147 238 CB
1094 1436 22 (_) 21 SB
gr
gs 1096 2851 147 238 CB
1095 1436 22 (_) 20 SB
gr
32 0 0 33 33 0 0 0 30 /Times-Roman /font32 ANSIFont font
gs 1096 2851 147 238 CB
1115 1502 29 (e||) 28 SB
gr
32 0 0 42 42 0 0 0 38 /Times-Roman /font32 ANSIFont font
gs 1096 2851 147 238 CB
222 1535 33 (in) 32 SB
254 1535 23 ( t) 22 SB
276 1535 40 (he) 39 SB
315 1535 11 ( ) 10 SB
325 1535 47 (est) 46 SB
371 1535 45 (im) 44 SB
415 1535 19 (a) 18 SB
433 1535 24 (ti) 23 SB
456 1535 53 (on ) 52 SB
508 1535 46 (of ) 45 SB
553 1535 12 (t) 11 SB
564 1535 40 (he) 39 SB
603 1535 32 ( n) 31 SB
634 1535 40 (eo) 39 SB
673 1535 31 (cl) 30 SB
703 1535 19 (a) 18 SB
721 1535 16 (s) 17 SB
738 1535 28 (si) 27 SB
765 1535 38 (ca) 37 SB
802 1535 12 (l) 11 SB
813 1535 23 ( t) 22 SB
835 1535 33 (ra) 32 SB
867 1535 93 (nspor) 92 SB
959 1535 23 (t ) 22 SB
981 1535 19 (c) 18 SB
999 1535 40 (oe) 39 SB
1038 1535 14 (f) 13 SB
1051 1535 26 (fi) 26 SB
gr
gs 1096 2851 147 238 CB
1077 1535 14 (-) 13 SB
gr
gs 1096 2851 147 238 CB
222 1585 19 (c) 18 SB
240 1585 31 (ie) 30 SB
270 1585 33 (nt) 32 SB
302 1585 16 (s) 17 SB
319 1585 11 ( ) 10 SB
329 1585 42 (wi) 41 SB
370 1585 44 (th ) 43 SB
gr
32 0 0 42 42 0 0 0 41 /NewCenturySchlbk-Italic /font22 ANSIFont font
gs 1096 2851 147 238 CB
413 1582 22 (v) 21 SB
gr
32 0 0 42 42 0 0 0 42 /Symbol font
gs 1096 2851 147 238 CB
413 1538 21 (_) 21 SB
gr
gs 1096 2851 147 238 CB
414 1538 21 (_) 20 SB
gr
32 0 0 35 35 0 0 0 31 /Times-Roman /font32 ANSIFont font
gs 1096 2851 147 238 CB
434 1601 23 (e|) 24 SB
458 1601 7 (|) 7 SB
gr
32 0 0 42 42 0 0 0 38 /Times-Roman /font32 ANSIFont font
gs 1096 2851 147 238 CB
462 1585 11 ( ) 10 SB
472 1585 52 (the) 51 SB
523 1585 11 ( ) 10 SB
533 1585 33 (m) 32 SB
565 1585 38 (ea) 37 SB
602 1585 32 (n ) 31 SB
633 1585 40 (pa) 39 SB
672 1585 33 (ra) 32 SB
704 1585 12 (l) 11 SB
715 1585 31 (le) 30 SB
745 1585 23 (l ) 22 SB
767 1585 19 (e) 18 SB
785 1585 31 (le) 30 SB
815 1585 19 (c) 18 SB
833 1585 26 (tr) 25 SB
858 1585 53 (on ) 52 SB
910 1585 40 (ve) 39 SB
949 1585 33 (lo) 32 SB
981 1585 31 (ci) 30 SB
1011 1585 33 (ty) 32 SB
1043 1585 30 ( a) 29 SB
1072 1585 16 (s) 16 SB
gr
gs 1096 2851 147 238 CB
222 1635 33 (m) 32 SB
254 1635 38 (ea) 37 SB
291 1635 70 (sure) 69 SB
360 1635 32 (d ) 31 SB
391 1635 21 (b) 20 SB
411 1635 32 (y ) 31 SB
442 1635 33 (th) 32 SB
474 1635 30 (e ) 29 SB
503 1635 28 (R) 27 SB
530 1635 90 (FEA.) 89 SB
619 1635 11 ( ) 10 SB
629 1635 42 (Al) 41 SB
670 1635 54 (tho) 53 SB
723 1635 74 (ugh ) 72 SB
795 1635 52 (the) 51 SB
846 1635 11 ( ) 10 SB
856 1635 19 (c) 18 SB
874 1635 12 (l) 11 SB
885 1635 82 (assic) 81 SB
966 1635 19 (a) 18 SB
984 1635 23 (l ) 21 SB
1005 1635 47 (dif) 46 SB
1051 1635 14 (f) 13 SB
1064 1635 91 (usion) 90 SB
gr
gs 1096 2851 147 238 CB
222 1685 19 (c) 18 SB
240 1685 40 (oe) 39 SB
279 1685 14 (f) 13 SB
292 1685 45 (fic) 44 SB
336 1685 12 (i) 11 SB
347 1685 40 (en) 39 SB
386 1685 23 (t ) 22 SB
408 1685 39 (is ) 38 SB
446 1685 33 (m) 32 SB
478 1685 40 (uc) 39 SB
517 1685 32 (h ) 31 SB
548 1685 31 (le) 30 SB
578 1685 32 (ss) 33 SB
611 1685 11 ( ) 10 SB
621 1685 12 (t) 11 SB
632 1685 40 (ha) 39 SB
671 1685 32 (n ) 31 SB
702 1685 52 (the) 51 SB
753 1685 11 ( ) 10 SB
763 1685 19 (e) 18 SB
781 1685 61 (xpe) 60 SB
841 1685 26 (ri) 25 SB
866 1685 52 (me) 51 SB
917 1685 33 (nt) 32 SB
949 1685 19 (a) 18 SB
967 1685 23 (l ) 22 SB
989 1685 33 (m) 32 SB
1021 1685 38 (ea) 37 SB
1058 1685 70 (sure) 69 SB
1127 1685 21 (d) 21 SB
gr
gs 1096 2851 147 238 CB
222 1735 40 (va) 39 SB
261 1735 33 (lu) 32 SB
293 1735 46 (es,) 45 SB
338 1735 11 ( ) 10 SB
348 1735 84 (good) 83 SB
431 1735 30 ( a) 29 SB
460 1735 35 (gr) 34 SB
494 1735 38 (ee) 37 SB
531 1735 33 (m) 32 SB
563 1735 19 (e) 18 SB
581 1735 44 (nt ) 43 SB
624 1735 49 (wa) 48 SB
672 1735 41 (s f) 40 SB
712 1735 84 (ound) 83 SB
795 1735 32 ( b) 31 SB
826 1735 31 (et) 30 SB
856 1735 49 (we) 48 SB
904 1735 40 (en) 39 SB
943 1735 23 ( t) 22 SB
965 1735 40 (he) 39 SB
1004 1735 11 ( ) 10 SB
1014 1735 40 (ne) 39 SB
1053 1735 40 (oc) 39 SB
1092 1735 31 (la) 30 SB
1122 1735 16 (s) 16 SB
gr
gs 1096 2851 147 238 CB
1138 1735 14 (-) 14 SB
gr
gs 1096 2851 147 238 CB
222 1785 47 (sic) 46 SB
268 1785 19 (a) 18 SB
286 1785 23 (l ) 22 SB
308 1785 26 (tr) 25 SB
333 1785 40 (an) 39 SB
372 1785 16 (s) 17 SB
389 1785 21 (p) 20 SB
409 1785 47 (ort) 46 SB
455 1785 30 ( a) 29 SB
484 1785 42 (nd) 41 SB
525 1785 44 ( m) 43 SB
568 1785 19 (e) 18 SB
586 1785 19 (a) 18 SB
604 1785 16 (s) 17 SB
621 1785 21 (u) 20 SB
641 1785 33 (re) 32 SB
673 1785 32 (d ) 31 SB
704 1785 40 (va) 39 SB
743 1785 52 (lue) 51 SB
794 1785 27 (s.) 26 SB
gr
gs 1096 2851 147 238 CB
222 1857 26 (T) 25 SB
247 1857 40 (he) 39 SB
286 1857 30 ( e) 29 SB
315 1857 12 (l) 11 SB
326 1857 38 (ec) 37 SB
363 1857 12 (t) 11 SB
374 1857 67 (ron ) 66 SB
440 1857 12 (t) 11 SB
451 1857 33 (ra) 32 SB
483 1857 105 (nsport) 104 SB
587 1857 11 ( ) 10 SB
597 1857 44 (in ) 43 SB
640 1857 12 (t) 11 SB
651 1857 40 (he) 39 SB
690 1857 39 ( st) 38 SB
728 1857 40 (oc) 39 SB
767 1857 40 (ha) 39 SB
806 1857 16 (s) 17 SB
823 1857 12 (t) 11 SB
834 1857 31 (ic) 30 SB
864 1857 11 ( ) 10 SB
874 1857 33 (m) 32 SB
906 1857 40 (ag) 39 SB
945 1857 40 (ne) 39 SB
984 1857 24 (ti) 23 SB
1007 1857 30 (c ) 29 SB
1036 1857 26 (fi) 25 SB
1061 1857 19 (e) 18 SB
1079 1857 44 (ld ) 43 SB
1122 1857 12 (i) 11 SB
1133 1857 16 (s) 17 SB
gr
gs 1096 2851 147 238 CB
222 1907 120 (shown ) 119 SB
341 1907 12 (i) 11 SB
352 1907 32 (n ) 31 SB
383 1907 67 (Fig.) 66 SB
449 1907 11 ( ) 10 SB
459 1907 32 (2,) 31 SB
490 1907 11 ( ) 10 SB
500 1907 70 (whe) 69 SB
569 1907 33 (re) 32 SB
601 1907 23 ( t) 22 SB
623 1907 40 (he) 39 SB
662 1907 11 ( ) 10 SB
672 1907 31 (el) 30 SB
702 1907 19 (e) 18 SB
720 1907 31 (ct) 30 SB
750 1907 56 (ron) 55 SB
805 1907 32 ( d) 31 SB
836 1907 26 (if) 25 SB
861 1907 35 (fu) 34 SB
895 1907 16 (s) 17 SB
912 1907 12 (i) 11 SB
923 1907 53 (on ) 52 SB
975 1907 19 (c) 18 SB
993 1907 40 (oe) 39 SB
1032 1907 14 (f) 13 SB
1045 1907 26 (fi) 25 SB
1070 1907 31 (ci) 30 SB
1100 1907 19 (e) 18 SB
1118 1907 33 (nt) 33 SB
gr
gs 1096 2851 147 238 CB
222 1957 39 (is ) 38 SB
260 1957 19 (a) 18 SB
278 1957 40 (ga) 39 SB
317 1957 44 (in ) 43 SB
360 1957 33 (pl) 32 SB
392 1957 33 (ot) 32 SB
424 1957 31 (te) 30 SB
454 1957 32 (d ) 31 SB
485 1957 19 (a) 18 SB
503 1957 16 (s) 17 SB
520 1957 11 ( ) 10 SB
530 1957 19 (a) 18 SB
548 1957 11 ( ) 10 SB
558 1957 75 (func) 74 SB
632 1957 12 (t) 11 SB
643 1957 54 (ion) 53 SB
696 1957 32 ( o) 31 SB
727 1957 25 (f ) 24 SB
751 1957 52 (the) 51 SB
802 1957 11 ( ) 10 SB
812 1957 19 (e) 18 SB
830 1957 31 (le) 30 SB
860 1957 19 (c) 18 SB
878 1957 47 (tro) 46 SB
924 1957 75 (n-ne) 74 SB
998 1957 33 (ut) 32 SB
1030 1957 33 (ra) 32 SB
1062 1957 23 (l ) 22 SB
1084 1957 19 (c) 18 SB
1102 1957 33 (ol) 32 SB
gr
gs 1096 2851 147 238 CB
1134 1957 14 (-) 14 SB
gr
gs 1096 2851 147 238 CB
222 2007 24 (li) 23 SB
245 2007 70 (sion) 69 SB
314 2007 31 (al) 30 SB
344 2007 24 (it) 23 SB
367 2007 32 (y ) 31 SB
398 2007 14 (\() 14 SB
gr
32 0 0 42 42 0 0 0 42 /Symbol font
gs 1096 2851 147 238 CB
412 2003 23 (l) 23 SB
gr
32 0 0 33 33 0 0 0 33 /Symbol font
gs 1096 2851 147 238 CB
435 2021 17 (n) 17 SB
gr
32 0 0 42 42 0 0 0 42 /Symbol font
gs 1096 2851 147 238 CB
452 2003 12 (/) 11 SB
463 2003 23 (p) 23 SB
gr
32 0 0 42 42 0 0 0 39 /Times-Italic /font31 ANSIFont font
gs 1096 2851 147 238 CB
486 2006 26 (R) 25 SB
gr
32 0 0 42 42 0 0 0 38 /Times-Roman /font32 ANSIFont font
gs 1096 2851 147 238 CB
511 2007 25 (\).) 24 SB
535 2007 37 ( T) 36 SB
571 2007 40 (he) 39 SB
610 2007 11 ( ) 10 SB
620 2007 40 (pe) 39 SB
659 2007 26 (rt) 25 SB
684 2007 75 (urba) 74 SB
758 2007 12 (t) 11 SB
769 2007 54 (ion) 53 SB
822 2007 25 ( f) 24 SB
846 2007 31 (ie) 30 SB
876 2007 33 (ld) 32 SB
908 2007 11 ( ) 10 SB
918 2007 40 (cu) 39 SB
957 2007 47 (rre) 46 SB
1003 2007 33 (nt) 32 SB
1035 2007 11 ( ) 10 SB
1045 2007 39 (is ) 38 SB
1083 2007 47 (set) 46 SB
gr
gs 1096 2851 147 238 CB
222 2057 19 (a) 18 SB
240 2057 23 (t ) 22 SB
gr
32 0 0 42 42 0 0 0 39 /Times-Italic /font31 ANSIFont font
gs 1096 2851 147 238 CB
262 2056 14 (I) 14 SB
gr
32 0 0 33 33 0 0 0 30 /Times-Roman /font32 ANSIFont font
gs 1096 2851 147 238 CB
276 2074 17 (h) 16 SB
gr
32 0 0 42 42 0 0 0 38 /Times-Roman /font32 ANSIFont font
gs 1096 2851 147 238 CB
292 2057 11 ( ) 10 SB
302 2057 35 (= ) 34 SB
336 2057 21 (1) 20 SB
356 2057 53 (00 ) 52 SB
408 2057 41 (A,) 40 SB
448 2057 11 ( ) 10 SB
458 2057 63 (whi) 62 SB
520 2057 40 (ch) 39 SB
559 2057 11 ( ) 10 SB
569 2057 52 (yie) 51 SB
620 2057 12 (l) 11 SB
631 2057 40 (de) 39 SB
670 2057 32 (d ) 31 SB
701 2057 19 (a) 18 SB
719 2057 32 (n ) 31 SB
750 2057 40 (isl) 39 SB
789 2057 40 (an) 39 SB
828 2057 32 (d ) 31 SB
859 2057 61 (ove) 60 SB
919 2057 26 (rl) 25 SB
944 2057 19 (a) 18 SB
962 2057 32 (p ) 31 SB
993 2057 40 (pa) 39 SB
1032 2057 33 (ra) 32 SB
1064 2057 33 (m) 32 SB
1096 2057 19 (e) 18 SB
gr
gs 1096 2851 147 238 CB
1114 2057 14 (-) 14 SB
gr
gs 1096 2851 147 238 CB
222 2107 31 (te) 30 SB
252 2107 25 (r ) 24 SB
gr
32 0 0 42 42 0 0 0 42 /Symbol font
gs 1096 2851 147 238 CB
276 2103 21 (d) 20 SB
gr
32 0 0 42 42 0 0 0 38 /Times-Roman /font32 ANSIFont font
gs 1096 2851 147 238 CB
296 2107 35 ( =) 34 SB
330 2107 11 ( ) 10 SB
340 2107 14 (\() 14 SB
gr
32 0 0 42 42 0 0 0 42 /Symbol font
gs 1096 2851 147 238 CB
354 2103 26 (D) 25 SB
gr
32 0 0 33 33 0 0 0 30 /Times-Roman /font32 ANSIFont font
gs 1096 2851 147 238 CB
379 2124 43 (mn) 43 SB
gr
32 0 0 42 42 0 0 0 38 /Times-Roman /font32 ANSIFont font
gs 1096 2851 147 238 CB
422 2107 11 ( ) 10 SB
432 2107 24 (+) 23 SB
455 2107 11 ( ) 10 SB
gr
32 0 0 42 42 0 0 0 42 /Symbol font
gs 1096 2851 147 238 CB
465 2103 26 (D) 26 SB
gr
32 0 0 33 33 0 0 0 30 /Times-Roman /font32 ANSIFont font
gs 1096 2851 147 238 CB
491 2124 26 (m) 26 SB
gr
32 0 0 33 33 0 0 0 33 /Symbol font
gs 1096 2851 147 238 CB
517 2121 8 (\242) 8 SB
gr
32 0 0 33 33 0 0 0 30 /Times-Roman /font32 ANSIFont font
gs 1096 2851 147 238 CB
525 2124 17 (n) 16 SB
gr
32 0 0 33 33 0 0 0 33 /Symbol font
gs 1096 2851 147 238 CB
541 2121 8 (\242) 8 SB
gr
32 0 0 42 42 0 0 0 38 /Times-Roman /font32 ANSIFont font
gs 1096 2851 147 238 CB
549 2107 26 (\)/) 25 SB
574 2107 43 (\(2|) 43 SB
gr
32 0 0 42 42 0 0 0 39 /Times-Italic /font31 ANSIFont font
gs 1096 2851 147 238 CB
617 2106 16 (r) 16 SB
gr
32 0 0 33 33 0 0 0 30 /Times-Roman /font32 ANSIFont font
gs 1096 2851 147 238 CB
633 2124 43 (mn) 43 SB
gr
32 0 0 42 42 0 0 0 38 /Times-Roman /font32 ANSIFont font
gs 1096 2851 147 238 CB
676 2107 11 ( ) 10 SB
686 2107 25 (- ) 24 SB
gr
32 0 0 42 42 0 0 0 39 /Times-Italic /font31 ANSIFont font
gs 1096 2851 147 238 CB
710 2106 16 (r) 16 SB
gr
32 0 0 33 33 0 0 0 30 /Times-Roman /font32 ANSIFont font
gs 1096 2851 147 238 CB
726 2124 26 (m) 26 SB
gr
32 0 0 33 33 0 0 0 33 /Symbol font
gs 1096 2851 147 238 CB
752 2121 8 (\242) 8 SB
gr
32 0 0 33 33 0 0 0 30 /Times-Roman /font32 ANSIFont font
gs 1096 2851 147 238 CB
760 2124 17 (n) 16 SB
gr
32 0 0 33 33 0 0 0 33 /Symbol font
gs 1096 2851 147 238 CB
776 2121 8 (\242) 8 SB
gr
32 0 0 42 42 0 0 0 38 /Times-Roman /font32 ANSIFont font
gs 1096 2851 147 238 CB
784 2107 8 (|) 9 SB
793 2107 14 (\)) 13 SB
806 2107 34 ( ~) 33 SB
839 2107 11 ( ) 10 SB
849 2107 32 (2.) 31 SB
880 2107 60 ( He) 59 SB
939 2107 33 (re) 32 SB
971 2107 11 ( ) 10 SB
gr
32 0 0 42 42 0 0 0 39 /Times-Italic /font31 ANSIFont font
gs 1096 2851 147 238 CB
981 2106 16 (r) 16 SB
gr
32 0 0 33 33 0 0 0 30 /Times-Roman /font32 ANSIFont font
gs 1096 2851 147 238 CB
997 2124 43 (mn) 43 SB
gr
32 0 0 42 42 0 0 0 38 /Times-Roman /font32 ANSIFont font
gs 1096 2851 147 238 CB
1040 2107 11 (,) 10 SB
1050 2107 11 ( ) 10 SB
gr
32 0 0 42 42 0 0 0 39 /Times-Italic /font31 ANSIFont font
gs 1096 2851 147 238 CB
1060 2106 16 (r) 16 SB
gr
32 0 0 33 33 0 0 0 30 /Times-Roman /font32 ANSIFont font
gs 1096 2851 147 238 CB
1076 2124 26 (m) 26 SB
gr
32 0 0 33 33 0 0 0 33 /Symbol font
gs 1096 2851 147 238 CB
1102 2121 8 (\242) 8 SB
gr
32 0 0 33 33 0 0 0 30 /Times-Roman /font32 ANSIFont font
gs 1096 2851 147 238 CB
1110 2124 17 (n) 17 SB
gr
32 0 0 33 33 0 0 0 33 /Symbol font
gs 1096 2851 147 238 CB
1127 2121 8 (\242) 8 SB
gr
32 0 0 42 42 0 0 0 38 /Times-Roman /font32 ANSIFont font
gs 1096 2851 147 238 CB
222 2157 19 (a) 18 SB
240 2157 33 (re) 32 SB
272 2157 23 ( t) 22 SB
294 2157 40 (he) 39 SB
333 2157 11 ( ) 10 SB
343 2157 33 (ra) 32 SB
375 2157 45 (dii) 44 SB
419 2157 11 ( ) 10 SB
429 2157 46 (of ) 45 SB
474 2157 33 (th) 32 SB
506 2157 30 (e ) 29 SB
535 2157 19 (a) 18 SB
553 2157 33 (dj) 32 SB
585 2157 38 (ac) 37 SB
622 2157 19 (e) 18 SB
640 2157 33 (nt) 32 SB
672 2157 23 ( i) 22 SB
694 2157 47 (sla) 46 SB
740 2157 69 (nds ) 68 SB
808 2157 19 (a) 18 SB
826 2157 53 (nd ) 52 SB
gr
32 0 0 42 42 0 0 0 42 /Symbol font
gs 1096 2851 147 238 CB
878 2153 26 (D) 26 SB
gr
32 0 0 33 33 0 0 0 30 /Times-Roman /font32 ANSIFont font
gs 1096 2851 147 238 CB
904 2174 26 (m) 25 SB
929 2174 17 (n) 17 SB
gr
32 0 0 42 42 0 0 0 38 /Times-Roman /font32 ANSIFont font
gs 1096 2851 147 238 CB
946 2157 11 (,) 10 SB
956 2157 11 ( ) 10 SB
gr
32 0 0 42 42 0 0 0 42 /Symbol font
gs 1096 2851 147 238 CB
966 2153 26 (D) 26 SB
gr
32 0 0 33 33 0 0 0 30 /Times-Roman /font32 ANSIFont font
gs 1096 2851 147 238 CB
992 2174 26 (m) 26 SB
gr
32 0 0 33 33 0 0 0 33 /Symbol font
gs 1096 2851 147 238 CB
1018 2171 8 (\242) 8 SB
gr
32 0 0 33 33 0 0 0 30 /Times-Roman /font32 ANSIFont font
gs 1096 2851 147 238 CB
1026 2174 17 (n) 16 SB
gr
32 0 0 33 33 0 0 0 33 /Symbol font
gs 1096 2851 147 238 CB
1042 2171 8 (\242) 8 SB
gr
32 0 0 42 42 0 0 0 38 /Times-Roman /font32 ANSIFont font
gs 1096 2851 147 238 CB
1050 2157 11 ( ) 10 SB
1060 2157 33 (ar) 32 SB
1092 2157 19 (e) 19 SB
gr
gs 1096 2851 147 238 CB
222 2207 33 (th) 32 SB
254 2207 30 (e ) 29 SB
283 2207 12 (i) 11 SB
294 2207 16 (s) 17 SB
311 2207 12 (l) 11 SB
322 2207 19 (a) 18 SB
340 2207 53 (nd ) 52 SB
392 2207 63 (wid) 62 SB
454 2207 60 (ths.) 59 SB
513 2207 11 ( ) 10 SB
523 2207 47 (Th) 46 SB
569 2207 30 (e ) 29 SB
598 2207 40 (da) 39 SB
637 2207 12 (t) 11 SB
648 2207 30 (a ) 29 SB
677 2207 19 (a) 18 SB
695 2207 33 (re) 32 SB
727 2207 11 ( ) 10 SB
737 2207 31 (ta) 30 SB
767 2207 40 (ke) 39 SB
806 2207 32 (n ) 31 SB
837 2207 33 (in) 32 SB
869 2207 11 ( ) 10 SB
879 2207 52 (the) 51 SB
930 2207 11 ( ) 10 SB
940 2207 33 (m) 32 SB
972 2207 33 (id) 32 SB
1004 2207 52 (dle) 51 SB
1055 2207 11 ( ) 10 SB
1065 2207 33 (re) 32 SB
gr
gs 1096 2851 147 238 CB
1097 2207 14 (-) 14 SB
gr
gs 1096 2851 147 238 CB
222 2257 33 (gi) 32 SB
254 2257 53 (on ) 52 SB
306 2257 40 (be) 39 SB
345 2257 61 (twe) 60 SB
405 2257 19 (e) 18 SB
423 2257 32 (n ) 31 SB
454 2257 33 (th) 32 SB
486 2257 30 (e ) 29 SB
515 2257 33 (1/) 32 SB
547 2257 32 (3 ) 31 SB
578 2257 19 (a) 18 SB
596 2257 53 (nd ) 52 SB
648 2257 33 (1/) 32 SB
680 2257 32 (4 ) 31 SB
711 2257 59 (isla) 58 SB
769 2257 42 (nd) 41 SB
810 2257 25 ( r) 24 SB
834 2257 40 (ad) 39 SB
873 2257 24 (ii) 23 SB
896 2257 62 ( wh) 61 SB
957 2257 33 (er) 32 SB
989 2257 30 (e ) 29 SB
1018 2257 12 (t) 11 SB
1029 2257 40 (he) 39 SB
1068 2257 25 ( f) 24 SB
1092 2257 31 (ie) 30 SB
1122 2257 33 (ld) 32 SB
gr
gs 1096 2851 147 238 CB
222 2307 24 (li) 23 SB
245 2307 40 (ne) 39 SB
284 2307 46 (s a) 45 SB
329 2307 33 (re) 32 SB
361 2307 11 ( ) 10 SB
371 2307 33 (m) 32 SB
403 2307 60 (ost ) 59 SB
462 2307 68 (stoc) 67 SB
529 2307 40 (ha) 39 SB
568 2307 40 (sti) 39 SB
607 2307 19 (c) 18 SB
625 2307 22 (. ) 21 SB
646 2307 26 (T) 25 SB
671 2307 40 (he) 39 SB
710 2307 11 ( ) 10 SB
720 2307 45 (tre) 44 SB
764 2307 42 (nd) 41 SB
805 2307 11 ( ) 10 SB
815 2307 46 (of ) 45 SB
860 2307 12 (t) 11 SB
871 2307 40 (he) 39 SB
910 2307 11 ( ) 10 SB
920 2307 40 (de) 39 SB
959 2307 33 (cr) 32 SB
991 2307 38 (ea) 37 SB
1028 2307 49 (sin) 48 SB
1076 2307 21 (g) 21 SB
gr
32 0 0 42 42 0 0 0 39 /Times-Italic /font31 ANSIFont font
gs 1096 2851 147 238 CB
222 2356 30 (D) 30 SB
gr
32 0 0 33 33 0 0 0 33 /Symbol font
gs 1096 2851 147 238 CB
252 2371 22 (^) 22 SB
gr
32 0 0 42 42 0 0 0 38 /Times-Roman /font32 ANSIFont font
gs 1096 2851 147 238 CB
274 2357 11 ( ) 10 SB
284 2357 54 (wit) 53 SB
337 2357 32 (h ) 31 SB
368 2357 33 (in) 32 SB
400 2357 33 (cr) 32 SB
432 2357 38 (ea) 37 SB
469 2357 49 (sin) 48 SB
517 2357 32 (g ) 31 SB
548 2357 40 (co) 39 SB
587 2357 24 (ll) 23 SB
610 2357 61 (isio) 60 SB
670 2357 32 (n ) 31 SB
701 2357 47 (fre) 46 SB
747 2357 61 (que) 60 SB
807 2357 40 (nc) 39 SB
846 2357 32 (y ) 31 SB
877 2357 12 (i) 11 SB
888 2357 16 (s) 17 SB
905 2357 11 ( ) 10 SB
915 2357 42 (du) 41 SB
956 2357 30 (e ) 29 SB
985 2357 12 (t) 11 SB
996 2357 32 (o ) 31 SB
1027 2357 54 (gra) 53 SB
1080 2357 33 (di) 32 SB
gr
gs 1096 2851 147 238 CB
1112 2357 14 (-) 14 SB
gr
gs 1096 2851 147 238 CB
222 2407 19 (e) 18 SB
240 2407 44 (nt ) 43 SB
283 2407 12 (i) 11 SB
294 2407 40 (nc) 39 SB
333 2407 33 (re) 32 SB
365 2407 47 (asi) 46 SB
411 2407 53 (ng ) 52 SB
463 2407 54 (wit) 53 SB
516 2407 32 (h ) 31 SB
547 2407 33 (in) 32 SB
579 2407 33 (cr) 32 SB
611 2407 38 (ea) 37 SB
648 2407 49 (sin) 48 SB
696 2407 32 (g ) 31 SB
727 2407 40 (co) 39 SB
766 2407 24 (ll) 23 SB
789 2407 61 (isio) 60 SB
849 2407 32 (n ) 31 SB
880 2407 47 (fre) 46 SB
926 2407 61 (que) 60 SB
986 2407 40 (nc) 39 SB
1025 2407 32 (y ) 31 SB
gr
32 0 0 42 42 0 0 0 42 /Symbol font
gs 1096 2851 147 238 CB
1056 2403 22 (n) 21 SB
gr
32 0 0 33 33 0 0 0 30 /Times-Roman /font32 ANSIFont font
gs 1096 2851 147 238 CB
1077 2424 32 (en) 32 SB
gr
32 0 0 42 42 0 0 0 38 /Times-Roman /font32 ANSIFont font
gs 1096 2851 147 238 CB
1109 2407 11 (.) 10 SB
1119 2407 11 ( ) 10 SB
gr
gs 1096 2851 147 238 CB
222 2480 26 (T) 25 SB
247 2480 40 (he) 39 SB
286 2480 30 ( e) 29 SB
315 2480 42 (xp) 41 SB
356 2480 33 (er) 32 SB
388 2480 45 (im) 44 SB
432 2480 19 (e) 18 SB
450 2480 52 (nta) 51 SB
501 2480 12 (l) 11 SB
512 2480 25 ( r) 24 SB
536 2480 68 (esul) 67 SB
603 2480 39 (ts ) 38 SB
641 2480 19 (a) 18 SB
659 2480 33 (re) 32 SB
691 2480 30 ( c) 29 SB
720 2480 54 (om) 53 SB
773 2480 40 (pa) 39 SB
812 2480 33 (re) 32 SB
844 2480 32 (d ) 31 SB
875 2480 42 (wi) 41 SB
916 2480 44 (th ) 43 SB
959 2480 12 (t) 11 SB
970 2480 40 (he) 39 SB
1009 2480 32 ( p) 31 SB
1040 2480 33 (re) 32 SB
1072 2480 52 (dic) 51 SB
gr
gs 1096 2851 147 238 CB
1123 2480 14 (-) 14 SB
gr
gs 1096 2851 147 238 CB
222 2530 24 (ti) 23 SB
245 2530 69 (ons ) 68 SB
313 2530 46 (of ) 45 SB
358 2530 12 (t) 11 SB
369 2530 40 (he) 39 SB
408 2530 25 ( f) 24 SB
432 2530 67 (our ) 66 SB
498 2530 19 (a) 18 SB
516 2530 40 (na) 39 SB
555 2530 33 (ly) 32 SB
587 2530 24 (ti) 23 SB
610 2530 30 (c ) 29 SB
639 2530 33 (m) 32 SB
671 2530 61 (ode) 60 SB
731 2530 12 (l) 11 SB
742 2530 16 (s) 17 SB
759 2530 11 (.) 10 SB
769 2530 11 ( ) 10 SB
779 2530 26 (T) 25 SB
804 2530 40 (he) 39 SB
843 2530 11 ( ) 10 SB
853 2530 70 (four) 69 SB
922 2530 11 ( ) 10 SB
932 2530 40 (an) 39 SB
971 2530 31 (al) 30 SB
1001 2530 33 (yt) 32 SB
1033 2530 31 (ic) 30 SB
1063 2530 11 ( ) 10 SB
1073 2530 33 (m) 32 SB
1105 2530 42 (od) 42 SB
gr
gs 1096 2851 147 238 CB
1147 2530 14 (-) 14 SB
gr
gs 1096 2851 147 238 CB
222 2580 19 (e) 18 SB
240 2580 39 (ls,) 38 SB
278 2580 23 ( i) 22 SB
300 2580 32 (n ) 31 SB
331 2580 56 (ord) 55 SB
386 2580 33 (er) 32 SB
418 2580 32 ( o) 31 SB
449 2580 25 (f ) 24 SB
473 2580 52 (the) 51 SB
524 2580 11 ( ) 10 SB
534 2580 33 (in) 32 SB
566 2580 33 (cr) 32 SB
598 2580 38 (ea) 37 SB
635 2580 49 (sin) 48 SB
683 2580 32 (g ) 31 SB
714 2580 40 (co) 39 SB
753 2580 24 (ll) 23 SB
776 2580 61 (isio) 60 SB
836 2580 40 (na) 39 SB
875 2580 24 (li) 23 SB
898 2580 33 (ty) 30 SB
928 2580 11 (,) 10 SB
938 2580 11 ( ) 10 SB
948 2580 19 (a) 18 SB
966 2580 33 (re) 32 SB
998 2580 23 ( t) 22 SB
1020 2580 40 (he) 39 SB
gr
gs 1096 2851 147 238 CB
222 2630 19 (c) 18 SB
240 2630 45 (oll) 44 SB
284 2630 40 (isi) 39 SB
323 2630 54 (onl) 53 SB
376 2630 62 (ess ) 61 SB
gr
32 0 0 42 42 0 0 0 39 /Times-Italic /font31 ANSIFont font
gs 1096 2851 147 238 CB
437 2629 30 (D) 30 SB
gr
32 0 0 33 33 0 0 0 30 /Times-Roman /font32 ANSIFont font
gs 1096 2851 147 238 CB
467 2618 22 (C) 22 SB
gr
32 0 0 42 42 0 0 0 38 /Times-Roman /font32 ANSIFont font
gs 1096 2851 147 238 CB
489 2630 22 (, ) 21 SB
510 2630 12 (t) 11 SB
521 2630 40 (he) 39 SB
560 2630 39 ( R) 38 SB
598 2630 38 (ec) 37 SB
635 2630 40 (he) 39 SB
674 2630 47 (ste) 46 SB
720 2630 14 (r) 13 SB
733 2630 42 (-R) 41 SB
774 2630 77 (osen) 76 SB
850 2630 54 (blu) 53 SB
903 2630 44 (th ) 43 SB
gr
32 0 0 42 42 0 0 0 39 /Times-Italic /font31 ANSIFont font
gs 1096 2851 147 238 CB
946 2629 30 (D) 30 SB
gr
32 0 0 33 33 0 0 0 30 /Times-Roman /font32 ANSIFont font
gs 1096 2851 147 238 CB
976 2618 44 (RR) 44 SB
gr
32 0 0 42 42 0 0 0 38 /Times-Roman /font32 ANSIFont font
gs 1096 2851 147 238 CB
1020 2630 11 (,) 10 SB
1030 2630 11 ( ) 10 SB
1040 2630 52 (the) 51 SB
gr
gs 1096 2851 147 238 CB
222 2680 49 (Ka) 48 SB
270 2680 75 (dom) 74 SB
344 2680 47 (tse) 46 SB
390 2680 79 (v-Po) 78 SB
468 2680 54 (gut) 53 SB
521 2680 16 (s) 17 SB
538 2680 19 (e) 18 SB
556 2680 11 ( ) 10 SB
gr
32 0 0 42 42 0 0 0 39 /Times-Italic /font31 ANSIFont font
gs 1096 2851 147 238 CB
566 2679 30 (D) 30 SB
gr
32 0 0 33 33 0 0 0 30 /Times-Roman /font32 ANSIFont font
gs 1096 2851 147 238 CB
596 2668 42 (KP) 43 SB
gr
32 0 0 42 42 0 0 0 38 /Times-Roman /font32 ANSIFont font
gs 1096 2851 147 238 CB
639 2680 11 (,) 10 SB
649 2680 11 ( ) 10 SB
659 2680 19 (a) 18 SB
677 2680 53 (nd ) 52 SB
729 2680 26 (fl) 25 SB
754 2680 54 (uid) 53 SB
807 2680 11 ( ) 11 SB
gr
32 0 0 42 42 0 0 0 39 /Times-Italic /font31 ANSIFont font
gs 1096 2851 147 238 CB
818 2679 30 (D) 30 SB
gr
32 0 0 33 33 0 0 0 30 /Times-Roman /font32 ANSIFont font
gs 1096 2851 147 238 CB
848 2668 18 (F) 18 SB
gr
32 0 0 42 42 0 0 0 38 /Times-Roman /font32 ANSIFont font
gs 1096 2851 147 238 CB
866 2680 11 ( ) 10 SB
876 2680 33 (m) 32 SB
908 2680 61 (ode) 60 SB
968 2680 39 (ls.) 38 SB
1006 2680 11 ( ) 10 SB
1016 2680 26 (T) 25 SB
1041 2680 40 (he) 39 SB
gr
gs 1096 2851 147 238 CB
222 2730 56 (hor) 55 SB
277 2730 31 (iz) 30 SB
307 2730 54 (ont) 53 SB
360 2730 19 (a) 18 SB
378 2730 23 (l ) 22 SB
400 2730 40 (ba) 39 SB
439 2730 41 (rs ) 40 SB
479 2730 40 (ab) 39 SB
518 2730 61 (ove) 60 SB
578 2730 11 ( ) 10 SB
588 2730 52 (the) 51 SB
639 2730 11 ( ) 10 SB
649 2730 26 (fi) 25 SB
674 2730 75 (gure) 74 SB
748 2730 11 ( ) 10 SB
758 2730 99 (show ) 98 SB
856 2730 33 (th) 32 SB
888 2730 30 (e ) 29 SB
917 2730 33 (re) 32 SB
949 2730 12 (l) 11 SB
960 2730 40 (ev) 39 SB
999 2730 40 (an) 39 SB
1038 2730 23 (t ) 22 SB
1060 2730 19 (a) 18 SB
1078 2730 54 (ppl) 53 SB
1131 2730 12 (i) 12 SB
gr
gs 1096 2851 147 238 CB
1143 2730 14 (-) 14 SB
gr
gs 1096 2851 147 238 CB
222 2780 19 (c) 18 SB
240 2780 40 (ab) 39 SB
279 2780 24 (il) 23 SB
302 2780 24 (it) 23 SB
325 2780 32 (y ) 31 SB
356 2780 33 (re) 32 SB
388 2780 66 (gim) 65 SB
453 2780 19 (e) 18 SB
471 2780 41 (s f) 40 SB
511 2780 46 (or ) 45 SB
556 2780 33 (th) 32 SB
588 2780 30 (e ) 29 SB
617 2780 35 (fo) 34 SB
651 2780 46 (ur ) 45 SB
696 2780 19 (a) 18 SB
714 2780 40 (na) 39 SB
753 2780 45 (lyt) 44 SB
797 2780 31 (ic) 30 SB
827 2780 11 ( ) 10 SB
837 2780 33 (m) 32 SB
869 2780 61 (ode) 60 SB
929 2780 12 (l) 11 SB
940 2780 16 (s) 17 SB
957 2780 11 (.) 10 SB
967 2780 11 ( ) 10 SB
977 2780 26 (T) 25 SB
1002 2780 40 (he) 39 SB
1041 2780 11 ( ) 10 SB
1051 2780 19 (e) 18 SB
1069 2780 21 (x) 21 SB
gr
gs 1096 2851 147 238 CB
1090 2780 14 (-) 14 SB
gr
gs 1096 2851 147 238 CB
222 2830 40 (pe) 39 SB
261 2830 26 (ri) 25 SB
286 2830 52 (me) 51 SB
337 2830 33 (nt) 32 SB
369 2830 19 (a) 18 SB
387 2830 23 (l ) 22 SB
409 2830 33 (m) 32 SB
441 2830 38 (ea) 37 SB
478 2830 70 (sure) 69 SB
547 2830 32 (d ) 31 SB
578 2830 40 (va) 39 SB
617 2830 12 (l) 11 SB
628 2830 40 (ue) 39 SB
667 2830 32 ( o) 31 SB
698 2830 25 (f ) 24 SB
722 2830 52 (the) 51 SB
773 2830 11 ( ) 10 SB
783 2830 26 (tr) 25 SB
808 2830 40 (an) 39 SB
847 2830 16 (s) 17 SB
864 2830 21 (p) 20 SB
884 2830 47 (ort) 46 SB
930 2830 30 ( c) 29 SB
959 2830 40 (oe) 39 SB
998 2830 14 (f) 13 SB
1011 2830 26 (fi) 25 SB
1036 2830 19 (c) 18 SB
1054 2830 31 (ie) 30 SB
1084 2830 33 (nt) 32 SB
1116 2830 23 ( i) 22 SB
1138 2830 16 (s) 16 SB
gr
gs 1096 2851 147 238 CB
222 2880 19 (a) 18 SB
240 2880 75 (bout) 74 SB
314 2880 11 ( ) 10 SB
324 2880 30 (a ) 29 SB
353 2880 33 (fa) 32 SB
385 2880 19 (c) 18 SB
403 2880 33 (to) 32 SB
435 2880 25 (r ) 24 SB
459 2880 46 (of ) 45 SB
504 2880 32 (3 ) 31 SB
535 2880 32 (\226 ) 31 SB
566 2880 32 (5 ) 31 SB
597 2880 31 (la) 30 SB
627 2880 14 (r) 13 SB
640 2880 40 (ge) 39 SB
679 2880 25 (r ) 24 SB
703 2880 12 (t) 11 SB
714 2880 40 (ha) 39 SB
753 2880 32 (n ) 31 SB
784 2880 52 (the) 51 SB
835 2880 11 ( ) 10 SB
845 2880 54 (pre) 53 SB
898 2880 33 (di) 32 SB
930 2880 19 (c) 18 SB
948 2880 24 (ti) 23 SB
971 2880 69 (ons ) 68 SB
1039 2880 46 (of ) 45 SB
1084 2880 33 (th) 32 SB
1116 2880 19 (e) 19 SB
gr
gs 1096 2851 147 238 CB
222 2930 56 (fou) 55 SB
277 2930 25 (r ) 24 SB
301 2930 54 (mo) 53 SB
354 2930 40 (de) 39 SB
393 2930 39 (ls.) 38 SB
431 2930 62 ( Ho) 61 SB
492 2930 70 (wev) 69 SB
561 2930 33 (er) 31 SB
592 2930 11 (,) 10 SB
602 2930 11 ( ) 10 SB
612 2930 63 (goo) 62 SB
674 2930 32 (d ) 31 SB
705 2930 61 (qua) 60 SB
765 2930 24 (li) 23 SB
788 2930 31 (ta) 30 SB
818 2930 12 (t) 11 SB
829 2930 52 (ive) 51 SB
880 2930 11 ( ) 10 SB
890 2930 19 (a) 18 SB
908 2930 54 (gre) 53 SB
961 2930 19 (e) 18 SB
979 2930 52 (me) 51 SB
1030 2930 33 (nt) 32 SB
1062 2930 11 ( ) 10 SB
1072 2930 40 (be) 39 SB
gr
gs 1096 2851 147 238 CB
1111 2930 14 (-) 14 SB
gr
gs 1096 1501 1152 238 CB
1227 312 61 (twe) 60 SB
1287 312 19 (e) 18 SB
1305 312 32 (n ) 31 SB
1336 312 33 (th) 32 SB
1368 312 30 (e ) 29 SB
1397 312 19 (e) 18 SB
1415 312 61 (xpe) 60 SB
1475 312 26 (ri) 25 SB
1500 312 33 (m) 32 SB
1532 312 40 (en) 39 SB
1571 312 31 (ta) 30 SB
1601 312 23 (l ) 22 SB
1623 312 33 (re) 32 SB
1655 312 49 (sul) 48 SB
1703 312 39 (ts ) 38 SB
1741 312 40 (an) 39 SB
1780 312 32 (d ) 31 SB
1811 312 52 (the) 51 SB
1862 312 11 ( ) 10 SB
1872 312 54 (pre) 53 SB
1925 312 33 (di) 32 SB
1957 312 19 (c) 18 SB
1975 312 24 (ti) 23 SB
1998 312 69 (ons ) 68 SB
2066 312 46 (of ) 45 SB
2111 312 33 (th) 32 SB
2143 312 19 (e) 19 SB
gr
gs 1096 1501 1152 238 CB
1227 362 56 (fou) 55 SB
1282 362 25 (r ) 24 SB
1306 362 40 (an) 39 SB
1345 362 31 (al) 30 SB
1375 362 33 (yt) 32 SB
1407 362 31 (ic) 30 SB
1437 362 11 ( ) 10 SB
1447 362 54 (mo) 53 SB
1500 362 40 (de) 39 SB
1539 362 39 (ls ) 38 SB
1577 362 38 (ca) 37 SB
1614 362 32 (n ) 31 SB
1645 362 40 (be) 39 SB
1684 362 11 ( ) 10 SB
1694 362 56 (fou) 55 SB
1749 362 53 (nd.) 52 SB
1801 362 11 ( ) 10 SB
1811 362 47 (Th) 46 SB
1857 362 30 (e ) 29 SB
1886 362 35 (sc) 34 SB
1920 362 19 (a) 18 SB
1938 362 24 (li) 23 SB
1961 362 53 (ng ) 52 SB
2013 362 35 (of) 34 SB
2047 362 11 ( ) 10 SB
2057 362 52 (the) 51 SB
gr
gs 1096 1501 1152 238 CB
1227 412 40 (da) 39 SB
1266 412 31 (ta) 30 SB
1296 412 11 ( ) 10 SB
1306 412 54 (wit) 53 SB
1359 412 32 (h ) 31 SB
1390 412 33 (th) 32 SB
1422 412 30 (e ) 29 SB
1451 412 19 (c) 18 SB
1469 412 33 (ol) 32 SB
1501 412 24 (li) 23 SB
1524 412 16 (s) 17 SB
1541 412 12 (i) 11 SB
1552 412 61 (ona) 60 SB
1612 412 12 (l) 11 SB
1623 412 24 (it) 23 SB
1646 412 32 (y ) 31 SB
1677 412 40 (ag) 39 SB
1716 412 33 (re) 32 SB
1748 412 46 (es ) 45 SB
1793 412 49 (we) 48 SB
1841 412 24 (ll) 23 SB
1864 412 53 ( wi) 52 SB
1916 412 33 (th) 32 SB
1948 412 23 ( t) 22 SB
1970 412 40 (he) 39 SB
2009 412 11 ( ) 10 SB
2019 412 54 (pre) 53 SB
2072 412 33 (di) 32 SB
2104 412 19 (c) 19 SB
gr
gs 1096 1501 1152 238 CB
2123 412 14 (-) 13 SB
gr
gs 1096 1501 1152 238 CB
1227 462 24 (ti) 23 SB
1250 462 69 (ons ) 68 SB
1318 462 46 (of ) 45 SB
1363 462 12 (t) 11 SB
1374 462 40 (he) 39 SB
1413 462 25 ( f) 24 SB
1437 462 67 (our ) 66 SB
1503 462 19 (a) 18 SB
1521 462 40 (na) 39 SB
1560 462 33 (ly) 32 SB
1592 462 24 (ti) 23 SB
1615 462 30 (c ) 29 SB
1644 462 33 (m) 32 SB
1676 462 61 (ode) 60 SB
1736 462 12 (l) 11 SB
1747 462 16 (s) 17 SB
1764 462 11 ( ) 10 SB
1774 462 12 (i) 11 SB
1785 462 32 (n ) 31 SB
1816 462 38 (ea) 37 SB
1853 462 19 (c) 18 SB
1871 462 32 (h ) 31 SB
1902 462 46 (of ) 45 SB
1947 462 12 (t) 11 SB
1958 462 40 (he) 39 SB
1997 462 37 (ir ) 36 SB
2033 462 19 (a) 18 SB
2051 462 54 (ppl) 53 SB
2104 462 31 (ic) 30 SB
2134 462 19 (a) 18 SB
gr
gs 1096 1501 1152 238 CB
2152 462 14 (-) 14 SB
gr
gs 1096 1501 1152 238 CB
1227 512 33 (bl) 32 SB
1259 512 30 (e ) 29 SB
1288 512 33 (re) 32 SB
1320 512 33 (gi) 32 SB
1352 512 33 (m) 32 SB
1384 512 46 (es.) 45 SB
gr
32 0 0 38 38 0 0 0 36 /Helvetica-Narrow /font15 ANSIFont font
gs 1096 1501 1152 238 CB
1257 577 40 (H. ) 39 SB
1296 577 59 (Lin, ) 58 SB
1354 577 40 (R. ) 39 SB
1393 577 37 (F. ) 36 SB
1429 577 91 (Gandy) 90 SB
1519 577 18 (, ) 17 SB
1536 577 30 (S.) 29 SB
1565 577 28 ( F) 27 SB
1592 577 18 (. ) 17 SB
1609 577 93 (Knowlt) 92 SB
1701 577 52 (on, ) 51 SB
1752 577 42 (G. ) 41 SB
1793 577 16 (J) 15 SB
1808 577 18 (. ) 17 SB
1825 577 22 (H) 23 SB
1848 577 36 (art) 35 SB
1883 577 22 (w) 23 SB
1906 577 24 (el) 23 SB
1929 577 25 (l, ) 24 SB
1953 577 40 (D. ) 39 SB
1992 577 54 (Pric) 53 SB
2045 577 70 (hard,) 70 SB
gr
gs 1096 1501 1152 238 CB
1257 622 33 (G.) 32 SB
1289 622 30 ( S) 29 SB
1318 622 49 (ass) 48 SB
1366 622 45 (er, ) 44 SB
1410 622 81 (and E) 80 SB
1490 622 18 (. ) 17 SB
1507 622 79 (Thom) 78 SB
1585 622 42 (as,) 41 SB
1626 622 25 ( J) 24 SB
1650 622 19 (r.) 19 SB
gr
gs 1096 1501 1152 238 CB
1257 667 54 (Phy) 53 SB
1310 667 16 (s) 15 SB
1325 667 23 (ic) 22 SB
1347 667 16 (s) 15 SB
1362 667 127 ( Departm) 126 SB
1488 667 52 (ent ) 51 SB
1539 667 30 ( A) 29 SB
1568 667 61 (ubur) 62 SB
1630 667 26 (n ) 25 SB
1655 667 62 (Univ) 61 SB
1716 667 27 (er) 28 SB
1744 667 16 (s) 15 SB
1759 667 16 (it) 15 SB
1774 667 16 (y) 15 SB
1789 667 9 ( ) 9 SB
gr
gs 1096 1501 1152 238 CB
1257 712 108 (Auburn,) 107 SB
1364 712 30 ( A) 29 SB
1393 712 120 (L 36849,) 119 SB
1512 712 73 ( USA) 72 SB
gr
gs 1096 1501 1152 238 CB
1257 793 98 (Phone:) 97 SB
1354 793 218 ( \(205\) 844-4126) 218 SB
gr
gs 1096 1501 1152 238 CB
1257 838 40 (FA) 39 SB
1296 838 30 (X:) 29 SB
1325 838 218 ( \(205\) 844-5864) 218 SB
gr
gs 1096 1501 1152 238 CB
1257 883 57 (E-M) 56 SB
1313 883 40 (ail:) 39 SB
1352 883 93 ( gandy) 92 SB
1444 883 32 (@) 31 SB
1475 883 66 (phys) 65 SB
1540 883 23 (ic) 22 SB
1562 883 16 (s) 15 SB
1577 883 113 (.auburn.) 112 SB
1689 883 51 (edu) 51 SB
gr
32 0 0 38 38 0 0 0 34 /Times-Roman /font32 ANSIFont font
gs 1096 1501 1152 238 CB
1227 978 55 (Ref) 54 SB
1281 978 30 (er) 29 SB
1310 978 36 (en) 35 SB
1345 978 34 (ce) 33 SB
1378 978 25 (s ) 24 SB
gr
gs 1096 1501 1152 238 CB
1227 1028 32 ([1) 31 SB
1258 1028 23 (].) 22 SB
gr
gs 1096 1501 1152 238 CB
1302 1028 25 (J.) 24 SB
1326 1028 10 ( ) 9 SB
1335 1028 47 (A. ) 46 SB
1381 1028 40 (Kr) 39 SB
1420 1028 49 (om) 48 SB
1468 1028 30 (m) 29 SB
1497 1028 32 (es) 31 SB
1528 1028 10 (,) 9 SB
1537 1028 45 ( C.) 44 SB
1581 1028 10 ( ) 9 SB
1590 1028 63 (Obe) 62 SB
1652 1028 43 (rm) 42 SB
1694 1028 17 (a) 16 SB
1710 1028 29 (n,) 28 SB
1738 1028 45 ( R.) 44 SB
1782 1028 10 ( ) 9 SB
1791 1028 37 (G.) 36 SB
1827 1028 48 ( Kl) 47 SB
1874 1028 17 (e) 16 SB
1890 1028 46 (va,) 45 SB
1935 1028 10 ( ) 9 SB
1944 1028 25 (J.) 24 SB
1968 1028 10 ( ) 9 SB
1977 1028 32 (Pl) 31 SB
2008 1028 32 (as) 31 SB
2039 1028 30 (m) 29 SB
2068 1028 27 (a ) 26 SB
2094 1028 40 (Ph) 39 SB
2133 1028 34 (ys) 34 SB
gr
gs 1096 1501 1152 238 CB
2167 1028 13 (-) 12 SB
gr
gs 1096 1501 1152 238 CB
1302 1078 11 (i) 10 SB
1312 1078 32 (cs) 31 SB
1343 1078 20 (, ) 19 SB
gr
32 0 0 38 38 0 0 0 37 /Times-Bold /font29 ANSIFont font
gs 1096 1501 1152 238 CB
1362 1075 38 (30) 37 SB
gr
32 0 0 38 38 0 0 0 34 /Times-Roman /font32 ANSIFont font
gs 1096 1501 1152 238 CB
1399 1078 23 ( \() 22 SB
1421 1078 57 (198) 56 SB
1477 1078 42 (3\) ) 41 SB
1518 1078 38 (11) 37 SB
1555 1078 10 (.) 10 SB
gr
gs 1096 1501 1152 238 CB
1227 1128 32 ([2) 31 SB
1258 1128 23 (].) 22 SB
gr
gs 1096 1501 1152 238 CB
1302 1128 47 ( A ) 46 SB
1348 1128 10 (.) 9 SB
1357 1128 37 (G.) 36 SB
1393 1128 48 ( Di) 47 SB
1440 1128 30 (ki) 29 SB
1469 1128 11 (i) 10 SB
1479 1128 10 (,) 9 SB
1488 1128 47 ( V.) 46 SB
1534 1128 10 ( ) 9 SB
1543 1128 34 (M) 33 SB
1576 1128 20 (. ) 19 SB
1595 1128 40 (Za) 39 SB
1634 1128 11 (l) 10 SB
1644 1128 30 (ki) 29 SB
1673 1128 48 (nd,) 47 SB
1720 1128 10 ( ) 9 SB
1729 1128 47 (G. ) 46 SB
1775 1128 37 (G.) 36 SB
1811 1128 10 ( ) 9 SB
1820 1128 55 (Les) 54 SB
1874 1128 55 (nya) 54 SB
1928 1128 57 (kov) 56 SB
1984 1128 20 (, ) 19 SB
2003 1128 37 (O.) 36 SB
2039 1128 10 ( ) 9 SB
2048 1128 31 (S.) 30 SB
gr
gs 1096 1501 1152 238 CB
1302 1178 38 (Pa) 37 SB
1339 1178 30 (vl) 29 SB
1368 1178 28 (ic) 27 SB
1395 1178 36 (he) 35 SB
1430 1178 67 (nko,) 66 SB
1496 1178 10 ( ) 9 SB
1505 1178 37 (A.) 36 SB
1541 1178 47 ( V.) 46 SB
1587 1178 10 ( ) 9 SB
1596 1178 38 (Pa) 37 SB
1633 1178 51 (shc) 50 SB
1683 1178 36 (he) 35 SB
1718 1178 67 (nko,) 66 SB
1784 1178 10 ( ) 9 SB
1793 1178 37 (V.) 36 SB
1829 1178 47 ( K.) 46 SB
1875 1178 10 ( ) 9 SB
1884 1178 38 (Pa) 37 SB
1921 1178 53 (shn) 52 SB
1973 1178 46 (ev,) 45 SB
2018 1178 10 ( ) 9 SB
2027 1178 37 (D.) 36 SB
2063 1178 31 ( P) 30 SB
2093 1178 10 (.) 10 SB
gr
gs 1096 1501 1152 238 CB
1302 1228 59 (Pog) 58 SB
1360 1228 55 (ozh) 54 SB
1414 1228 46 (ev,) 45 SB
1459 1228 10 ( ) 9 SB
1468 1228 36 (an) 35 SB
1503 1228 29 (d ) 28 SB
1531 1228 71 (V.M) 70 SB
1601 1228 10 (.) 9 SB
1610 1228 10 ( ) 9 SB
1619 1228 80 (Tonk) 79 SB
1698 1228 51 (opr) 50 SB
1748 1228 55 (yad) 54 SB
1802 1228 20 (, ) 19 SB
1821 1228 40 (So) 39 SB
1860 1228 29 (v.) 28 SB
1888 1228 25 ( J) 24 SB
1912 1228 10 (.) 9 SB
1921 1228 31 ( P) 30 SB
1951 1228 28 (la) 27 SB
1978 1228 45 (sm) 44 SB
2022 1228 17 (a) 16 SB
2038 1228 10 ( ) 9 SB
2047 1228 74 (Phys) 73 SB
2120 1228 10 (.) 9 SB
gr
32 0 0 38 38 0 0 0 37 /Times-Bold /font29 ANSIFont font
gs 1096 1501 1152 238 CB
1302 1275 38 (14) 38 SB
gr
32 0 0 38 38 0 0 0 34 /Times-Roman /font32 ANSIFont font
gs 1096 1501 1152 238 CB
1340 1278 10 (,) 9 SB
1349 1278 10 ( ) 9 SB
1358 1278 57 (160) 56 SB
1414 1278 23 ( \() 22 SB
1436 1278 57 (198) 56 SB
1492 1278 42 (8\).) 41 SB
1533 1278 10 ( ) 9 SB
gr
gs 1096 1501 1152 238 CB
1227 1328 32 ([3) 31 SB
1258 1328 23 (].) 22 SB
gr
gs 1096 1501 1152 238 CB
1302 1328 47 (K. ) 46 SB
1348 1328 34 (M) 33 SB
1381 1328 11 (i) 10 SB
1391 1328 66 (yam) 65 SB
1456 1328 19 (o) 18 SB
1474 1328 30 (to) 29 SB
1503 1328 20 (, ) 19 SB
gr
32 0 0 38 38 0 0 0 35 /Times-Italic /font31 ANSIFont font
gs 1096 1501 1152 238 CB
1522 1327 34 (Pl) 33 SB
1555 1327 34 (as) 33 SB
1588 1327 56 (ma ) 55 SB
1643 1327 59 (Phy) 58 SB
1701 1327 26 (si) 25 SB
1726 1327 17 (c) 16 SB
1742 1327 25 (s ) 24 SB
1766 1327 11 (f) 10 SB
1776 1327 44 (or ) 43 SB
1819 1327 61 (Nuc) 60 SB
1879 1327 11 (l) 10 SB
1889 1327 51 (ear) 50 SB
1939 1327 10 ( ) 9 SB
1948 1327 57 (Fus) 56 SB
2004 1327 30 (io) 29 SB
2033 1327 19 (n) 19 SB
gr
32 0 0 38 38 0 0 0 34 /Times-Roman /font32 ANSIFont font
gs 1096 1501 1152 238 CB
2052 1328 10 (,) 9 SB
2061 1328 29 ( p) 28 SB
2089 1328 29 (.1) 28 SB
2117 1328 48 (58,) 47 SB
gr
gs 1096 1501 1152 238 CB
1302 1378 34 (M) 33 SB
1335 1378 46 (IT ) 45 SB
1380 1378 34 (Pr) 33 SB
1413 1378 32 (es) 31 SB
1444 1378 25 (s,) 24 SB
1468 1378 10 ( ) 9 SB
1477 1378 72 (Cam) 71 SB
1548 1378 19 (b) 18 SB
1566 1378 24 (ri) 23 SB
1589 1378 55 (dge) 54 SB
1643 1378 10 (,) 9 SB
1652 1378 23 ( \() 22 SB
1674 1378 57 (198) 56 SB
1730 1378 42 (9\).) 41 SB
gr
gs 1096 1501 1152 238 CB
1227 1428 32 ([4) 31 SB
1258 1428 23 (].) 22 SB
gr
gs 1096 1501 1152 238 CB
1302 1428 45 (B. ) 44 SB
1346 1428 34 (Li) 33 SB
1379 1428 34 (ps) 33 SB
1412 1428 36 (ch) 35 SB
1447 1428 30 (ul) 29 SB
1476 1428 28 (tz) 27 SB
1503 1428 10 (,) 9 SB
1512 1428 23 ( I) 22 SB
1534 1428 10 (.) 9 SB
1543 1428 56 ( Hu) 55 SB
1598 1428 28 (tc) 27 SB
1625 1428 30 (hi) 29 SB
1654 1428 34 (ns) 33 SB
1687 1428 48 (on,) 47 SB
1734 1428 10 ( ) 9 SB
1743 1428 45 (B. ) 44 SB
1787 1428 40 (La) 39 SB
1826 1428 74 (Bom) 73 SB
1899 1428 36 (ba) 35 SB
1934 1428 32 (rd) 31 SB
1965 1428 20 (, ) 19 SB
1984 1428 17 (a) 16 SB
2000 1428 48 (nd ) 47 SB
2047 1428 37 (A.) 36 SB
2083 1428 46 ( W) 45 SB
2128 1428 17 (a) 16 SB
2144 1428 29 (n.) 28 SB
2172 1428 10 (,) 10 SB
gr
gs 1096 1501 1152 238 CB
1302 1478 25 (J.) 24 SB
1326 1478 10 ( ) 9 SB
1335 1478 61 (Vac) 60 SB
1395 1478 10 (.) 9 SB
1404 1478 31 ( S) 30 SB
1434 1478 28 (ci) 27 SB
1461 1478 10 (.) 9 SB
1470 1478 33 ( T) 32 SB
1502 1478 34 (ec) 33 SB
1535 1478 68 (hnol) 67 SB
1602 1478 10 (.) 9 SB
1611 1478 10 ( ) 9 SB
gr
32 0 0 38 38 0 0 0 37 /Times-Bold /font29 ANSIFont font
gs 1096 1501 1152 238 CB
1620 1475 46 (A4) 46 SB
gr
32 0 0 38 38 0 0 0 34 /Times-Roman /font32 ANSIFont font
gs 1096 1501 1152 238 CB
1666 1478 32 (\(3) 31 SB
1697 1478 23 (\),) 22 SB
1719 1478 10 ( ) 9 SB
1728 1478 51 (Ma) 50 SB
1778 1478 30 (y/) 29 SB
1807 1478 34 (Ju) 33 SB
1840 1478 29 (n ) 28 SB
1868 1478 51 (\(19) 50 SB
1918 1478 51 (86\)) 50 SB
1968 1478 29 ( 1) 28 SB
1996 1478 67 (810.) 66 SB
gr
/PPT_ProcessAll true def

userdict /VPsave save put
userdict begin
/showpage{}def
1002 1588 1002 2920 2398 2920 2398 1588 newpath moveto lineto lineto lineto clip newpath
1227 2638 translate 300 72 div dup neg scale
946 300 div 358 72 div div 771 300 div 292 72 div div scale
-35 -45 translate
/AutoFlatness true def
/wCorel5Dict 300 dict def wCorel5Dict begin/bd{bind def}bind def/ld{load def}
bd/xd{exch def}bd/_ null def/rp{{pop}repeat}bd/@cp/closepath ld/@gs/gsave ld
/@gr/grestore ld/@np/newpath ld/Tl/translate ld/$sv 0 def/@sv{/$sv save def}bd
/@rs{$sv restore}bd/spg/showpage ld/showpage{}bd currentscreen/@dsp xd/$dsp
/@dsp def/$dsa xd/$dsf xd/$sdf false def/$SDF false def/$Scra 0 def/SetScr
/setscreen ld/setscreen{3 rp}bd/@ss{2 index 0 eq{$dsf 3 1 roll 4 -1 roll pop}
if exch $Scra add exch load SetScr}bd/SepMode_5 where{pop}{/SepMode_5 0 def}
ifelse/CurrentInkName_5 where{pop}{/CurrentInkName_5(Composite)def}ifelse
/$ink_5 where{pop}{/$ink_5 -1 def}ifelse/$c 0 def/$m 0 def/$y 0 def/$k 0 def
/$t 1 def/$n _ def/$o 0 def/$fil 0 def/$C 0 def/$M 0 def/$Y 0 def/$K 0 def/$T 1
def/$N _ def/$O 0 def/$PF false def/s1c 0 def/s1m 0 def/s1y 0 def/s1k 0 def
/s1t 0 def/s1n _ def/$bkg false def/SK 0 def/SM 0 def/SY 0 def/SC 0 def/$op
false def matrix currentmatrix/$ctm xd/$ptm matrix def/$ttm matrix def/$stm
matrix def/$fst 128 def/$pad 0 def/$rox 0 def/$roy 0 def/$ffpnt true def
/CorelDrawReencodeVect[16#0/grave 16#5/breve 16#6/dotaccent 16#8/ring
16#A/hungarumlaut 16#B/ogonek 16#C/caron 16#D/dotlessi 16#27/quotesingle
16#60/grave 16#7C/bar
16#82/quotesinglbase/florin/quotedblbase/ellipsis/dagger/daggerdbl
16#88/circumflex/perthousand/Scaron/guilsinglleft/OE
16#91/quoteleft/quoteright/quotedblleft/quotedblright/bullet/endash/emdash
16#98/tilde/trademark/scaron/guilsinglright/oe 16#9F/Ydieresis
16#A1/exclamdown/cent/sterling/currency/yen/brokenbar/section
16#a8/dieresis/copyright/ordfeminine/guillemotleft/logicalnot/minus/registered/macron
16#b0/degree/plusminus/twosuperior/threesuperior/acute/mu/paragraph/periodcentered
16#b8/cedilla/onesuperior/ordmasculine/guillemotright/onequarter/onehalf/threequarters/questiondown
16#c0/Agrave/Aacute/Acircumflex/Atilde/Adieresis/Aring/AE/Ccedilla
16#c8/Egrave/Eacute/Ecircumflex/Edieresis/Igrave/Iacute/Icircumflex/Idieresis
16#d0/Eth/Ntilde/Ograve/Oacute/Ocircumflex/Otilde/Odieresis/multiply
16#d8/Oslash/Ugrave/Uacute/Ucircumflex/Udieresis/Yacute/Thorn/germandbls
16#e0/agrave/aacute/acircumflex/atilde/adieresis/aring/ae/ccedilla
16#e8/egrave/eacute/ecircumflex/edieresis/igrave/iacute/icircumflex/idieresis
16#f0/eth/ntilde/ograve/oacute/ocircumflex/otilde/odieresis/divide
16#f8/oslash/ugrave/uacute/ucircumflex/udieresis/yacute/thorn/ydieresis]def
/@BeginSysCorelDict{systemdict/Corel20Dict known{systemdict/Corel20Dict get
exec}if}bd/@EndSysCorelDict{systemdict/Corel20Dict known{end}if}bd AutoFlatness
{/@ifl{dup currentflat exch sub 10 gt{
([Error: PathTooComplex; OffendingCommand: AnyPaintingOperator]\n)print flush
@np exit}{currentflat 2 add setflat}ifelse}bd/@fill/fill ld/fill{currentflat{
{@fill}stopped{@ifl}{exit}ifelse}bind loop setflat}bd/@eofill/eofill ld/eofill
{currentflat{{@eofill}stopped{@ifl}{exit}ifelse}bind loop setflat}bd/@clip
/clip ld/clip{currentflat{{@clip}stopped{@ifl}{exit}ifelse}bind loop setflat}
bd/@eoclip/eoclip ld/eoclip{currentflat{{@eoclip}stopped{@ifl}{exit}ifelse}
bind loop setflat}bd/@stroke/stroke ld/stroke{currentflat{{@stroke}stopped
{@ifl}{exit}ifelse}bind loop setflat}bd}if/d/setdash ld/j/setlinejoin ld/J
/setlinecap ld/M/setmiterlimit ld/w/setlinewidth ld/O{/$o xd}bd/R{/$O xd}bd/W
/eoclip ld/c/curveto ld/C/c ld/l/lineto ld/L/l ld/rl/rlineto ld/m/moveto ld/n
/newpath ld/N/newpath ld/P{11 rp}bd/u{}bd/U{}bd/A{pop}bd/q/@gs ld/Q/@gr ld/`
{}bd/~{}bd/@{}bd/&{}bd/@j{@sv @np}bd/@J{@rs}bd/g{1 exch sub/$k xd/$c 0 def/$m 0
def/$y 0 def/$t 1 def/$n _ def/$fil 0 def}bd/G{1 sub neg/$K xd _ 1 0 0 0/$C xd
/$M xd/$Y xd/$T xd/$N xd}bd/k{1 index type/stringtype eq{/$t xd/$n xd}{/$t 0
def/$n _ def}ifelse/$k xd/$y xd/$m xd/$c xd/$fil 0 def}bd/K{1 index type
/stringtype eq{/$T xd/$N xd}{/$T 0 def/$N _ def}ifelse/$K xd/$Y xd/$M xd/$C xd
}bd/x/k ld/X/K ld/sf{1 index type/stringtype eq{/s1t xd/s1n xd}{/s1t 0 def/s1n
_ def}ifelse/s1k xd/s1y xd/s1m xd/s1c xd}bd/i{dup 0 ne{setflat}{pop}ifelse}bd
/v{4 -2 roll 2 copy 6 -2 roll c}bd/V/v ld/y{2 copy c}bd/Y/y ld/@w{matrix rotate
/$ptm xd matrix scale $ptm dup concatmatrix/$ptm xd 1 eq{$ptm exch dup
concatmatrix/$ptm xd}if 1 w}bd/@g{1 eq dup/$sdf xd{/$scp xd/$sca xd/$scf xd}if
}bd/@G{1 eq dup/$SDF xd{/$SCP xd/$SCA xd/$SCF xd}if}bd/@D{2 index 0 eq{$dsf 3 1
roll 4 -1 roll pop}if 3 copy exch $Scra add exch load SetScr/$dsp xd/$dsa xd
/$dsf xd}bd/$ngx{$SDF{$SCF SepMode_5 0 eq{$SCA}{$dsa}ifelse $SCP @ss}if}bd/p{
/$pm xd 7 rp/$pyf xd/$pxf xd/$pn xd/$fil 1 def}bd/@MN{2 copy le{pop}{exch pop}
ifelse}bd/@MX{2 copy ge{pop}{exch pop}ifelse}bd/InRange{3 -1 roll @MN @MX}bd
/wDstChck{2 1 roll dup 3 -1 roll eq{1 add}if}bd/@dot{dup mul exch dup mul add 1
exch sub}bd/@lin{exch pop abs 1 exch sub}bd/cmyk2rgb{3{dup 5 -1 roll add 1 exch
sub dup 0 lt{pop 0}if exch}repeat pop}bd/rgb2cmyk{3{1 exch sub 3 1 roll}repeat
3 copy @MN @MN 3{dup 5 -1 roll sub neg exch}repeat}bd/rgb2g{2 index .299 mul 2
index .587 mul add 1 index .114 mul add 4 1 roll 3 rp}bd/WaldoColor_5 where{
pop}{/SetRgb/setrgbcolor ld/GetRgb/currentrgbcolor ld/SetGry/setgray ld/GetGry
/currentgray ld/SetRgb2 systemdict/setrgbcolor get def/GetRgb2 systemdict
/currentrgbcolor get def/SetHsb systemdict/sethsbcolor get def/GetHsb
systemdict/currenthsbcolor get def/rgb2hsb{SetRgb2 GetHsb}bd/hsb2rgb{3 -1 roll
dup floor sub 3 1 roll SetHsb GetRgb2}bd/setcmykcolor where{pop/SetCmyk_5
/setcmykcolor ld}{/SetCmyk_5{cmyk2rgb SetRgb}bd}ifelse/currentcmykcolor where{
pop/GetCmyk/currentcmykcolor ld}{/GetCmyk{GetRgb rgb2cmyk}bd}ifelse
/setoverprint where{pop}{/setoverprint{/$op xd}bd}ifelse/currentoverprint where
{pop}{/currentoverprint{$op}bd}ifelse/@tc_5{5 -1 roll dup 1 ge{pop}{4{dup 6 -1
roll mul exch}repeat pop}ifelse}bd/@trp{exch pop 5 1 roll @tc_5}bd
/setprocesscolor_5{SepMode_5 0 eq{SetCmyk_5}{0 4 $ink_5 sub index exch pop 5 1
roll 4 rp SepsColor true eq{$ink_5 3 gt{1 sub neg SetGry}{0 0 0 4 $ink_5 roll
SetCmyk_5}ifelse}{1 sub neg SetGry}ifelse}ifelse}bd/findcmykcustomcolor where
{pop}{/findcmykcustomcolor{5 array astore}bd}ifelse/setcustomcolor where{pop}{
/setcustomcolor{exch aload pop SepMode_5 0 eq{pop @tc_5 setprocesscolor_5}{
CurrentInkName_5 eq{4 index}{0}ifelse 6 1 roll 5 rp 1 sub neg SetGry}ifelse}bd
}ifelse/@scc_5{dup type/booleantype eq{setoverprint}{1 eq setoverprint}ifelse
dup _ eq{pop setprocesscolor_5 pop}{findcmykcustomcolor exch setcustomcolor}
ifelse SepMode_5 0 eq{true}{GetGry 1 eq currentoverprint and not}ifelse}bd
/colorimage where{pop/ColorImage/colorimage ld}{/ColorImage{/ncolors xd pop
/dataaq xd{dataaq ncolors dup 3 eq{/$dat xd 0 1 $dat length 3 div 1 sub{dup 3
mul $dat 1 index get 255 div $dat 2 index 1 add get 255 div $dat 3 index 2 add
get 255 div rgb2g 255 mul cvi exch pop $dat 3 1 roll put}for $dat 0 $dat length
3 idiv getinterval pop}{4 eq{/$dat xd 0 1 $dat length 4 div 1 sub{dup 4 mul
$dat 1 index get 255 div $dat 2 index 1 add get 255 div $dat 3 index 2 add get
255 div $dat 4 index 3 add get 255 div cmyk2rgb rgb2g 255 mul cvi exch pop $dat
3 1 roll put}for $dat 0 $dat length ncolors idiv getinterval}if}ifelse}image}
bd}ifelse/setcmykcolor{1 5 1 roll _ currentoverprint @scc_5/$ffpnt xd}bd
/currentcmykcolor{0 0 0 0}bd/setrgbcolor{rgb2cmyk setcmykcolor}bd
/currentrgbcolor{currentcmykcolor cmyk2rgb}bd/sethsbcolor{hsb2rgb setrgbcolor}
bd/currenthsbcolor{currentrgbcolor rgb2hsb}bd/setgray{dup dup setrgbcolor}bd
/currentgray{currentrgbcolor rgb2g}bd}ifelse/WaldoColor_5 true def/@sft{$tllx
$pxf add dup $tllx gt{$pwid sub}if/$tx xd $tury $pyf sub dup $tury lt{$phei
add}if/$ty xd}bd/@stb{pathbbox/$ury xd/$urx xd/$lly xd/$llx xd}bd/@ep{{cvx exec
}forall}bd/@tp{@sv/$in true def 2 copy dup $lly le{/$in false def}if $phei sub
$ury ge{/$in false def}if dup $urx ge{/$in false def}if $pwid add $llx le{/$in
false def}if $in{@np 2 copy m $pwid 0 rl 0 $phei neg rl $pwid neg 0 rl 0 $phei
rl clip @np $pn cvlit load aload pop 7 -1 roll 5 index sub 7 -1 roll 3 index
sub Tl matrix currentmatrix/$ctm xd @ep 4 rp}{2 rp}ifelse @rs}bd/@th{@sft 0 1
$tly 1 sub{dup $psx mul $tx add{dup $llx gt{$pwid sub}{exit}ifelse}loop exch
$phei mul $ty exch sub 0 1 $tlx 1 sub{$pwid mul 3 copy 3 -1 roll add exch @tp
pop}for 2 rp}for}bd/@tv{@sft 0 1 $tlx 1 sub{dup $pwid mul $tx add exch $psy mul
$ty exch sub{dup $ury lt{$phei add}{exit}ifelse}loop 0 1 $tly 1 sub{$phei mul 3
copy sub @tp pop}for 2 rp}for}bd/@pf{@gs $ctm setmatrix $pm concat @stb eoclip
Bburx Bbury $pm itransform/$tury xd/$turx xd Bbllx Bblly $pm itransform/$tlly
xd/$tllx xd/$wid $turx $tllx sub def/$hei $tury $tlly sub def @gs $vectpat{1 0
0 0 0 _ $o @scc_5{eofill}if}{$t $c $m $y $k $n $o @scc_5{SepMode_5 0 eq $pfrg
or{$tllx $tlly Tl $wid $hei scale <00> 8 1 false[8 0 0 1 0 0]{}imagemask}{
/$bkg true def}ifelse}if}ifelse @gr $wid 0 gt $hei 0 gt and{$pn cvlit load
aload pop/$pd xd 3 -1 roll sub/$phei xd exch sub/$pwid xd $wid $pwid div
ceiling 1 add/$tlx xd $hei $phei div ceiling 1 add/$tly xd $psx 0 eq{@tv}{@th}
ifelse}if @gr @np/$bkg false def}bd/@dlt{$fse $fss sub/nff xd $frb dup 1 eq
exch 2 eq or{$frt dup $frc $frm $fry $frk @tc_5 4 copy cmyk2rgb rgb2hsb 3 copy
/myb xd/mys xd/myh xd $tot $toc $tom $toy $tok @tc_5 cmyk2rgb rgb2hsb 3 1 roll
4 1 roll 5 1 roll sub neg nff div/kdb xd sub neg nff div/kds xd sub neg dup 0
eq{pop $frb 2 eq{.99}{-.99}ifelse}if dup $frb 2 eq exch 0 lt and{1 add}if dup
$frb 1 eq exch 0 gt and{1 sub}if nff div/kdh xd}{$frt dup $frc $frm $fry $frk
@tc_5 5 copy $tot dup $toc $tom $toy $tok @tc_5 5 1 roll 6 1 roll 7 1 roll 8 1
roll 9 1 roll sub neg nff dup 1 gt{1 sub}if div/$dk xd sub neg nff dup 1 gt{1
sub}if div/$dy xd sub neg nff dup 1 gt{1 sub}if div/$dm xd sub neg nff dup 1 gt
{1 sub}if div/$dc xd sub neg nff dup 1 gt{1 sub}if div/$dt xd}ifelse}bd/ffcol{
5 copy $fsit 0 eq{setcmykcolor pop}{SepMode_5 0 ne{$frn findcmykcustomcolor
exch setcustomcolor}{4 rp $frc $frm $fry $frk $frn findcmykcustomcolor exch
setcustomcolor}ifelse}ifelse}bd/@ftl{1 index 4 index sub dup $pad mul dup/$pdw
xd 2 mul sub $fst div/$wid xd 2 index sub/$hei xd pop Tl @dlt $fss 0 eq{ffcol n
0 0 m 0 $hei l $pdw $hei l $pdw 0 l @cp $ffpnt{fill}{@np}ifelse}if $fss $wid
mul $pdw add 0 Tl nff{ffcol n 0 0 m 0 $hei l $wid $hei l $wid 0 l @cp $ffpnt
{fill}{@np}ifelse $wid 0 Tl $frb dup 1 eq exch 2 eq or{4 rp myh mys myb kdb add
3 1 roll kds add 3 1 roll kdh add 3 1 roll 3 copy/myb xd/mys xd/myh xd hsb2rgb
rgb2cmyk}{$dk add 5 1 roll $dy add 5 1 roll $dm add 5 1 roll $dc add 5 1 roll
$dt add 5 1 roll}ifelse}repeat 5 rp $tot dup $toc $tom $toy $tok @tc_5 ffcol n
0 0 m 0 $hei l $pdw $hei l $pdw 0 l @cp $ffpnt{fill}{@np}ifelse 5 rp}bd/@ftrs{
1 index 4 index sub dup $rox mul/$row xd 2 div 1 index 4 index sub dup $roy mul
/$roh xd 2 div 2 copy dup mul exch dup mul add sqrt $row dup mul $roh dup mul
add sqrt add dup/$hei xd $fst div/$wid xd 4 index add $roh add exch 5 index add
$row add exch Tl $fan rotate 4 rp @dlt $fss 0 eq{ffcol $fty 3 eq{$hei dup neg
dup m 2 mul @sqr}{0 0 m 0 0 $hei 0 360 arc}ifelse $ffpnt{fill}{@np}ifelse}if
1.0 $pad 2 mul sub dup scale $hei $fss $wid mul sub/$hei xd nff{ffcol $fty 3 eq
{n $hei dup neg dup m 2 mul @sqr}{n 0 0 m 0 0 $hei 0 360 arc}ifelse $ffpnt
{fill}{@np}ifelse/$hei $hei $wid sub def $frb dup 1 eq exch 2 eq or{4 rp myh
mys myb kdb add 3 1 roll kds add 3 1 roll kdh add 3 1 roll 3 copy/myb xd/mys xd
/myh xd hsb2rgb rgb2cmyk}{$dk add 5 1 roll $dy add 5 1 roll $dm add 5 1 roll
$dc add 5 1 roll $dt add 5 1 roll}ifelse}repeat 5 rp}bd/@ftc{1 index 4 index
sub dup $rox mul/$row xd 2 div 1 index 4 index sub dup $roy mul/$roh xd 2 div 2
copy dup mul exch dup mul add sqrt $row dup mul $roh dup mul add sqrt add dup
/$hei xd $fst div/$wid xd 4 index add $roh add exch 5 index add $row add exch
Tl 4 rp @dlt $fss 0 eq{ffcol $ffpnt{fill}{@np}ifelse}{n}ifelse/$dang 180 $fst 1
sub div def/$sang $dang -2 div 180 add def/$eang $dang 2 div 180 add def/$sang
$sang $dang $fss mul add def/$eang $eang $dang $fss mul add def/$sang $eang
$dang sub def nff{ffcol n 0 0 m 0 0 $hei $sang $fan add $eang $fan add arc
$ffpnt{fill}{@np}ifelse 0 0 m 0 0 $hei $eang neg $fan add $sang neg $fan add
arc $ffpnt{fill}{@np}ifelse/$sang $eang def/$eang $eang $dang add def $frb dup
1 eq exch 2 eq or{4 rp myh mys myb kdb add 3 1 roll kds add 3 1 roll kdh add 3
1 roll 3 copy/myb xd/mys xd/myh xd hsb2rgb rgb2cmyk}{$dk add 5 1 roll $dy add 5
1 roll $dm add 5 1 roll $dc add 5 1 roll $dt add 5 1 roll}ifelse}repeat 5 rp}
bd/@ff{/$fss 0 def $o 1 eq setoverprint 1 1 $fsc 1 sub{dup 1 sub $fsit 0 eq{
$fsa exch 5 mul 5 getinterval aload 2 rp/$frk xd/$fry xd/$frm xd/$frc xd/$frn _
def/$frt 1 def $fsa exch 5 mul 5 getinterval aload pop $fss add/$fse xd/$tok xd
/$toy xd/$tom xd/$toc xd/$ton _ def/$tot 1 def}{$fsa exch 7 mul 7 getinterval
aload 2 rp/$frt xd/$frn xd/$frk xd/$fry xd/$frm xd/$frc xd $fsa exch 7 mul 7
getinterval aload pop $fss add/$fse xd/$tot xd/$ton xd/$tok xd/$toy xd/$tom xd
/$toc xd}ifelse $fsit 0 eq SepMode_5 0 eq or dup not CurrentInkName_5 $frn eq
and or{@sv $ctm setmatrix eoclip Bbllx Bblly Bburx Bbury $fty 2 eq{@ftc}{1
index 3 index m 2 copy l 3 index 1 index l 3 index 3 index l @cp $fty dup 1 eq
exch 3 eq or{@ftrs}{4 rp $fan rotate pathbbox @ftl}ifelse}ifelse @rs/$fss $fse
def}{1 0 0 0 0 _ $o @scc_5{fill}if}ifelse}for @np}bd/@Pf{@sv SepMode_5 0 eq
$ink_5 3 eq or{0 J 0 j[]0 d $t $c $m $y $k $n $o @scc_5 pop $ctm setmatrix 72
1000 div dup matrix scale dup concat dup Bburx exch Bbury exch itransform
ceiling cvi/Bbury xd ceiling cvi/Bburx xd Bbllx exch Bblly exch itransform
floor cvi/Bblly xd floor cvi/Bbllx xd $Prm aload pop $Psn load exec}{1 SetGry
eofill}ifelse @rs @np}bd/F{matrix currentmatrix $sdf{$scf $sca $scp @ss}if $fil
1 eq{@pf}{$fil 2 eq{@ff}{$fil 3 eq{@Pf}{$t $c $m $y $k $n $o @scc_5{eofill}
{@np}ifelse}ifelse}ifelse}ifelse $sdf{$dsf $dsa $dsp @ss}if setmatrix}bd/f{@cp
F}bd/S{matrix currentmatrix $ctm setmatrix $SDF{$SCF $SCA $SCP @ss}if $T $C $M
$Y $K $N $O @scc_5{matrix currentmatrix $ptm concat stroke setmatrix}
{@np}ifelse $SDF{$dsf $dsa $dsp @ss}if setmatrix}bd/s{@cp S}bd/B{@gs F @gr S}
bd/b{@cp B}bd/E{5 array astore exch cvlit xd}bd/@cc{currentfile $dat
readhexstring pop}bd/@sm{/$ctm $ctm currentmatrix def}bd/@E{/Bbury xd/Bburx xd
/Bblly xd/Bbllx xd}bd/@c{@cp}bd/@p{/$fil 1 def 1 eq dup/$vectpat xd{/$pfrg true
def}{@gs $t $c $m $y $k $n $o @scc_5/$pfrg xd @gr}ifelse/$pm xd/$psy xd/$psx xd
/$pyf xd/$pxf xd/$pn xd}bd/@P{/$fil 3 def/$Psn xd array astore/$Prm xd}bd/@k{
/$fil 2 def/$roy xd/$rox xd/$pad xd/$fty xd/$fan xd $fty 1 eq{/$fan 0 def}if
/$frb xd/$fst xd/$fsc xd/$fsa xd/$fsit 0 def}bd/@x{/$fil 2 def/$roy xd/$rox xd
/$pad xd/$fty xd/$fan xd $fty 1 eq{/$fan 0 def}if/$frb xd/$fst xd/$fsc xd/$fsa
xd/$fsit 1 def}bd/@ii{concat 3 index 3 index m 3 index 1 index l 2 copy l 1
index 3 index l 3 index 3 index l clip 4 rp}bd/tcc{@cc}def/@i{@sm @gs @ii 6
index 1 ne{/$frg true def 2 rp}{1 eq{s1t s1c s1m s1y s1k s1n $O @scc_5/$frg xd
}{/$frg false def}ifelse 1 eq{@gs $ctm setmatrix F @gr}if}ifelse @np/$ury xd
/$urx xd/$lly xd/$llx xd/$bts xd/$hei xd/$wid xd/$dat $wid $bts mul 8 div
ceiling cvi string def $bkg $frg or{$SDF{$SCF $SCA $SCP @ss}if $llx $lly Tl
$urx $llx sub $ury $lly sub scale $bkg{$t $c $m $y $k $n $o @scc_5 pop}if $wid
$hei abs $bts 1 eq{$bkg}{$bts}ifelse[$wid 0 0 $hei neg 0 $hei 0
gt{$hei}{0}ifelse]/tcc load $bts 1 eq{imagemask}{image}ifelse $SDF{$dsf $dsa
$dsp @ss}if}{$hei abs{tcc pop}repeat}ifelse @gr $ctm setmatrix}bd/@M{@sv}bd/@N
{/@cc{}def 1 eq{12 -1 roll neg 12 1 roll @I}{13 -1 roll neg 13 1 roll @i}
ifelse @rs}bd/@I{@sm @gs @ii @np/$ury xd/$urx xd/$lly xd/$llx xd/$ncl xd/$bts
xd/$hei xd/$wid xd/$dat $wid $bts mul $ncl mul 8 div ceiling cvi string def
$ngx $llx $lly Tl $urx $llx sub $ury $lly sub scale $wid $hei abs $bts[$wid 0 0
$hei neg 0 $hei 0 gt{$hei}{0}ifelse]/@cc load false $ncl ColorImage $SDF{$dsf
$dsa $dsp @ss}if @gr $ctm setmatrix}bd/z{exch findfont exch scalefont setfont}
bd/ZB{9 dict dup begin 4 1 roll/FontType 3 def/FontMatrix xd/FontBBox xd
/Encoding 256 array def 0 1 255{Encoding exch/.notdef put}for/CharStrings 256
dict def CharStrings/.notdef{}put/Metrics 256 dict def Metrics/.notdef 3 -1
roll put/BuildChar{exch dup/$char exch/Encoding get 3 index get def dup
/Metrics get $char get aload pop setcachedevice begin Encoding exch get
CharStrings exch get end exec}def end definefont pop}bd/ZBAddChar{findfont
begin dup 4 1 roll dup 6 1 roll Encoding 3 1 roll put CharStrings 3 1 roll put
Metrics 3 1 roll put end}bd/Z{findfont dup maxlength 2 add dict exch dup{1
index/FID ne{3 index 3 1 roll put}{2 rp}ifelse}forall pop dup dup/Encoding get
256 array copy dup/$fe xd/Encoding exch put dup/Fontname 3 index put 3 -1 roll
dup length 0 ne{0 exch{dup type 0 type eq{exch pop}{$fe exch 2 index exch put 1
add}ifelse}forall pop}if dup 256 dict dup/$met xd/Metrics exch put dup
/FontMatrix get 0 get 1000 mul 1 exch div 3 index length 256 eq{0 1 255{dup $fe
exch get dup/.notdef eq{2 rp}{5 index 3 -1 roll get 2 index mul $met 3 1 roll
put}ifelse}for}if pop definefont pop pop}bd/@ftx{{currentpoint 3 -1 roll(0)dup
3 -1 roll 0 exch put dup @gs true charpath $ctm setmatrix @@txt @gr @np
stringwidth pop 3 -1 roll add exch moveto}forall}bd/@ft{matrix currentmatrix
exch $sdf{$scf $sca $scp @ss}if $fil 1 eq{/@@txt/@pf ld @ftx}{$fil 2 eq{/@@txt
/@ff ld @ftx}{$fil 3 eq{/@@txt/@Pf ld @ftx}{$t $c $m $y $k $n $o @scc_5{show}
{pop}ifelse}ifelse}ifelse}ifelse $sdf{$dsf $dsa $dsp @ss}if setmatrix}bd/@st{
matrix currentmatrix exch $SDF{$SCF $SCA $SCP @ss}if $T $C $M $Y $K $N $O
@scc_5{{currentpoint 3 -1 roll(0)dup 3 -1 roll 0 exch put dup @gs true charpath
$ctm setmatrix $ptm concat stroke @gr @np stringwidth pop 3 -1 roll add exch
moveto}forall}{pop}ifelse $SDF{$dsf $dsa $dsp @ss}if setmatrix}bd/@te{@ft}bd
/@tr{@st}bd/@ta{dup @gs @ft @gr @st}bd/@t@a{dup @gs @st @gr @ft}bd/@tm{@sm
concat}bd/e{/t{@te}def}bd/r{/t{@tr}def}bd/o{/t{pop}def}bd/a{/t{@ta}def}bd/@a{
/t{@t@a}def}bd/t{@te}def/T{@np $ctm setmatrix/$ttm matrix def}bd/ddt{t}def/@t{
/$stm $stm currentmatrix def 3 1 roll moveto $ttm concat ddt $stm setmatrix}bd
/@n{/$ttm exch matrix rotate def}bd/@s{}bd/@l{}bd/@B{@gs S @gr F}bd/@b{@cp @B}
bd/@sep{CurrentInkName_5(Composite)eq{/$ink_5 -1 def}{CurrentInkName_5(Cyan)eq
{/$ink_5 0 def}{CurrentInkName_5(Magenta)eq{/$ink_5 1 def}{CurrentInkName_5
(Yellow)eq{/$ink_5 2 def}{CurrentInkName_5(Black)eq{/$ink_5 3 def}{/$ink_5 4
def}ifelse}ifelse}ifelse}ifelse}ifelse}bd/@whi{@gs -72000 dup moveto -72000
72000 lineto 72000 dup lineto 72000 -72000 lineto @cp 1 SetGry fill @gr}bd
/@neg{[{1 exch sub}/exec cvx currenttransfer/exec cvx]cvx settransfer @whi}bd
/currentscale{1 0 dtransform matrix defaultmatrix idtransform dup mul exch dup
mul add sqrt 0 1 dtransform matrix defaultmatrix idtransform dup mul exch dup
mul add sqrt}bd/@unscale{currentscale 1 exch div exch 1 exch div exch scale}bd
/@sqr{dup 0 rlineto dup 0 exch rlineto neg 0 rlineto @cp}bd/corelsym{@gs @np Tl
-90 rotate 7{45 rotate -.75 2 moveto 1.5 @sqr fill}repeat @gr}bd/@reg_cor{@gs
@np Tl -6 -6 moveto 12 @sqr @gs 1 GetGry sub SetGry fill @gr 4{90 rotate 0 4 m
0 4 rl}repeat stroke 0 0 corelsym @gr}bd/@reg_std{@gs @np Tl .3 w 0 0 5 0 360
arc @cp @gs 1 GetGry sub SetGry fill @gr 4{90 rotate 0 0 m 0 8 rl}repeat stroke
@gr}bd/@reg_inv{@gs @np Tl .3 w 0 0 5 0 360 arc @cp @gs 1 GetGry sub SetGry
fill @gr 4{90 rotate 0 0 m 0 8 rl}repeat stroke 0 0 m 0 0 5 90 180 arc @cp 0 0
5 270 360 arc @cp GetGry fill @gr}bd/$corelmeter[1 .95 .75 .50 .25 .05 0]def
/@colormeter{@gs @np 0 SetGry 0.3 w/Courier findfont 5 scalefont setfont/yy xd
/xx xd 0 1 6{dup xx 20 sub yy m 20 @sqr @gs $corelmeter exch get dup SetGry
fill @gr stroke xx 18 sub yy 2 add m exch dup 3 ge{1 SetGry}{0 SetGry}ifelse 3
eq{pop}{100 mul 100 exch sub cvi 20 string cvs show}ifelse/yy yy 20 add def}
for @gr}bd/@calbar{@gs Tl @gs @np 0 0 m @gs 20 @sqr 1 1 0 0 0 _ 0 @scc_5 pop
fill @gr 20 0 Tl 0 0 m @gs 20 @sqr 1 1 0 1 0 _ 0 @scc_5 pop fill @gr 20 0 Tl 0
0 m @gs 20 @sqr 1 0 0 1 0 _ 0 @scc_5 pop fill @gr 20 0 Tl 0 0 m @gs 20 @sqr 1 0
1 1 0 _ 0 @scc_5 pop fill @gr 20 0 Tl 0 0 m @gs 20 @sqr 1 0 1 0 0 _ 0 @scc_5
pop fill @gr 20 0 Tl 0 0 m @gs 20 @sqr 1 1 1 0 0 _ 0 @scc_5 pop fill @gr 20 0
Tl 0 0 m @gs 20 @sqr 1 1 1 1 0 _ 0 @scc_5 pop fill @gr @gr @np -84 0 Tl @gs 0 0
m 20 @sqr clip 1 1 0 0 0 _ 0 @scc_5 pop @gain @gr 20 0 Tl @gs 0 0 m 20 @sqr
clip 1 0 1 0 0 _ 0 @scc_5 pop @gain @gr 20 0 Tl @gs 0 0 m 20 @sqr clip 1 0 0 1
0 _ 0 @scc_5 pop @gain @gr 20 0 Tl @gs 0 0 m 20 @sqr clip 1 0 0 0 1 _ 0 @scc_5
pop @gain @gr @gr}bd/@gain{10 10 Tl @np 0 0 m 0 10 360{0 0 15 4 -1 roll dup 5
add arc @cp}for fill}bd/@crop{@gs 10 div/$croplen xd .3 w 0 SetGry Tl rotate 0
0 m 0 $croplen neg rl stroke @gr}bd/@colorbox{@gs @np Tl 100 exch sub 100 div
SetGry -8 -8 moveto 16 @sqr fill 0 SetGry 10 -2 moveto show @gr}bd/deflevel 0
def/@sax{/deflevel deflevel 1 add def}bd/@eax{/deflevel deflevel dup 0 gt{1
sub}if def deflevel 0 gt{/eax load}{eax}ifelse}bd/eax{{exec}forall}bd/@rax{
deflevel 0 eq{@rs @sv}if}bd/@daq{dup type/arraytype eq{{}forall}if}bd/@BMP{
/@cc xd 11 index 1 eq{12 -1 roll pop @i}{7 -2 roll 2 rp @I}ifelse}bd systemdict
/pdfmark known not{/pdfmark/cleartomark ld}if end
wCorel5Dict begin
@BeginSysCorelDict
2.6131 setmiterlimit
1.00 setflat
/$fst 14 def
[ 0 0 0 0 0 0 0 0 0 0 0 0 0 0 0 0 0 
0 0 0 0 0 0 0 0 0 0 0 0 0 0 0 278 
278 355 556 556 889 667 191 333 333 389 584 278 333 278 278 556 
556 556 556 556 556 556 556 556 556 278 278 584 584 584 556 1015 
667 667 722 722 667 611 778 722 278 500 667 556 833 722 778 667 
778 722 667 611 722 667 944 667 667 611 278 278 278 469 556 333 
556 556 500 556 556 278 556 556 222 222 500 222 833 556 556 556 
556 333 500 278 556 500 722 500 500 500 334 260 334 584 350 350 
350 222 556 333 1000 556 556 333 1000 667 333 1000 350 350 350 350 
222 222 333 333 350 556 1000 333 1000 500 333 944 350 350 667 278 
333 556 556 556 556 260 556 333 737 370 556 584 584 737 333 400 
584 333 333 333 556 537 278 333 333 365 556 834 834 834 611 667 
667 667 667 667 667 1000 722 667 667 667 667 278 278 278 278 722 
722 778 778 778 778 778 584 778 722 722 722 722 667 667 611 556 
556 556 556 556 556 889 500 556 556 556 556 278 278 278 278 556 
556 556 556 556 556 556 584 611 556 556 556 556 500 556 500 ]
CorelDrawReencodeVect /_R44-Helvetica /Helvetica Z

@sv
@sm
@sv
@sm @sv @sv
@rax 
329.33 181.51 329.62 188.57 @E
0 J 2 j [] 0 d 0 R 0 @G
0.00 0.00 0.00 1.00 K
0 1.008 1.008 0.000 @w
329.47 181.51 m
329.47 188.57 L
S

@rax 
327.46 188.42 331.49 188.71 @E
0 J 2 j [] 0 d 0 R 0 @G
0.00 0.00 0.00 1.00 K
0 1.008 1.008 0.000 @w
327.46 188.57 m
331.49 188.57 L
S

@rax 
329.33 168.55 329.62 181.51 @E
0 J 2 j [] 0 d 0 R 0 @G
0.00 0.00 0.00 1.00 K
0 1.008 1.008 0.000 @w
329.47 181.51 m
329.47 168.55 L
S

@rax 
327.46 168.41 331.49 168.70 @E
0 J 2 j [] 0 d 0 R 0 @G
0.00 0.00 0.00 1.00 K
0 1.008 1.008 0.000 @w
327.46 168.55 m
331.49 168.55 L
S

@rax 
313.34 182.52 313.63 190.51 @E
0 J 2 j [] 0 d 0 R 0 @G
0.00 0.00 0.00 1.00 K
0 1.008 1.008 0.000 @w
313.49 182.52 m
313.49 190.51 L
S

@rax 
311.47 190.37 315.43 190.66 @E
0 J 2 j [] 0 d 0 R 0 @G
0.00 0.00 0.00 1.00 K
0 1.008 1.008 0.000 @w
311.47 190.51 m
315.43 190.51 L
S

@rax 
313.34 170.57 313.63 182.52 @E
0 J 2 j [] 0 d 0 R 0 @G
0.00 0.00 0.00 1.00 K
0 1.008 1.008 0.000 @w
313.49 182.52 m
313.49 170.57 L
S

@rax 
311.47 170.42 315.43 170.71 @E
0 J 2 j [] 0 d 0 R 0 @G
0.00 0.00 0.00 1.00 K
0 1.008 1.008 0.000 @w
311.47 170.57 m
315.43 170.57 L
S

@rax 
300.38 184.54 300.67 191.52 @E
0 J 2 j [] 0 d 0 R 0 @G
0.00 0.00 0.00 1.00 K
0 1.008 1.008 0.000 @w
300.53 184.54 m
300.53 191.52 L
S

@rax 
298.51 191.38 302.54 191.66 @E
0 J 2 j [] 0 d 0 R 0 @G
0.00 0.00 0.00 1.00 K
0 1.008 1.008 0.000 @w
298.51 191.52 m
302.54 191.52 L
S

@rax 
300.38 171.50 300.67 184.54 @E
0 J 2 j [] 0 d 0 R 0 @G
0.00 0.00 0.00 1.00 K
0 1.008 1.008 0.000 @w
300.53 184.54 m
300.53 171.50 L
S

@rax 
298.51 171.36 302.54 171.65 @E
0 J 2 j [] 0 d 0 R 0 @G
0.00 0.00 0.00 1.00 K
0 1.008 1.008 0.000 @w
298.51 171.50 m
302.54 171.50 L
S

@rax 
288.36 182.52 288.65 189.50 @E
0 J 2 j [] 0 d 0 R 0 @G
0.00 0.00 0.00 1.00 K
0 1.008 1.008 0.000 @w
288.50 182.52 m
288.50 189.50 L
S

@rax 
286.56 189.36 290.52 189.65 @E
0 J 2 j [] 0 d 0 R 0 @G
0.00 0.00 0.00 1.00 K
0 1.008 1.008 0.000 @w
286.56 189.50 m
290.52 189.50 L
S

@rax 
288.36 169.56 288.65 182.52 @E
0 J 2 j [] 0 d 0 R 0 @G
0.00 0.00 0.00 1.00 K
0 1.008 1.008 0.000 @w
288.50 182.52 m
288.50 169.56 L
S

@rax 
286.56 169.42 290.52 169.70 @E
0 J 2 j [] 0 d 0 R 0 @G
0.00 0.00 0.00 1.00 K
0 1.008 1.008 0.000 @w
286.56 169.56 m
290.52 169.56 L
S

@rax 
276.41 183.53 276.70 190.51 @E
0 J 2 j [] 0 d 0 R 0 @G
0.00 0.00 0.00 1.00 K
0 1.008 1.008 0.000 @w
276.55 183.53 m
276.55 190.51 L
S

@rax 
274.54 190.37 278.57 190.66 @E
0 J 2 j [] 0 d 0 R 0 @G
0.00 0.00 0.00 1.00 K
0 1.008 1.008 0.000 @w
274.54 190.51 m
278.57 190.51 L
S

@rax 
276.41 170.57 276.70 183.53 @E
0 J 2 j [] 0 d 0 R 0 @G
0.00 0.00 0.00 1.00 K
0 1.008 1.008 0.000 @w
276.55 183.53 m
276.55 170.57 L
S

@rax 
274.54 170.42 278.57 170.71 @E
0 J 2 j [] 0 d 0 R 0 @G
0.00 0.00 0.00 1.00 K
0 1.008 1.008 0.000 @w
274.54 170.57 m
278.57 170.57 L
S

@rax 
264.38 187.56 264.67 194.54 @E
0 J 2 j [] 0 d 0 R 0 @G
0.00 0.00 0.00 1.00 K
0 1.008 1.008 0.000 @w
264.53 187.56 m
264.53 194.54 L
S

@rax 
262.51 194.40 266.54 194.69 @E
0 J 2 j [] 0 d 0 R 0 @G
0.00 0.00 0.00 1.00 K
0 1.008 1.008 0.000 @w
262.51 194.54 m
266.54 194.54 L
S

@rax 
264.38 174.53 264.67 187.56 @E
0 J 2 j [] 0 d 0 R 0 @G
0.00 0.00 0.00 1.00 K
0 1.008 1.008 0.000 @w
264.53 187.56 m
264.53 174.53 L
S

@rax 
262.51 174.38 266.54 174.67 @E
0 J 2 j [] 0 d 0 R 0 @G
0.00 0.00 0.00 1.00 K
0 1.008 1.008 0.000 @w
262.51 174.53 m
266.54 174.53 L
S

@rax 
247.39 189.50 247.68 196.56 @E
0 J 2 j [] 0 d 0 R 0 @G
0.00 0.00 0.00 1.00 K
0 1.008 1.008 0.000 @w
247.54 189.50 m
247.54 196.56 L
S

@rax 
245.52 196.42 249.55 196.70 @E
0 J 2 j [] 0 d 0 R 0 @G
0.00 0.00 0.00 1.00 K
0 1.008 1.008 0.000 @w
245.52 196.56 m
249.55 196.56 L
S

@rax 
247.39 176.54 247.68 189.50 @E
0 J 2 j [] 0 d 0 R 0 @G
0.00 0.00 0.00 1.00 K
0 1.008 1.008 0.000 @w
247.54 189.50 m
247.54 176.54 L
S

@rax 
245.52 176.40 249.55 176.69 @E
0 J 2 j [] 0 d 0 R 0 @G
0.00 0.00 0.00 1.00 K
0 1.008 1.008 0.000 @w
245.52 176.54 m
249.55 176.54 L
S

@rax 
231.41 190.51 231.70 198.50 @E
0 J 2 j [] 0 d 0 R 0 @G
0.00 0.00 0.00 1.00 K
0 1.008 1.008 0.000 @w
231.55 190.51 m
231.55 198.50 L
S

@rax 
229.54 198.36 233.57 198.65 @E
0 J 2 j [] 0 d 0 R 0 @G
0.00 0.00 0.00 1.00 K
0 1.008 1.008 0.000 @w
229.54 198.50 m
233.57 198.50 L
S

@rax 
231.41 177.55 231.70 190.51 @E
0 J 2 j [] 0 d 0 R 0 @G
0.00 0.00 0.00 1.00 K
0 1.008 1.008 0.000 @w
231.55 190.51 m
231.55 177.55 L
S

@rax 
229.54 177.41 233.57 177.70 @E
0 J 2 j [] 0 d 0 R 0 @G
0.00 0.00 0.00 1.00 K
0 1.008 1.008 0.000 @w
229.54 177.55 m
233.57 177.55 L
S

@rax 
216.36 191.52 216.65 198.50 @E
0 J 2 j [] 0 d 0 R 0 @G
0.00 0.00 0.00 1.00 K
0 1.008 1.008 0.000 @w
216.50 191.52 m
216.50 198.50 L
S

@rax 
214.56 198.36 218.52 198.65 @E
0 J 2 j [] 0 d 0 R 0 @G
0.00 0.00 0.00 1.00 K
0 1.008 1.008 0.000 @w
214.56 198.50 m
218.52 198.50 L
S

@rax 
216.36 178.56 216.65 191.52 @E
0 J 2 j [] 0 d 0 R 0 @G
0.00 0.00 0.00 1.00 K
0 1.008 1.008 0.000 @w
216.50 191.52 m
216.50 178.56 L
S

@rax 
214.56 178.42 218.52 178.70 @E
0 J 2 j [] 0 d 0 R 0 @G
0.00 0.00 0.00 1.00 K
0 1.008 1.008 0.000 @w
214.56 178.56 m
218.52 178.56 L
S

@rax 
204.41 194.54 204.70 201.53 @E
0 J 2 j [] 0 d 0 R 0 @G
0.00 0.00 0.00 1.00 K
0 1.008 1.008 0.000 @w
204.55 194.54 m
204.55 201.53 L
S

@rax 
202.54 201.38 206.57 201.67 @E
0 J 2 j [] 0 d 0 R 0 @G
0.00 0.00 0.00 1.00 K
0 1.008 1.008 0.000 @w
202.54 201.53 m
206.57 201.53 L
S

@rax 
204.41 181.51 204.70 194.54 @E
0 J 2 j [] 0 d 0 R 0 @G
0.00 0.00 0.00 1.00 K
0 1.008 1.008 0.000 @w
204.55 194.54 m
204.55 181.51 L
S

@rax 
202.54 181.37 206.57 181.66 @E
0 J 2 j [] 0 d 0 R 0 @G
0.00 0.00 0.00 1.00 K
0 1.008 1.008 0.000 @w
202.54 181.51 m
206.57 181.51 L
S

@rax 
180.36 194.54 180.65 201.53 @E
0 J 2 j [] 0 d 0 R 0 @G
0.00 0.00 0.00 1.00 K
0 1.008 1.008 0.000 @w
180.50 194.54 m
180.50 201.53 L
S

@rax 
178.56 201.38 182.52 201.67 @E
0 J 2 j [] 0 d 0 R 0 @G
0.00 0.00 0.00 1.00 K
0 1.008 1.008 0.000 @w
178.56 201.53 m
182.52 201.53 L
S

@rax 
180.36 181.51 180.65 194.54 @E
0 J 2 j [] 0 d 0 R 0 @G
0.00 0.00 0.00 1.00 K
0 1.008 1.008 0.000 @w
180.50 194.54 m
180.50 181.51 L
S

@rax 
178.56 181.37 182.52 181.66 @E
0 J 2 j [] 0 d 0 R 0 @G
0.00 0.00 0.00 1.00 K
0 1.008 1.008 0.000 @w
178.56 181.51 m
182.52 181.51 L
S

@rax 
165.38 195.55 165.67 203.54 @E
0 J 2 j [] 0 d 0 R 0 @G
0.00 0.00 0.00 1.00 K
0 1.008 1.008 0.000 @w
165.53 195.55 m
165.53 203.54 L
S

@rax 
163.51 203.40 167.54 203.69 @E
0 J 2 j [] 0 d 0 R 0 @G
0.00 0.00 0.00 1.00 K
0 1.008 1.008 0.000 @w
163.51 203.54 m
167.54 203.54 L
S

@rax 
165.38 182.52 165.67 195.55 @E
0 J 2 j [] 0 d 0 R 0 @G
0.00 0.00 0.00 1.00 K
0 1.008 1.008 0.000 @w
165.53 195.55 m
165.53 182.52 L
S

@rax 
163.51 182.38 167.54 182.66 @E
0 J 2 j [] 0 d 0 R 0 @G
0.00 0.00 0.00 1.00 K
0 1.008 1.008 0.000 @w
163.51 182.52 m
167.54 182.52 L
S

@rax 
152.42 193.54 152.71 200.52 @E
0 J 2 j [] 0 d 0 R 0 @G
0.00 0.00 0.00 1.00 K
0 1.008 1.008 0.000 @w
152.57 193.54 m
152.57 200.52 L
S

@rax 
150.55 200.38 154.51 200.66 @E
0 J 2 j [] 0 d 0 R 0 @G
0.00 0.00 0.00 1.00 K
0 1.008 1.008 0.000 @w
150.55 200.52 m
154.51 200.52 L
S

@rax 
152.42 180.50 152.71 193.54 @E
0 J 2 j [] 0 d 0 R 0 @G
0.00 0.00 0.00 1.00 K
0 1.008 1.008 0.000 @w
152.57 193.54 m
152.57 180.50 L
S

@rax 
150.55 180.36 154.51 180.65 @E
0 J 2 j [] 0 d 0 R 0 @G
0.00 0.00 0.00 1.00 K
0 1.008 1.008 0.000 @w
150.55 180.50 m
154.51 180.50 L
S

@rax 
130.39 194.54 130.68 201.53 @E
0 J 2 j [] 0 d 0 R 0 @G
0.00 0.00 0.00 1.00 K
0 1.008 1.008 0.000 @w
130.54 194.54 m
130.54 201.53 L
S

@rax 
128.52 201.38 132.55 201.67 @E
0 J 2 j [] 0 d 0 R 0 @G
0.00 0.00 0.00 1.00 K
0 1.008 1.008 0.000 @w
128.52 201.53 m
132.55 201.53 L
S

@rax 
130.39 181.51 130.68 194.54 @E
0 J 2 j [] 0 d 0 R 0 @G
0.00 0.00 0.00 1.00 K
0 1.008 1.008 0.000 @w
130.54 194.54 m
130.54 181.51 L
S

@rax 
128.52 181.37 132.55 181.66 @E
0 J 2 j [] 0 d 0 R 0 @G
0.00 0.00 0.00 1.00 K
0 1.008 1.008 0.000 @w
128.52 181.51 m
132.55 181.51 L
S

@rax 
130.54 181.51 329.47 195.55 @E
0 J 0 j [] 0 d 0 R 0 @G
0.00 0.00 0.00 1.00 K
0 1.008 1.008 0.000 @w
329.47 181.51 m
327.46 181.51 L
325.44 181.51 L
323.50 182.52 L
321.48 182.52 L
319.46 182.52 L
317.45 182.52 L
315.43 182.52 L
313.49 182.52 L
311.47 183.53 L
309.46 183.53 L
307.44 183.53 L
305.57 183.53 L
303.55 183.53 L
301.54 183.53 L
299.52 183.53 L
297.50 183.53 L
295.56 183.53 L
293.54 183.53 L
291.53 183.53 L
289.51 183.53 L
287.57 183.53 L
285.55 183.53 L
283.54 183.53 L
281.52 183.53 L
279.50 183.53 L
277.56 184.54 L
275.54 184.54 L
273.53 184.54 L
271.51 185.54 L
269.57 185.54 L
267.55 186.55 L
265.54 186.55 L
263.52 187.56 L
261.50 187.56 L
259.56 187.56 L
257.54 188.57 L
255.53 188.57 L
253.51 188.57 L
251.57 188.57 L
249.55 189.50 L
247.54 189.50 L
245.52 189.50 L
243.50 189.50 L
241.56 189.50 L
238.54 190.51 L
237.53 190.51 L
234.50 190.51 L
232.56 190.51 L
231.55 190.51 L
228.53 190.51 L
227.52 191.52 L
224.57 191.52 L
222.55 191.52 L
221.54 191.52 L
218.52 191.52 L
216.50 191.52 L
214.56 192.53 L
212.54 192.53 L
210.53 192.53 L
208.51 193.54 L
206.57 193.54 L
204.55 194.54 L
202.54 194.54 L
200.52 194.54 L
198.50 194.54 L
196.56 194.54 L
194.54 194.54 L
192.53 194.54 L
190.51 194.54 L
188.57 194.54 L
186.55 194.54 L
184.54 194.54 L
182.52 194.54 L
180.50 194.54 L
178.56 194.54 L
176.54 195.55 L
174.53 195.55 L
172.51 195.55 L
170.57 195.55 L
168.55 195.55 L
166.54 195.55 L
164.52 195.55 L
162.50 195.55 L
160.56 194.54 L
158.54 194.54 L
156.53 194.54 L
154.51 194.54 L
152.57 194.54 L
150.55 194.54 L
148.54 194.54 L
146.52 194.54 L
144.50 194.54 L
142.56 194.54 L
140.54 194.54 L
138.53 194.54 L
136.51 194.54 L
134.57 194.54 L
132.55 194.54 L
130.54 194.54 L
S

@rax 
328.46 173.38 329.47 173.66 @E
0 J 2 j [] 0 d 0 R 0 @G
0.00 0.00 0.00 1.00 K
0 1.008 1.008 0.000 @w
329.47 173.52 m
328.46 173.52 L
S

@rax 
324.43 173.38 325.44 173.66 @E
0 J 2 j [] 0 d 0 R 0 @G
0.00 0.00 0.00 1.00 K
0 1.008 1.008 0.000 @w
325.44 173.52 m
324.43 173.52 L
S

@rax 
320.47 173.38 321.48 173.66 @E
0 J 2 j [] 0 d 0 R 0 @G
0.00 0.00 0.00 1.00 K
0 1.008 1.008 0.000 @w
321.48 173.52 m
320.47 173.52 L
S

@rax 
316.44 173.38 317.45 173.66 @E
0 J 2 j [] 0 d 0 R 0 @G
0.00 0.00 0.00 1.00 K
0 1.008 1.008 0.000 @w
317.45 173.52 m
316.44 173.52 L
S

@rax 
312.48 173.38 313.49 173.66 @E
0 J 2 j [] 0 d 0 R 0 @G
0.00 0.00 0.00 1.00 K
0 1.008 1.008 0.000 @w
313.49 173.52 m
312.48 173.52 L
S

@rax 
308.45 173.38 309.46 173.66 @E
0 J 2 j [] 0 d 0 R 0 @G
0.00 0.00 0.00 1.00 K
0 1.008 1.008 0.000 @w
309.46 173.52 m
308.45 173.52 L
S

@rax 
304.56 173.38 305.57 173.66 @E
0 J 2 j [] 0 d 0 R 0 @G
0.00 0.00 0.00 1.00 K
0 1.008 1.008 0.000 @w
305.57 173.52 m
304.56 173.52 L
S

@rax 
300.53 173.38 301.54 173.66 @E
0 J 2 j [] 0 d 0 R 0 @G
0.00 0.00 0.00 1.00 K
0 1.008 1.008 0.000 @w
301.54 173.52 m
300.53 173.52 L
S

@rax 
296.57 173.38 297.50 173.66 @E
0 J 2 j [] 0 d 0 R 0 @G
0.00 0.00 0.00 1.00 K
0 1.008 1.008 0.000 @w
297.50 173.52 m
296.57 173.52 L
S

@rax 
292.54 173.38 293.54 173.66 @E
0 J 2 j [] 0 d 0 R 0 @G
0.00 0.00 0.00 1.00 K
0 1.008 1.008 0.000 @w
293.54 173.52 m
292.54 173.52 L
S

@rax 
288.50 173.38 289.51 173.66 @E
0 J 2 j [] 0 d 0 R 0 @G
0.00 0.00 0.00 1.00 K
0 1.008 1.008 0.000 @w
289.51 173.52 m
288.50 173.52 L
S

@rax 
284.54 173.38 285.55 173.66 @E
0 J 2 j [] 0 d 0 R 0 @G
0.00 0.00 0.00 1.00 K
0 1.008 1.008 0.000 @w
285.55 173.52 m
284.54 173.52 L
S

@rax 
280.51 173.38 281.52 173.66 @E
0 J 2 j [] 0 d 0 R 0 @G
0.00 0.00 0.00 1.00 K
0 1.008 1.008 0.000 @w
281.52 173.52 m
280.51 173.52 L
S

@rax 
276.55 173.38 277.56 173.66 @E
0 J 2 j [] 0 d 0 R 0 @G
0.00 0.00 0.00 1.00 K
0 1.008 1.008 0.000 @w
277.56 173.52 m
276.55 173.52 L
S

@rax 
272.52 173.38 273.53 173.66 @E
0 J 2 j [] 0 d 0 R 0 @G
0.00 0.00 0.00 1.00 K
0 1.008 1.008 0.000 @w
273.53 173.52 m
272.52 173.52 L
S

@rax 
268.56 173.38 269.57 173.66 @E
0 J 2 j [] 0 d 0 R 0 @G
0.00 0.00 0.00 1.00 K
0 1.008 1.008 0.000 @w
269.57 173.52 m
268.56 173.52 L
S

@rax 
264.53 173.38 266.54 173.66 @E
0 J 2 j [] 0 d 0 R 0 @G
0.00 0.00 0.00 1.00 K
0 1.008 1.008 0.000 @w
266.54 173.52 m
264.53 173.52 L
S

@rax 
260.57 173.38 261.50 173.66 @E
0 J 2 j [] 0 d 0 R 0 @G
0.00 0.00 0.00 1.00 K
0 1.008 1.008 0.000 @w
261.50 173.52 m
260.57 173.52 L
S

@rax 
256.54 173.38 257.54 173.66 @E
0 J 2 j [] 0 d 0 R 0 @G
0.00 0.00 0.00 1.00 K
0 1.008 1.008 0.000 @w
257.54 173.52 m
256.54 173.52 L
S

@rax 
252.50 173.38 253.51 173.66 @E
0 J 2 j [] 0 d 0 R 0 @G
0.00 0.00 0.00 1.00 K
0 1.008 1.008 0.000 @w
253.51 173.52 m
252.50 173.52 L
S

@rax 
248.54 173.38 249.55 173.66 @E
0 J 2 j [] 0 d 0 R 0 @G
0.00 0.00 0.00 1.00 K
0 1.008 1.008 0.000 @w
249.55 173.52 m
248.54 173.52 L
S

@rax 
244.51 173.38 245.52 173.66 @E
0 J 2 j [] 0 d 0 R 0 @G
0.00 0.00 0.00 1.00 K
0 1.008 1.008 0.000 @w
245.52 173.52 m
244.51 173.52 L
S

@rax 
241.56 173.38 242.57 173.66 @E
0 J 2 j [] 0 d 0 R 0 @G
0.00 0.00 0.00 1.00 K
0 1.008 1.008 0.000 @w
242.57 173.52 m
241.56 173.52 L
S

@rax 
237.53 173.38 238.54 173.66 @E
0 J 2 j [] 0 d 0 R 0 @G
0.00 0.00 0.00 1.00 K
0 1.008 1.008 0.000 @w
238.54 173.52 m
237.53 173.52 L
S

@rax 
232.56 173.38 234.50 173.66 @E
0 J 2 j [] 0 d 0 R 0 @G
0.00 0.00 0.00 1.00 K
0 1.008 1.008 0.000 @w
234.50 173.52 m
232.56 173.52 L
S

@rax 
228.53 173.38 229.54 173.66 @E
0 J 2 j [] 0 d 0 R 0 @G
0.00 0.00 0.00 1.00 K
0 1.008 1.008 0.000 @w
229.54 173.52 m
228.53 173.52 L
S

@rax 
224.57 173.38 225.50 173.66 @E
0 J 2 j [] 0 d 0 R 0 @G
0.00 0.00 0.00 1.00 K
0 1.008 1.008 0.000 @w
225.50 173.52 m
224.57 173.52 L
S

@rax 
220.54 173.38 221.54 173.66 @E
0 J 2 j [] 0 d 0 R 0 @G
0.00 0.00 0.00 1.00 K
0 1.008 1.008 0.000 @w
221.54 173.52 m
220.54 173.52 L
S

@rax 
216.50 173.38 217.51 173.66 @E
0 J 2 j [] 0 d 0 R 0 @G
0.00 0.00 0.00 1.00 K
0 1.008 1.008 0.000 @w
217.51 173.52 m
216.50 173.52 L
S

@rax 
212.54 173.38 213.55 173.66 @E
0 J 2 j [] 0 d 0 R 0 @G
0.00 0.00 0.00 1.00 K
0 1.008 1.008 0.000 @w
213.55 173.52 m
212.54 173.52 L
S

@rax 
208.51 173.38 209.52 173.66 @E
0 J 2 j [] 0 d 0 R 0 @G
0.00 0.00 0.00 1.00 K
0 1.008 1.008 0.000 @w
209.52 173.52 m
208.51 173.52 L
S

@rax 
204.55 173.38 205.56 173.66 @E
0 J 2 j [] 0 d 0 R 0 @G
0.00 0.00 0.00 1.00 K
0 1.008 1.008 0.000 @w
205.56 173.52 m
204.55 173.52 L
S

@rax 
201.53 173.38 202.54 173.66 @E
0 J 2 j [] 0 d 0 R 0 @G
0.00 0.00 0.00 1.00 K
0 1.008 1.008 0.000 @w
202.54 173.52 m
201.53 173.52 L
S

@rax 
197.57 173.38 198.50 173.66 @E
0 J 2 j [] 0 d 0 R 0 @G
0.00 0.00 0.00 1.00 K
0 1.008 1.008 0.000 @w
198.50 173.52 m
197.57 173.52 L
S

@rax 
193.54 173.38 194.54 173.66 @E
0 J 2 j [] 0 d 0 R 0 @G
0.00 0.00 0.00 1.00 K
0 1.008 1.008 0.000 @w
194.54 173.52 m
193.54 173.52 L
S

@rax 
189.50 173.38 190.51 173.66 @E
0 J 2 j [] 0 d 0 R 0 @G
0.00 0.00 0.00 1.00 K
0 1.008 1.008 0.000 @w
190.51 173.52 m
189.50 173.52 L
S

@rax 
184.54 173.38 185.54 173.66 @E
0 J 2 j [] 0 d 0 R 0 @G
0.00 0.00 0.00 1.00 K
0 1.008 1.008 0.000 @w
185.54 173.52 m
184.54 173.52 L
S

@rax 
180.50 173.38 181.51 173.66 @E
0 J 2 j [] 0 d 0 R 0 @G
0.00 0.00 0.00 1.00 K
0 1.008 1.008 0.000 @w
181.51 173.52 m
180.50 173.52 L
S

@rax 
176.54 173.38 177.55 173.66 @E
0 J 2 j [] 0 d 0 R 0 @G
0.00 0.00 0.00 1.00 K
0 1.008 1.008 0.000 @w
177.55 173.52 m
176.54 173.52 L
S

@rax 
173.52 173.38 174.53 173.66 @E
0 J 2 j [] 0 d 0 R 0 @G
0.00 0.00 0.00 1.00 K
0 1.008 1.008 0.000 @w
174.53 173.52 m
173.52 173.52 L
S

@rax 
169.56 173.38 170.57 173.66 @E
0 J 2 j [] 0 d 0 R 0 @G
0.00 0.00 0.00 1.00 K
0 1.008 1.008 0.000 @w
170.57 173.52 m
169.56 173.52 L
S

@rax 
165.53 173.38 166.54 173.66 @E
0 J 2 j [] 0 d 0 R 0 @G
0.00 0.00 0.00 1.00 K
0 1.008 1.008 0.000 @w
166.54 173.52 m
165.53 173.52 L
S

@rax 
161.57 173.38 162.50 173.66 @E
0 J 2 j [] 0 d 0 R 0 @G
0.00 0.00 0.00 1.00 K
0 1.008 1.008 0.000 @w
162.50 173.52 m
161.57 173.52 L
S

@rax 
157.54 173.38 158.54 173.66 @E
0 J 2 j [] 0 d 0 R 0 @G
0.00 0.00 0.00 1.00 K
0 1.008 1.008 0.000 @w
158.54 173.52 m
157.54 173.52 L
S

@rax 
152.57 173.38 153.50 173.66 @E
0 J 2 j [] 0 d 0 R 0 @G
0.00 0.00 0.00 1.00 K
0 1.008 1.008 0.000 @w
153.50 173.52 m
152.57 173.52 L
S

@rax 
148.54 173.38 149.54 173.66 @E
0 J 2 j [] 0 d 0 R 0 @G
0.00 0.00 0.00 1.00 K
0 1.008 1.008 0.000 @w
149.54 173.52 m
148.54 173.52 L
S

@rax 
145.51 173.38 146.52 173.66 @E
0 J 2 j [] 0 d 0 R 0 @G
0.00 0.00 0.00 1.00 K
0 1.008 1.008 0.000 @w
146.52 173.52 m
145.51 173.52 L
S

@rax 
141.55 173.38 142.56 173.66 @E
0 J 2 j [] 0 d 0 R 0 @G
0.00 0.00 0.00 1.00 K
0 1.008 1.008 0.000 @w
142.56 173.52 m
141.55 173.52 L
S

@rax 
137.52 173.38 138.53 173.66 @E
0 J 2 j [] 0 d 0 R 0 @G
0.00 0.00 0.00 1.00 K
0 1.008 1.008 0.000 @w
138.53 173.52 m
137.52 173.52 L
S

@rax 
133.56 173.38 134.57 173.66 @E
0 J 2 j [] 0 d 0 R 0 @G
0.00 0.00 0.00 1.00 K
0 1.008 1.008 0.000 @w
134.57 173.52 m
133.56 173.52 L
S

@rax 
321.48 99.50 329.47 104.54 @E
0 J 2 j [] 0 d 0 R 0 @G
0.00 0.00 0.00 1.00 K
0 1.008 1.008 0.000 @w
329.47 99.50 m
321.48 104.54 L
S

@rax 
305.57 109.51 313.49 113.54 @E
0 J 2 j [] 0 d 0 R 0 @G
0.00 0.00 0.00 1.00 K
0 1.008 1.008 0.000 @w
313.49 109.51 m
305.57 113.54 L
S

@rax 
289.51 117.50 297.50 122.54 @E
0 J 2 j [] 0 d 0 R 0 @G
0.00 0.00 0.00 1.00 K
0 1.008 1.008 0.000 @w
297.50 117.50 m
289.51 122.54 L
S

@rax 
273.53 127.51 281.52 131.54 @E
0 J 0 j [] 0 d 0 R 0 @G
0.00 0.00 0.00 1.00 K
0 1.008 1.008 0.000 @w
281.52 127.51 m
276.55 129.53 L
273.53 131.54 L
S

@rax 
257.54 135.50 266.54 140.54 @E
0 J 0 j [] 0 d 0 R 0 @G
0.00 0.00 0.00 1.00 K
0 1.008 1.008 0.000 @w
266.54 135.50 m
264.53 136.51 L
257.54 140.54 L
S

@rax 
242.57 145.51 249.55 149.54 @E
0 J 0 j [] 0 d 0 R 0 @G
0.00 0.00 0.00 1.00 K
0 1.008 1.008 0.000 @w
249.55 145.51 m
247.54 146.52 L
242.57 149.54 L
S

@rax 
225.50 154.51 234.50 159.55 @E
0 J 0 j [] 0 d 0 R 0 @G
0.00 0.00 0.00 1.00 K
0 1.008 1.008 0.000 @w
234.50 154.51 m
231.55 155.52 L
225.50 159.55 L
S

@rax 
209.52 163.51 217.51 168.55 @E
0 J 0 j [] 0 d 0 R 0 @G
0.00 0.00 0.00 1.00 K
0 1.008 1.008 0.000 @w
217.51 163.51 m
216.50 164.52 L
209.52 168.55 L
S

@rax 
193.54 172.51 202.54 176.54 @E
0 J 2 j [] 0 d 0 R 0 @G
0.00 0.00 0.00 1.00 K
0 1.008 1.008 0.000 @w
202.54 172.51 m
193.54 176.54 L
S

@rax 
177.55 181.51 185.54 186.55 @E
0 J 0 j [] 0 d 0 R 0 @G
0.00 0.00 0.00 1.00 K
0 1.008 1.008 0.000 @w
185.54 181.51 m
180.50 184.54 L
177.55 186.55 L
S

@rax 
162.50 190.51 170.57 195.55 @E
0 J 0 j [] 0 d 0 R 0 @G
0.00 0.00 0.00 1.00 K
0 1.008 1.008 0.000 @w
170.57 190.51 m
165.53 193.54 L
162.50 195.55 L
S

@rax 
146.52 199.51 153.50 204.55 @E
0 J 0 j [] 0 d 0 R 0 @G
0.00 0.00 0.00 1.00 K
0 1.008 1.008 0.000 @w
153.50 199.51 m
152.57 200.52 L
146.52 204.55 L
S

@rax 
130.54 208.51 138.53 213.55 @E
0 J 2 j [] 0 d 0 R 0 @G
0.00 0.00 0.00 1.00 K
0 1.008 1.008 0.000 @w
138.53 208.51 m
130.54 213.55 L
S

@rax 
130.54 139.54 329.47 253.51 @E
0 J 0 j [] 0 d 0 R 0 @G
0.00 0.00 0.00 1.00 K
0 1.728 1.728 0.000 @w
329.47 139.54 m
313.49 148.54 L
300.53 156.53 L
288.50 163.51 L
276.55 169.56 L
264.53 176.54 L
247.54 186.55 L
231.55 195.55 L
216.50 203.54 L
204.55 210.53 L
180.50 224.57 L
165.53 233.57 L
152.57 240.55 L
130.54 253.51 L
S

@rax 
324.43 151.56 329.47 152.57 @E
0 J 2 j [] 0 d 0 R 0 @G
0.00 0.00 0.00 1.00 K
0 1.008 1.008 0.000 @w
329.47 151.56 m
324.43 152.57 L
S

@rax 
318.46 153.36 319.46 153.65 @E
0 J 2 j [] 0 d 0 R 0 @G
0.00 0.00 0.00 1.00 K
0 1.008 1.008 0.000 @w
319.46 153.50 m
318.46 153.50 L
S

@rax 
308.45 154.51 313.49 155.52 @E
0 J 2 j [] 0 d 0 R 0 @G
0.00 0.00 0.00 1.00 K
0 1.008 1.008 0.000 @w
313.49 154.51 m
308.45 155.52 L
S

@rax 
302.54 155.52 303.55 156.53 @E
0 J 2 j [] 0 d 0 R 0 @G
0.00 0.00 0.00 1.00 K
0 1.008 1.008 0.000 @w
303.55 155.52 m
302.54 156.53 L
S

@rax 
292.54 156.53 297.50 157.54 @E
0 J 2 j [] 0 d 0 R 0 @G
0.00 0.00 0.00 1.00 K
0 1.008 1.008 0.000 @w
297.50 156.53 m
292.54 157.54 L
S

@rax 
286.56 158.40 287.57 158.69 @E
0 J 2 j [] 0 d 0 R 0 @G
0.00 0.00 0.00 1.00 K
0 1.008 1.008 0.000 @w
287.57 158.54 m
286.56 158.54 L
S

@rax 
276.55 159.41 281.52 159.70 @E
0 J 2 j [] 0 d 0 R 0 @G
0.00 0.00 0.00 1.00 K
0 1.008 1.008 0.000 @w
281.52 159.55 m
276.55 159.55 L
S

@rax 
270.50 160.42 271.51 160.70 @E
0 J 2 j [] 0 d 0 R 0 @G
0.00 0.00 0.00 1.00 K
0 1.008 1.008 0.000 @w
271.51 160.56 m
270.50 160.56 L
S

@rax 
260.57 161.57 266.54 162.50 @E
0 J 0 j [] 0 d 0 R 0 @G
0.00 0.00 0.00 1.00 K
0 1.008 1.008 0.000 @w
266.54 161.57 m
264.53 161.57 L
260.57 162.50 L
S

@rax 
254.52 162.36 255.53 162.65 @E
0 J 2 j [] 0 d 0 R 0 @G
0.00 0.00 0.00 1.00 K
0 1.008 1.008 0.000 @w
255.53 162.50 m
254.52 162.50 L
S

@rax 
244.51 162.50 249.55 163.51 @E
0 J 0 j [] 0 d 0 R 0 @G
0.00 0.00 0.00 1.00 K
0 1.008 1.008 0.000 @w
249.55 162.50 m
247.54 162.50 L
244.51 163.51 L
S

@rax 
239.54 163.51 240.55 164.52 @E
0 J 2 j [] 0 d 0 R 0 @G
0.00 0.00 0.00 1.00 K
0 1.008 1.008 0.000 @w
240.55 163.51 m
239.54 164.52 L
S

@rax 
228.53 164.38 234.50 164.66 @E
0 J 0 j [] 0 d 0 R 0 @G
0.00 0.00 0.00 1.00 K
0 1.008 1.008 0.000 @w
234.50 164.52 m
231.55 164.52 L
228.53 164.52 L
S

@rax 
222.55 164.38 223.56 164.66 @E
0 J 2 j [] 0 d 0 R 0 @G
0.00 0.00 0.00 1.00 K
0 1.008 1.008 0.000 @w
223.56 164.52 m
222.55 164.52 L
S

@rax 
212.54 164.38 217.51 164.66 @E
0 J 0 j [] 0 d 0 R 0 @G
0.00 0.00 0.00 1.00 K
0 1.008 1.008 0.000 @w
217.51 164.52 m
216.50 164.52 L
212.54 164.52 L
S

@rax 
206.57 165.38 207.50 165.67 @E
0 J 2 j [] 0 d 0 R 0 @G
0.00 0.00 0.00 1.00 K
0 1.008 1.008 0.000 @w
207.50 165.53 m
206.57 165.53 L
S

@rax 
197.57 165.53 202.54 166.54 @E
0 J 2 j [] 0 d 0 R 0 @G
0.00 0.00 0.00 1.00 K
0 1.008 1.008 0.000 @w
202.54 165.53 m
197.57 166.54 L
S

@rax 
190.51 166.39 191.52 166.68 @E
0 J 2 j [] 0 d 0 R 0 @G
0.00 0.00 0.00 1.00 K
0 1.008 1.008 0.000 @w
191.52 166.54 m
190.51 166.54 L
S

@rax 
180.50 167.40 185.54 167.69 @E
0 J 2 j [] 0 d 0 R 0 @G
0.00 0.00 0.00 1.00 K
0 1.008 1.008 0.000 @w
185.54 167.54 m
180.50 167.54 L
S

@rax 
175.39 167.40 175.68 167.69 @E
0 J 2 j [] 0 d 0 R 0 @G
0.00 0.00 0.00 1.00 K
0 1.008 1.008 0.000 @w
175.54 167.54 m
175.54 167.54 L
S

@rax 
165.53 167.40 170.57 167.69 @E
0 J 2 j [] 0 d 0 R 0 @G
0.00 0.00 0.00 1.00 K
0 1.008 1.008 0.000 @w
170.57 167.54 m
165.53 167.54 L
S

@rax 
158.54 168.41 159.55 168.70 @E
0 J 2 j [] 0 d 0 R 0 @G
0.00 0.00 0.00 1.00 K
0 1.008 1.008 0.000 @w
159.55 168.55 m
158.54 168.55 L
S

@rax 
148.54 168.55 153.50 169.56 @E
0 J 0 j [] 0 d 0 R 0 @G
0.00 0.00 0.00 1.00 K
0 1.008 1.008 0.000 @w
153.50 168.55 m
152.57 168.55 L
148.54 169.56 L
S

@rax 
143.57 169.42 144.50 169.70 @E
0 J 2 j [] 0 d 0 R 0 @G
0.00 0.00 0.00 1.00 K
0 1.008 1.008 0.000 @w
144.50 169.56 m
143.57 169.56 L
S

@rax 
133.56 169.56 138.53 170.57 @E
0 J 2 j [] 0 d 0 R 0 @G
0.00 0.00 0.00 1.00 K
0 1.008 1.008 0.000 @w
138.53 169.56 m
133.56 170.57 L
S

@rax 
92.52 87.41 98.57 87.70 @E
0 J 2 j [] 0 d 0 R 0 @G
0.00 0.00 0.00 1.00 K
0 1.008 1.008 0.000 @w
92.52 87.55 m
98.57 87.55 L
S

@rax 
92.52 125.42 98.57 125.71 @E
0 J 2 j [] 0 d 0 R 0 @G
0.00 0.00 0.00 1.00 K
0 1.008 1.008 0.000 @w
92.52 125.57 m
98.57 125.57 L
S

@rax 
92.52 162.36 98.57 162.65 @E
0 J 2 j [] 0 d 0 R 0 @G
0.00 0.00 0.00 1.00 K
0 1.008 1.008 0.000 @w
92.52 162.50 m
98.57 162.50 L
S

@rax 
92.52 200.38 98.57 200.66 @E
0 J 2 j [] 0 d 0 R 0 @G
0.00 0.00 0.00 1.00 K
0 1.008 1.008 0.000 @w
92.52 200.52 m
98.57 200.52 L
S

@rax 
92.52 237.38 98.57 237.67 @E
0 J 2 j [] 0 d 0 R 0 @G
0.00 0.00 0.00 1.00 K
0 1.008 1.008 0.000 @w
92.52 237.53 m
98.57 237.53 L
S

@rax 
92.52 275.40 98.57 275.69 @E
0 J 2 j [] 0 d 0 R 0 @G
0.00 0.00 0.00 1.00 K
0 1.008 1.008 0.000 @w
92.52 275.54 m
98.57 275.54 L
S

@rax 
92.52 98.42 95.54 98.71 @E
0 J 2 j [] 0 d 0 R 0 @G
0.00 0.00 0.00 1.00 K
0 1.008 1.008 0.000 @w
92.52 98.57 m
95.54 98.57 L
S

@rax 
92.52 105.41 95.54 105.70 @E
0 J 2 j [] 0 d 0 R 0 @G
0.00 0.00 0.00 1.00 K
0 1.008 1.008 0.000 @w
92.52 105.55 m
95.54 105.55 L
S

@rax 
92.52 110.38 95.54 110.66 @E
0 J 2 j [] 0 d 0 R 0 @G
0.00 0.00 0.00 1.00 K
0 1.008 1.008 0.000 @w
92.52 110.52 m
95.54 110.52 L
S

@rax 
92.52 113.40 95.54 113.69 @E
0 J 2 j [] 0 d 0 R 0 @G
0.00 0.00 0.00 1.00 K
0 1.008 1.008 0.000 @w
92.52 113.54 m
95.54 113.54 L
S

@rax 
92.52 116.42 95.54 116.71 @E
0 J 2 j [] 0 d 0 R 0 @G
0.00 0.00 0.00 1.00 K
0 1.008 1.008 0.000 @w
92.52 116.57 m
95.54 116.57 L
S

@rax 
92.52 119.38 95.54 119.66 @E
0 J 2 j [] 0 d 0 R 0 @G
0.00 0.00 0.00 1.00 K
0 1.008 1.008 0.000 @w
92.52 119.52 m
95.54 119.52 L
S

@rax 
92.52 121.39 95.54 121.68 @E
0 J 2 j [] 0 d 0 R 0 @G
0.00 0.00 0.00 1.00 K
0 1.008 1.008 0.000 @w
92.52 121.54 m
95.54 121.54 L
S

@rax 
92.52 123.41 95.54 123.70 @E
0 J 2 j [] 0 d 0 R 0 @G
0.00 0.00 0.00 1.00 K
0 1.008 1.008 0.000 @w
92.52 123.55 m
95.54 123.55 L
S

@rax 
92.52 136.37 95.54 136.66 @E
0 J 2 j [] 0 d 0 R 0 @G
0.00 0.00 0.00 1.00 K
0 1.008 1.008 0.000 @w
92.52 136.51 m
95.54 136.51 L
S

@rax 
92.52 143.42 95.54 143.71 @E
0 J 2 j [] 0 d 0 R 0 @G
0.00 0.00 0.00 1.00 K
0 1.008 1.008 0.000 @w
92.52 143.57 m
95.54 143.57 L
S

@rax 
92.52 147.38 95.54 147.67 @E
0 J 2 j [] 0 d 0 R 0 @G
0.00 0.00 0.00 1.00 K
0 1.008 1.008 0.000 @w
92.52 147.53 m
95.54 147.53 L
S

@rax 
92.52 151.42 95.54 151.70 @E
0 J 2 j [] 0 d 0 R 0 @G
0.00 0.00 0.00 1.00 K
0 1.008 1.008 0.000 @w
92.52 151.56 m
95.54 151.56 L
S

@rax 
92.52 154.37 95.54 154.66 @E
0 J 2 j [] 0 d 0 R 0 @G
0.00 0.00 0.00 1.00 K
0 1.008 1.008 0.000 @w
92.52 154.51 m
95.54 154.51 L
S

@rax 
92.52 156.38 95.54 156.67 @E
0 J 2 j [] 0 d 0 R 0 @G
0.00 0.00 0.00 1.00 K
0 1.008 1.008 0.000 @w
92.52 156.53 m
95.54 156.53 L
S

@rax 
92.52 158.40 95.54 158.69 @E
0 J 2 j [] 0 d 0 R 0 @G
0.00 0.00 0.00 1.00 K
0 1.008 1.008 0.000 @w
92.52 158.54 m
95.54 158.54 L
S

@rax 
92.52 160.42 95.54 160.70 @E
0 J 2 j [] 0 d 0 R 0 @G
0.00 0.00 0.00 1.00 K
0 1.008 1.008 0.000 @w
92.52 160.56 m
95.54 160.56 L
S

@rax 
92.52 174.38 95.54 174.67 @E
0 J 2 j [] 0 d 0 R 0 @G
0.00 0.00 0.00 1.00 K
0 1.008 1.008 0.000 @w
92.52 174.53 m
95.54 174.53 L
S

@rax 
92.52 180.36 95.54 180.65 @E
0 J 2 j [] 0 d 0 R 0 @G
0.00 0.00 0.00 1.00 K
0 1.008 1.008 0.000 @w
92.52 180.50 m
95.54 180.50 L
S

@rax 
92.52 185.40 95.54 185.69 @E
0 J 2 j [] 0 d 0 R 0 @G
0.00 0.00 0.00 1.00 K
0 1.008 1.008 0.000 @w
92.52 185.54 m
95.54 185.54 L
S

@rax 
92.52 188.42 95.54 188.71 @E
0 J 2 j [] 0 d 0 R 0 @G
0.00 0.00 0.00 1.00 K
0 1.008 1.008 0.000 @w
92.52 188.57 m
95.54 188.57 L
S

@rax 
92.52 191.38 95.54 191.66 @E
0 J 2 j [] 0 d 0 R 0 @G
0.00 0.00 0.00 1.00 K
0 1.008 1.008 0.000 @w
92.52 191.52 m
95.54 191.52 L
S

@rax 
92.52 194.40 95.54 194.69 @E
0 J 2 j [] 0 d 0 R 0 @G
0.00 0.00 0.00 1.00 K
0 1.008 1.008 0.000 @w
92.52 194.54 m
95.54 194.54 L
S

@rax 
92.52 196.42 95.54 196.70 @E
0 J 2 j [] 0 d 0 R 0 @G
0.00 0.00 0.00 1.00 K
0 1.008 1.008 0.000 @w
92.52 196.56 m
95.54 196.56 L
S

@rax 
92.52 198.36 95.54 198.65 @E
0 J 2 j [] 0 d 0 R 0 @G
0.00 0.00 0.00 1.00 K
0 1.008 1.008 0.000 @w
92.52 198.50 m
95.54 198.50 L
S

@rax 
92.52 211.39 95.54 211.68 @E
0 J 2 j [] 0 d 0 R 0 @G
0.00 0.00 0.00 1.00 K
0 1.008 1.008 0.000 @w
92.52 211.54 m
95.54 211.54 L
S

@rax 
92.52 218.38 95.54 218.66 @E
0 J 2 j [] 0 d 0 R 0 @G
0.00 0.00 0.00 1.00 K
0 1.008 1.008 0.000 @w
92.52 218.52 m
95.54 218.52 L
S

@rax 
92.52 222.41 95.54 222.70 @E
0 J 2 j [] 0 d 0 R 0 @G
0.00 0.00 0.00 1.00 K
0 1.008 1.008 0.000 @w
92.52 222.55 m
95.54 222.55 L
S

@rax 
92.52 226.37 95.54 226.66 @E
0 J 2 j [] 0 d 0 R 0 @G
0.00 0.00 0.00 1.00 K
0 1.008 1.008 0.000 @w
92.52 226.51 m
95.54 226.51 L
S

@rax 
92.52 229.39 95.54 229.68 @E
0 J 2 j [] 0 d 0 R 0 @G
0.00 0.00 0.00 1.00 K
0 1.008 1.008 0.000 @w
92.52 229.54 m
95.54 229.54 L
S

@rax 
92.52 231.41 95.54 231.70 @E
0 J 2 j [] 0 d 0 R 0 @G
0.00 0.00 0.00 1.00 K
0 1.008 1.008 0.000 @w
92.52 231.55 m
95.54 231.55 L
S

@rax 
92.52 234.36 95.54 234.65 @E
0 J 2 j [] 0 d 0 R 0 @G
0.00 0.00 0.00 1.00 K
0 1.008 1.008 0.000 @w
92.52 234.50 m
95.54 234.50 L
S

@rax 
92.52 236.38 95.54 236.66 @E
0 J 2 j [] 0 d 0 R 0 @G
0.00 0.00 0.00 1.00 K
0 1.008 1.008 0.000 @w
92.52 236.52 m
95.54 236.52 L
S

@rax 
92.52 249.41 95.54 249.70 @E
0 J 2 j [] 0 d 0 R 0 @G
0.00 0.00 0.00 1.00 K
0 1.008 1.008 0.000 @w
92.52 249.55 m
95.54 249.55 L
S

@rax 
92.52 255.38 95.54 255.67 @E
0 J 2 j [] 0 d 0 R 0 @G
0.00 0.00 0.00 1.00 K
0 1.008 1.008 0.000 @w
92.52 255.53 m
95.54 255.53 L
S

@rax 
92.52 260.42 95.54 260.71 @E
0 J 2 j [] 0 d 0 R 0 @G
0.00 0.00 0.00 1.00 K
0 1.008 1.008 0.000 @w
92.52 260.57 m
95.54 260.57 L
S

@rax 
92.52 264.38 95.54 264.67 @E
0 J 2 j [] 0 d 0 R 0 @G
0.00 0.00 0.00 1.00 K
0 1.008 1.008 0.000 @w
92.52 264.53 m
95.54 264.53 L
S

@rax 
92.52 267.41 95.54 267.70 @E
0 J 2 j [] 0 d 0 R 0 @G
0.00 0.00 0.00 1.00 K
0 1.008 1.008 0.000 @w
92.52 267.55 m
95.54 267.55 L
S

@rax 
92.52 269.42 95.54 269.71 @E
0 J 2 j [] 0 d 0 R 0 @G
0.00 0.00 0.00 1.00 K
0 1.008 1.008 0.000 @w
92.52 269.57 m
95.54 269.57 L
S

@rax 
92.52 271.37 95.54 271.66 @E
0 J 2 j [] 0 d 0 R 0 @G
0.00 0.00 0.00 1.00 K
0 1.008 1.008 0.000 @w
92.52 271.51 m
95.54 271.51 L
S

@rax 
92.52 273.38 95.54 273.67 @E
0 J 2 j [] 0 d 0 R 0 @G
0.00 0.00 0.00 1.00 K
0 1.008 1.008 0.000 @w
92.52 273.53 m
95.54 273.53 L
S

@rax 
350.50 87.41 356.47 87.70 @E
0 J 2 j [] 0 d 0 R 0 @G
0.00 0.00 0.00 1.00 K
0 1.008 1.008 0.000 @w
356.47 87.55 m
350.50 87.55 L
S

@rax 
350.50 125.42 356.47 125.71 @E
0 J 2 j [] 0 d 0 R 0 @G
0.00 0.00 0.00 1.00 K
0 1.008 1.008 0.000 @w
356.47 125.57 m
350.50 125.57 L
S

@rax 
350.50 162.36 356.47 162.65 @E
0 J 2 j [] 0 d 0 R 0 @G
0.00 0.00 0.00 1.00 K
0 1.008 1.008 0.000 @w
356.47 162.50 m
350.50 162.50 L
S

@rax 
350.50 200.38 356.47 200.66 @E
0 J 2 j [] 0 d 0 R 0 @G
0.00 0.00 0.00 1.00 K
0 1.008 1.008 0.000 @w
356.47 200.52 m
350.50 200.52 L
S

@rax 
350.50 237.38 356.47 237.67 @E
0 J 2 j [] 0 d 0 R 0 @G
0.00 0.00 0.00 1.00 K
0 1.008 1.008 0.000 @w
356.47 237.53 m
350.50 237.53 L
S

@rax 
350.50 275.40 356.47 275.69 @E
0 J 2 j [] 0 d 0 R 0 @G
0.00 0.00 0.00 1.00 K
0 1.008 1.008 0.000 @w
356.47 275.54 m
350.50 275.54 L
S

@rax 
353.45 98.42 356.47 98.71 @E
0 J 2 j [] 0 d 0 R 0 @G
0.00 0.00 0.00 1.00 K
0 1.008 1.008 0.000 @w
356.47 98.57 m
353.45 98.57 L
S

@rax 
353.45 105.41 356.47 105.70 @E
0 J 2 j [] 0 d 0 R 0 @G
0.00 0.00 0.00 1.00 K
0 1.008 1.008 0.000 @w
356.47 105.55 m
353.45 105.55 L
S

@rax 
353.45 110.38 356.47 110.66 @E
0 J 2 j [] 0 d 0 R 0 @G
0.00 0.00 0.00 1.00 K
0 1.008 1.008 0.000 @w
356.47 110.52 m
353.45 110.52 L
S

@rax 
353.45 113.40 356.47 113.69 @E
0 J 2 j [] 0 d 0 R 0 @G
0.00 0.00 0.00 1.00 K
0 1.008 1.008 0.000 @w
356.47 113.54 m
353.45 113.54 L
S

@rax 
353.45 116.42 356.47 116.71 @E
0 J 2 j [] 0 d 0 R 0 @G
0.00 0.00 0.00 1.00 K
0 1.008 1.008 0.000 @w
356.47 116.57 m
353.45 116.57 L
S

@rax 
353.45 119.38 356.47 119.66 @E
0 J 2 j [] 0 d 0 R 0 @G
0.00 0.00 0.00 1.00 K
0 1.008 1.008 0.000 @w
356.47 119.52 m
353.45 119.52 L
S

@rax 
353.45 121.39 356.47 121.68 @E
0 J 2 j [] 0 d 0 R 0 @G
0.00 0.00 0.00 1.00 K
0 1.008 1.008 0.000 @w
356.47 121.54 m
353.45 121.54 L
S

@rax 
353.45 123.41 356.47 123.70 @E
0 J 2 j [] 0 d 0 R 0 @G
0.00 0.00 0.00 1.00 K
0 1.008 1.008 0.000 @w
356.47 123.55 m
353.45 123.55 L
S

@rax 
353.45 136.37 356.47 136.66 @E
0 J 2 j [] 0 d 0 R 0 @G
0.00 0.00 0.00 1.00 K
0 1.008 1.008 0.000 @w
356.47 136.51 m
353.45 136.51 L
S

@rax 
353.45 143.42 356.47 143.71 @E
0 J 2 j [] 0 d 0 R 0 @G
0.00 0.00 0.00 1.00 K
0 1.008 1.008 0.000 @w
356.47 143.57 m
353.45 143.57 L
S

@rax 
353.45 147.38 356.47 147.67 @E
0 J 2 j [] 0 d 0 R 0 @G
0.00 0.00 0.00 1.00 K
0 1.008 1.008 0.000 @w
356.47 147.53 m
353.45 147.53 L
S

@rax 
353.45 151.42 356.47 151.70 @E
0 J 2 j [] 0 d 0 R 0 @G
0.00 0.00 0.00 1.00 K
0 1.008 1.008 0.000 @w
356.47 151.56 m
353.45 151.56 L
S

@rax 
353.45 154.37 356.47 154.66 @E
0 J 2 j [] 0 d 0 R 0 @G
0.00 0.00 0.00 1.00 K
0 1.008 1.008 0.000 @w
356.47 154.51 m
353.45 154.51 L
S

@rax 
353.45 156.38 356.47 156.67 @E
0 J 2 j [] 0 d 0 R 0 @G
0.00 0.00 0.00 1.00 K
0 1.008 1.008 0.000 @w
356.47 156.53 m
353.45 156.53 L
S

@rax 
353.45 158.40 356.47 158.69 @E
0 J 2 j [] 0 d 0 R 0 @G
0.00 0.00 0.00 1.00 K
0 1.008 1.008 0.000 @w
356.47 158.54 m
353.45 158.54 L
S

@rax 
353.45 160.42 356.47 160.70 @E
0 J 2 j [] 0 d 0 R 0 @G
0.00 0.00 0.00 1.00 K
0 1.008 1.008 0.000 @w
356.47 160.56 m
353.45 160.56 L
S

@rax 
353.45 174.38 356.47 174.67 @E
0 J 2 j [] 0 d 0 R 0 @G
0.00 0.00 0.00 1.00 K
0 1.008 1.008 0.000 @w
356.47 174.53 m
353.45 174.53 L
S

@rax 
353.45 180.36 356.47 180.65 @E
0 J 2 j [] 0 d 0 R 0 @G
0.00 0.00 0.00 1.00 K
0 1.008 1.008 0.000 @w
356.47 180.50 m
353.45 180.50 L
S

@rax 
353.45 185.40 356.47 185.69 @E
0 J 2 j [] 0 d 0 R 0 @G
0.00 0.00 0.00 1.00 K
0 1.008 1.008 0.000 @w
356.47 185.54 m
353.45 185.54 L
S

@rax 
353.45 188.42 356.47 188.71 @E
0 J 2 j [] 0 d 0 R 0 @G
0.00 0.00 0.00 1.00 K
0 1.008 1.008 0.000 @w
356.47 188.57 m
353.45 188.57 L
S

@rax 
353.45 191.38 356.47 191.66 @E
0 J 2 j [] 0 d 0 R 0 @G
0.00 0.00 0.00 1.00 K
0 1.008 1.008 0.000 @w
356.47 191.52 m
353.45 191.52 L
S

@rax 
353.45 194.40 356.47 194.69 @E
0 J 2 j [] 0 d 0 R 0 @G
0.00 0.00 0.00 1.00 K
0 1.008 1.008 0.000 @w
356.47 194.54 m
353.45 194.54 L
S

@rax 
353.45 196.42 356.47 196.70 @E
0 J 2 j [] 0 d 0 R 0 @G
0.00 0.00 0.00 1.00 K
0 1.008 1.008 0.000 @w
356.47 196.56 m
353.45 196.56 L
S

@rax 
353.45 198.36 356.47 198.65 @E
0 J 2 j [] 0 d 0 R 0 @G
0.00 0.00 0.00 1.00 K
0 1.008 1.008 0.000 @w
356.47 198.50 m
353.45 198.50 L
S

@rax 
353.45 211.39 356.47 211.68 @E
0 J 2 j [] 0 d 0 R 0 @G
0.00 0.00 0.00 1.00 K
0 1.008 1.008 0.000 @w
356.47 211.54 m
353.45 211.54 L
S

@rax 
353.45 218.38 356.47 218.66 @E
0 J 2 j [] 0 d 0 R 0 @G
0.00 0.00 0.00 1.00 K
0 1.008 1.008 0.000 @w
356.47 218.52 m
353.45 218.52 L
S

@rax 
353.45 222.41 356.47 222.70 @E
0 J 2 j [] 0 d 0 R 0 @G
0.00 0.00 0.00 1.00 K
0 1.008 1.008 0.000 @w
356.47 222.55 m
353.45 222.55 L
S

@rax 
353.45 226.37 356.47 226.66 @E
0 J 2 j [] 0 d 0 R 0 @G
0.00 0.00 0.00 1.00 K
0 1.008 1.008 0.000 @w
356.47 226.51 m
353.45 226.51 L
S

@rax 
353.45 229.39 356.47 229.68 @E
0 J 2 j [] 0 d 0 R 0 @G
0.00 0.00 0.00 1.00 K
0 1.008 1.008 0.000 @w
356.47 229.54 m
353.45 229.54 L
S

@rax 
353.45 231.41 356.47 231.70 @E
0 J 2 j [] 0 d 0 R 0 @G
0.00 0.00 0.00 1.00 K
0 1.008 1.008 0.000 @w
356.47 231.55 m
353.45 231.55 L
S

@rax 
353.45 234.36 356.47 234.65 @E
0 J 2 j [] 0 d 0 R 0 @G
0.00 0.00 0.00 1.00 K
0 1.008 1.008 0.000 @w
356.47 234.50 m
353.45 234.50 L
S

@rax 
353.45 236.38 356.47 236.66 @E
0 J 2 j [] 0 d 0 R 0 @G
0.00 0.00 0.00 1.00 K
0 1.008 1.008 0.000 @w
356.47 236.52 m
353.45 236.52 L
S

@rax 
353.45 249.41 356.47 249.70 @E
0 J 2 j [] 0 d 0 R 0 @G
0.00 0.00 0.00 1.00 K
0 1.008 1.008 0.000 @w
356.47 249.55 m
353.45 249.55 L
S

@rax 
353.45 255.38 356.47 255.67 @E
0 J 2 j [] 0 d 0 R 0 @G
0.00 0.00 0.00 1.00 K
0 1.008 1.008 0.000 @w
356.47 255.53 m
353.45 255.53 L
S

@rax 
353.45 260.42 356.47 260.71 @E
0 J 2 j [] 0 d 0 R 0 @G
0.00 0.00 0.00 1.00 K
0 1.008 1.008 0.000 @w
356.47 260.57 m
353.45 260.57 L
S

@rax 
353.45 264.38 356.47 264.67 @E
0 J 2 j [] 0 d 0 R 0 @G
0.00 0.00 0.00 1.00 K
0 1.008 1.008 0.000 @w
356.47 264.53 m
353.45 264.53 L
S

@rax 
353.45 267.41 356.47 267.70 @E
0 J 2 j [] 0 d 0 R 0 @G
0.00 0.00 0.00 1.00 K
0 1.008 1.008 0.000 @w
356.47 267.55 m
353.45 267.55 L
S

@rax 
353.45 269.42 356.47 269.71 @E
0 J 2 j [] 0 d 0 R 0 @G
0.00 0.00 0.00 1.00 K
0 1.008 1.008 0.000 @w
356.47 269.57 m
353.45 269.57 L
S

@rax 
353.45 271.37 356.47 271.66 @E
0 J 2 j [] 0 d 0 R 0 @G
0.00 0.00 0.00 1.00 K
0 1.008 1.008 0.000 @w
356.47 271.51 m
353.45 271.51 L
S

@rax 
353.45 273.38 356.47 273.67 @E
0 J 2 j [] 0 d 0 R 0 @G
0.00 0.00 0.00 1.00 K
0 1.008 1.008 0.000 @w
356.47 273.53 m
353.45 273.53 L
S

@rax 
336.31 87.55 336.60 93.53 @E
0 J 2 j [] 0 d 0 R 0 @G
0.00 0.00 0.00 1.00 K
0 1.008 1.008 0.000 @w
336.46 87.55 m
336.46 93.53 L
S

@rax 
270.36 87.55 270.65 93.53 @E
0 J 2 j [] 0 d 0 R 0 @G
0.00 0.00 0.00 1.00 K
0 1.008 1.008 0.000 @w
270.50 87.55 m
270.50 93.53 L
S

@rax 
204.41 87.55 204.70 93.53 @E
0 J 2 j [] 0 d 0 R 0 @G
0.00 0.00 0.00 1.00 K
0 1.008 1.008 0.000 @w
204.55 87.55 m
204.55 93.53 L
S

@rax 
138.38 87.55 138.67 93.53 @E
0 J 2 j [] 0 d 0 R 0 @G
0.00 0.00 0.00 1.00 K
0 1.008 1.008 0.000 @w
138.53 87.55 m
138.53 93.53 L
S

@rax 
356.33 87.55 356.62 90.50 @E
0 J 2 j [] 0 d 0 R 0 @G
0.00 0.00 0.00 1.00 K
0 1.008 1.008 0.000 @w
356.47 87.55 m
356.47 90.50 L
S

@rax 
351.29 87.55 351.58 90.50 @E
0 J 2 j [] 0 d 0 R 0 @G
0.00 0.00 0.00 1.00 K
0 1.008 1.008 0.000 @w
351.43 87.55 m
351.43 90.50 L
S

@rax 
346.32 87.55 346.61 90.50 @E
0 J 2 j [] 0 d 0 R 0 @G
0.00 0.00 0.00 1.00 K
0 1.008 1.008 0.000 @w
346.46 87.55 m
346.46 90.50 L
S

@rax 
343.30 87.55 343.58 90.50 @E
0 J 2 j [] 0 d 0 R 0 @G
0.00 0.00 0.00 1.00 K
0 1.008 1.008 0.000 @w
343.44 87.55 m
343.44 90.50 L
S

@rax 
339.34 87.55 339.62 90.50 @E
0 J 2 j [] 0 d 0 R 0 @G
0.00 0.00 0.00 1.00 K
0 1.008 1.008 0.000 @w
339.48 87.55 m
339.48 90.50 L
S

@rax 
316.30 87.55 316.58 90.50 @E
0 J 2 j [] 0 d 0 R 0 @G
0.00 0.00 0.00 1.00 K
0 1.008 1.008 0.000 @w
316.44 87.55 m
316.44 90.50 L
S

@rax 
305.42 87.55 305.71 90.50 @E
0 J 2 j [] 0 d 0 R 0 @G
0.00 0.00 0.00 1.00 K
0 1.008 1.008 0.000 @w
305.57 87.55 m
305.57 90.50 L
S

@rax 
296.42 87.55 296.71 90.50 @E
0 J 2 j [] 0 d 0 R 0 @G
0.00 0.00 0.00 1.00 K
0 1.008 1.008 0.000 @w
296.57 87.55 m
296.57 90.50 L
S

@rax 
290.38 87.55 290.66 90.50 @E
0 J 2 j [] 0 d 0 R 0 @G
0.00 0.00 0.00 1.00 K
0 1.008 1.008 0.000 @w
290.52 87.55 m
290.52 90.50 L
S

@rax 
285.41 87.55 285.70 90.50 @E
0 J 2 j [] 0 d 0 R 0 @G
0.00 0.00 0.00 1.00 K
0 1.008 1.008 0.000 @w
285.55 87.55 m
285.55 90.50 L
S

@rax 
280.37 87.55 280.66 90.50 @E
0 J 2 j [] 0 d 0 R 0 @G
0.00 0.00 0.00 1.00 K
0 1.008 1.008 0.000 @w
280.51 87.55 m
280.51 90.50 L
S

@rax 
277.42 87.55 277.70 90.50 @E
0 J 2 j [] 0 d 0 R 0 @G
0.00 0.00 0.00 1.00 K
0 1.008 1.008 0.000 @w
277.56 87.55 m
277.56 90.50 L
S

@rax 
273.38 87.55 273.67 90.50 @E
0 J 2 j [] 0 d 0 R 0 @G
0.00 0.00 0.00 1.00 K
0 1.008 1.008 0.000 @w
273.53 87.55 m
273.53 90.50 L
S

@rax 
250.42 87.55 250.70 90.50 @E
0 J 2 j [] 0 d 0 R 0 @G
0.00 0.00 0.00 1.00 K
0 1.008 1.008 0.000 @w
250.56 87.55 m
250.56 90.50 L
S

@rax 
239.40 87.55 239.69 90.50 @E
0 J 2 j [] 0 d 0 R 0 @G
0.00 0.00 0.00 1.00 K
0 1.008 1.008 0.000 @w
239.54 87.55 m
239.54 90.50 L
S

@rax 
230.40 87.55 230.69 90.50 @E
0 J 2 j [] 0 d 0 R 0 @G
0.00 0.00 0.00 1.00 K
0 1.008 1.008 0.000 @w
230.54 87.55 m
230.54 90.50 L
S

@rax 
224.42 87.55 224.71 90.50 @E
0 J 2 j [] 0 d 0 R 0 @G
0.00 0.00 0.00 1.00 K
0 1.008 1.008 0.000 @w
224.57 87.55 m
224.57 90.50 L
S

@rax 
219.38 87.55 219.67 90.50 @E
0 J 2 j [] 0 d 0 R 0 @G
0.00 0.00 0.00 1.00 K
0 1.008 1.008 0.000 @w
219.53 87.55 m
219.53 90.50 L
S

@rax 
214.42 87.55 214.70 90.50 @E
0 J 2 j [] 0 d 0 R 0 @G
0.00 0.00 0.00 1.00 K
0 1.008 1.008 0.000 @w
214.56 87.55 m
214.56 90.50 L
S

@rax 
210.38 87.55 210.67 90.50 @E
0 J 2 j [] 0 d 0 R 0 @G
0.00 0.00 0.00 1.00 K
0 1.008 1.008 0.000 @w
210.53 87.55 m
210.53 90.50 L
S

@rax 
207.36 87.55 207.65 90.50 @E
0 J 2 j [] 0 d 0 R 0 @G
0.00 0.00 0.00 1.00 K
0 1.008 1.008 0.000 @w
207.50 87.55 m
207.50 90.50 L
S

@rax 
184.39 87.55 184.68 90.50 @E
0 J 2 j [] 0 d 0 R 0 @G
0.00 0.00 0.00 1.00 K
0 1.008 1.008 0.000 @w
184.54 87.55 m
184.54 90.50 L
S

@rax 
173.38 87.55 173.66 90.50 @E
0 J 2 j [] 0 d 0 R 0 @G
0.00 0.00 0.00 1.00 K
0 1.008 1.008 0.000 @w
173.52 87.55 m
173.52 90.50 L
S

@rax 
164.38 87.55 164.66 90.50 @E
0 J 2 j [] 0 d 0 R 0 @G
0.00 0.00 0.00 1.00 K
0 1.008 1.008 0.000 @w
164.52 87.55 m
164.52 90.50 L
S

@rax 
158.40 87.55 158.69 90.50 @E
0 J 2 j [] 0 d 0 R 0 @G
0.00 0.00 0.00 1.00 K
0 1.008 1.008 0.000 @w
158.54 87.55 m
158.54 90.50 L
S

@rax 
153.36 87.55 153.65 90.50 @E
0 J 2 j [] 0 d 0 R 0 @G
0.00 0.00 0.00 1.00 K
0 1.008 1.008 0.000 @w
153.50 87.55 m
153.50 90.50 L
S

@rax 
148.39 87.55 148.68 90.50 @E
0 J 2 j [] 0 d 0 R 0 @G
0.00 0.00 0.00 1.00 K
0 1.008 1.008 0.000 @w
148.54 87.55 m
148.54 90.50 L
S

@rax 
144.36 87.55 144.65 90.50 @E
0 J 2 j [] 0 d 0 R 0 @G
0.00 0.00 0.00 1.00 K
0 1.008 1.008 0.000 @w
144.50 87.55 m
144.50 90.50 L
S

@rax 
141.41 87.55 141.70 90.50 @E
0 J 2 j [] 0 d 0 R 0 @G
0.00 0.00 0.00 1.00 K
0 1.008 1.008 0.000 @w
141.55 87.55 m
141.55 90.50 L
S

@rax 
118.37 87.55 118.66 90.50 @E
0 J 2 j [] 0 d 0 R 0 @G
0.00 0.00 0.00 1.00 K
0 1.008 1.008 0.000 @w
118.51 87.55 m
118.51 90.50 L
S

@rax 
107.42 87.55 107.71 90.50 @E
0 J 2 j [] 0 d 0 R 0 @G
0.00 0.00 0.00 1.00 K
0 1.008 1.008 0.000 @w
107.57 87.55 m
107.57 90.50 L
S

@rax 
98.42 87.55 98.71 90.50 @E
0 J 2 j [] 0 d 0 R 0 @G
0.00 0.00 0.00 1.00 K
0 1.008 1.008 0.000 @w
98.57 87.55 m
98.57 90.50 L
S

@rax 
336.31 269.57 336.60 275.54 @E
0 J 2 j [] 0 d 0 R 0 @G
0.00 0.00 0.00 1.00 K
0 1.008 1.008 0.000 @w
336.46 275.54 m
336.46 269.57 L
S

@rax 
270.36 269.57 270.65 275.54 @E
0 J 2 j [] 0 d 0 R 0 @G
0.00 0.00 0.00 1.00 K
0 1.008 1.008 0.000 @w
270.50 275.54 m
270.50 269.57 L
S

@rax 
204.41 269.57 204.70 275.54 @E
0 J 2 j [] 0 d 0 R 0 @G
0.00 0.00 0.00 1.00 K
0 1.008 1.008 0.000 @w
204.55 275.54 m
204.55 269.57 L
S

@rax 
138.38 269.57 138.67 275.54 @E
0 J 2 j [] 0 d 0 R 0 @G
0.00 0.00 0.00 1.00 K
0 1.008 1.008 0.000 @w
138.53 275.54 m
138.53 269.57 L
S

@rax 
356.33 272.52 356.62 275.54 @E
0 J 2 j [] 0 d 0 R 0 @G
0.00 0.00 0.00 1.00 K
0 1.008 1.008 0.000 @w
356.47 275.54 m
356.47 272.52 L
S

@rax 
351.29 272.52 351.58 275.54 @E
0 J 2 j [] 0 d 0 R 0 @G
0.00 0.00 0.00 1.00 K
0 1.008 1.008 0.000 @w
351.43 275.54 m
351.43 272.52 L
S

@rax 
346.32 272.52 346.61 275.54 @E
0 J 2 j [] 0 d 0 R 0 @G
0.00 0.00 0.00 1.00 K
0 1.008 1.008 0.000 @w
346.46 275.54 m
346.46 272.52 L
S

@rax 
343.30 272.52 343.58 275.54 @E
0 J 2 j [] 0 d 0 R 0 @G
0.00 0.00 0.00 1.00 K
0 1.008 1.008 0.000 @w
343.44 275.54 m
343.44 272.52 L
S

@rax 
339.34 272.52 339.62 275.54 @E
0 J 2 j [] 0 d 0 R 0 @G
0.00 0.00 0.00 1.00 K
0 1.008 1.008 0.000 @w
339.48 275.54 m
339.48 272.52 L
S

@rax 
316.30 272.52 316.58 275.54 @E
0 J 2 j [] 0 d 0 R 0 @G
0.00 0.00 0.00 1.00 K
0 1.008 1.008 0.000 @w
316.44 275.54 m
316.44 272.52 L
S

@rax 
305.42 272.52 305.71 275.54 @E
0 J 2 j [] 0 d 0 R 0 @G
0.00 0.00 0.00 1.00 K
0 1.008 1.008 0.000 @w
305.57 275.54 m
305.57 272.52 L
S

@rax 
296.42 272.52 296.71 275.54 @E
0 J 2 j [] 0 d 0 R 0 @G
0.00 0.00 0.00 1.00 K
0 1.008 1.008 0.000 @w
296.57 275.54 m
296.57 272.52 L
S

@rax 
290.38 272.52 290.66 275.54 @E
0 J 2 j [] 0 d 0 R 0 @G
0.00 0.00 0.00 1.00 K
0 1.008 1.008 0.000 @w
290.52 275.54 m
290.52 272.52 L
S

@rax 
285.41 272.52 285.70 275.54 @E
0 J 2 j [] 0 d 0 R 0 @G
0.00 0.00 0.00 1.00 K
0 1.008 1.008 0.000 @w
285.55 275.54 m
285.55 272.52 L
S

@rax 
280.37 272.52 280.66 275.54 @E
0 J 2 j [] 0 d 0 R 0 @G
0.00 0.00 0.00 1.00 K
0 1.008 1.008 0.000 @w
280.51 275.54 m
280.51 272.52 L
S

@rax 
277.42 272.52 277.70 275.54 @E
0 J 2 j [] 0 d 0 R 0 @G
0.00 0.00 0.00 1.00 K
0 1.008 1.008 0.000 @w
277.56 275.54 m
277.56 272.52 L
S

@rax 
273.38 272.52 273.67 275.54 @E
0 J 2 j [] 0 d 0 R 0 @G
0.00 0.00 0.00 1.00 K
0 1.008 1.008 0.000 @w
273.53 275.54 m
273.53 272.52 L
S

@rax 
250.42 272.52 250.70 275.54 @E
0 J 2 j [] 0 d 0 R 0 @G
0.00 0.00 0.00 1.00 K
0 1.008 1.008 0.000 @w
250.56 275.54 m
250.56 272.52 L
S

@rax 
239.40 272.52 239.69 275.54 @E
0 J 2 j [] 0 d 0 R 0 @G
0.00 0.00 0.00 1.00 K
0 1.008 1.008 0.000 @w
239.54 275.54 m
239.54 272.52 L
S

@rax 
230.40 272.52 230.69 275.54 @E
0 J 2 j [] 0 d 0 R 0 @G
0.00 0.00 0.00 1.00 K
0 1.008 1.008 0.000 @w
230.54 275.54 m
230.54 272.52 L
S

@rax 
224.42 272.52 224.71 275.54 @E
0 J 2 j [] 0 d 0 R 0 @G
0.00 0.00 0.00 1.00 K
0 1.008 1.008 0.000 @w
224.57 275.54 m
224.57 272.52 L
S

@rax 
219.38 272.52 219.67 275.54 @E
0 J 2 j [] 0 d 0 R 0 @G
0.00 0.00 0.00 1.00 K
0 1.008 1.008 0.000 @w
219.53 275.54 m
219.53 272.52 L
S

@rax 
214.42 272.52 214.70 275.54 @E
0 J 2 j [] 0 d 0 R 0 @G
0.00 0.00 0.00 1.00 K
0 1.008 1.008 0.000 @w
214.56 275.54 m
214.56 272.52 L
S

@rax 
210.38 272.52 210.67 275.54 @E
0 J 2 j [] 0 d 0 R 0 @G
0.00 0.00 0.00 1.00 K
0 1.008 1.008 0.000 @w
210.53 275.54 m
210.53 272.52 L
S

@rax 
207.36 272.52 207.65 275.54 @E
0 J 2 j [] 0 d 0 R 0 @G
0.00 0.00 0.00 1.00 K
0 1.008 1.008 0.000 @w
207.50 275.54 m
207.50 272.52 L
S

@rax 
184.39 272.52 184.68 275.54 @E
0 J 2 j [] 0 d 0 R 0 @G
0.00 0.00 0.00 1.00 K
0 1.008 1.008 0.000 @w
184.54 275.54 m
184.54 272.52 L
S

@rax 
173.38 272.52 173.66 275.54 @E
0 J 2 j [] 0 d 0 R 0 @G
0.00 0.00 0.00 1.00 K
0 1.008 1.008 0.000 @w
173.52 275.54 m
173.52 272.52 L
S

@rax 
164.38 272.52 164.66 275.54 @E
0 J 2 j [] 0 d 0 R 0 @G
0.00 0.00 0.00 1.00 K
0 1.008 1.008 0.000 @w
164.52 275.54 m
164.52 272.52 L
S

@rax 
158.40 272.52 158.69 275.54 @E
0 J 2 j [] 0 d 0 R 0 @G
0.00 0.00 0.00 1.00 K
0 1.008 1.008 0.000 @w
158.54 275.54 m
158.54 272.52 L
S

@rax 
153.36 272.52 153.65 275.54 @E
0 J 2 j [] 0 d 0 R 0 @G
0.00 0.00 0.00 1.00 K
0 1.008 1.008 0.000 @w
153.50 275.54 m
153.50 272.52 L
S

@rax 
148.39 272.52 148.68 275.54 @E
0 J 2 j [] 0 d 0 R 0 @G
0.00 0.00 0.00 1.00 K
0 1.008 1.008 0.000 @w
148.54 275.54 m
148.54 272.52 L
S

@rax 
144.36 272.52 144.65 275.54 @E
0 J 2 j [] 0 d 0 R 0 @G
0.00 0.00 0.00 1.00 K
0 1.008 1.008 0.000 @w
144.50 275.54 m
144.50 272.52 L
S

@rax 
141.41 272.52 141.70 275.54 @E
0 J 2 j [] 0 d 0 R 0 @G
0.00 0.00 0.00 1.00 K
0 1.008 1.008 0.000 @w
141.55 275.54 m
141.55 272.52 L
S

@rax 
118.37 272.52 118.66 275.54 @E
0 J 2 j [] 0 d 0 R 0 @G
0.00 0.00 0.00 1.00 K
0 1.008 1.008 0.000 @w
118.51 275.54 m
118.51 272.52 L
S

@rax 
107.42 272.52 107.71 275.54 @E
0 J 2 j [] 0 d 0 R 0 @G
0.00 0.00 0.00 1.00 K
0 1.008 1.008 0.000 @w
107.57 275.54 m
107.57 272.52 L
S

@rax 
98.42 272.52 98.71 275.54 @E
0 J 2 j [] 0 d 0 R 0 @G
0.00 0.00 0.00 1.00 K
0 1.008 1.008 0.000 @w
98.57 275.54 m
98.57 272.52 L
S

@rax 
92.38 87.55 92.66 275.54 @E
0 J 2 j [] 0 d 0 R 0 @G
0.00 0.00 0.00 1.00 K
0 1.440 1.440 0.000 @w
92.52 87.55 m
92.52 275.54 L
S

@rax 
356.33 87.55 356.62 275.54 @E
0 J 2 j [] 0 d 0 R 0 @G
0.00 0.00 0.00 1.00 K
0 1.440 1.440 0.000 @w
356.47 87.55 m
356.47 275.54 L
S

@rax 
92.52 87.41 356.47 87.70 @E
0 J 2 j [] 0 d 0 R 0 @G
0.00 0.00 0.00 1.00 K
0 1.440 1.440 0.000 @w
92.52 87.55 m
356.47 87.55 L
S

@rax 
92.52 275.40 356.47 275.69 @E
0 J 2 j [] 0 d 0 R 0 @G
0.00 0.00 0.00 1.00 K
0 1.440 1.440 0.000 @w
92.52 275.54 m
356.47 275.54 L
S

@rax 
327.46 180.50 330.48 183.53 @E
0 J 0 j [] 0 d 0 R 0 @G
0.00 0.00 0.00 1.00 K
0 1.008 1.008 0.000 @w
330.48 182.02 m
330.48 181.22 329.76 180.50 328.97 180.50 c
328.18 180.50 327.46 181.22 327.46 182.02 c
327.46 182.88 328.18 183.53 328.97 183.53 c
329.76 183.53 330.48 182.88 330.48 182.02 c
@c
S

@rax 
311.47 181.51 314.50 184.54 @E
0 J 0 j [] 0 d 0 R 0 @G
0.00 0.00 0.00 1.00 K
0 1.008 1.008 0.000 @w
314.50 183.02 m
314.50 182.23 313.78 181.51 312.98 181.51 c
312.12 181.51 311.47 182.23 311.47 183.02 c
311.47 183.89 312.12 184.54 312.98 184.54 c
313.78 184.54 314.50 183.89 314.50 183.02 c
@c
S

@rax 
298.51 183.53 301.54 186.55 @E
0 J 0 j [] 0 d 0 R 0 @G
0.00 0.00 0.00 1.00 K
0 1.008 1.008 0.000 @w
301.54 185.04 m
301.54 184.25 300.89 183.53 300.02 183.53 c
299.23 183.53 298.51 184.25 298.51 185.04 c
298.51 185.83 299.23 186.55 300.02 186.55 c
300.89 186.55 301.54 185.83 301.54 185.04 c
@c
S

@rax 
286.56 181.51 289.51 184.54 @E
0 J 0 j [] 0 d 0 R 0 @G
0.00 0.00 0.00 1.00 K
0 1.008 1.008 0.000 @w
289.51 183.02 m
289.51 182.23 288.86 181.51 288.00 181.51 c
287.21 181.51 286.56 182.23 286.56 183.02 c
286.56 183.89 287.21 184.54 288.00 184.54 c
288.86 184.54 289.51 183.89 289.51 183.02 c
@c
S

@rax 
274.54 182.52 277.56 185.54 @E
0 J 0 j [] 0 d 0 R 0 @G
0.00 0.00 0.00 1.00 K
0 1.008 1.008 0.000 @w
277.56 184.03 m
277.56 183.24 276.84 182.52 276.05 182.52 c
275.18 182.52 274.54 183.24 274.54 184.03 c
274.54 184.82 275.18 185.54 276.05 185.54 c
276.84 185.54 277.56 184.82 277.56 184.03 c
@c
S

@rax 
262.51 186.55 265.54 189.50 @E
0 J 0 j [] 0 d 0 R 0 @G
0.00 0.00 0.00 1.00 K
0 1.008 1.008 0.000 @w
265.54 188.06 m
265.54 187.20 264.89 186.55 264.02 186.55 c
263.23 186.55 262.51 187.20 262.51 188.06 c
262.51 188.86 263.23 189.50 264.02 189.50 c
264.89 189.50 265.54 188.86 265.54 188.06 c
@c
S

@rax 
245.52 188.57 248.54 191.52 @E
0 J 0 j [] 0 d 0 R 0 @G
0.00 0.00 0.00 1.00 K
0 1.008 1.008 0.000 @w
248.54 190.01 m
248.54 189.22 247.82 188.57 247.03 188.57 c
246.24 188.57 245.52 189.22 245.52 190.01 c
245.52 190.87 246.24 191.52 247.03 191.52 c
247.82 191.52 248.54 190.87 248.54 190.01 c
@c
S

@rax 
229.54 189.50 232.56 192.53 @E
0 J 0 j [] 0 d 0 R 0 @G
0.00 0.00 0.00 1.00 K
0 1.008 1.008 0.000 @w
232.56 191.02 m
232.56 190.22 231.84 189.50 231.05 189.50 c
230.18 189.50 229.54 190.22 229.54 191.02 c
229.54 191.88 230.18 192.53 231.05 192.53 c
231.84 192.53 232.56 191.88 232.56 191.02 c
@c
S

@rax 
214.56 190.51 217.51 193.54 @E
0 J 0 j [] 0 d 0 R 0 @G
0.00 0.00 0.00 1.00 K
0 1.008 1.008 0.000 @w
217.51 192.02 m
217.51 191.23 216.86 190.51 216.00 190.51 c
215.21 190.51 214.56 191.23 214.56 192.02 c
214.56 192.89 215.21 193.54 216.00 193.54 c
216.86 193.54 217.51 192.89 217.51 192.02 c
@c
S

@rax 
202.54 193.54 205.56 196.56 @E
0 J 0 j [] 0 d 0 R 0 @G
0.00 0.00 0.00 1.00 K
0 1.008 1.008 0.000 @w
205.56 195.05 m
205.56 194.18 204.84 193.54 204.05 193.54 c
203.18 193.54 202.54 194.18 202.54 195.05 c
202.54 195.84 203.18 196.56 204.05 196.56 c
204.84 196.56 205.56 195.84 205.56 195.05 c
@c
S

@rax 
178.56 193.54 181.51 196.56 @E
0 J 0 j [] 0 d 0 R 0 @G
0.00 0.00 0.00 1.00 K
0 1.008 1.008 0.000 @w
181.51 195.05 m
181.51 194.18 180.86 193.54 180.00 193.54 c
179.21 193.54 178.56 194.18 178.56 195.05 c
178.56 195.84 179.21 196.56 180.00 196.56 c
180.86 196.56 181.51 195.84 181.51 195.05 c
@c
S

@rax 
163.51 194.54 166.54 197.57 @E
0 J 0 j [] 0 d 0 R 0 @G
0.00 0.00 0.00 1.00 K
0 1.008 1.008 0.000 @w
166.54 196.06 m
166.54 195.19 165.89 194.54 165.02 194.54 c
164.23 194.54 163.51 195.19 163.51 196.06 c
163.51 196.85 164.23 197.57 165.02 197.57 c
165.89 197.57 166.54 196.85 166.54 196.06 c
@c
S

@rax 
150.55 192.53 153.50 195.55 @E
0 J 0 j [] 0 d 0 R 0 @G
0.00 0.00 0.00 1.00 K
0 1.008 1.008 0.000 @w
153.50 194.04 m
153.50 193.25 152.86 192.53 152.06 192.53 c
151.20 192.53 150.55 193.25 150.55 194.04 c
150.55 194.83 151.20 195.55 152.06 195.55 c
152.86 195.55 153.50 194.83 153.50 194.04 c
@c
S

@rax 
128.52 193.54 131.54 196.56 @E
0 J 0 j [] 0 d 0 R 0 @G
0.00 0.00 0.00 1.00 K
0 1.008 1.008 0.000 @w
131.54 195.05 m
131.54 194.18 130.82 193.54 130.03 193.54 c
129.24 193.54 128.52 194.18 128.52 195.05 c
128.52 195.84 129.24 196.56 130.03 196.56 c
130.82 196.56 131.54 195.84 131.54 195.05 c
@c
S

@rax 70.56 84.74 83.38 93.46 @E
[0.07200 0.00000 0.00000 0.07200 70.56000 85.03200] @tm
 0 O 0 @g
0.00 0.00 0.00 1.00 k
e
/_R44-Helvetica 166.00 z
0 0 (10) @t
T
@rax 70.56 84.74 83.38 93.46 @E
[0.07200 0.00000 0.00000 0.07200 70.56000 85.03200] @tm
0 J 0 j [] 0 d 0 R 0 @G
0.00 0.00 0.00 1.00 K
0 0.216 0.216 0.000 @w
r
/_R44-Helvetica 166.00 z
0 0 (10) @t
T
@rax 84.53 91.80 88.63 97.56 @E
[0.07200 0.00000 0.00000 0.07200 84.52800 92.01600] @tm
 0 O 0 @g
0.00 0.00 0.00 1.00 k
e
/_R44-Helvetica 110.00 z
0 0 (3) @t
T
@rax 84.53 91.80 88.63 97.56 @E
[0.07200 0.00000 0.00000 0.07200 84.52800 92.01600] @tm
0 J 0 j [] 0 d 0 R 0 @G
0.00 0.00 0.00 1.00 K
0 0.216 0.216 0.000 @w
r
/_R44-Helvetica 110.00 z
0 0 (3) @t
T
@rax 70.56 122.76 83.38 131.47 @E
[0.07200 0.00000 0.00000 0.07200 70.56000 123.04799] @tm
 0 O 0 @g
0.00 0.00 0.00 1.00 k
e
/_R44-Helvetica 166.00 z
0 0 (10) @t
T
@rax 70.56 122.76 83.38 131.47 @E
[0.07200 0.00000 0.00000 0.07200 70.56000 123.04799] @tm
0 J 0 j [] 0 d 0 R 0 @G
0.00 0.00 0.00 1.00 K
0 0.216 0.216 0.000 @w
r
/_R44-Helvetica 166.00 z
0 0 (10) @t
T
@rax 84.53 129.02 88.70 134.57 @E
[0.07200 0.00000 0.00000 0.07200 84.52800 129.02399] @tm
 0 O 0 @g
0.00 0.00 0.00 1.00 k
e
/_R44-Helvetica 110.00 z
0 0 (4) @t
T
@rax 84.53 129.02 88.70 134.57 @E
[0.07200 0.00000 0.00000 0.07200 84.52800 129.02399] @tm
0 J 0 j [] 0 d 0 R 0 @G
0.00 0.00 0.00 1.00 K
0 0.216 0.216 0.000 @w
r
/_R44-Helvetica 110.00 z
0 0 (4) @t
T
@rax 70.56 160.78 83.38 169.49 @E
[0.07200 0.00000 0.00000 0.07200 70.56000 161.06399] @tm
 0 O 0 @g
0.00 0.00 0.00 1.00 k
e
/_R44-Helvetica 166.00 z
0 0 (10) @t
T
@rax 70.56 160.78 83.38 169.49 @E
[0.07200 0.00000 0.00000 0.07200 70.56000 161.06399] @tm
0 J 0 j [] 0 d 0 R 0 @G
0.00 0.00 0.00 1.00 K
0 0.216 0.216 0.000 @w
r
/_R44-Helvetica 166.00 z
0 0 (10) @t
T
@rax 84.53 166.82 88.63 172.51 @E
[0.07200 0.00000 0.00000 0.07200 84.52800 167.03999] @tm
 0 O 0 @g
0.00 0.00 0.00 1.00 k
e
/_R44-Helvetica 110.00 z
0 0 (5) @t
T
@rax 84.53 166.82 88.63 172.51 @E
[0.07200 0.00000 0.00000 0.07200 84.52800 167.03999] @tm
0 J 0 j [] 0 d 0 R 0 @G
0.00 0.00 0.00 1.00 K
0 0.216 0.216 0.000 @w
r
/_R44-Helvetica 110.00 z
0 0 (5) @t
T
@rax 70.56 197.71 83.38 206.42 @E
[0.07200 0.00000 0.00000 0.07200 70.56000 197.99999] @tm
 0 O 0 @g
0.00 0.00 0.00 1.00 k
e
/_R44-Helvetica 166.00 z
0 0 (10) @t
T
@rax 70.56 197.71 83.38 206.42 @E
[0.07200 0.00000 0.00000 0.07200 70.56000 197.99999] @tm
0 J 0 j [] 0 d 0 R 0 @G
0.00 0.00 0.00 1.00 K
0 0.216 0.216 0.000 @w
r
/_R44-Helvetica 166.00 z
0 0 (10) @t
T
@rax 84.53 203.83 88.63 209.59 @E
[0.07200 0.00000 0.00000 0.07200 84.52800 204.04799] @tm
 0 O 0 @g
0.00 0.00 0.00 1.00 k
e
/_R44-Helvetica 110.00 z
0 0 (6) @t
T
@rax 84.53 203.83 88.63 209.59 @E
[0.07200 0.00000 0.00000 0.07200 84.52800 204.04799] @tm
0 J 0 j [] 0 d 0 R 0 @G
0.00 0.00 0.00 1.00 K
0 0.216 0.216 0.000 @w
r
/_R44-Helvetica 110.00 z
0 0 (6) @t
T
@rax 70.56 235.73 83.38 244.44 @E
[0.07200 0.00000 0.00000 0.07200 70.56000 236.01599] @tm
 0 O 0 @g
0.00 0.00 0.00 1.00 k
e
/_R44-Helvetica 166.00 z
0 0 (10) @t
T
@rax 70.56 235.73 83.38 244.44 @E
[0.07200 0.00000 0.00000 0.07200 70.56000 236.01599] @tm
0 J 0 j [] 0 d 0 R 0 @G
0.00 0.00 0.00 1.00 K
0 0.216 0.216 0.000 @w
r
/_R44-Helvetica 166.00 z
0 0 (10) @t
T
@rax 84.53 242.06 88.70 247.54 @E
[0.07200 0.00000 0.00000 0.07200 84.52800 242.06399] @tm
 0 O 0 @g
0.00 0.00 0.00 1.00 k
e
/_R44-Helvetica 110.00 z
0 0 (7) @t
T
@rax 84.53 242.06 88.70 247.54 @E
[0.07200 0.00000 0.00000 0.07200 84.52800 242.06399] @tm
0 J 0 j [] 0 d 0 R 0 @G
0.00 0.00 0.00 1.00 K
0 0.216 0.216 0.000 @w
r
/_R44-Helvetica 110.00 z
0 0 (7) @t
T
@rax 70.56 272.74 83.38 281.45 @E
[0.07200 0.00000 0.00000 0.07200 70.56000 273.02399] @tm
 0 O 0 @g
0.00 0.00 0.00 1.00 k
e
/_R44-Helvetica 166.00 z
0 0 (10) @t
T
@rax 70.56 272.74 83.38 281.45 @E
[0.07200 0.00000 0.00000 0.07200 70.56000 273.02399] @tm
0 J 0 j [] 0 d 0 R 0 @G
0.00 0.00 0.00 1.00 K
0 0.216 0.216 0.000 @w
r
/_R44-Helvetica 166.00 z
0 0 (10) @t
T
@rax 84.53 279.79 88.63 285.55 @E
[0.07200 0.00000 0.00000 0.07200 84.52800 280.00799] @tm
 0 O 0 @g
0.00 0.00 0.00 1.00 k
e
/_R44-Helvetica 110.00 z
0 0 (8) @t
T
@rax 84.53 279.79 88.63 285.55 @E
[0.07200 0.00000 0.00000 0.07200 84.52800 280.00799] @tm
0 J 0 j [] 0 d 0 R 0 @G
0.00 0.00 0.00 1.00 K
0 0.216 0.216 0.000 @w
r
/_R44-Helvetica 110.00 z
0 0 (8) @t
T
@rax 329.47 73.73 343.73 82.44 @E
[0.07200 0.00000 0.00000 0.07200 329.47199 74.01600] @tm
 0 O 0 @g
0.00 0.00 0.00 1.00 k
e
/_R44-Helvetica 166.00 z
0 0 (0.1) @t
T
@rax 329.47 73.73 343.73 82.44 @E
[0.07200 0.00000 0.00000 0.07200 329.47199 74.01600] @tm
0 J 0 j [] 0 d 0 R 0 @G
0.00 0.00 0.00 1.00 K
0 0.216 0.216 0.000 @w
r
/_R44-Helvetica 166.00 z
0 0 (0.1) @t
T
@rax 268.56 74.02 272.88 82.44 @E
[0.07200 0.00000 0.00000 0.07200 268.55999 74.01600] @tm
 0 O 0 @g
0.00 0.00 0.00 1.00 k
e
/_R44-Helvetica 166.00 z
0 0 (1) @t
T
@rax 268.56 74.02 272.88 82.44 @E
[0.07200 0.00000 0.00000 0.07200 268.55999 74.01600] @tm
0 J 0 j [] 0 d 0 R 0 @G
0.00 0.00 0.00 1.00 K
0 0.216 0.216 0.000 @w
r
/_R44-Helvetica 166.00 z
0 0 (1) @t
T
@rax 198.50 73.73 211.32 82.44 @E
[0.07200 0.00000 0.00000 0.07200 198.50399 74.01600] @tm
 0 O 0 @g
0.00 0.00 0.00 1.00 k
e
/_R44-Helvetica 166.00 z
0 0 (10) @t
T
@rax 198.50 73.73 211.32 82.44 @E
[0.07200 0.00000 0.00000 0.07200 198.50399 74.01600] @tm
0 J 0 j [] 0 d 0 R 0 @G
0.00 0.00 0.00 1.00 K
0 0.216 0.216 0.000 @w
r
/_R44-Helvetica 166.00 z
0 0 (10) @t
T
@rax 128.52 73.73 147.96 82.44 @E
[0.07200 0.00000 0.00000 0.07200 128.51999 74.01600] @tm
 0 O 0 @g
0.00 0.00 0.00 1.00 k
e
/_R44-Helvetica 166.00 z
0 0 (100) @t
T
@rax 128.52 73.73 147.96 82.44 @E
[0.07200 0.00000 0.00000 0.07200 128.51999 74.01600] @tm
0 J 0 j [] 0 d 0 R 0 @G
0.00 0.00 0.00 1.00 K
0 0.216 0.216 0.000 @w
r
/_R44-Helvetica 166.00 z
0 0 (100) @t
T
@rax 199.01 49.90 233.14 68.04 @E
[0.07200 0.00000 0.00000 0.07200 206.63999 54.07200] @tm
 0 O 0 @g
0.00 0.00 0.00 1.00 k
e
/Symbol 193.00 z
-52 0 (l) @t
154 0 (/p) @t
T
@rax 199.01 49.90 233.14 68.04 @E
[0.07200 0.00000 0.00000 0.07200 206.63999 54.07200] @tm
0 J 0 j [] 0 d 0 R 0 @G
0.00 0.00 0.00 1.00 K
0 0.216 0.216 0.000 @w
r
/Symbol 193.00 z
-52 0 (l) @t
154 0 (/p) @t
T
@rax 232.06 55.15 241.56 65.16 @E
[0.07200 0.00000 0.00000 0.07200 232.05599 55.15200] @tm
 0 O 0 @g
0.00 0.00 0.00 1.00 k
e
/_R44-Helvetica 193.00 z
0 0 (R) @t
T
@rax 232.06 55.15 241.56 65.16 @E
[0.07200 0.00000 0.00000 0.07200 232.05599 55.15200] @tm
0 J 0 j [] 0 d 0 R 0 @G
0.00 0.00 0.00 1.00 K
0 0.216 0.216 0.000 @w
r
/_R44-Helvetica 193.00 z
0 0 (R) @t
T
@rax 210.10 46.66 215.93 53.06 @E
[0.07200 0.00000 0.00000 0.07200 210.23999 46.94400] @tm
 0 O 0 @g
0.00 0.00 0.00 1.00 k
e
/Symbol 166.00 z
0 0 (n) @t
T
@rax 210.10 46.66 215.93 53.06 @E
[0.07200 0.00000 0.00000 0.07200 210.23999 46.94400] @tm
0 J 0 j [] 0 d 0 R 0 @G
0.00 0.00 0.00 1.00 K
0 0.216 0.216 0.000 @w
r
/Symbol 166.00 z
0 0 (n) @t
T
@rax 
91.01 279.00 355.97 287.06 @E
 0 O 0 @g
0.00 0.00 0.00 0.00 k
0 J 0 j [] 0 d 0 R 0 @G
0.00 0.00 0.00 1.00 K
0 0.216 0.216 0.000 @w
91.01 287.06 m
355.97 287.06 L
355.97 279.00 L
91.01 279.00 L
91.01 287.06 L
@c
B

@rax 
91.51 278.57 356.47 286.56 @E
0 J 0 j [] 0 d 0 R 0 @G
0.00 0.00 0.00 1.00 K
0 1.008 1.008 0.000 @w
91.51 286.56 m
91.51 278.57 L
356.47 278.57 L
356.47 286.56 L
91.51 286.56 L
@c
S

@rax 
91.01 292.03 355.97 300.02 @E
 0 O 0 @g
0.00 0.00 0.00 0.00 k
0 J 0 j [] 0 d 0 R 0 @G
0.00 0.00 0.00 1.00 K
0 0.216 0.216 0.000 @w
91.01 300.02 m
355.97 300.02 L
355.97 292.03 L
91.01 292.03 L
91.01 300.02 L
@c
B

@rax 
91.51 291.53 356.47 299.52 @E
0 J 0 j [] 0 d 0 R 0 @G
0.00 0.00 0.00 1.00 K
0 1.008 1.008 0.000 @w
91.51 299.52 m
91.51 291.53 L
356.47 291.53 L
356.47 299.52 L
91.51 299.52 L
@c
S

@rax 
91.01 315.00 355.97 323.06 @E
 0 O 0 @g
0.00 0.00 0.00 0.00 k
0 J 0 j [] 0 d 0 R 0 @G
0.00 0.00 0.00 1.00 K
0 0.216 0.216 0.000 @w
91.01 323.06 m
355.97 323.06 L
355.97 315.00 L
91.01 315.00 L
91.01 323.06 L
@c
B

@rax 
91.51 314.57 356.47 322.56 @E
0 J 0 j [] 0 d 0 R 0 @G
0.00 0.00 0.00 1.00 K
0 1.008 1.008 0.000 @w
91.51 322.56 m
91.51 314.57 L
356.47 314.57 L
356.47 322.56 L
91.51 322.56 L
@c
S

@rax 
91.01 304.06 355.97 312.05 @E
 0 O 0 @g
0.00 0.00 0.00 0.00 k
0 J 0 j [] 0 d 0 R 0 @G
0.00 0.00 0.00 1.00 K
0 0.216 0.216 0.000 @w
91.01 312.05 m
355.97 312.05 L
355.97 304.06 L
91.01 304.06 L
91.01 312.05 L
@c
B

@rax 
91.51 303.55 356.47 311.54 @E
0 J 0 j [] 0 d 0 R 0 @G
0.00 0.00 0.00 1.00 K
0 1.008 1.008 0.000 @w
91.51 311.54 m
91.51 303.55 L
356.47 303.55 L
356.47 311.54 L
91.51 311.54 L
@c
S

@rax 371.45 277.06 379.51 285.62 @E
[0.07200 0.00000 0.00000 0.07200 371.44798 277.05599] @tm
 0 O 0 @g
0.00 0.00 0.00 1.00 k
e
/_R44-Helvetica 166.00 z
0 0 (D) @t
T
@rax 371.45 277.06 379.51 285.62 @E
[0.07200 0.00000 0.00000 0.07200 371.44798 277.05599] @tm
0 J 0 j [] 0 d 0 R 0 @G
0.00 0.00 0.00 1.00 K
0 0.216 0.216 0.000 @w
r
/_R44-Helvetica 166.00 z
0 0 (D) @t
T
@rax 380.45 283.82 384.19 288.29 @E
[0.07200 0.00000 0.00000 0.07200 380.44798 284.03999] @tm
 0 O 0 @g
0.00 0.00 0.00 1.00 k
e
/_R44-Helvetica 110.00 z
0 0 (c) @t
T
@rax 380.45 283.82 384.19 288.29 @E
[0.07200 0.00000 0.00000 0.07200 380.44798 284.03999] @tm
0 J 0 j [] 0 d 0 R 0 @G
0.00 0.00 0.00 1.00 K
0 0.216 0.216 0.000 @w
r
/_R44-Helvetica 110.00 z
0 0 (c) @t
T
@rax 371.45 290.02 379.51 298.58 @E
[0.07200 0.00000 0.00000 0.07200 371.44798 290.01599] @tm
 0 O 0 @g
0.00 0.00 0.00 1.00 k
e
/_R44-Helvetica 166.00 z
0 0 (D) @t
T
@rax 371.45 290.02 379.51 298.58 @E
[0.07200 0.00000 0.00000 0.07200 371.44798 290.01599] @tm
0 J 0 j [] 0 d 0 R 0 @G
0.00 0.00 0.00 1.00 K
0 0.216 0.216 0.000 @w
r
/_R44-Helvetica 166.00 z
0 0 (D) @t
T
@rax 380.45 297.00 391.54 302.69 @E
[0.07200 0.00000 0.00000 0.07200 380.44798 296.99999] @tm
 0 O 0 @g
0.00 0.00 0.00 1.00 k
e
/_R44-Helvetica 110.00 z
0 0 (RR) @t
T
@rax 380.45 297.00 391.54 302.69 @E
[0.07200 0.00000 0.00000 0.07200 380.44798 296.99999] @tm
0 J 0 j [] 0 d 0 R 0 @G
0.00 0.00 0.00 1.00 K
0 0.216 0.216 0.000 @w
r
/_R44-Helvetica 110.00 z
0 0 (RR) @t
T
@rax 371.45 303.05 379.51 311.62 @E
[0.07200 0.00000 0.00000 0.07200 371.44798 303.04799] @tm
 0 O 0 @g
0.00 0.00 0.00 1.00 k
e
/_R44-Helvetica 166.00 z
0 0 (D) @t
T
@rax 371.45 303.05 379.51 311.62 @E
[0.07200 0.00000 0.00000 0.07200 371.44798 303.04799] @tm
0 J 0 j [] 0 d 0 R 0 @G
0.00 0.00 0.00 1.00 K
0 0.216 0.216 0.000 @w
r
/_R44-Helvetica 166.00 z
0 0 (D) @t
T
@rax 380.45 310.03 390.60 315.72 @E
[0.07200 0.00000 0.00000 0.07200 380.44798 310.03199] @tm
 0 O 0 @g
0.00 0.00 0.00 1.00 k
e
/_R44-Helvetica 110.00 z
0 0 (KP) @t
T
@rax 380.45 310.03 390.60 315.72 @E
[0.07200 0.00000 0.00000 0.07200 380.44798 310.03199] @tm
0 J 0 j [] 0 d 0 R 0 @G
0.00 0.00 0.00 1.00 K
0 0.216 0.216 0.000 @w
r
/_R44-Helvetica 110.00 z
0 0 (KP) @t
T
@rax 371.45 315.00 379.51 323.57 @E
[0.07200 0.00000 0.00000 0.07200 371.44798 314.99999] @tm
 0 O 0 @g
0.00 0.00 0.00 1.00 k
e
/_R44-Helvetica 166.00 z
0 0 (D) @t
T
@rax 371.45 315.00 379.51 323.57 @E
[0.07200 0.00000 0.00000 0.07200 371.44798 314.99999] @tm
0 J 0 j [] 0 d 0 R 0 @G
0.00 0.00 0.00 1.00 K
0 0.216 0.216 0.000 @w
r
/_R44-Helvetica 166.00 z
0 0 (D) @t
T
@rax 380.45 322.06 385.06 327.74 @E
[0.07200 0.00000 0.00000 0.07200 380.44798 322.05599] @tm
 0 O 0 @g
0.00 0.00 0.00 1.00 k
e
/_R44-Helvetica 110.00 z
0 0 (F) @t
T
@rax 380.45 322.06 385.06 327.74 @E
[0.07200 0.00000 0.00000 0.07200 380.44798 322.05599] @tm
0 J 0 j [] 0 d 0 R 0 @G
0.00 0.00 0.00 1.00 K
0 0.216 0.216 0.000 @w
r
/_R44-Helvetica 110.00 z
0 0 (F) @t
T
@rax 
91.01 327.02 156.02 335.02 @E
 0 O 0 @g
0.00 0.00 0.00 0.25 k
0 J 0 j [] 0 d 0 R 0 @G
0.00 0.00 0.00 1.00 K
0 0.216 0.216 0.000 @w
91.01 335.02 m
156.02 335.02 L
156.02 327.02 L
91.01 327.02 L
91.01 335.02 L
@c
B

@rax 
91.51 326.52 156.53 334.51 @E
0 J 0 j [] 0 d 0 R 0 @G
0.00 0.00 0.00 1.00 K
0 1.008 1.008 0.000 @w
91.51 334.51 m
91.51 326.52 L
156.53 326.52 L
156.53 334.51 L
91.51 334.51 L
@c
S

@rax 
232.06 327.02 296.06 335.02 @E
 0 O 0 @g
0.00 0.00 0.00 0.00 k
0 J 0 j [] 0 d 0 R 0 @G
0.00 0.00 0.00 1.00 K
0 0.216 0.216 0.000 @w
232.06 335.02 m
296.06 335.02 L
296.06 327.02 L
232.06 327.02 L
232.06 335.02 L
@c
B

@rax 
232.56 326.52 296.57 334.51 @E
0 J 0 j [] 0 d 0 R 0 @G
0.00 0.00 0.00 1.00 K
0 1.008 1.008 0.000 @w
232.56 334.51 m
232.56 326.52 L
296.57 326.52 L
296.57 334.51 L
232.56 334.51 L
@c
S

@rax 173.52 326.81 198.58 335.59 @E
[0.07200 0.00000 0.00000 0.07200 173.51999 327.02399] @tm
 0 O 0 @g
0.00 0.00 0.00 1.00 k
e
/_R44-Helvetica 166.00 z
0 0 (V) @t
99 0 (alid) @t
T
@rax 173.52 326.81 198.58 335.59 @E
[0.07200 0.00000 0.00000 0.07200 173.51999 327.02399] @tm
0 J 0 j [] 0 d 0 R 0 @G
0.00 0.00 0.00 1.00 K
0 0.216 0.216 0.000 @w
r
/_R44-Helvetica 166.00 z
0 0 (V) @t
99 0 (alid) @t
T
@rax 311.47 326.81 344.81 335.59 @E
[0.07200 0.00000 0.00000 0.07200 311.47199 327.02399] @tm
 0 O 0 @g
0.00 0.00 0.00 1.00 k
e
/_R44-Helvetica 166.00 z
0 0 (In) @t
135 0 (v) @t
214 0 (alid) @t
T
@rax 311.47 326.81 344.81 335.59 @E
[0.07200 0.00000 0.00000 0.07200 311.47199 327.02399] @tm
0 J 0 j [] 0 d 0 R 0 @G
0.00 0.00 0.00 1.00 K
0 0.216 0.216 0.000 @w
r
/_R44-Helvetica 166.00 z
0 0 (In) @t
135 0 (v) @t
214 0 (alid) @t
T
@rax 
91.01 279.00 201.02 287.06 @E
 0 O 0 @g
0.00 0.00 0.00 0.25 k
0 J 0 j [] 0 d 0 R 0 @G
0.00 0.00 0.00 1.00 K
0 0.216 0.216 0.000 @w
91.01 287.06 m
201.02 287.06 L
201.02 279.00 L
91.01 279.00 L
91.01 287.06 L
@c
B

@rax 
91.51 278.57 201.53 286.56 @E
0 J 0 j [] 0 d 0 R 0 @G
0.00 0.00 0.00 1.00 K
0 1.008 1.008 0.000 @w
91.51 286.56 m
91.51 278.57 L
201.53 278.57 L
201.53 286.56 L
91.51 286.56 L
@c
S

@rax 
229.03 304.06 318.96 312.05 @E
 0 O 0 @g
0.00 0.00 0.00 0.25 k
0 J 0 j [] 0 d 0 R 0 @G
0.00 0.00 0.00 1.00 K
0 0.216 0.216 0.000 @w
229.03 312.05 m
318.96 312.05 L
318.96 304.06 L
229.03 304.06 L
229.03 312.05 L
@c
B

@rax 
229.54 303.55 319.46 311.54 @E
0 J 0 j [] 0 d 0 R 0 @G
0.00 0.00 0.00 1.00 K
0 1.008 1.008 0.000 @w
229.54 311.54 m
229.54 303.55 L
319.46 303.55 L
319.46 311.54 L
229.54 311.54 L
@c
S

@rax 
277.06 315.00 356.98 323.06 @E
 0 O 0 @g
0.00 0.00 0.00 0.25 k
0 J 0 j [] 0 d 0 R 0 @G
0.00 0.00 0.00 1.00 K
0 0.216 0.216 0.000 @w
277.06 323.06 m
356.98 323.06 L
356.98 315.00 L
277.06 315.00 L
277.06 323.06 L
@c
B

@rax 
277.56 314.57 357.48 322.56 @E
0 J 0 j [] 0 d 0 R 0 @G
0.00 0.00 0.00 1.00 K
0 1.008 1.008 0.000 @w
277.56 322.56 m
277.56 314.57 L
357.48 314.57 L
357.48 322.56 L
277.56 322.56 L
@c
S

@rax 
172.01 292.03 262.01 300.02 @E
 0 O 0 @g
0.00 0.00 0.00 0.25 k
0 J 0 j [] 0 d 0 R 0 @G
0.00 0.00 0.00 1.00 K
0 0.216 0.216 0.000 @w
172.01 300.02 m
262.01 300.02 L
262.01 292.03 L
172.01 292.03 L
172.01 300.02 L
@c
B

@rax 
172.51 291.53 262.51 299.52 @E
0 J 0 j [] 0 d 0 R 0 @G
0.00 0.00 0.00 1.00 K
0 1.008 1.008 0.000 @w
172.51 299.52 m
172.51 291.53 L
262.51 291.53 L
262.51 299.52 L
172.51 299.52 L
@c
S

@rax 105.55 251.06 113.62 259.63 @E
[0.07200 0.00000 0.00000 0.07200 105.55200 251.06399] @tm
 0 O 0 @g
0.00 0.00 0.00 1.00 k
e
/_R44-Helvetica 166.00 z
0 0 (D) @t
T
@rax 105.55 251.06 113.62 259.63 @E
[0.07200 0.00000 0.00000 0.07200 105.55200 251.06399] @tm
0 J 0 j [] 0 d 0 R 0 @G
0.00 0.00 0.00 1.00 K
0 0.216 0.216 0.000 @w
r
/_R44-Helvetica 166.00 z
0 0 (D) @t
T
@rax 115.56 258.05 120.17 263.74 @E
[0.07200 0.00000 0.00000 0.07200 115.56000 258.04799] @tm
 0 O 0 @g
0.00 0.00 0.00 1.00 k
e
/_R44-Helvetica 110.00 z
0 0 (F) @t
T
@rax 115.56 258.05 120.17 263.74 @E
[0.07200 0.00000 0.00000 0.07200 115.56000 258.04799] @tm
0 J 0 j [] 0 d 0 R 0 @G
0.00 0.00 0.00 1.00 K
0 0.216 0.216 0.000 @w
r
/_R44-Helvetica 110.00 z
0 0 (F) @t
T
@rax 101.52 213.05 109.58 221.62 @E
[0.07200 0.00000 0.00000 0.07200 101.52000 213.04799] @tm
 0 O 0 @g
0.00 0.00 0.00 1.00 k
e
/_R44-Helvetica 166.00 z
0 0 (D) @t
T
@rax 101.52 213.05 109.58 221.62 @E
[0.07200 0.00000 0.00000 0.07200 101.52000 213.04799] @tm
0 J 0 j [] 0 d 0 R 0 @G
0.00 0.00 0.00 1.00 K
0 0.216 0.216 0.000 @w
r
/_R44-Helvetica 166.00 z
0 0 (D) @t
T
@rax 110.52 220.03 121.61 225.72 @E
[0.07200 0.00000 0.00000 0.07200 110.52000 220.03199] @tm
 0 O 0 @g
0.00 0.00 0.00 1.00 k
e
/_R44-Helvetica 110.00 z
0 0 (RR) @t
T
@rax 110.52 220.03 121.61 225.72 @E
[0.07200 0.00000 0.00000 0.07200 110.52000 220.03199] @tm
0 J 0 j [] 0 d 0 R 0 @G
0.00 0.00 0.00 1.00 K
0 0.216 0.216 0.000 @w
r
/_R44-Helvetica 110.00 z
0 0 (RR) @t
T
@rax 100.51 186.05 108.58 194.62 @E
[0.07200 0.00000 0.00000 0.07200 100.51200 186.04799] @tm
 0 O 0 @g
0.00 0.00 0.00 1.00 k
e
/_R44-Helvetica 166.00 z
0 0 (D) @t
T
@rax 100.51 186.05 108.58 194.62 @E
[0.07200 0.00000 0.00000 0.07200 100.51200 186.04799] @tm
0 J 0 j [] 0 d 0 R 0 @G
0.00 0.00 0.00 1.00 K
0 0.216 0.216 0.000 @w
r
/_R44-Helvetica 166.00 z
0 0 (D) @t
T
@rax 110.52 193.03 125.93 198.72 @E
[0.07200 0.00000 0.00000 0.07200 110.52000 193.03199] @tm
 0 O 0 @g
0.00 0.00 0.00 1.00 k
e
/_R44-Helvetica 110.00 z
0 0 (EXP) @t
T
@rax 110.52 193.03 125.93 198.72 @E
[0.07200 0.00000 0.00000 0.07200 110.52000 193.03199] @tm
0 J 0 j [] 0 d 0 R 0 @G
0.00 0.00 0.00 1.00 K
0 0.216 0.216 0.000 @w
r
/_R44-Helvetica 110.00 z
0 0 (EXP) @t
T
@rax 108.50 170.06 116.57 178.63 @E
[0.07200 0.00000 0.00000 0.07200 108.50400 170.06399] @tm
 0 O 0 @g
0.00 0.00 0.00 1.00 k
e
/_R44-Helvetica 166.00 z
0 0 (D) @t
T
@rax 108.50 170.06 116.57 178.63 @E
[0.07200 0.00000 0.00000 0.07200 108.50400 170.06399] @tm
0 J 0 j [] 0 d 0 R 0 @G
0.00 0.00 0.00 1.00 K
0 0.216 0.216 0.000 @w
r
/_R44-Helvetica 166.00 z
0 0 (D) @t
T
@rax 117.50 176.83 122.90 182.88 @E
[0.07200 0.00000 0.00000 0.07200 117.50399 177.04799] @tm
 0 O 0 @g
0.00 0.00 0.00 1.00 k
e
/_R44-Helvetica 110.00 z
0 0 (C) @t
T
@rax 117.50 176.83 122.90 182.88 @E
[0.07200 0.00000 0.00000 0.07200 117.50399 177.04799] @tm
0 J 0 j [] 0 d 0 R 0 @G
0.00 0.00 0.00 1.00 K
0 0.216 0.216 0.000 @w
r
/_R44-Helvetica 110.00 z
0 0 (C) @t
T
@rax 108.50 158.04 116.57 166.61 @E
[0.07200 0.00000 0.00000 0.07200 108.50400 158.03999] @tm
 0 O 0 @g
0.00 0.00 0.00 1.00 k
e
/_R44-Helvetica 166.00 z
0 0 (D) @t
T
@rax 108.50 158.04 116.57 166.61 @E
[0.07200 0.00000 0.00000 0.07200 108.50400 158.03999] @tm
0 J 0 j [] 0 d 0 R 0 @G
0.00 0.00 0.00 1.00 K
0 0.216 0.216 0.000 @w
r
/_R44-Helvetica 166.00 z
0 0 (D) @t
T
@rax 117.50 165.02 127.66 170.71 @E
[0.07200 0.00000 0.00000 0.07200 117.50399 165.02399] @tm
 0 O 0 @g
0.00 0.00 0.00 1.00 k
e
/_R44-Helvetica 110.00 z
0 0 (KP) @t
T
@rax 117.50 165.02 127.66 170.71 @E
[0.07200 0.00000 0.00000 0.07200 117.50399 165.02399] @tm
0 J 0 j [] 0 d 0 R 0 @G
0.00 0.00 0.00 1.00 K
0 0.216 0.216 0.000 @w
r
/_R44-Helvetica 110.00 z
0 0 (KP) @t
T
@rax 36.36 84.10 65.23 285.77 @E
[0.00001 0.07200 -0.07200 0.00001 60.40800 89.92800] @tm
 0 O 0 @g
0.00 0.00 0.00 1.00 k
e
/_R44-Helvetica 222.00 z
0 0 (Diffusion Coefficient \133cm /s\135) @t
T
@rax 36.36 84.10 65.23 285.77 @E
[0.00001 0.07200 -0.07200 0.00001 60.40800 89.92800] @tm
 0 O 0 @g
0.00 0.00 0.00 1.00 k
e
/_R44-Helvetica 111.00 z
2391 111 (2) @t
T
@rs @rs @sm
@rs
@rs
@EndSysCorelDict
end

end
clear
userdict /VPsave get restore

/PPT_ProcessAll false def
32 0 0 38 38 0 0 0 36 /Helvetica-Bold /font13 ANSIFont font
gs 1096 470 1152 2619 CB
1227 2693 34 (Fi) 33 SB
1260 2693 34 (g.) 33 SB
1293 2693 11 ( ) 10 SB
1303 2693 32 (2.) 31 SB
gr
32 0 0 38 38 0 0 0 34 /Helvetica /font12 ANSIFont font
gs 1096 470 1152 2619 CB
1334 2695 45 ( El) 44 SB
1378 2695 20 (e) 21 SB
1399 2695 19 (c) 18 SB
1417 2695 24 (tr) 23 SB
1440 2695 53 (on ) 52 SB
1492 2695 30 (di) 29 SB
1521 2695 11 (f) 10 SB
1531 2695 11 (f) 10 SB
1541 2695 49 (usi) 48 SB
1589 2695 53 (on ) 52 SB
1641 2695 40 (co) 39 SB
1680 2695 20 (e) 21 SB
1701 2695 11 (f) 10 SB
1711 2695 20 (fi) 19 SB
1730 2695 28 (ci) 27 SB
1757 2695 20 (e) 21 SB
1778 2695 21 (n) 20 SB
1798 2695 22 (t ) 21 SB
1819 2695 9 (i) 8 SB
1827 2695 32 (n ) 31 SB
1858 2695 11 (t) 10 SB
1868 2695 41 (he) 42 SB
1910 2695 11 ( ) 10 SB
1920 2695 30 (st) 29 SB
1949 2695 61 (och) 60 SB
2009 2695 51 (ast) 50 SB
2059 2695 9 (i) 8 SB
2067 2695 19 (c) 19 SB
gr
gs 1096 470 1152 2619 CB
1227 2740 33 (m) 31 SB
1258 2740 103 (agneti) 102 SB
1360 2740 30 (c ) 29 SB
1389 2740 11 (f) 10 SB
1399 2740 9 (i) 8 SB
1407 2740 20 (e) 21 SB
1428 2740 9 (l) 8 SB
1436 2740 51 (ds ) 50 SB
1486 2740 51 (as ) 50 SB
1536 2740 32 (a ) 31 SB
1567 2740 11 (f) 10 SB
1577 2740 61 (unc) 60 SB
1637 2740 20 (ti) 19 SB
1656 2740 42 (on) 41 SB
1697 2740 32 ( o) 31 SB
1728 2740 22 (f ) 21 SB
1749 2740 11 (t) 10 SB
1759 2740 41 (he) 42 SB
1801 2740 11 ( ) 10 SB
1811 2740 20 (e) 21 SB
1832 2740 9 (l) 8 SB
1840 2740 20 (e) 21 SB
1861 2740 19 (c) 18 SB
1879 2740 24 (tr) 23 SB
1902 2740 55 (on-) 54 SB
1956 2740 41 (ne) 42 SB
1998 2740 21 (u) 20 SB
2018 2740 24 (tr) 23 SB
2041 2740 30 (al) 29 SB
2070 2740 11 ( ) 10 SB
2080 2740 49 (col) 48 SB
2128 2740 9 (l) 8 SB
2136 2740 9 (i) 9 SB
gr
gs 1096 470 1152 2619 CB
2145 2740 13 (-) 12 SB
gr
gs 1096 470 1152 2619 CB
1227 2785 28 (si) 27 SB
1254 2785 63 (ona) 62 SB
1316 2785 18 (li) 17 SB
1333 2785 11 (t) 10 SB
1343 2785 30 (y.) 29 SB
1372 2785 11 ( ) 10 SB
1382 2785 64 (The) 65 SB
1447 2785 11 ( ) 10 SB
1457 2785 20 (e) 21 SB
1478 2785 59 (xpe) 60 SB
1538 2785 13 (r) 12 SB
1550 2785 42 (im) 40 SB
1590 2785 20 (e) 21 SB
1611 2785 21 (n) 20 SB
1631 2785 32 (ta) 31 SB
1662 2785 20 (l ) 19 SB
1681 2785 13 (r) 12 SB
1693 2785 20 (e) 21 SB
1714 2785 49 (sul) 48 SB
1762 2785 11 (t) 10 SB
1772 2785 30 (s ) 29 SB
1801 2785 34 (ar) 33 SB
1834 2785 20 (e) 21 SB
1855 2785 11 ( ) 10 SB
1865 2785 73 (com) 71 SB
1936 2785 55 (par) 54 SB
1990 2785 20 (e) 21 SB
2011 2785 32 (d ) 31 SB
2042 2785 36 (wi) 35 SB
2077 2785 11 (t) 10 SB
2087 2785 21 (h) 21 SB
gr
gs 1096 470 1152 2619 CB
1227 2830 11 (t) 10 SB
1237 2830 41 (he) 42 SB
1279 2830 11 ( ) 10 SB
1289 2830 34 (pr) 33 SB
1322 2830 20 (e) 21 SB
1343 2830 30 (di) 29 SB
1372 2830 30 (ct) 29 SB
1401 2830 9 (i) 8 SB
1409 2830 72 (ons ) 71 SB
1480 2830 32 (of) 31 SB
1511 2830 11 ( ) 10 SB
1521 2830 11 (f) 10 SB
1531 2830 55 (our) 54 SB
1585 2830 32 ( a) 31 SB
1616 2830 51 (nal) 50 SB
1666 2830 30 (yt) 29 SB
1695 2830 9 (i) 8 SB
1703 2830 30 (c ) 29 SB
1732 2830 33 (m) 32 SB
1764 2830 21 (o) 20 SB
1784 2830 41 (de) 42 SB
1826 2830 9 (l) 8 SB
1834 2830 30 (s.) 29 SB
1863 2830 11 ( ) 10 SB
1873 2830 64 (The) 65 SB
1938 2830 11 ( ) 10 SB
1948 2830 42 (ho) 41 SB
1989 2830 22 (ri) 21 SB
2010 2830 18 (z) 19 SB
2029 2830 42 (on) 41 SB
2070 2830 32 (ta) 31 SB
2101 2830 9 (l) 9 SB
gr
gs 1096 470 1152 2619 CB
1227 2875 55 (bar) 54 SB
1281 2875 30 (s ) 29 SB
1310 2875 63 (abo) 62 SB
1372 2875 39 (ve) 40 SB
1412 2875 11 ( ) 10 SB
1422 2875 11 (t) 10 SB
1432 2875 41 (he) 42 SB
1474 2875 11 ( ) 10 SB
1484 2875 11 (f) 10 SB
1494 2875 30 (ig) 29 SB
1523 2875 78 (ure r) 77 SB
1600 2875 20 (e) 21 SB
1621 2875 34 (pr) 33 SB
1654 2875 20 (e) 21 SB
1675 2875 71 (sent) 70 SB
1745 2875 22 ( t) 21 SB
1766 2875 65 (he r) 64 SB
1830 2875 20 (e) 21 SB
1851 2875 9 (l) 8 SB
1859 2875 20 (e) 21 SB
1880 2875 40 (va) 39 SB
1919 2875 32 (nt) 31 SB
1950 2875 32 ( a) 31 SB
1981 2875 51 (ppl) 50 SB
2031 2875 9 (i) 8 SB
2039 2875 70 (cabi) 69 SB
2108 2875 9 (l) 8 SB
2116 2875 9 (i) 8 SB
2124 2875 30 (ty) 29 SB
gr
gs 1096 470 1152 2619 CB
1227 2920 63 (regi) 62 SB
1289 2920 33 (m) 32 SB
1321 2920 52 (e o) 51 SB
1372 2920 22 (f ) 21 SB
1393 2920 20 (e) 21 SB
1414 2920 21 (a) 20 SB
1434 2920 51 (ch ) 50 SB
1484 2920 33 (m) 31 SB
1515 2920 62 (ode) 63 SB
1578 2920 9 (l) 8 SB
1586 2920 11 (.) 10 SB
gr
32 0 0 42 42 0 0 0 38 /Helvetica /font12 ANSIFont font
gs 2394 239 0 2938 CB
1172 3134 51 (-2-) 51 SB
gr
1 #C
statusdict begin /manualfeed false store end
EJ RS
SVDoc restore
end

